\documentclass[preprint2]{aastex}

\usepackage{morefloats}
\usepackage{afterpage}
\usepackage{color}

\newcommand{\myemail}{adecia@eso.org}
\newcommand{\mgii}{\mbox{Mg\,{\sc ii}}} 
\newcommand{\oiii}{\mbox{O\,{\sc iii}}}
\newcommand{\oii}{\mbox{O\,{\sc ii}}}

\slugcomment{\today}

\shorttitle{PTF SLSN-I light curves}
\shortauthors{De Cia et al.}

\begin{document}

\title{Light curves of hydrogen-poor Superluminous Supernovae\\ from the Palomar Transient Factory}

\author{Annalisa De Cia\altaffilmark{1,2}\footnote{ESO fellow}, A. Gal-Yam\altaffilmark{1}, A. Rubin\altaffilmark{1}, G. Leloudas\altaffilmark{1,3}, P. Vreeswijk\altaffilmark{1}, D. A. Perley\altaffilmark{4}, R. Quimby\altaffilmark{5,6}, Lin Yan\altaffilmark{7,8}, M. Sullivan\altaffilmark{9}, A. Fl\"ors\altaffilmark{2}, J. Sollerman\altaffilmark{10}, D. Bersier\altaffilmark{4}, S. B. Cenko\altaffilmark{11,12}, M. Gal-Yam\altaffilmark{13}, K. Maguire\altaffilmark{14}, E. O. Ofek\altaffilmark{1}, S. Prentice\altaffilmark{4}, S. Schulze\altaffilmark{1}, J. Spyromilio\altaffilmark{2}, S. Valenti\altaffilmark{20}, I. Arcavi\altaffilmark{15,16}\footnote{Einstein Fellow}, A. Corsi\altaffilmark{17}, D. A. Howell\altaffilmark{15,16}, P. Mazzali\altaffilmark{4,18}, M. M. Kasliwal\altaffilmark{19}, F. Taddia\altaffilmark{10}, O. Yaron\altaffilmark{1}
}

\begin{center}
\altaffiltext{1}{ Department of Particle Physics and Astrophysics, Weizmann Institute of Science, Rehovot 76100, Israel}
\altaffiltext{2}{ European Southern Observatory, Karl-Schwarzschild Str. 2, D-85748 Garching bei M\"{u}nchen, Germany, \myemail}
\altaffiltext{3}{ Dark Cosmology Centre, Niels Bohr Institute, University of Copenhagen, Juliane Maries Vej 30, DK-2100, Copenhagen, Denmark}
\altaffiltext{4}{ Astrophysics Research Institute, Liverpool John Moores University, Liverpool Science Park, 146 Brownlow Hill, Liverpool L35RF, UK}
\altaffiltext{5}{ Department of Astronomy, San Diego State University, San Diego, CA 92182, USA}
\altaffiltext{6}{ Kavli IPMU (WPI), UTIAS, The University of Tokyo, Kashiwa, Chiba 277-8583, Japan}
\altaffiltext{7}{ MS100-22, Caltech/IPAC, California Institute of Technology, Pasadena, CA 91125, USA}
\altaffiltext{8}{ Caltech Optical Observatories, California Institute of Technology, Pasadena, CA 91125, USA}
\altaffiltext{9}{ Department of Physics and Astronomy, University of Southampton, Southampton SO17 1BJ, UK}
\altaffiltext{10}{ The Oskar Klein Centre, Department of Astronomy, Stockholm University, AlbaNova, SE-10691, Stockholm, Sweden}
\altaffiltext{11}{ Astrophysics Science Division, NASA Goddard Space Flight Center, Mail Code 661, Greenbelt, MD 20771, USA}
\altaffiltext{12}{ Joint Space-Science Institute, University of Maryland, College Park, MD 20742, USA}
\altaffiltext{13}{ The Schwartz/Reisman Science Education Center, Weizmann Institute of Science, Rehovot, Israel} 
\altaffiltext{14}{ Astrophysics Research Centre, School of Mathematics and Physics, Queen’s University Belfast, Belfast BT7 1NN, UK}
\altaffiltext{15}{ Las Cumbres Observatory, 6740 Cortona Dr. Suite 102, Goleta, CA 93117, USA}
\altaffiltext{16}{ Department of Physics, University of California, Santa Barbara, CA 93106-9530, USA}
\altaffiltext{17}{ Department of Physics and Astronomy, Texas Tech University, Box 1051, Lubbock, TX 79409-1051, USA}
\altaffiltext{18}{ Max-Planck-Institut f\"ur Astrophysik, Karl-Schwarzschild-Str. 1, D-85748 Garching, Germany}
\altaffiltext{19}{ Astronomy Department, California Institute of Technology, Pasadena, CA 91125, USA}
\altaffiltext{20}{ Department of Physics, University of California, Davis, CA 95616, USA}
\end{center}

\begin{abstract}
We investigate the light-curve properties of a sample of 26 spectroscopically confirmed hydrogen-poor superluminous supernovae (SLSNe-I) in the Palomar Transient Factory (PTF) survey. These events are brighter than SNe Ib/c and SNe Ic-BL, on average, by about 4 and 2~mag, respectively. The peak absolute magnitudes of SLSNe-I in rest-frame $g$ band span $-22\lesssim M_g \lesssim-20$~mag, and these peaks are not powered by radioactive $^{56}$Ni, unless strong asymmetries are at play. The rise timescales are longer for SLSNe than for normal SNe Ib/c, by roughly 10 days, for events with similar decay times. Thus, SLSNe-I can be considered as a separate population based on photometric properties. After peak, SLSNe-I decay with a wide range of slopes, with no obvious gap between rapidly declining and slowly declining events. The latter events show more irregularities (bumps) in the light curves at all times. At late times, the SLSN-I light curves slow down and cluster around the $^{56}$Co radioactive decay rate. Powering the late-time light curves with radioactive decay would require between 1 and 10${\rm M}_\odot$ of Ni masses. Alternatively, a simple magnetar model can reasonably fit the majority of SLSNe-I light curves, with four exceptions, and can mimic the radioactive decay of $^{56}$Co, up to $\sim400$ days from explosion. The resulting spin values do not correlate with the host-galaxy metallicities. Finally, the analysis of our sample cannot strengthen the case for using SLSNe-I for cosmology.      
\end{abstract}

\section{Introduction}

Since the advent of wide-field untargeted transient surveys, a class of superluminous supernovae (SLSNe) that are over 10 times more luminous than regular SNe \citep[see][for a review]{Gal-Yam12}, with absolute magnitudes $\lesssim-21$~mag, has emerged. The first few objects showed a striking diversity, e.g. SN~2005ap \citep{Quimby07}, SN~2006gy \citep{Ofek07,Smith07}, and SN~2007bi \citep{Gal-Yam09}, leading to a natural division between H-rich events (SLSNe-II) and H-poor events (SLSNe-I). Most SLSNe-II show narrow lines (SLSN-IIn) and are powered by the interaction of the SN ejecta with the circumstellar medium \citep[CSM; e.g.,][]{Chugai94,Chevalier94,Chevalier11,Ofek13b,Inserra18}. SLSNe-I are less well-understood, and the physical processes that dominate these explosions are still under debate.

\citet{Quimby11} inspected the first sample of SLSNe-I, and found that they have UV-bright light curves over extended periods of time. Quimby et al. also showed similarity in their spectral features, and suggested that their progenitors may have initial masses $90 < M< 130$ ${\rm M}_\odot$, perhaps exploding as core-collapse SNe with massive ejecta interacting with a H-poor CSM. Asymmetry in the ejecta can hide signatures of hydrogen or helium in SLSN-I spectra \citep{Kozyreva15}, as well as ionization \citep{Mazzali16}. \citet{Gal-Yam12} proposed that a group of slowly declining events (SLSNe-R), similar to SN~2007bi, have late-time light curves that are powered by radioactivity and could be associated with pair-instability SNe \citep[PISN;][]{Barkat67,Heger02,Gal-Yam09}, but this is widely debated \citep{Dessart12,Nicholl13}. \citet{Inserra13} showed that the late-time decay of a few SLSNe slows down to a 'tail' that could be explained if the light curves were powered by the spin-down of a newly born magnetar \citep{Kasen10}. \citet{Nicholl15} studied a sample of SLSNe and suggested that the ejecta mass is the main driver of the observed diversity. More recently, \citet{Nicholl17} fitted a magnetar model to the literature sample of SLSNe-I.

Early-time bumps (pre-peak or double peaks, or excess emission) have been observed in some SLSNe light curves, such as SN~2006oz \citep{Leloudas12}, LSQ~14bdq \citep{Nicholl15b}, PTF~12dam and iPTF~13dcc \citep{Vreeswijk17}, and DES~14X3taz \citep{Smith16}. These early bumps can be explained by shock-cooling or CSM interaction models \citep[e.g][]{Nakar10,Rabinak11,Chatzopoulos12,Piro15}. Such early bumps or double-peaked light curves may in fact be common among SLSNe \citep{Nicholl16}. Late-time bumps (postpeak) have also been observed in a few cases, such as for SN~2007bi \citep{Gal-Yam09}, iPTF~13ehe \citep{Yan15}, PS1-14bj \citep{Lunnan16}, and SN~2015bn \citep{Nicholl16b}. Wiggles in the late-time decay have often been observed in a handful of slowly declining SLSNe by \citet{Inserra17}. Such late-time bumps cannot be explained by magnetar and radioactive decay models. Late-time emergence of hydrogen emission has been detected in a few cases \citep[e.g., iPTF~15esb;][]{Yan15,Yan17}, and in these cases it was explained with substantial mass loss that occurred shortly before the progenitors of the SLSNe exploded \citep{Yan15,Yan17}. Indeed, \citet{Liu17} showed that the light curves of iPTF~15esb could be explained with a multiple-shell CSM interaction model. The diversity observed so far in H-poor SLSNe seems to indicate that multiple processes may contribute to powering their light curves.

In this paper, we present a sample of 26 SLSNe-I from the Palomar Transient Factory \citep[PTF;][]{Law09,Rau09} and its successor the intermediate Palomar Transient Factory (iPTF). This is the largest sample of SLSNe-I homogeneously selected from a single survey available so far. Here we characterize and discuss the properties of the light curves of these events, and compare them to a large PTF sample of SNe Ib/c and Ic-BL (with broad lines). We address the question on how luminous SLSNe-I are, and whether they can be considered a separate population based on their light-curve properties. We investigate whether Type SLSNe-R are a separate class of events, and whether we can use SLSNe-I for cosmology. The spectra of (i)PTF SLSNe, and the host galaxies of PTF SLSNe up to 2012 are studied in \citet{Quimby18} and \citet{Perley16}, respectively. 

The paper is organized as follows. In Sect. \ref{sec sample} and \ref{sec observations} we describe the SLSN sample and observations, respectively. We characterize the SN light curves in Sect. \ref{sec characterization}, discuss our results in Sect. \ref{sec discussion}, and present our conclusions in Sect. \ref{sec conclusions}. We adopt the cosmological parameters $H_0 = 70$~km~s$^{-1}$~Mpc$^{-1}$, $\Omega_m = 0.3$, and $\Omega_\Lambda = 0.7$ throughout the paper.

\section{The PTF sample of SLSNe}
\label{sec sample}

PTF was a wide-field (7.26 deg$^2$ field of view), nontargeted survey designed to investigate the optical transient and variable sky \citep{Law09,Rau09}, carried out using the refurbished $CFH12k$ camera \citep{Rahmer08}, mounted on the Palomar Observatory 48-inch Samuel Oschin Telescope (P48), in California. The PTF survey is optimized for the discovery of SNe of different types. Since its start in 2009, PTF has discovered and classified over 3000 SNe. The classification and follow-up observations of these SNe are performed through a wide network of telescopes \citep{Gal-Yam11}, as described below for our sample. The selection of SN candidates for spectroscopic classification within the PTF survey is not free of biases. For example, SLSN searches may have given more weight to candidates that were brighter than their host galaxies. Nevertheless, PTF has discovered a large number of SLSNe-II as well, which can explode in normal host galaxies \citep[e.g.][]{Perley16}, reassuring us that such selection biases are not dominant. 

The 26 SLSNe discussed in this paper are all the hydrogen-poor SLSNe discovered between 2009 and 2013 by the (i)PTF survey. The sample is shown in Table~\ref{table sample}. This significantly increases the sample of about 50 currently known H-poor SLSNe with reported spectral classification in the literature \citep[either published or with spectra reported in Astronomer's Telegrams; e.g.][]{Nicholl15,Schulze18,Lunnan18}, 10 of which have been discovered by PTF and are part of this work as well.  

The PTF SLSNe in our sample have been spectroscopically classified as SLSNe-I by \citet{Quimby18}. This sample is thus spectroscopically selected and assumes no luminosity thresholds. The sample of PTF SLSN host galaxies of \citet{Perley16} is slightly different because that work also applied a luminosity cut, while PTF~12hni and PTF~12gty are presented here for the first time. PTF~12hni is classified as an SLSN-I by \citet{Quimby18} with some uncertainty, and having possible matches to SN Ia and SN Ic. In addition, three other events are reported as possible SLSNe by \citet{Quimby18}, namely PTF~09q, PTF~10gvb, and PTF~11mnb, but are most likely not SLSNe, and therefore we do not include these in our sample.\footnote{For PTF~09q there is a single spectrum available, which is well consistent with an SN Ic, and its host galaxy is a massive galaxy \citep{Quimby18}. Three spectra are available for PTF~10gvb, but one is mostly featureless and lacks the typical SLSN \oii{} features, and the the other two are well-matched with an SN Ic-BL \citep{Quimby18}. PTF~11mnb is most likely a SN Ic, as studied in detail by \citet{Taddia18a}.} 

What makes the PTF sample unique is not only the fact that it is homogeneously selected from a single survey, but also that its average redshift is low ($<z>\, = 0.27$; see Sect. \ref{sec redshift}). A higher-$z$ ($0.3<z<1.6$) sample of 17 SLSNe-I from the Pan-STARRS1 Medium Deep Survey \citep[PS1;][]{Kaiser10} is presented by \citet{Lunnan18}. In addition, the light-curve coverage of the PTF sample often extends to late times, beyond 100 days after the peak for half of the sample. The currently known H-poor SLSNe in the literature typically lack photometry later than 120 days after peak \citep[e.g.][]{Nicholl15}. \citet{Jerkstrand17} and \citet{Inserra17} have studied a small sample of slow-evolving SLSNe, with data coverage up to 400 days after peak.

As a comparison sample, we also select all Type Ib, Ib/c, Ic, and Ic-BL SNe discovered between 2009 and 2013 by PTF. These SNe are studied in more detail in \citet{Arcavi10}, \citet{Corsi16}, \citet{Prentice16}, and \citet{Taddia18b}, and will be presented in full in forthcoming publications (Barbarino et al. 2018, in preparation; Fremling et al. 2018, in preparation; Huang et al. 2018, in preparation; Karamehmetoglu et al. 2018, in preparation; Schulze et al. 2018, in preparation).

\begin{figure*}
\includegraphics[angle=90,width=18cm]{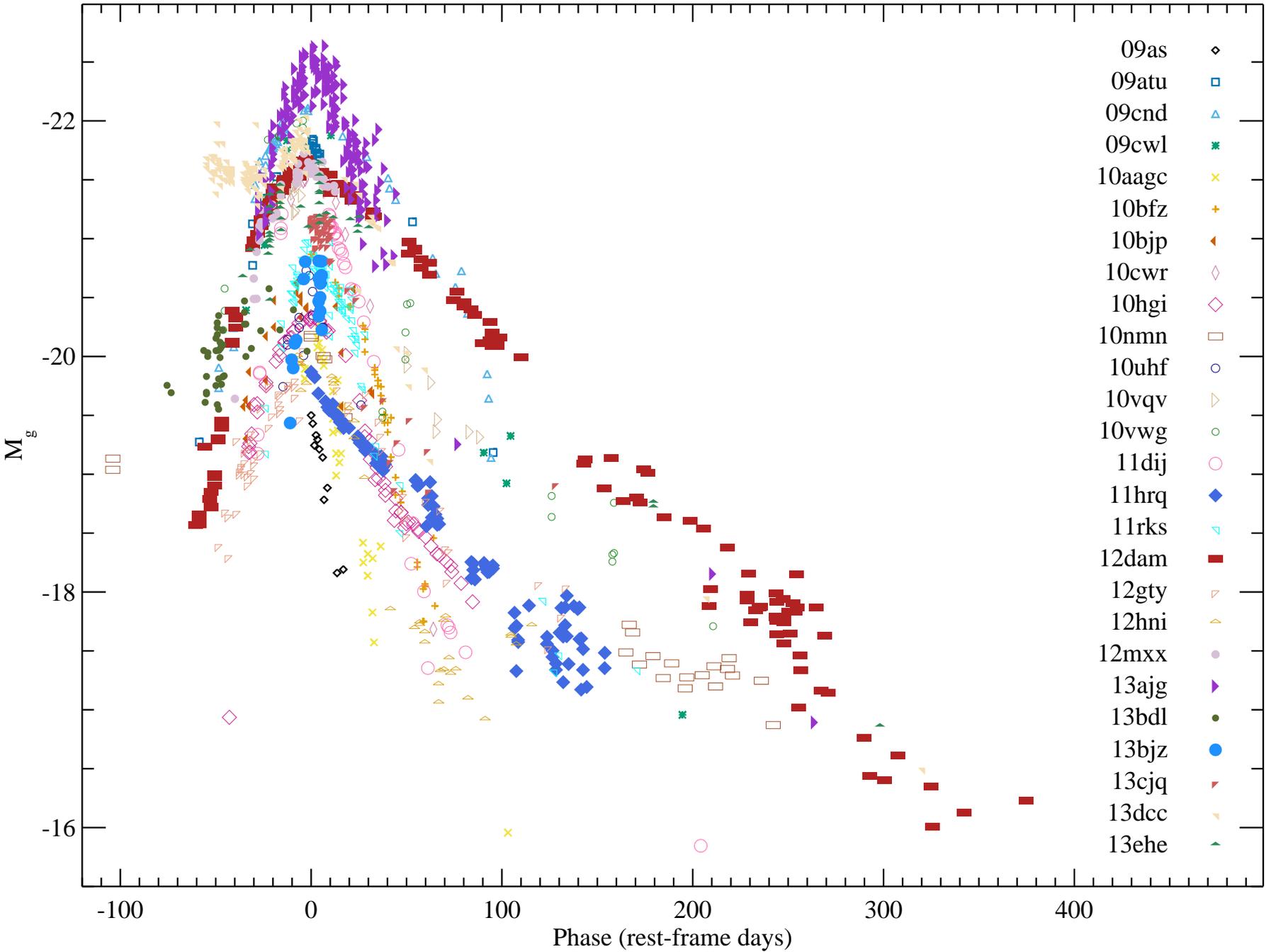}
\caption{Rest-frame $M_g$ light curves of the 26 H-poor SLSNe in our sample. The error bars are omitted here for readability, but are shown in Figs. \ref{fig ind abs g mag 1} to \ref{fig ind abs g mag 5}.\label{fig all lc}}
\end{figure*}

We derive the rest-frame $g$-band absolute magnitudes, $M_g$, from the apparent $r$ magnitudes $m_r$ including the $k$-correction term as $K_{gr}$ described in Sect. \ref{sec k-corr} and listed in Table~\ref{table k-corr} ($M_g = m_r - DM(z) -K_{gr}$, where $DM(z)$ is the distance modulus for a given redshift $z$, and $m_r$ is corrected for foreground Galactic extinction, reported in Table~\ref{table sample}). Figure~\ref{fig all lc} shows the rest-frame $M_g$ light curves of all the 26 SLSNe in our sample. 

\begin{table}
\begin{center}
\caption{The PTF sample of 26 hydrogen-poor SLSNe.\label{table sample}}
\begin{tabular}{@{}l @{\hspace{3mm}} l @{\hspace{3mm}} l @{\hspace{3mm}} l @{\hspace{3mm}} l @{\hspace{3mm}} l@{} }
\tableline\tableline
PTF & R.A. & Decl. & $z$ & Type & $A_V^a$\\
ID & (hh:mm:ss) & (\degr\ :\arcmin\ :\arcsec) &   &   & (mag)\\
\tableline

09as        &   12:59:15.78      &   $+$27:16:38.5  &   0.1864    &   I   & 0.02\\
09atu       &   16:30:24.55      &   $+$23:38:25.0  &   0.5014    &   I   & 0.13\\
09cnd       &   16:12:08.94      &   $+$51:29:16.2  &   0.2585    &   I   & 0.06\\
09cwl       &   14:49:10.08      &   $+$29:25:11.4  &   0.3502    &   I   & 0.04\\
10aagc      &   09:39:56.93      &   $+$21:43:16.9  &   0.2067    &   I   & 0.07\\
10bfz       &   12:54:41.27      &   $+$15:24:17.0  &   0.1699    &   I   & 0.05\\
10bjp       &   10:06:34.30      &   $+$67:59:19.0  &   0.3585    &   I   & 0.17\\
10cwr       &   11:25:46.67      &   $-$08:49:41.2  &   0.2301    &   I   & 0.10\\
10hgi       &   16:37:47.04      &   $+$06:12:32.3  &   0.0982    &   I   & 0.22\\
10nmn       &   15:50:02.79      &   $-$07:24:42.1  &   0.1236    &   I/R & 0.42\\
10uhf       &   16:52:46.68      &   $+$47:36:22.0  &   0.2879    &   I   & 0.05\\
10vqv       &   03:03:06.84      &   $-$01:32:34.9  &   0.4520    &   I   & 0.17\\
10vwg       &   18:59:32.86      &   $+$19:24:25.7  &   0.1901    &   I/R & 1.41\\
11dij       &   13:50:57.77      &   $+$26:16:42.8  &   0.1429    &   I   & 0.03\\
11hrq       &   00:51:47.22      &   $-$26:25:10.0  &   0.0571    &   I/R & 0.04\\
11rks       &   01:39:45.51      &   $+$29:55:27.0  &   0.1924    &   I   & 0.11\\
12dam       &   14:24:46.20      &   $+$46:13:48.3  &   0.1075    &   I/R & 0.03\\
12gty       &   16:01:15.23      &   $+$21:23:17.4  &   0.1768    &   I   & 0.18\\
12hni       &   22:31:55.86      &   $-$06:47:49.0  &   0.1056    &   I   & 0.16\\
12mxx       &   22:30:16.68      &   $+$27:58:21.9  &   0.3274    &   I   & 0.12\\
13ajg       &   16:39:03.95      &   $+$37:01:38.4  &   0.7403    &   I   & 0.04\\
13bdl       &  12:36:56.14       &   $+$13:07:45.5  &   0.4030    &   I   & 0.13\\
13bjz       &  10:38:19.83       &   $+$24:24:51.0  &   0.2712    &   I   & 0.06\\
13cjq       &  00:14:27.18       &   $+$24:17:08.8  &   0.3962    &   I   & 0.13\\
13dcc       &  02:57:02.50       &   $-$00:18:44.0  &   0.4308    &   I/R & 0.18\\
13ehe       &  06:53:21.50       &   $+$67:07:56.0  &   0.3434    &   I/R & 0.14\\

\tableline
\end{tabular}
\tablenotetext{a}{Galactic foreground extinction.}
\end{center}
\end{table}

\subsection{The redshift distribution}
\label{sec redshift}
Although most normal SNe are observed in the nearby universe ($z\lesssim0.2$), the most luminous ones can be detected out to higher redshift. SNe Type Ia, for instance, are currently discovered out to $z\sim2$ in deep imaging \citep{Jones13}. A few SLSNe have been studied out to $z\sim4$ in the deepest surveys \citep{Cooke12}, although in limited detail compared to nearby targets.Recently, a small sample of $z\sim2$ SLSNe has been studied by \citet{Moriya18} and \citet{Curtin18}. In the future, the James Webb Space Telescope is expected to be able to detect SLSNe out to $z\sim20$ \citep{Abbott17}. The PTF survey typically discovers SLSNe below $z\lesssim1$. 

The redshifts in our SLSN sample are all measured spectroscopically and normally measured from narrow \mgii{} absorption lines in the SN spectra. The typical uncertainties on the redshift estimates are of the order of 0.0005, given the typical resolution of the follow-up spectra \citep[described in][]{Quimby18}. The redshifts in Table \ref{table sample} are taken from \citet{Quimby18} for all events up to 2012, except for PTF~10vwg for which we adopt the slightly more accurate redshift of \citet{Perley16}. We adopt the redshifts of \citet{Vreeswijk14} for PTF~13ajg and of \citet{Yan15} for PTF~13ehe. For the other 2013 events, we directly measure the redshifts from the \mgii{} narrow absorption lines in the spectra. In PTF~13ehe, the most common spectral features are very weak. The redshift measurement is based on a weak \oiii{} 5007 emission line, and its uncertainty is of the order of 0.001.

\begin{figure}
\epsscale{1.2}
\plotone{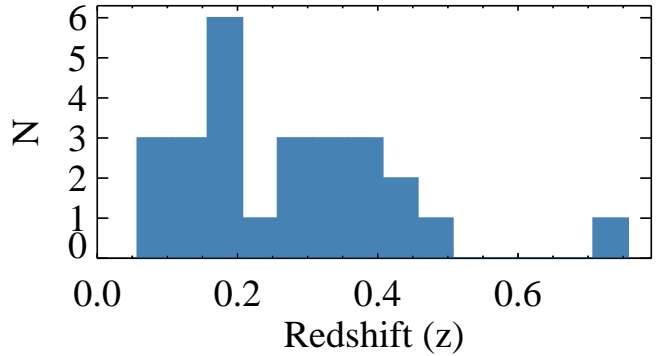}
\caption{Redshift distribution of the sample of hydrogen-poor SLSNe presented in Table~\ref{table sample}.\label{fig z distr}}
\end{figure}

Figure.~\ref{fig z distr} shows the redshift distribution of our sample, where the mean redshift is $<z> = 0.27$ with standard deviation $\sigma_z = 0.15$. The volume-weighted mean is $<z> = 0.33$.\footnote{A Gaussian fit through the redshift distribution data prefers a mean redshift $<z> = 0.16$ with a standard deviation $\sigma_z = 0.20$ for a $z$ bin size of 0.05. The mean $<z>$ is 0.11 and 0.17 for bin sizes of 0.01 and 0.1, respectively.} 

The mean redshift of the PTF H-poor SLSN sample presented here is comparable to the ''golden'' SLSN sample of \citet[][$<z>=0.22$, while $<z>=0.63$ for their ''silver'' sample]{Nicholl15}, and the SLSN host sample of \citet[][$<z>=0.34$ with a standard deviation of 0.2]{Leloudas15}. On the other hand, SLSNe discovered by PS1 tend to be at higher redshifts, typically $z>0.5$ \citep{McCrum15}, and in particular $0.3<z<1.6$ \citep{Lunnan18}.

The drop of the redshift distribution above $z\sim0.5$ in our sample is an observational selection effect due to the limiting magnitude of the PTF survey \citep[m$_{\rm r, lim}\sim 20.5$ mag,][]{Cao16}. This limit hampers further investigations of the evolution of the sample properties with redshift.

\section{Observations and data processing}
\label{sec observations}

\subsection{Palomar P48 discovery and imaging}
\label{sec P48}

As part of the standard PTF operations, SN candidates are discovered in P48 images using image subtraction in Mould-R ($r$) or the Sloan Digital Sky Survey \citep[SDSS,][]{York00,sdss16} $g'$ filter. The best SN candidates are then classified spectroscopically and followed-up with other telescopes. The raw P48 images are initially processed by the Infrared Processing and Analysis Center \citep[IPAC,][]{Laher14}. The photometric calibration and system are described in \citet{Ofek12b}. Image-subtraction point-spread function (PSF) photometry is performed with a custom routine \citep[a pipeline written by one of us (M.S.) and used extensively in PTF; e.g.][]{Sullivan06,Ofek14,Firth15,Dimitriadis17}. This pipeline constructs deep reference images - either before the SN explosion or after the SN has faded - and astrometrically aligns the images using the Automated Astrometry described in \citet{Hogg08} and the Naval Observatory Mergered Astrometric Dataset \citep[NOMAD,][]{Zacharias04}. The image PSFs are then matched in order to perform the image subtraction and then to extract the PSF photometry of the SN only (where the contribution of the reference image has been subtracted). The fluxes are calibrated against the SDSS Data Release 10 \citep{Ahn14} when available, and otherwise against the photometric catalog of \citet{Ofek12b}, and making no assumption on the SLSN colors.

The formal uncertainties derived with the MS pipeline only include statistical uncertainties, but not uncertainties from poor image subtraction or calibration. As a result, the formal uncertainties are slightly underestimated. For example, under excellent data coverage, we can observe a larger scatter than that accounted for by the formal uncertainties. The best example is for iPTF~13ajg, which shows a scatter of $\sim0.5$~mag around peak. We quantify the additional source of uncertainty by assuming a $\chi^2_\nu=1$ for the light-curve fit around the peak of iPTF~13ajg (see Sect. \ref{sec peak mag}). The additional required uncertainty is 0.05 mag, which we add to all formal errors derived with the MS pipeline (i.e. for the data taken with the P48, P60, and LT telescopes; see below), to account for poor image subtraction or calibration.

Nondetections, and in particular the last nondetection limits before the SN discoveries, are not included in our analysis. The reason for this is that nondetections are largely dominated by noisy data and are uninformative. In addition, in most cases the SLSNe-I were discovered long after explosions. For the case of PTF~12dam co-adds of the prediscovery nondetections are presented by \citet{Vreeswijk17}. The analysis presented in this paper is independent of the nondetection limits. Thus, we leave the treatment of prediscovery limits, which is beyond the scope of this paper, for future case-by-case studies.

\subsection{Palomar P60 imaging}
Follow-up imaging was obtained with the Palomar 60-inch telescope \citep[P60;][]{Cenko06}. The filters employed for our observations are Johnson $B$ \citep{Bessel90}, Kron $R$ \citep[similar to Cousins $R_C$,][]{Bessel90}, Sloan $i'$ and $z'$ \citep{Fukugita96}, and Gunn $g$ \citep{Thuan76}. The SN photometry is extracted with the same routine described above for the P48 data processing, but calibrated using the AAVSO Photometric All-Sky Survey \citep[APASS;][]{Henden09} for the $B$ filter or for fields that are not covered by the SDSS footprint.

\subsection{Keck/LRIS imaging}
We observed PTF SLSNe at late times using the Low-Resolution Imaging Spectrometer (LRIS) on the Keck I telescope to monitor the late-time evolution of the light curve or to produce a deep reference image (after the SN has faded) for image subtraction or host galaxy study \citep{Perley16}. Images were processed using standard techniques via the custom pipeline LPIPE\footnote{\texttt{www.astro.caltech.edu/$\sim$dperley/programs/lpipe.html}} and co-added using SWarp. Photometry was performed after image subtraction of the reference image \citep[taken from][]{Perley16}, with the custom-made IRAF routine \texttt{mkdifflc} \citep{Gal-Yam04,Gal-Yam08}. 

\subsection{Liverpool Telescope imaging}
Follow-up imaging was also obtained with the 2m robotic Liverpool Telescope \citep[LT;][]{Steele04} at the Roque de los Muchachos Observatory on La Palma, Spain, with the RATCAM and IO:O optical imagers in $g$, $r$, and $i$ filters (similar to SDSS). The images were processed following \citet{Maguire14} and using the image-subtraction PSF photometry custom routine described above for the P48 Telescope. 

\subsection{Las Cumbres Observatory imaging}
The LCO \citep[previously known as LCOGT;][]{Brown13} data have been reduced using a custom pipeline \citep{Valenti16}. The pipeline employs standard procedures (PYRAF, DAOPHOT) in a \texttt{Python} framework. Host galaxy flux was removed using image subtraction technique (High Order Transform of PSF ANd Template Subtraction, HOTPANTS\footnote{\texttt{www.astro.washington.edu/users/becker/v2.0/hotpants.html}}). PSF magnitudes were computed on the subtracted images and transformed to the standard SDSS filter system (for $gri$) via standard star observations taken during clear nights.

\subsection{Discovery Channel Telescope imaging}
We imaged several of the SLSNe in our sample with the Large Monolithic Imager (LMI) mounted on the 4.3\,m Discovery Channel Telescope (DCT) in Happy Jack, AZ. The LMI images were processed using a custom IRAF pipeline for basic detrending (bias subtraction and flat fielding), and individual dithered images were combined using \texttt{SWarp} \citep{Bertin02}. SN magnitudes were measured using aperture photometry with the inclusion radius matched to the FWHM of the image PSF. Photometric calibration was performed relative to point sources from the SDSS \citep{York00,sdss16}. 

No DCT reference images are available for image subtraction, so we account for the contribution of the host galaxies by subtracting the host magnitudes from the observed fluxes, and we include this into the photometric uncertainty budget. The SLSNe iPTF~13dcc and 13ehe were observed with the DCT. The host galaxy of iPTF~13dcc has $B = 26.3 \pm 0.2$ and $i=25.0 \pm 0.2$ (Perley et al. 2018, in preparation), and we assume $g = 26.3 \pm 0.2$ and $r=25.0 \pm 0.2$ for the subtraction of the host galaxy contribution to the $r$-band data point, which is a reasonable assumption given typical host galaxy colors \citep{Perley16}. The host galaxy of iPTF~13ehe has $B = 25.0 \pm 0.1$ and $R = 24.0 \pm 0.1.$  (Perley et al. 2018, in preparation), which we use to subtract the host-galaxy contribution. For both SLSNe, this host-galaxy correction affects significantly (by 0.2~mag) only the last $r$-band epoch of their light curves. In both cases, the DCT data points are consistent with the photometry from other facilities, including late-time \textit{Hubble Space Telescope} (HST) photometry (Sec. \ref{sec hst}).

\subsection{Swift/UVOT imaging}
\label{sec swift}
A number of supernovae in our sample were observed with the UltraViolet/Optical Telescope (UVOT; \citealt{Roming05}) on board the \textit{Swift} Gamma-ray Burst Explorer \citep{Gehrels04}.  Data were processed using the standard UVOT pipeline, and photometry was extracted at the supernova location using a 3\arcsec\ radius. Photometric calibration was calculated using the zero point measurements from \citet{Poole08} and \citet{Breeveld10}.  The magnitudes reported in Tables \ref{table phot09as} to \ref{table phot13ehe} are all on the AB system. No attempt has been made to correct for underlying contributions from the host galaxy emission. At these redshifts, the host galaxy contribution should not significantly affect the observed UV flux in most cases. This may not be true for some cases, in particular for PTF~12dam and its luminous underlying starburst host galaxy \citep{Chen15,Thoene15,Perley16,Cikota17}. However, even in this case, the host galaxy brightness in the $F225W$ filter is $19.94 \pm 0.17$ \citep{Perley16}, which is 1--2 mag fainter that the unsubtracted SN photometry \citep{Nicholl13,Chen15}.

\subsection{Palomar P200/Large Format Camera (LFC) imaging}
Follow-up imaging was obtained with the Palomar 200-inch Hale Telescope with the LFC\footnote{\texttt{www.astro.caltech.edu/palomar/about/telescopes/hale.html}}. The data reduction was performed with standard IRAF tasks. The SN magnitudes were derived by extracting aperture photometry at different radii, for the SN images and the reference image, and subtracting the host contribution. For the case of PTF~09cnd, we measure the photometry using both the image-subtraction routine \texttt{mkdifflc} and aperture photometry, and take the average between the two results. For the case of PTF~10cwr no reference image is available, so we extract aperture photometry with a 3\arcsec\ radius and subtract the host magnitude reported by \citet{Perley16} and include this into the photometric uncertainty budget.

\subsection{HST}
\label{sec hst}
The SLSNe iPTF 13dcc and iPTF 13ehe were observed with the Advanced Camera for Surveys (ACS) in the Wide Field Channel on board the HST, with the \textit{F625W} filter, as part of the GO-13858 program (PI A. De Cia). The data were reduced using the CALACS software, which contains corrections for degradation of the charge transfer efficiency and electronic artifacts (bias-shift and -striping effects). Cosmic rays were removed using the LA Cosmic routine \citep{vanDokkum01}. The images were then processed with DrizzlePac 2.0,\footnote{\texttt{ http://drizzlepac.stsci.edu}} with inverse variance map (IVM) weighting and assuming a pixel scale of 0.033\arcsec\ and a pixel fraction of 0.6. 

The SN PSF is resolved from the more extended host galaxies. For iPTF 13dcc, the PSF of the host has an FWHM of 3.2 pixels (0.11\arcsec\ ), while field stars have 2.4 pixels (0.08\arcsec\ ). iPTF 13ehe is separated from its host galaxy. The SN PSFs were fitted and thus isolated from their host galaxies using a custom \texttt{IDL} routine. The PSF SN photometry was then extracted assuming HST zeropoints and applying a correction for an aperture of 0.5\arcsec\ radius.

\subsection{Literature data collection}
We complement the photometric dataset of the SLSNe in our sample with the data published in \citet{Quimby11}, \citet{Pastorello10}, \citet{Inserra13}, \citet{Nicholl13}, \citet{Chen15}, and \citet{Vreeswijk14}. The literature photometry is showed in Figs. \ref{fig ind abs g mag 1} to \ref{fig ind abs g mag 5}. The purpose of including the literature data in this paper is to collect the most complete available light curves for the SLSNe in our sample. The $r$-band photometry is used to calculate the rest-frame $g$-band photometry, which is reported in Tables \ref{table g phot09as} to \ref{table g phot13ehe}. Because the sources of our observations are already diverse, the inclusion of literature data does not affect significantly the quality of our dataset. 

The full light curve of PTF~10nmn will be presented by Yaron et al. (2018, in preparation), including a wider coverage of the SN peak, which is not presented in this paper. 

We exclude from the analysis a couple of published photometry datapoints in cases of disagreement with the photometry secured with other (multiple) telescopes, namely for iPTF~13ajg \citep[P60 $R$-band data at MJD 56429 from][excluded]{Vreeswijk14} and for PTF~09cnd \citep[\textit{Wise} $R$-band data at MJD 55089 from][excluded]{Quimby11}. The new measurements supersede the earlier ones.
     
\subsection{On the diversity of the dataset}

The dataset used in this paper was collected from a diversity of facilities, and the photometry is derived with different pipelines and methods. The quoted uncertainties assess the quality of the photometry for each facility or measurement method. The contribution of the host galaxy light to the SN measured flux is taken into account and reflected in the quoted uncertainties. An exception to this is for the UV photometry (\textit{Swift}), for which the contribution from the host galaxy is not subtracted, but should be minimal (Sect. \ref{sec swift}). Often the filter transmission curves of similar filters are different for different facilities or catalogs for calibration, such as $r$ , $R$, and $R_c$ for example. However, we did not correct for these differences because they depend on the source spectra and their evolution, and these differences are typically very small, normally well below $0.1$~mag. 

In Figs. \ref{fig ind abs g mag 1} to \ref{fig ind abs g mag 5}, all photometry ar shown together. When enough data are available, the photometry from different telescopes can be directly cross-checked, and we do not find evident discrepancies. Further corrections to the photometry, such as foreground extinction and $k$-corrections, and their uncertainties, are described below.

\subsection{Foreground dust extinction and $k$-corrections}
\label{sec k-corr}

We derive Galactic foreground optical extinction $A_V$ using the maps of \citet{Schlafly11} through the Galactic Dust Reddening and Extinction Service at the NASA/IPAC Infrared Science Archive, assuming a standard extinction law and an extinction to reddening ratio $A_V / E(B-V) = 3.1$.\footnote{The background and further cautionary notes are reported at \texttt{http://irsa.ipac.caltech.edu/ applications/DUST/docs/background.html}} The mean uncertainty in the Galactic foreground extinction $A_V$ for our sample is of 0.009~mag, and we do not include this uncertainty in the photometric budget. The adopted $A_V$ values are listed in Table~\ref{table sample}. We calculate the extinction $A_\lambda$ at the central wavelength of each filter using the reddening curve of \citet{Cardelli89} and including the update for the near-UV given by \citet{ODonnel94}. Both apparent and absolute magnitudes reported in this paper are corrected for Galactic foreground extinction.

Host-galaxy extinction is not considered. SLSN host galaxies tend to be faint and have low metallicity \citep{Neill11,Leloudas15,Lunnan15,Perley16,Chen17a,Schulze18}, and therefore we expect them to have negligible dust extinction in the red bands, with possible regions that may be locally more dusty, affecting mostly the UV \citep[e.g.][]{Cikota17}. On the other hand, SN Ib/c host galaxies can show significant extinction \citep[mean $<E(B-V)> \sim 0.2$ and $<0.6$~mag for $\sim$80\% of SNe Type Ic, Ib, and Ic-BL;][]{Taddia15, Prentice16}, but determining it case by case for our comparison sample is often not possible and is beyond the scope of this paper.

We calculate the $k$-corrections $K_{gr}$ for the SLSN sample from observed PTF $r$ to rest-frame $g$ (SDSS filter system) using spectral series of PTF~12dam and iPTF~13ajg \citep{Vreeswijk14,Vreeswijk17,Quimby18} and following \citet{Hogg02}. Using individual spectra for each SLSNe was not possible here, due to the paucity of sufficient spectral coverage at all epochs for the SLSNe in our sample. The spectral coverage of PTF~12 dam is frequently sampled and spans from $-25$ to $321$ rest-frame days after the peak, while the spectra of iPTF~13ajg are reliable until 60 days after peak. In the overlapping interval, there is good agreement between the $k$-corrections calculated from the two series of spectra. This indicates some level of similarity between the spectra, which is also confirmed by the spectral analysis of the PTF SLSN-I sample \citep{Quimby18}. The spectra of the more slowly evolving SLSNe-I change more slowly. However, the $k$-corrections based on PTF~12dam and iPTF~13ajg are similar, so the differences in $k$-corrections for faster and slower SLSNe-I should be small. Given this similarity, and due to the general lack of spectral series as complete as those for PTF~12dam, we apply the $k$-corrections derived from the spectra of PTF~12dam and their evolution to all of the SLSNe in our sample.

The spectra were not warped to match the observed photometry of the individual SLSNe. This could have led to more accurate $k$-corrections. However, the uncertainty from the fact that we use the spectrum of PTF~12dam as a reference for the $k$-correction for all individual SLSNe is likely larger than the precision that could be gained by such a procedure. In addition, to make a reliable warping, photometry in at least two bands (and much preferably three) would be necessary, and this was not always available. The spectra of PTF~12dam themselves were carefully flux calibrated. To ensure a smooth evolution of the $k$-correction with time, we interpolate the individual $k$-correction values and obtain a smooth $k$-correction evolution in time for each SLSN, through a third-order polynomial fit of the individual $k$-correction values. 

The residuals from the third-order polynomial fit of the $k$-correction values with time can be used to estimate the uncertainties on the $k$-corrections, which are between 0.004 and 0.05 mag in our sample, with an average of 0.02~mag. These values show the scatter around the best-fit $k$-correction curve. Fitting the $k$-correction through individual points ensures that potential outliers, e.g. due to inaccurate flux calibration of the reference spectra, become negligible because the fit is driven by the majority of the points.  The dominant source of uncertainty on the $k$-correction is likely the fact that we use the spectrum of PTF~12dam as a reference for the $k$-correction for all SLSNe, but this cannot be trivially estimated. Using different SLSN spectra as a reference for the $k$-correction in a different but comparable sample of SLSNe, the uncertainties on the $k$-corrections are 0.01--0.1~mag (Wiis, private communication, MSc thesis, Table 4.1.1), although these are potentially slightly overestimated because they are derived with linear fits to the $k$-correction data. We do not include the uncertainties on the $k$-correction in our photometry uncertainty budget.

Table~\ref{table k-corr} lists the values of the adopted smooth $k$-correction at the epochs of the PTF~12dam spectra, applied to the redshifts of the SLSNe of our sample. Our $k$-correction values are in agreement with those of \citet[][$K_{gr}=-0.3$ before peak for LSQ~14bdq at $z=0.347$]{Nicholl15b}. The $k-$corrections that we apply rely on the assumption that the spectra of our SLSN sample are similar to those of PTF~12dam out to late epochs. At late times, the assumption of similarity among the spectra is less certain. We therefore recommend exercising caution in trusting our $k$-corrections at late epochs.

For the comparison with the Ib/c sample at maximum light, we calculate the $k$-correction from the observed $r$ to rest-frame $r$ using the spectrum at the peak of PTF~10tqv and following \citet{Hogg02}. We then derive the rest-frame $g$ by applying a color correction from the observed mean $g-r = 0.36$ mag of a large sample of Type Ib/c SNe of \citet{Prentice16}. Where necessary, we convert $B-V$ measurements from \citet{Prentice16} at $V$ peak to $g-r$ and adopt the weighted average, and otherwise directly use the observed $g-r$ at the $g$ peak. The standard deviation on the $g-r$ distribution is $0.34$ mag ($0.25$ and $0.23$ mag for the $g-r$ and $B-V$ distributions). Since $g-r$ evolves significantly for SNe Ib/c \citep[e.g.][]{Taddia15,Prentice16}, the $r$ to $g$ conversion used here for SNe Ib/c is most reliable around SNe peaks.

\section{Characterizing the light curves}
\label{sec characterization}

The light curves of PTF H-poor SLSNe sometimes show complex features, such as bumps/plateau, double peaks, and a change of the decay rate. Besides, the data are often sparse. We use the following independent diagnostics to characterize different properties of the light curves.

\begin{enumerate}
\item \textit{SN peak magnitudes} -- derived with a second-order polynomial fit to the data around the peak (see Sect. \ref{sec peak mag}).
\item \textit{Early- and late-time decay rates} -- derived with two independent linear fits to the data at early and late times after peak (see Sect. \ref{sec decay}).
\item \textit{Rise and fall times by 1 mag, $t^{\rm \Delta 1mag}_{\rm rise}$ and $t^{\rm \Delta 1mag}_{\rm fall}$}\\ -- the times that the SN takes to rise and fall by 1 mag from the peak (see Sect. \ref{sec rise fall 1mag}), measured on light curves which have been smoothed using interpolation (see Sect. \ref{sec smooth}).
\item \textit{Half-flux rise and fall times $t_{\rm rise, 1/2}$, $t_{\rm fall, 1/2}$} -- the times for the SN fluxes to rise from half-flux to peak, and to fall form peak to half-flux (see Sect. \ref{sec rise fall t12}), measured on light curves that have been smoothed using interpolation (see Sect. \ref{sec smooth}).
\end{enumerate}

The derived quantities are listed in Table~\ref{table derived} and the details for each diagnostic are reported below.

\subsection{Peak-magnitude distribution}
\label{sec peak mag}

We calculate the absolute magnitudes using the distance modulus for a given $z$ \citep[e.g.][]{Hogg00}. At the $z$ considered here ($\sim0.3$), the difference in distance modulus obtained from different cosmology models is negligible compared to the uncertainties in the observed apparent magnitudes. The redshifts of the SLSNe are derived in most cases to three decimal digits (see Table~\ref{table sample}). In fact, here we are interested only in the relative luminosity distances between different SNe, and the relative uncertainties are even smaller. The uncertainties on the absolute magnitudes are therefore largely dominated by the uncertainties on the observed apparent magnitudes, and we do not make any attempt to include uncertainties due to the distance estimate.

We determine the peak times and magnitudes by fitting a second-order polynomial to the rest-frame $g$-band magnitudes around the maximum brightness, typically between $-30$ and $30$ days around the approximate peak, or adjusting this interval to adapt to the data coverage. The fitted curves and the relevant time intervals are shown in Figures \ref{fig ind abs g mag 1} to \ref{fig ind abs g mag 5}.

\begin{figure}
\epsscale{1.2}
\plotone{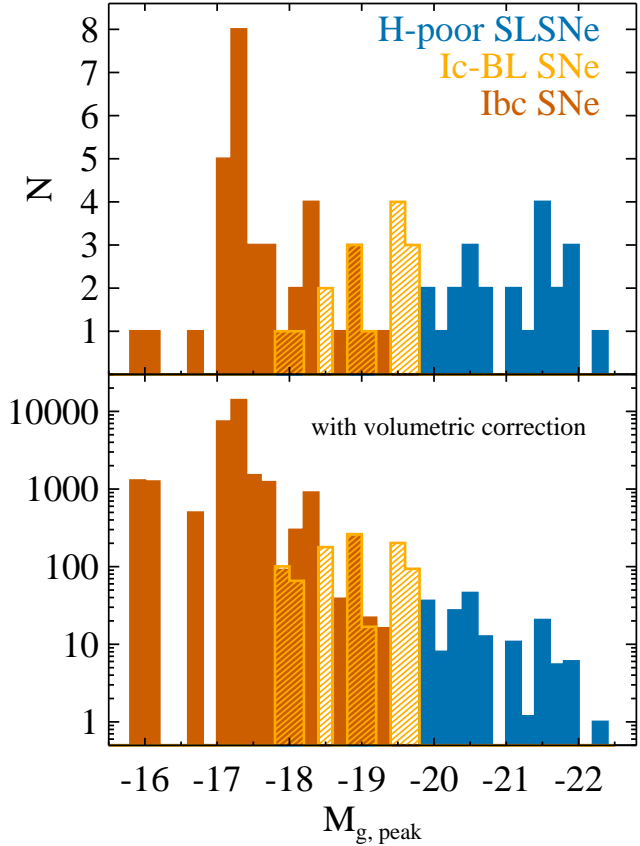}
\caption{Peak-magnitude distribution of the labelled types of PTF SNe, for rest-frame $g$ absolute magnitudes. The bottom panel shows the peak-magnitude distribution after volumetric correction (i.e. corrected for Malmquist bias).\label{fig peak mag}}
\end{figure}

Figure~\ref{fig peak mag}, top panel, shows the peak-magnitude distribution of H-poor SLSNe (solid blue), Type Ic-BL SNe (shaded yellow), and Type Ib, Ic, and Ib/c SNe (solid orange), all from the PTF survey, for a brightness bin of 0.2 mag. Note that only SNe where the peak could be observed and constrained are included in this plot. When calculating the number of SNe for each brightness bin, it is important to consider the observational biases. Although SLSNe are bright enough to be observed at larger distances, many normal SNe could actually be exploding at those distances, but be too faint to be detected (Malmquist bias). To compare the numbers of SNe in a fair way, it is therefore necessary to normalize the numbers to the same comoving volume. We calculate the volumetric correction $V_c$ for each SN as the ratio between the volume probed by the most luminous SLSN in our sample ($M_{g, \rm max}=-22.42$~mag at peak) and the volume probed by the individual SN, given the limiting magnitude of the PTF survey of $m_{\rm lim}=20.5$~mag \citep{Cao16}, i.e., the maximum luminosity distance at which each SN would have been observed with this limiting magnitude. The volumetric correction factor $V_c$ is then expressed as follows:
\begin{equation}
V_c = V_{\rm max} / V_{\rm max, i}  =   \left( \frac{D_{L, \rm max}}{1+z_{\rm max}}\right) ^3 /  \left( \frac{D_{L, \rm max, i}}{1+z_i}\right) ^3 \mbox{,}
\end{equation}
where the luminosity distance of the brightest SN in the sample is $D_{L, \rm max} = 10^{(((m_{\rm lim} - M_{g, \rm max}) +5.)/5.)}$, and the luminosity distance for each individual SN is $D_{L, \rm max, i} =10^{(((m_{\rm lim} - M_{g, i}) +5.)/5.)}$. Figure~\ref{fig peak mag}, bottom panel, shows the peak-magnitude distribution after the volumetric correction.
  The mean peak magnitude of the SLSN sample is $<M_{g,\rm peak}> = -21.14$~mag with a standard deviation of 0.75~mag.

\subsection{The postpeak early- and late-time decay rates}
\label{sec decay}

The light curves of H-poor SLSNe can show a change in decay rate \citep[e.g.][]{Inserra13}. Here we independently characterize the postpeak early- and late-time decay rates of H-poor SLSNe with linear fits to the early-time and late-time data separately. We study the early decay with a linear fit to the rest-frame $g$ magnitudes in a time interval between the peak and typically 60 days after peak. In some cases, this interval was adjusted to the data coverage, or to avoid changes of slope. The selected time intervals and the resulting linear fits to the data are displayed in Figs. \ref{fig ind abs g mag 1} to \ref{fig ind abs g mag 5} (solid curves). We define the late-time decay as typically beyond 60 days after peak and characterize the decay rate with a linear fit to the data, in the same way as we did for the early-time decay. The linear fits to the late-time decays are displayed in Figs. \ref{fig ind abs g mag 1} to \ref{fig ind abs g mag 5} (solid curves, typically beyond 60 days after peak).

\begin{figure}
\epsscale{1.2}
\plotone{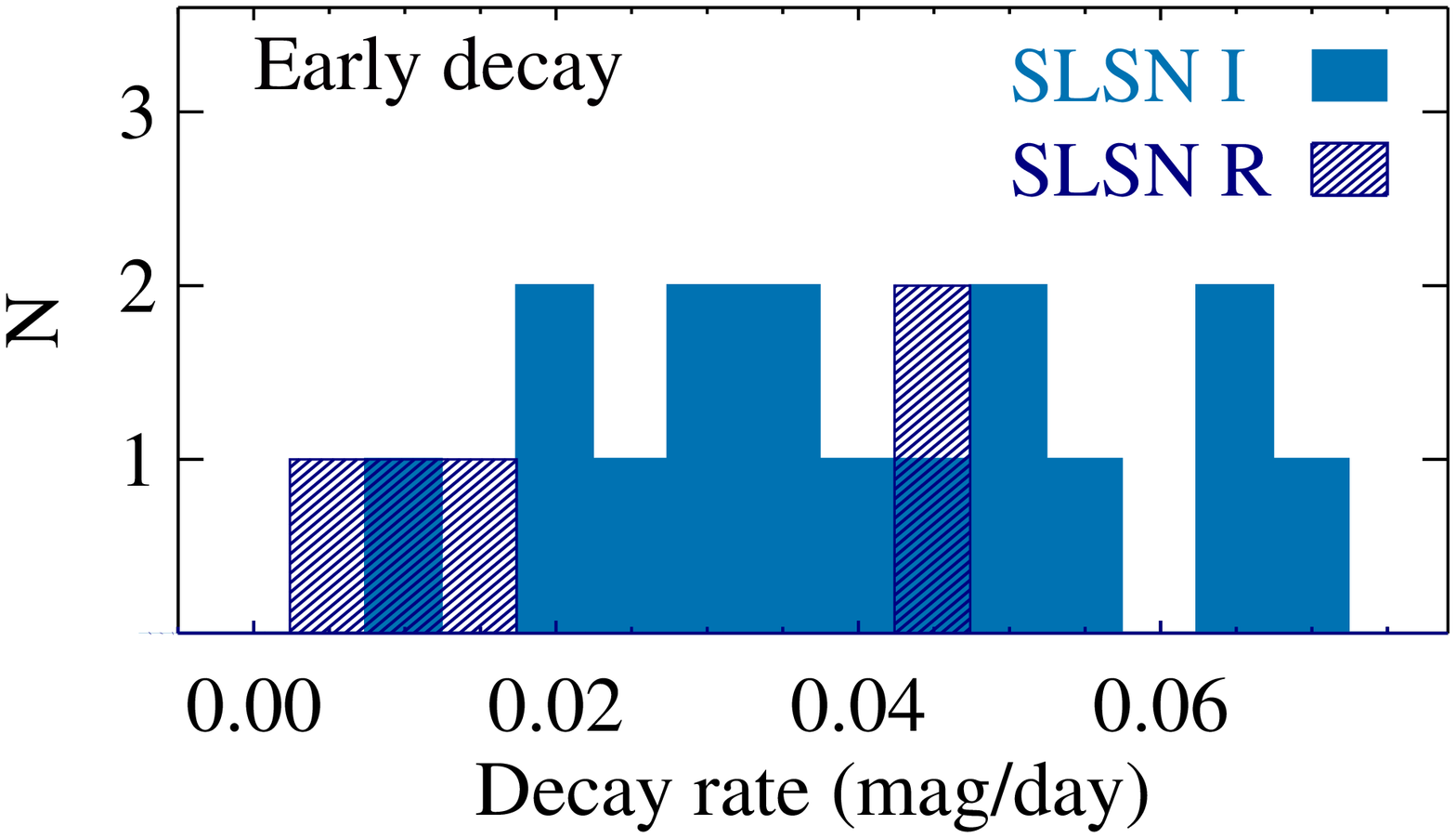}
\plotone{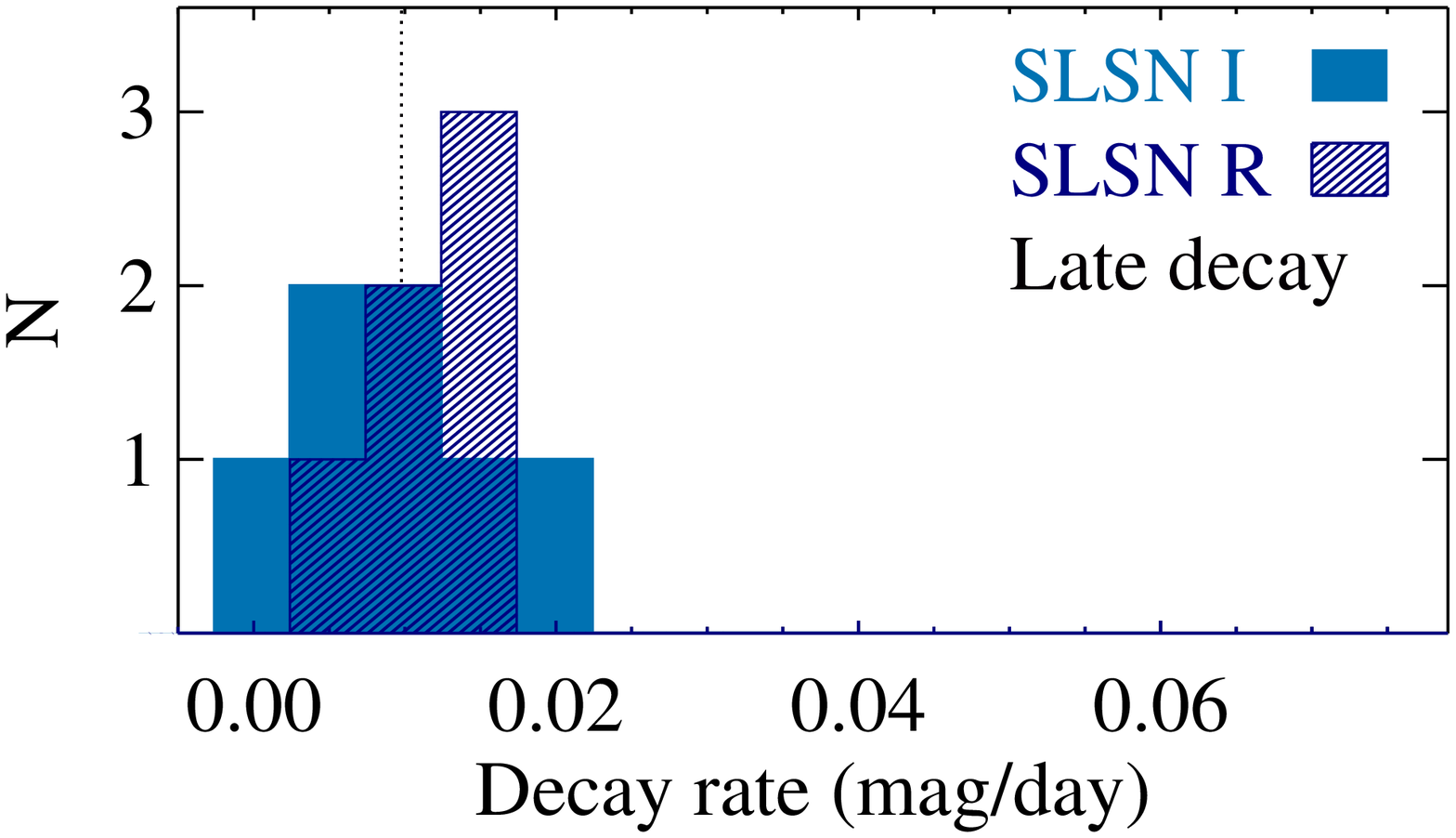}
\caption{Postpeak decay slope distribution at early times (typically below 60 days, top panel) and at late times (typically beyond 60 days, bottom panel). At late times, all observed SLSNe cluster around the $^{56}$Co to $^{56}$Fe decay rate of $0.0098$ mag day$^{-1}$ (dotted vertical line). \label{fig decay distr}}
\end{figure}

Figure~\ref{fig decay distr} shows the distribution of the early-time decay slopes (top panel) and the the late-time decay slopes (bottom panel). SLSNe that were originally classified as sub-type R within the PTF survey are marked separately in this figure and compared to the rest of the sample. The original criterion for being classified as an SLSN-R was either a slow decline or spectral similarity with SN~2007bi, with no quantitative threshold. We do not intend to use these criteria as a meaningful classification, but rather to test this classification scheme, because it is often used in the literature \citep[e.g.][]{Gal-Yam12,Inserra17}.

The decay rate of most SLSNe slows down from early to late times. The decay rates and the times of transition from a faster to a slower decay (the intersections between the early- and late-time linear fits) are reported in Table~\ref{table derived}.

\subsection{Light-curve smoothing}
\label{sec smooth}
We smooth the SN light curves to be able to further measure the rise and decay times more easily. We model the observed light curves with a nonparametric model, as follows. We first fit a first-order polynomial to the rest-frame $g$-band flux light curves locally. Then, we consider a fitting interval of 5 days (at phases until 5 days after peak), 10 days (at phases beyond 50 days after peak), and proportional to the phase (0.2 times) otherwise. For the interpolations, we use a Gaussian smoothing kernel that weights the fluxes according to their phase distance to each interpolated point. The smoothing algorithm also uses the uncertainties on the photometry to weight the data points. In order to avoid mathematical artifacts, a few auxiliary points are added to the observations. The light-curve smoothing algorithm is described in more detail in \citet{Rubin16}. In a few cases, to avoid unphysical wiggles in the smoothed light curves for poorly sampled regions, we binned scattered data during small time intervals. Namely, we binned the data for PTF~10aagc between 32 and 44 rest-frame days after peak; PTF~09cwl between 122 and 141; PTF~10vwg between 44 and 62; PTF~11rks at 55, and between 144 and 154; and PTF~12gty between 141 and 158. We adopt the formal error on the smoothed fluxes computed by the smoothing algorithm, and assume a minimum uncertainty of 10\% of the flux in those cases where the formal errors are smaller.

\begin{figure*}
\epsscale{1.2}
\plotone{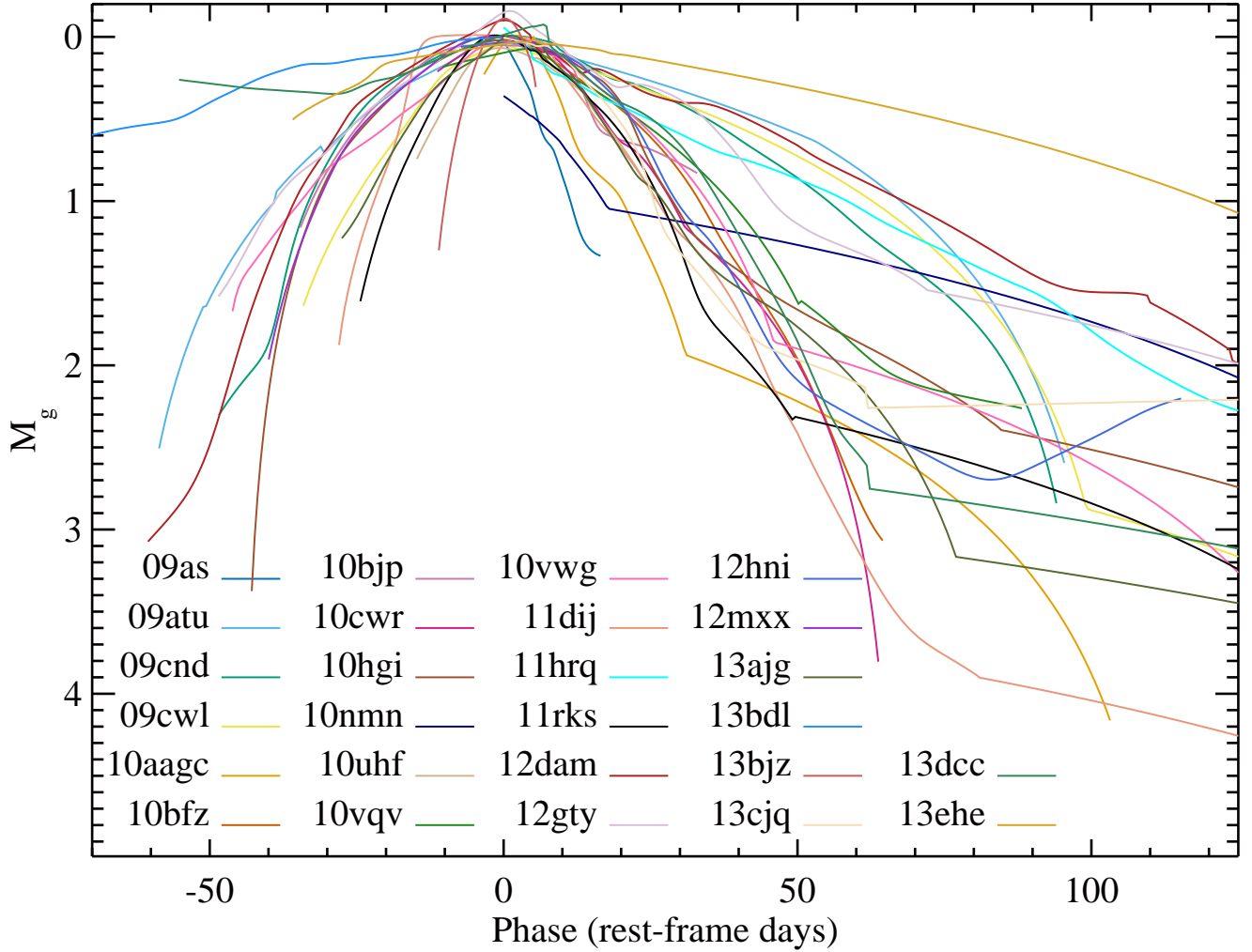}
\caption{Rest-frame $g$-band smoothed light curves of the SLSNe in our sample, normalized at peak. PTF~10nmn is normalized with respect to the peak magnitude, which is taken from Yaron et al. (2018, in preparation) and is not presented in this paper. The peak magnitudes are derived with a second-order polynomial fit to the data (Figs. \ref{fig ind abs g mag 1} to \ref{fig ind abs g mag 5}) and reported in Table \ref{table derived}. \label{fig all smooth norm}}
\end{figure*}

The light-curve smoothing fits to the data in flux space, including the auxiliary points, are shown in Figs. \ref{fig ind smooth flux 1} to \ref{fig ind smooth flux 4}. The collection of all smoothed light curves, normalized by the peak magnitude, is shown in Fig.~\ref{fig all smooth norm}. Even when normalized to the peak, there is a wide variety of light-curve behaviors among H-poor SLSNe, and the scatter is too large to reduce them to a single template. Remarkably, there is no clear gap between fast- and slow-decaying SLSNe. 

\subsection{Times to rise and decay by 1~mag from the peak}
\label{sec rise fall 1mag}

We derive the times to rise (and decay) by 1~mag to (from) the peak, $t^{\rm \Delta1mag}_{\rm rise}$ ($t^{\rm \Delta1mag}_{\rm fall}$) by inspecting the smooth light curves (Sect. \ref{sec smooth}). Figures \ref{fig ind smooth mag 1} to \ref{fig ind smooth mag 4} display the time intervals within 1~mag from the peak. Table~\ref{table derived} lists the resulting rise and decay times. The errors are estimated starting from the errors on the smoothed light curves (Sect. \ref{sec smooth}). We create a pseudo-random normal distribution of the smoothed flux errors around the smoothed light curves, through $n$ Monte Carlo realizations, and we estimate $n$ rise (and decay) times. We finally derive the uncertainties on the rise (decay) times from the standard deviation of the distribution of rise (and decay) times and assuming a minimum uncertainty of 2 days. We test for convergence of our results by varying the number of Monte Carlo realizations $n$ between 10, 100, 1000, and 10,000, and eventually use $n=1000$.  

\begin{figure*}
\epsscale{1.15}
\plottwo{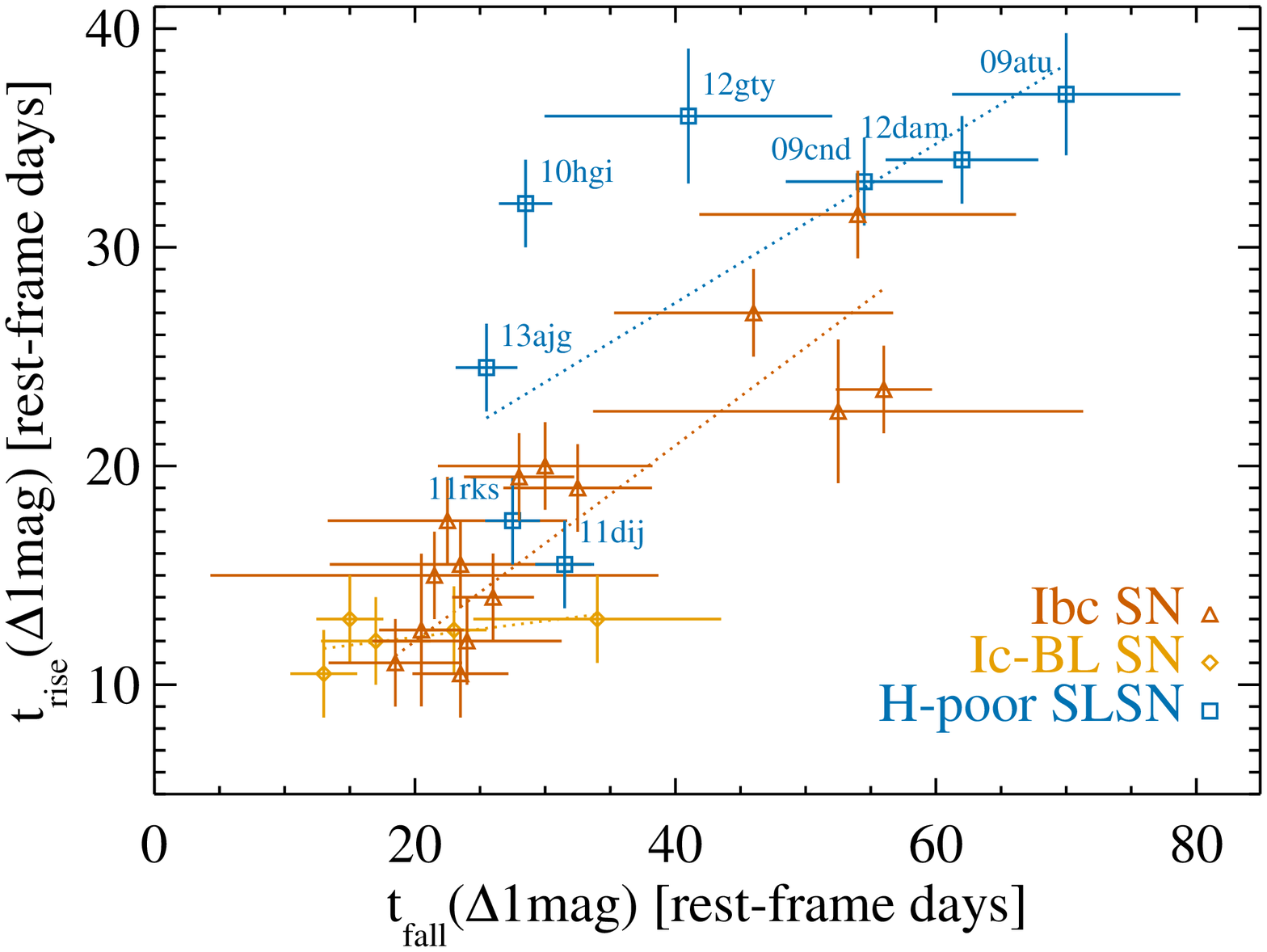}{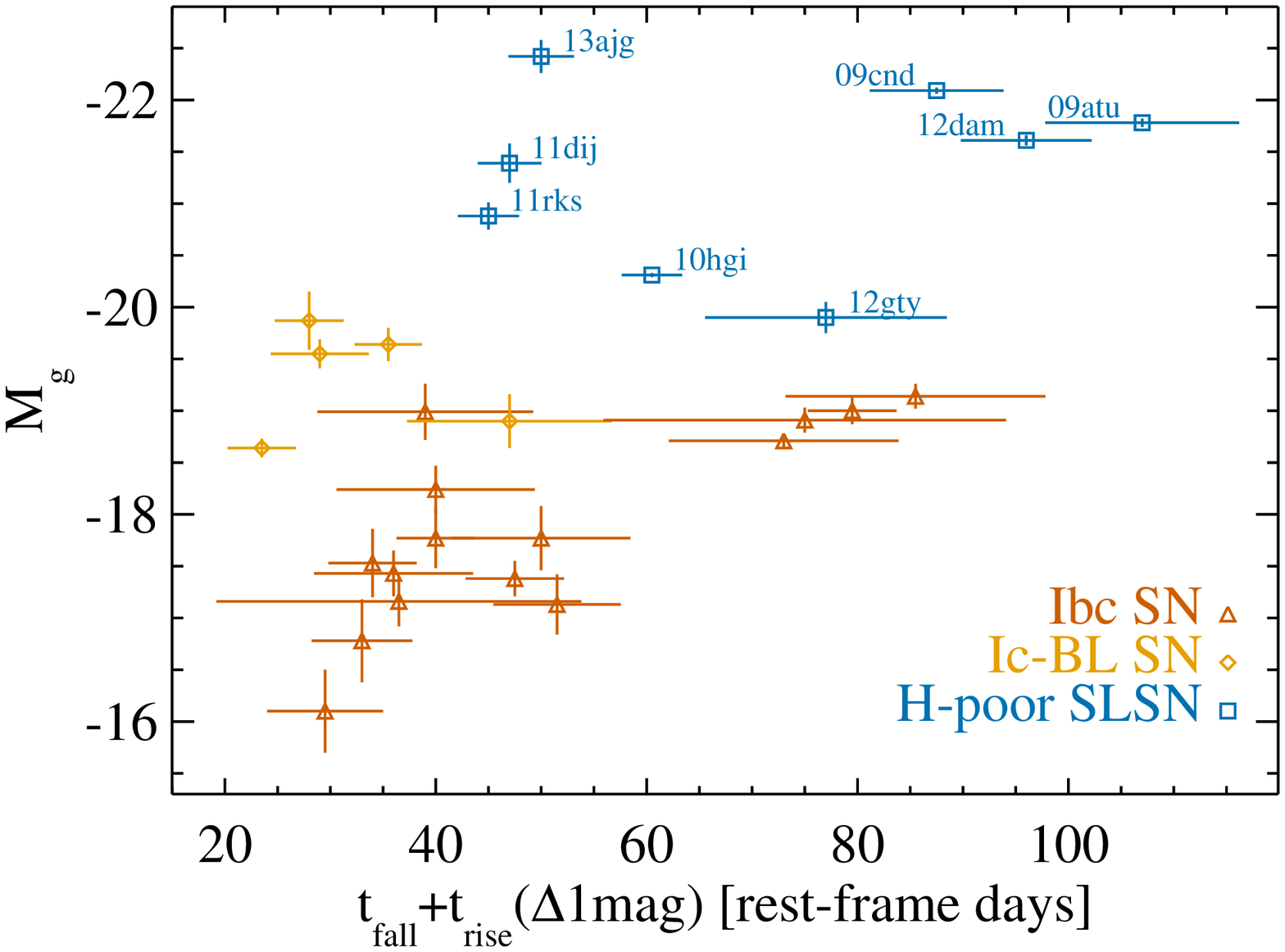}
\plottwo{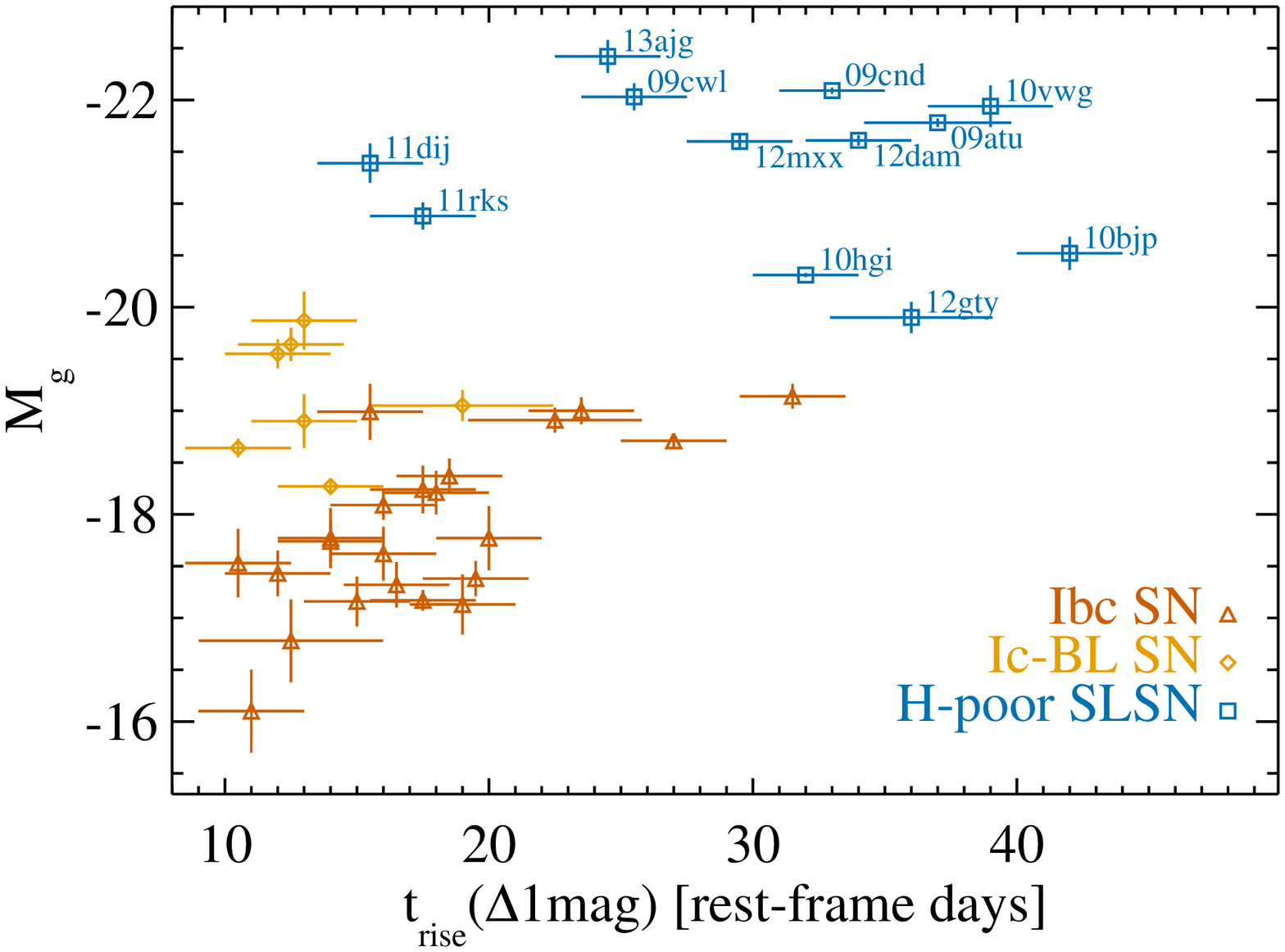}{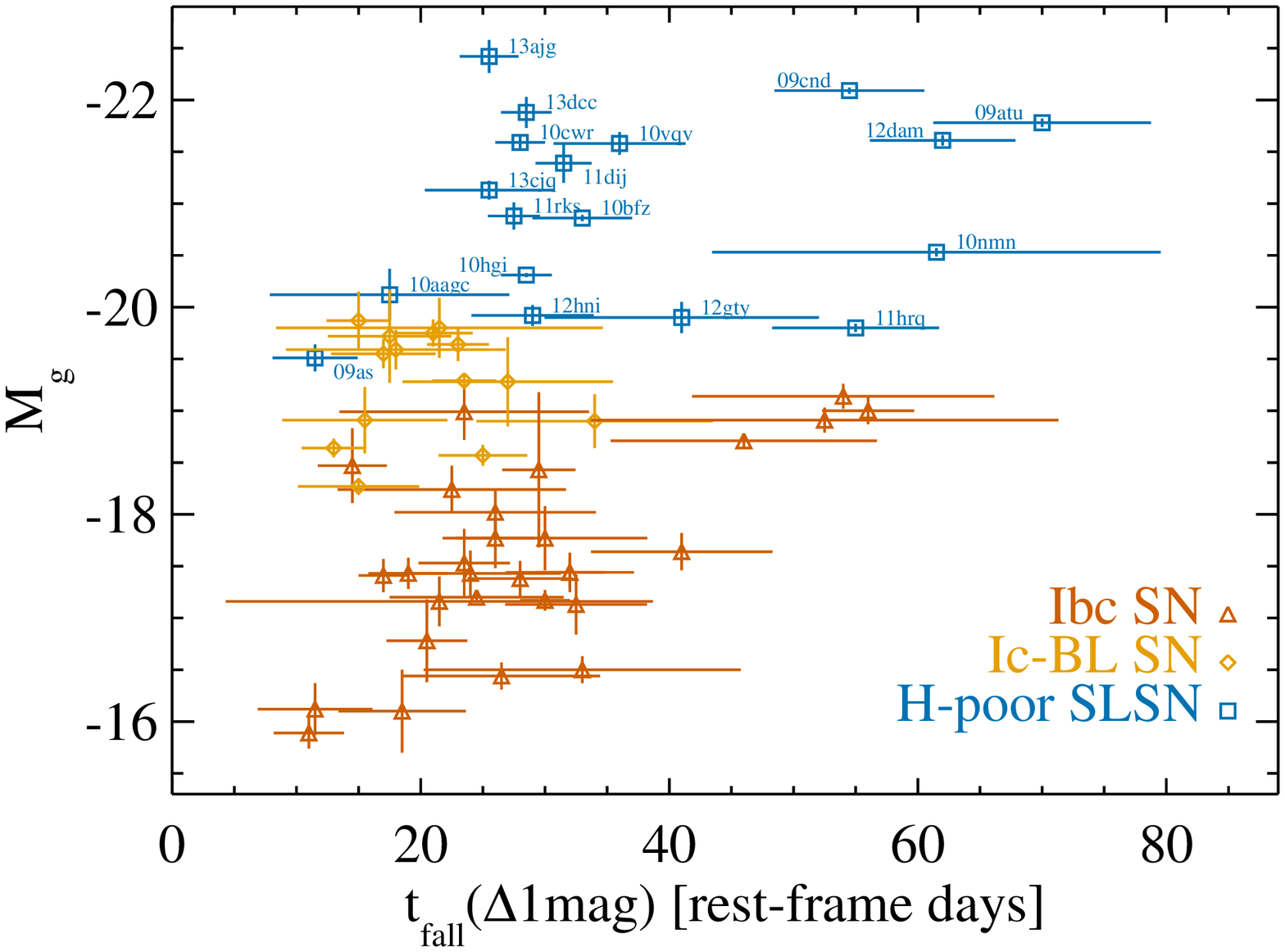}
\caption{Times to rise and decay by 1 mag to and from the peak, and peak magnitudes (rest-frame $g$). The dotted curves show linear fits to the data in our sample. \label{fig rise decay 1mag}}
\end{figure*}

In Fig.~\ref{fig rise decay 1mag} we investigate the cross-correlations between $t^{\rm \Delta1mag}_{\rm fall}$, $t^{\rm \Delta1mag}_{\rm rise}$, their sum, and the peak magnitude. There is a clear correlation between $t^{\rm \Delta1mag}_{\rm fall}$ and $t^{\rm \Delta1mag}_{\rm rise}$

We fit this correlation linearly, assuming $t^{\rm \Delta1mag}_{\rm rise} = A + B \times t^{\rm \Delta1mag}_{\rm fall}$ and including the observed uncertainties in both $x$ and $y$ axes, for each SN type. SNe where the data are not sufficient to constrain $t^{\rm \Delta1mag}_{\rm rise}$ and $t^{\rm \Delta1mag}_{\rm fall}$ are excluded from this fit, as reported in Table~\ref{table derived}. The results of this fit are shown in Fig.~\ref{fig rise decay 1mag} (dotted curves) and reported in Table~\ref{table risefall fit}. We also compute the Pearson and Spearman correlation coefficients, which measure the strength (tightness and monotonicity) of a correlation not taking the observed uncertainty into account, and their null probabilities. These are listed in Table~\ref{table risefall fit}. 
\begin{table}
\begin{center}
\caption{Normalizations and slopes of the linear fits of the correlations between rise times and decay times (Figs. \ref{fig rise decay 1mag} and \ref{fig rise decay t12}). Note. $r$ and $\rho$ are the Pearson and Spearman correlation coefficients, respectively, and are listed with their respective null probability ($p_r$ and $p_\rho$).\label{table risefall fit}}
\begin{tabular}{@{}l l l l l l l@{} }
\tableline\tableline
Type & A & B & $r$ & $p_r$ & $\rho$ & $p_\rho$\\
\tableline
\multicolumn{7}{c}{$t^{\rm \Delta1mag}_{\rm rise} = A + B \times t^{\rm \Delta1mag}_{\rm fall}$}\\
\tableline
SLSN  & $12.93\pm 6.55$ & $ 0.36\pm 0.15$ & $ 0.69$ & $0.057$ & $ 0.76$ & $0.028$ \\
\tableline
\multicolumn{7}{c}{$t_{\rm rise,1/2} = A + B \times t_{\rm fall, 1/2}$}\\
\tableline
SLSN  & $17.32\pm 6.22$ & $ 0.21\pm 2.06$ & $ 0.39$ & $0.270$ & $ 0.44$ & $0.206$ \\
\tableline
\end{tabular}
\end{center}
\end{table}

We also find a trend between the peak magnitudes and $t^{\rm \Delta1mag}_{\rm rise}$, $t^{\rm \Delta1mag}_{\rm fall}$, and the peak width.

\subsection{Times to rise and decay by half-flux}
\label{sec rise fall t12}

We further characterize the rise and decay times using the method of \citet{Prentice16}, who measured the time required to double or halve the flux with respect to the peak flux of a sample of stripped-envelope SNe (Ib/c). We derive $t_{\rm rise, 1/2}$ and $t_{\rm fall, 1/2}$, the time to double and halve the flux, respectively, for the PTF SN sample considered in this paper, using the smoothed flux light curves (Sect. \ref{sec smooth}). The resulting $t_{\rm rise, 1/2}$ and $t_{\rm fall, 1/2}$ are reported in Table \ref{table derived}. We calculated the uncertainties in the same way as for the rise and decay times by 1 mag (Sect. \ref{sec rise fall 1mag}). 

\begin{figure}
\epsscale{1.2}
\plotone{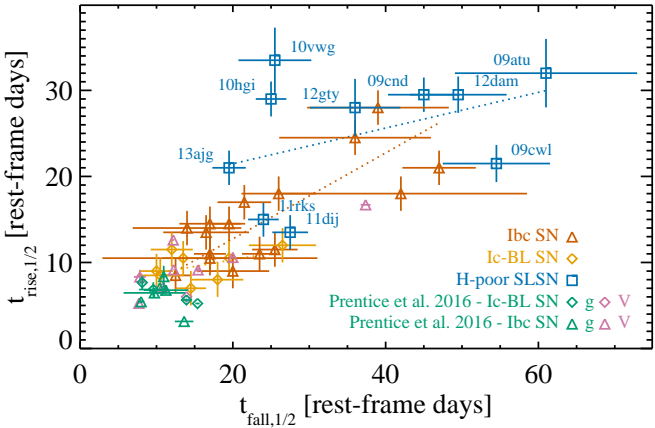}
\caption{Times to rise and decay by half-flux. The dotted curves show the linear fit to the data in our sample. \label{fig rise decay t12}}
\end{figure}

Figure~\ref{fig rise decay t12} shows the comparison of $t_{\rm rise, 1/2}$ and $t_{\rm fall, 1/2}$ among the different samples and SN types, and compares it with the results of \citet{Prentice16}. We linearly fit the correlations between rise and decay times for each SN type in the same way as for the rise and decay times by 1 mag (Sect. \ref{sec rise fall 1mag}). The results of the fit are shown in Fig.~\ref{fig rise decay t12} (dotted curves) and reported in Table~\ref{table risefall fit}.

\section{Discussion}
\label{sec discussion}

In this paper, we study 26 hydrogen-poor SLSNe at ''low'' redshift ($<z> = 0.27$), all spectroscopically classified as an SLSN-I and discovered by the (i)PTF survey. Here we characterize their light curves and discuss their (dis)similarity to SNe Ib/c and Ic-BL.

\subsection{Peak-magnitude distribution}
\label{sec disc peak}

Figure~\ref{fig peak mag}, top panel, shows the distribution of the rest-frame $g$-band absolute peak magnitudes of SLSNe, SNe Ib/c, and Ic-BL. These SNe are all discovered by the PTF survey and separated into these three classes through spectroscopic classification \citep{Quimby18, Schulze18}. The $k$-correction has been applied as described in Sect. \ref{sec k-corr}. Clearly, not all peak magnitudes of SLSNe are brighter than $<-21$ mag. This threshold was an operational definition that was used to start characterizing SLSNe in the early days of their discovery \citep{Quimby11,Gal-Yam12}. In fact, spectroscopically classified SLSNe-I from PTF span a wider range in absolute peak magnitudes, $-22.5\lesssim M_{g, \rm peak}\lesssim-20$~mag. The mean absolute peak magnitude in the PTF sample is $<M_{g, \rm peak}>\,=-21.14$~mag, with a standard deviation of 0.75 mag, which is about 2 and 4 mag brighter than the mean for SNe Ic-BL and SN Ib/c in our sample, respectively. The SLSN mean peak magnitude in the PTF sample is similar to what \citet{Lunnan18} found for the higher-$z$ sample from PS1, and thus we confirm no evidence for evolution of the SLSN peak luminosities with $z$ on the currently available data.

Furthermore, the peak magnitudes of SLSNe-I are all brighter than SNe Ib/c. The gap between the brightest SN Ibc and the faintest SLSN-I is of about 0.5~mag, although somewhat uncertain given the limited size of the samples and the uncertainty on the host-galaxy extinction for the SNe Ib/c. SNe Ic-BL are typically brighter than SNe Ib/c and fill this gap. The distribution of peak magnitudes is continuous from SNe Ib/c to SNe Ic-BL and SLSNe. There is very little overlap between the SLSN population and SNe Ic-BL.

It is crucial to take into consideration the fact that fainter SNe can be observed and counted only out to smaller distances. When applying the volumetric correction to compensate for this bias (Fig.~\ref{fig peak mag}, bottom panel), the peak magnitude distribution decays smoothly and exponentially from SNe Ib/c to Ic-BL and to SLSNe. Another important bias to keep in mind is spectroscopic completeness. Because SNe Ib/c exist in the same parameter space as Type Ia or IIp SNe, some of them may be not selected for spectroscopic classification and therefore missing from those that we sample.

We conclude that the peak magnitudes of SLSNe are brighter than those of SNe Ib/c and Ic-BL. However, there is no evidence for SLSNe being drawn from a separate population when considering only the peak-magnitude distribution and taking the volumetric correction into consideration. Further evidence for the difference between SLSNe and SNe Ib/c comes from the rise and decay timescales, which we discuss in Sect. \ref{sect disc rise decay times}.

\subsection{Observed colors}
\label{sec disc colors}

\begin{figure*}
\epsscale{1.}
\plotone{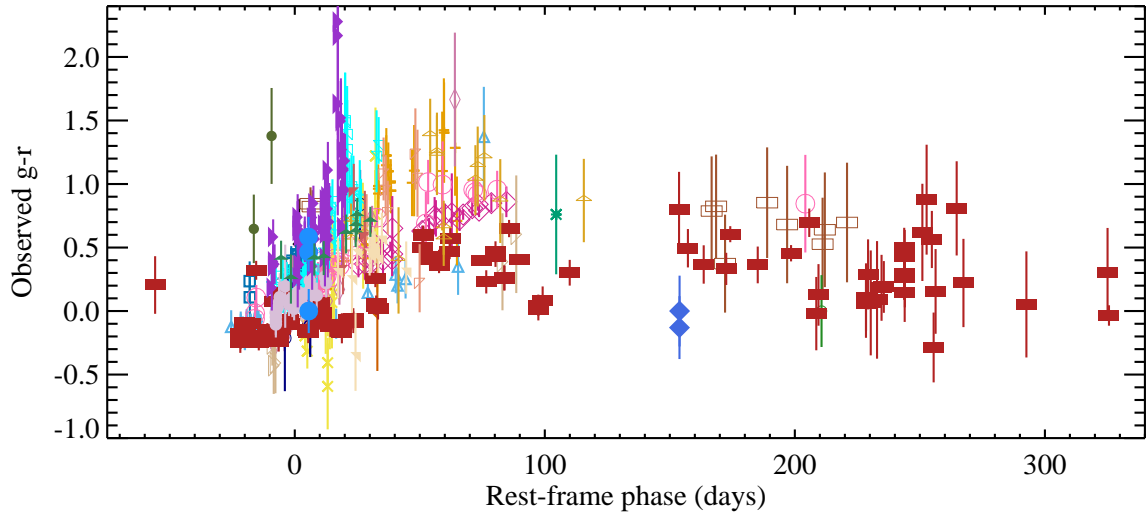}
\caption{Observed $g-r$ color of the SLSNe in our sample. The symbols and colors are as in Fig.~\ref{fig all lc}. \label{fig all g-r}}
\end{figure*}
Figure~\ref{fig all g-r} shows the evolution of the observed $g-r$ color for individual SLSNe. The $g-r$ color seems to increase at early epochs, until a few tens of days after peak. The mean observed $<g-r>$ at peak is 0.24 mag, with a standard deviation of 0.37 mag. At later times, the $g-r$ color evolution seems to stabilize at around $\sim0.5$ mag, and perhaps higher. 

This $g-r$ evolution in SLSNe is overall similar to that in SN Ib/c \citep[see Figures 22--25 of][]{Prentice16}. The $g-r$ color in SLSNe may, however, rise for a longer time (up to $\sim50$ days after peak, while SNe Ib/c reach a plateau at 10--20 days after peak), and to lower $g-r$ (the color plateau in SNe Ib/c spans roughly between 0.5--1 mag). However, we caution against a direct comparison of the observed $g-r$ between nearby SNe and SLSN, given their higher redshift. Indeed, after $k$-correction, the rest-frame $g-r$ colors at peak in SLSNe span roughly between $-0.6$ and $0.0$ mag. The observed-frame colors are reported here only as an observational reference. In Sect. \ref{sec cosmo} we further discuss rest-frame $g-r$ colors at peak in SLSNe.

\subsection{Decay rates}
\label{sec disc slopes}

Figure~\ref{fig decay distr} shows the postpeak decay rates at early times (top panel, typically before 60 days after peak) and late times (bottom panel, typically beyond 60 days after peak), as derived in Sect. \ref{sec decay}. The two distributions are quite different, indicating that at early times, SLSNe decay faster than at late times. The mean SLSN decay rates are 0.04 and 0.013 mag day$^{-1}$ at early and late times, respectively, with standard deviations of 0.02 and 0.005 mag day$^{-1}$. All SLSNe with available data in our sample slow down their decay rate from early to late times. Moreover, their late-time decay rate settles around $\sim1$~mag decay per 100~days. This rate is similar to the radioactive decay of $^{56}$Co to stable $^{56}$Fe \citep[more specifically, 102.3 days mag$^{-1}$;][]{Wheeler00,Nadyozhin94}. While at late times all SLSNe-I with available data show this slow decay, some selection biases may be present, because fast decays at late times may fall below the detection thresholds and not be measurable.

The late-time decays expected within the magnetar scenario can, under certain circumstances, mimic the radioactive $^{56}$Co decay \citep[e.g.][]{Moriya17}. We further discuss this in Sect. \ref{sec magnetar}.

The SLSN PTF~12hni is excluded from the late-time decay distribution because it exhibits a clear rebrightening (in all covered filters) and thus a negative decay rate, starting at about 75 days after peak, as reported in Table~\ref{table derived}. This could represent a case where interaction with the CSM re-energizes the light curve at late times, through the transformation of kinetic energy into luminosity. This typically requires a high optical depth, and one may naively not expect to observe broad lines and absorption features in this case \citep[see, however,][]{Moriya12}. The rebrightening of PTF~12hni was not covered by spectral observations \citep{Quimby18}.

In Fig.~\ref{fig decay distr}, we distinguish between classical H-poor SLSNe and those SNe originally classified as SLSNe-R within the PTF survey due to their slow decay early after the peak (consistent with radioactive decay, or spectrally similar to SN~2007bi). We stress again that we do not intend to use these criteria as a meaningful classification, but rather to test this classification scheme. At late times, the decay rates of SLSNe-R indeed cluster around the radioactive nickel decay rate, which is expected given the way SLSNe-R were originally selected. However, at early times, a couple of SLSNe-R have steeper decay slopes. Moreover, there is no evidence for a bimodal distribution in the decay properties in Fig.~\ref{fig all smooth norm}, where all smoothed light curves are plotted together, after being normalized around the peak. Therefore, there is no clear separation between SLSNe-R and classical SLSNe-I. This casts doubt on the existence of SLSN-R as a separate class, as initially suggested by \citet{Gal-Yam12}. 

\begin{figure*}
\epsscale{1.18}
\plotone{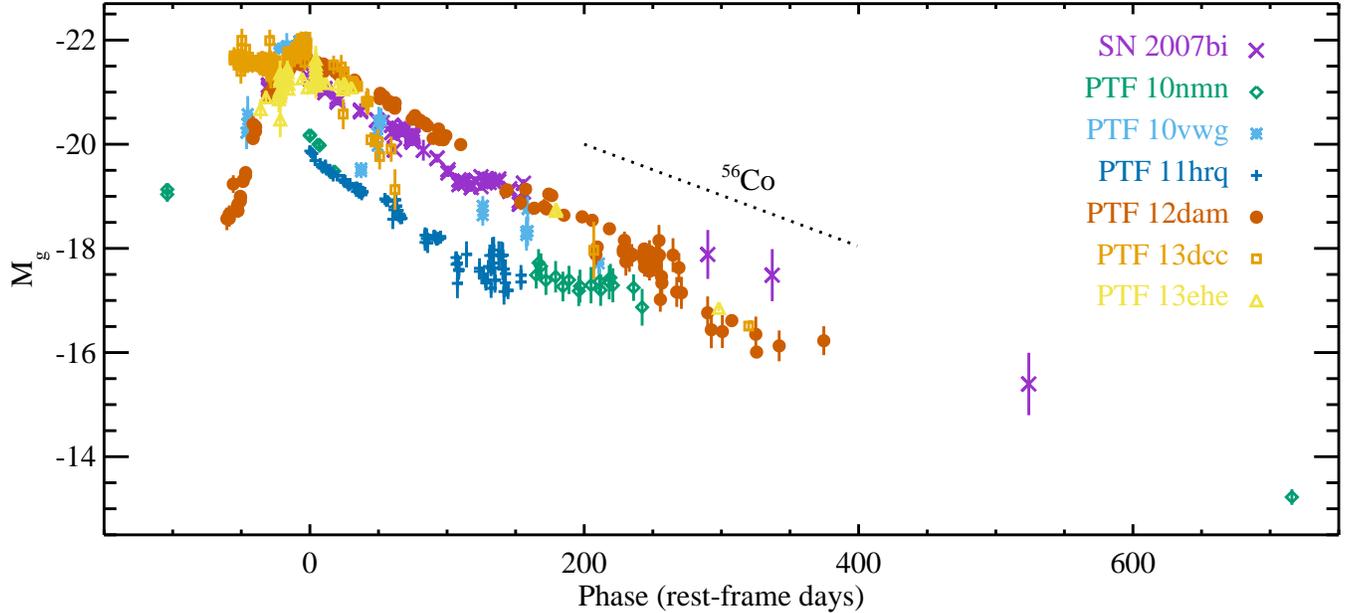}
\caption{Rest-frame $M_g$ light curves of the SLSNe originally classified as SLSNe of the R type. \label{fig R}}
\end{figure*}

Nevertheless, we note that early-time light-curve features are more common in SLSNe-R than in classical SLSNe-I. While none of the classical SLSNe-I show these features, three out of five SLSNe-R with early-time coverage \citep[10nmn, Yaron et al. 2018 in preparation; 12dam and 13dcc,][]{Vreeswijk17} and possibly a fourth case \citep[13ehe,][]{Yan15} have an early plateau or bumps of different strengths. At late times, the light-curve decay in SLSNe-R shows wiggles and bumps in virtually all events. Similar conclusions have also been drawn by \citet{Nicholl16} and \citet{Inserra17}. In the case of the hybrid SLSN iPTF~15esb (late-time emergence of H emission), \citet{Liu17} showed that the double peak of the light curves could be explained with a multiple-shell CSM interaction model. On the other hand, classical SLSNe-I might show fewer late-time features, the only clear example being PTF~12hni and perhaps PTF~12gty, which are the least luminous among our sample. However, the paucity of late-time data for such events prevents us from drawing firm conclusions. For this reason, we keep open the possibility that two separate subclasses of SLSNe-I exist (slowly/rapidly declining), until further evidence is collected. The light curves of all potential SLSNe-R from PTF are shown together in Fig.~\ref{fig R}. The presence of bumps in the light curves indicate that either CSM interaction or multiple sources are responsible for powering the light curve, as also found by \citet{Vreeswijk17} and \citet{Inserra17}. 

\subsection{Rise and decay times}
\label{sect disc rise decay times}

In Fig.~\ref{fig rise decay 1mag}, we compare the rise and decay timescales ($t^{\rm \Delta1mag}_{\rm rise}$ and $t^{\rm \Delta1mag}_{\rm fall}$; Sect. \ref{sec rise fall 1mag}) and peak magnitudes for PTF SLSNe-I and PTF SNe Type Ib/c and Ic-BL SNe. 

The peak magnitudes are brighter for SLSNe-I than for SNe Ib/c and Ic-BL, as discussed in Sect. \ref{sec disc peak}. We find possible mild trends between $M_g$ and $t^{\rm \Delta1mag}_{\rm rise}$ and between $M_g$ and $t^{\rm \Delta1mag}_{\rm fall}$, albeit with a very large scatter, and mostly when all SNe are considered. Indeed, neither luminous and fast-evolving events nor faint and slow-evolving ones are observed. A correlation between peak luminosity and rise time was observed for SNe IIn by \citet{Ofek14} and is consistent with the explanation of CSM interaction. The predictions for this correlation are in \citet{Ofek14c}.

The SLSNe in our sample tend to have longer rise timescales than SNe Ib/c; see below. Most SNe Ib/c and Ic-BL also tend to decay faster than SLSNe, although there are a few exceptions of slow-decaying, long-lived SNe Ib/c \citep[e.g., PTF~11bov, also known as SN~2011bm;][]{Valenti12}.

We find a correlation between $t^{\rm \Delta1mag}_{\rm rise}$ and $t^{\rm \Delta1mag}_{\rm fall}$ for SLSNe (Fig.~\ref{fig rise decay 1mag}, top left panel). The parameters and strength of this correlation are reported in Table \ref{table risefall fit}. Such a correlation is expected for both magnetar and nickel decay models \citep[e.g.][]{Nicholl15}. This correlation is continuous and does not show two separate classes of SLSNe-I, in contrast to the findings of \citet{Nicholl15}. Interestingly, SLSNe and ''normal'' SNe Ibc follow separate $t^{\rm \Delta1mag}_{\rm rise}$ -- $t^{\rm \Delta1mag}_{\rm fall}$ correlations. Although the correlation is not strong, there is an evident offset toward longer rise times for SLSNe with respect to SNe Ib/c. The offset is roughly a 10 day longer rise timescale for SLSNe. Finally, this correlation has a large scatter, and moreover, in some cases (PTF~11dij and PTF~11rks), the measured rise and decay times are similar to those of SNe Ib/c. Therefore, the above-mentioned correlation cannot be used to distinguish between SLSNe-I and SNe Ib/c, but only as an indicator of the average properties of the two populations.

As a sanity check, we further study the rise and decay times with a different approach, by considering the time to rise and decay by half-flux from the peak ($t_{\rm rise, 1/2}$ and $t_{\rm fall, 1/2}$, Sect. \ref{sec rise fall t12}). Again, the rise and decay times correlate continuously for SLSNe-I and the correlations are different for the three different classes of SNe. (The parameters and strength of this correlation are reported in Table \ref{table risefall fit}.)

The difference between the correlations of the rise with the decay timescales for SLSNe and SNe Ib/c (and Ic-BL) is evident using both independent methods ($\Delta1$ mag and half-flux). This suggests that SLSNe have longer rise timescales than SNe Ib/c, even for similar decay timescales. Observationally, we conclude that SLSNe show overall different light-curve properties from SNe Ib/c and Ic-BL. Therefore, SLSNe can be considered as a separate population, not only from a spectroscopic \citep{Quimby18} but also from a photometric perspective.

\subsection{Bolometric correction} 

We use the spectral information derived from a well-observed event to estimate the bolometric luminosity $L_{\rm bol}$ from single-band photometry. We adopt the conversion from the rest-frame $g$-band to the bolometric luminosity derived for PTF~12dam by \citet{Vreeswijk17} up to 334 rest-frame days after peak. The bolometric light curve of PTF~12dam was constructed from the observed spectral series and blackbody models of the UV/optical data. More details on this derivation are explained in \citet{Vreeswijk17}. We apply the bolometric correction from absolute magnitudes to the bolometric luminosity of PTF~12dam to all SLSNe in our sample, i.e. by basically adding a constant to the rest-frame $g$-band absolute magnitudes, where this constant evolves with the SN phase. This is valid under the assumption of spectral similarity among SLSNe-I. While there are strong indications for such similarity in our sample \citep{Quimby18}, this is not always guaranteed, especially at late times when the spectral coverage is typically poorer than around peak. A solid case-by-case bolometric correction can in principle only be attempted for the few best observed cases with sufficient spectral coverage. Due to the paucity of such complete datasets, this cannot be done for the full sample and is beyond the scope of the current paper. Nevertheless, given the overall similarity among the spectra in our sample, it is still informative, as a first approximation for the study of the energetics, to use a simple bolometric correction to derive the bolometric luminosities. Because the relative shapes of the light curves does not change between different SLSNe in bolometric luminosity, we do not show the individual light curves. We report the bolometric luminosities at peak in Table~\ref{table energy all}. The total radiated energy is then derived by integrating the bolometric light curves. Because the bolometric light curves are defined over a limited time interval, the derived radiated energies are lower limits.

\subsection{Ni masses} 
\label{sec Ni masses}

We investigate whether the peaks and light curve decays of SLSNe-I could be powered by Ni decay, using two independent methods. First, we derive a very rough estimate of what the required nickel masses would be if the SN peaks were completely powered by nickel. We use the relation $L_{\rm peak} = \alpha E_{\rm Ni} = \left( 6.45\times10^{43} e^{-(t_{\rm peak}/\tau_{\rm Ni})} + 1.43\times10^{43} e^{-(t_{\rm peak}/\tau_{\rm Co})}\right)  \times M_{\rm Ni}/M_\odot $ \citep{Nadyozhin94,Stritzinger05}, where $\tau_{\rm Ni}=8.8$~days and $\tau_{\rm Co}=111.3$~days, and assume no deviation from the Arnett rule \citep[$\alpha=1$,][]{Arnett79}. The time of explosion is quite uncertain in our sample, because the rise times are often not well-covered.\footnote{In most cases, the SN empirical-model fit of \citet{Bazin11} does not provide satisfactory results.} Thus, we use a representative $t_{\rm peak}=70$~days. For a few SLSNe-I that show indications for a longer rise time, we assume an explosion time of 100~days before peak (namely for PTF~10nmn, PTF~11hrq, and PTF~13dcc), and we assume the literature explosion time of 66 days before the peak for PTF 12dam \citep{Vreeswijk17}. The uncertainties in this Ni mass calculation are of about 20\% for an uncertainty in explosion date of about 30\%. The Ni mass that we derive with this method for PTF~12dam is similar to what has been derived by \citet{Vreeswijk17} with a more detailed Ni decay model of the full light curve. In this exercise, the main assumption is that the light curves are totally powered by radioactive Ni decay, while in fact there may be a significant contribution from CSM interaction, magnetars, or other sources. The derived nickel masses are therefore upper limits of the true values, for a peak time of 70~days after explosion. The results are reported in Table~\ref{table energy all}. 

\begin{figure*}
\epsscale{1.2}
\plotone{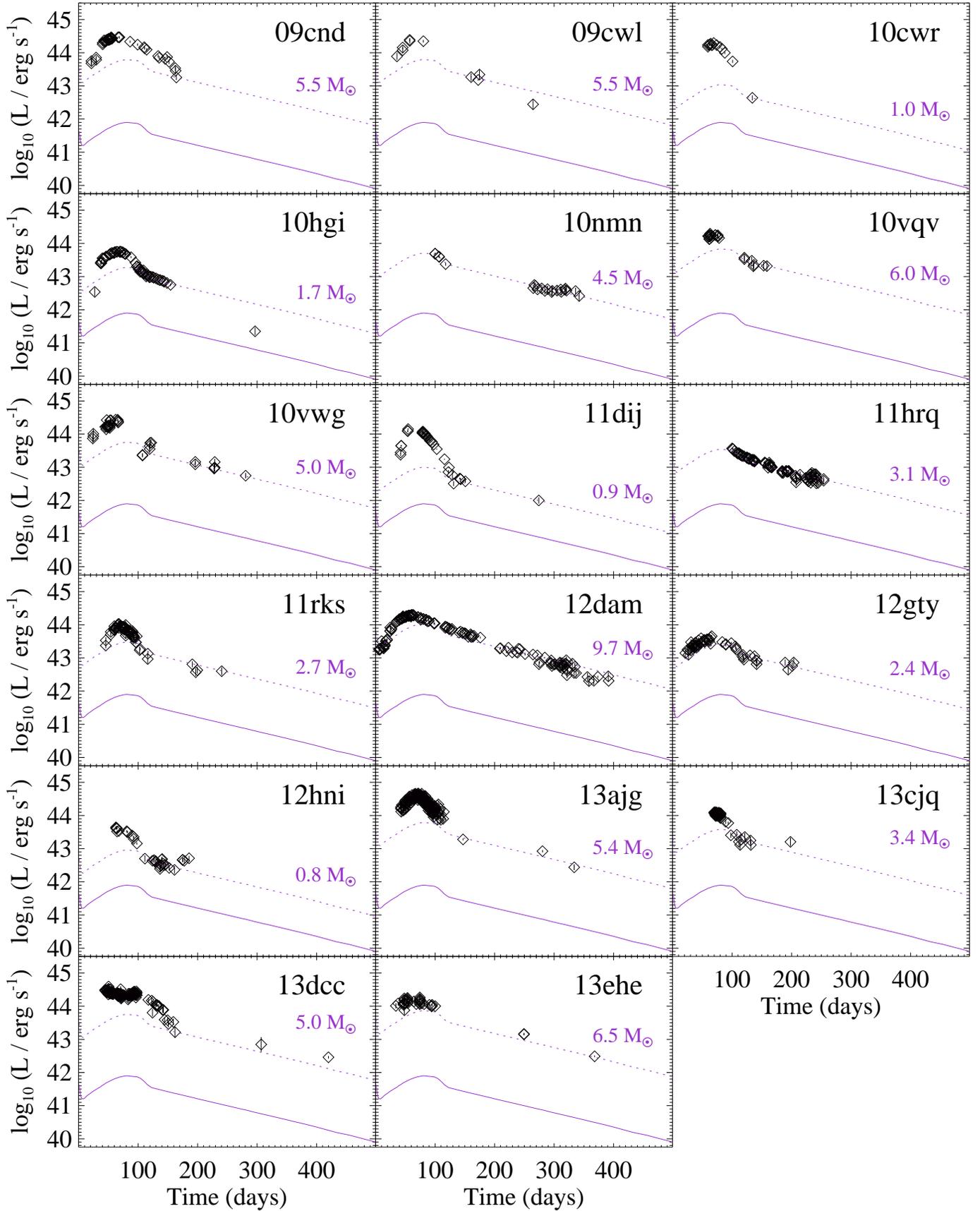}
\caption{Bolometric light curve of SN~1987A (solid curve) is scaled (dashed curves) to match the decay of SLSNe-I (black diamonds) at late times, after bolometric correction. The $M_{\rm Ni}$ roughly estimated from this comparison are labeled. The time reference is in rest-frame days after explosion. \label{fig MNi from decay}}
\end{figure*}

Second, we derive what the required nickel masses would roughly be if the SN late-time decay would be completely powered by nickel. In this case, we compare the SLSN decay, if sufficient photometry is available, to the decay of SN 1987A. For this SN, the well-studied decay is thought to be powered by the radioactive decay of 0.07 ${M}_\odot$ of $^{56}\rm Ni$ \citep{Fransson02,Seitenzahl14}. To compare the SLSNe light curves with SN~1987A, we assume an explosion date for the SLSNe and shift the bolometric light curves to that of SN 1987A \citep[taken from][]{Pun95}, using the same method as \citet{Gal-Yam09}. We shifted the light curve of SN~1987A to match the potential transition from the diffusive phase to the radioactive decay in the SLSN light curves or to the late-time decay. The explosion dates are quite uncertain. We assume the same explosion dates as discussed above. While the assumption on the explosion dates are not secure, here we are only interested in a zero-order estimate of the Ni masses from the tails. Figure~\ref{fig MNi from decay} shows the comparison between the SLSN light curves and SN~1987A. The Ni masses derived from the light-curve decay are labeled on the figure and reported in Table~\ref{table energy all}. These masses are upper limits, because they are calculated assuming that the late-time light curves are powered only by radioactive Ni, with no other contribution. The typical uncertainties on these nickel masses are large, roughly of the order of 50\% (accounting for a shift in explosion date of up to 100 days). Despite the large uncertainties, these estimates are useful for the comparison with the nickel masses derived from the peaks. 

In Table~\ref{table energy all}, we compare the nickel masses that we derived above with the two methods, one based purely on the peak luminosity and one based purely on the late-time decay. It is evident that the nickel masses derived from the peak luminosities are much higher than the nickel masses derived from the SN decay, both for fast- and slow-evolving SLSNe-I. This suggests that the SLSN peaks are not powered by nickel. This confirms the results of \citet{Inserra13}. In addition, the large ejecta masses required for powering the SLSN-I peaks with Ni radioactivity would increase the diffusion times, and therefore the light curves would show broader peaks than what is observed. A different component, such as magnetar spin-down or CSM interaction, is likely causing the high peak luminosities of most SLSNe-I. A factor of 5 discrepancy between the Ni masses required for the peak and the late-time decay was also found for SN~1998bw, which was possible to reconcile only with asymmetry of the ejecta \citep{Dessart17}. Strong evidence for asphericity of this event, based on the spectra, was found by \citet{Mazzali01}, \citet{Maeda02}, and \citet{Maeda06}. If asymmetry is relevant for SLSNe-I as well, the discrepancy between the nickel masses derived from the peak and the late-time decay may be partly mitigated (perhaps by a factor of 5 for asymmetries levels similar to SN~1998bw). 

When data coverage is available, we observe a late-time decay of SLSNe-I, which is close to the radioactive decay of $^{56}$Co to stable $^{56}$Fe, as observed from the late-time decay slopes (Sect. \ref{sec decay}). The nickel masses derived from this late-time decay are between $\leq1$ and $\leq10\,{M}_\odot$. This suggests that while nickel production is not the main source powering the light curve peaks, a nickel component could be important, and perhaps dominant, at late times. While the derived Ni masses are upper limits, producing up to $10\,{M}_\odot$ of Ni is challenging in classical SN models. The PISN model can produce 1--10$\,{M}_\odot$ of nickel from progenitor stars with cores of 90--105$\,{M}_\odot$ \citep{Heger02}.

In the case of PTF~10hgi, the only data point at late times seems fainter than what would be predicted from the decay rate of $^{56}$Co (Fig.~\ref{fig MNi from decay}). The current $^{56}$Ni mass estimated from the light-curve decay is $M_{\rm Ni}\sim2\,{M}_\odot$. However, estimating the Ni mass directly by scaling the SN~1987A light curve to the fainter data point at late times would provide $M_{\rm Ni}\sim0.2\,{M}_\odot$.

\begin{table}
\begin{center}
\caption{Radiated energy and nickel mass estimates from the peak luminosity and the late-time decay.
\label{table energy all}}
\begin{tabular}{@{}l l l l l@{} }
\tableline\tableline

PTF ID & $\log L_{\rm bol, peak}$ & $\log E_{\rm rad}$ & $M_{\rm Ni, peak}$ & $M_{\rm Ni, decay}$  \\

        &       (erg s$^{-1}$)  &     (erg)   & [${\rm M}_\odot$] & [${\rm M}_\odot$] \\
\tableline
   09as &  43.4 &$\geq 49.3 $ &$\leq  3.4$ &--\\
  09atu &  44.3 &$\geq 51.2 $ &$\leq 28.3$ &--\\
  09cnd &  44.5 &$\geq 51.3 $ &$\leq 37.3$ & $\leq  5.5$\\
  09cwl &  44.4 &$\geq 51.2 $ &$\leq 33.7$ & $\leq  5.5$\\
 10aagc &  43.7 &$\geq 50.0 $ &$\leq  6.5$ &--\\
  10bfz &  44.0 &$\geq 50.4 $ &$\leq 12.1$ &--\\
  10bjp &  43.8 &$\geq 50.5 $ &$\leq  8.8$ &--\\
  10cwr &  44.3 &$\geq 50.8 $ &$\leq 23.5$ & $\leq  1.0$\\
  10hgi &  43.7 &$\geq 50.5 $ &$\leq  7.2$ & $\leq  1.7$\\
  10nmn &  43.7 &$\geq 50.6 $ &$\leq  8.3$ & $\leq  4.5$\\
  10uhf &  43.9 &$\geq 50.3 $ &$\leq  9.8$ &--\\
  10vqv &  44.2 &$\geq 50.8 $ &$\leq 21.4$ & $\leq  6.0$\\
  10vwg &  44.4 &$\geq 51.2 $ &$\leq 32.0$ & $\leq  5.0$\\
  11dij &  44.2 &$\geq 50.8 $ &$\leq 21.0$ & $\leq  0.9$\\
  11hrq &  43.6 &$\geq 50.2 $ &$\leq  6.2$ & $\leq  3.1$\\
  11rks &  43.9 &$\geq 50.6 $ &$\leq 10.7$ & $\leq  2.7$\\
  12dam &  44.3 &$\geq 51.2 $ &$\leq 24.6$ & $\leq  9.7$\\
  12gty &  43.7 &$\geq 50.5 $ &$\leq  5.9$ & $\leq  2.4$\\
  12hni &  43.6 &$\geq 50.1 $ &$\leq  4.8$ & $\leq  0.8$\\
  12mxx &  44.2 &$\geq 50.8 $ &$\leq 22.6$ &--\\
  13ajg &  44.6 &$\geq 51.2 $ &$\leq 49.5$ & $\leq  5.4$\\
  13bdl &  43.9 &$\geq 50.6 $ &$\leq  9.6$ &--\\
  13bjz &  44.0 &$\geq 50.0 $ &$\leq 12.1$ &--\\
  13cjq &  44.1 &$\geq 50.7 $ &$\leq 15.2$ & $\leq  3.4$\\
  13dcc &  44.5 &$\geq 51.3 $ &$\leq 48.6$ & $\leq  5.0$\\
  13ehe &  44.2 &$\geq 51.2 $ &$\leq 18.3$ & $\leq  6.5$\\

\tableline
\end{tabular}
\end{center}
\end{table}

\subsection{Radioactive decay}
\label{sec RDE}

\begin{figure*}
\epsscale{1.1}
\plotone{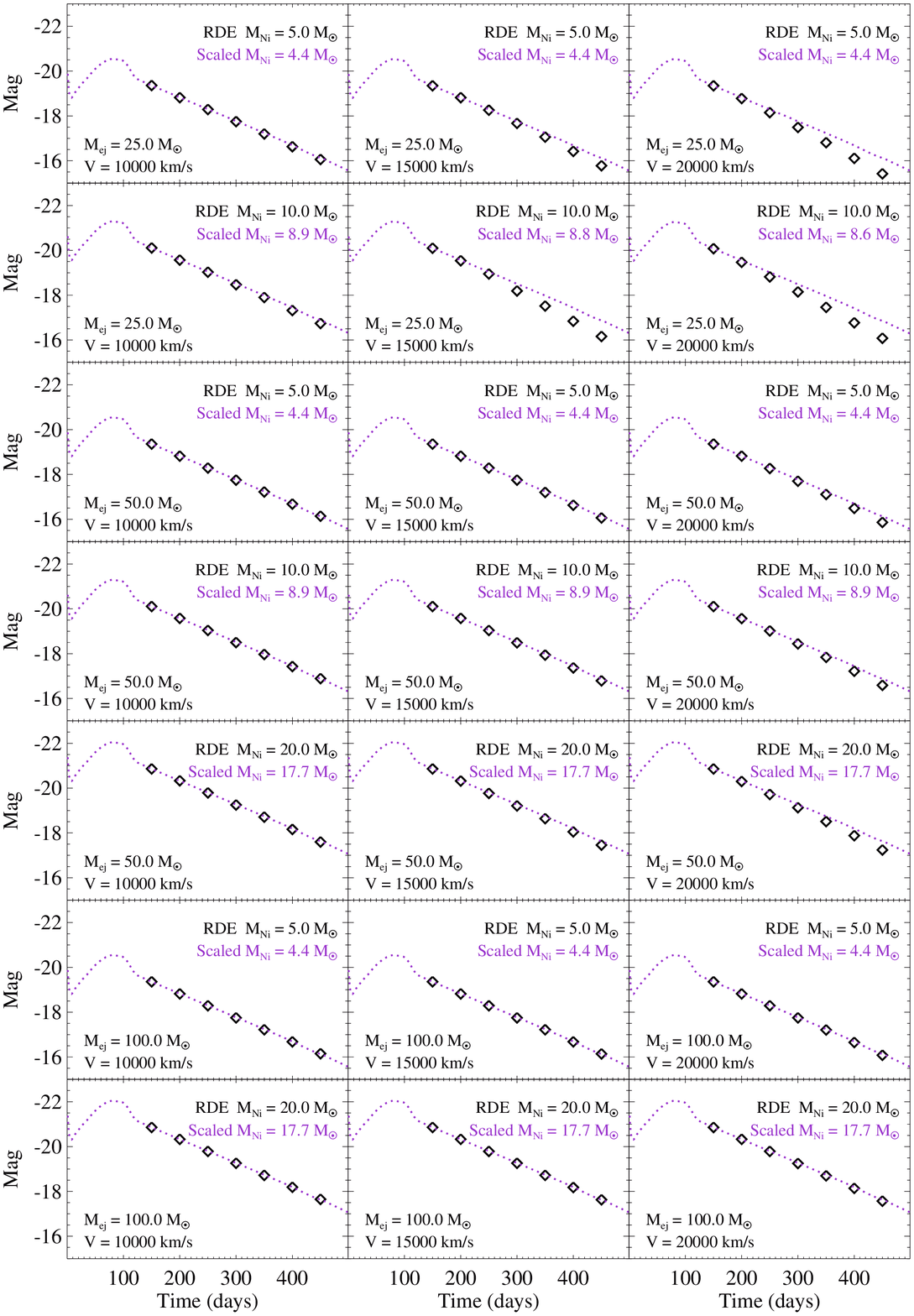}
\caption{Simulated light curves using RDE deposition (black diamonds) for stars with different initial parameters (black labels). A comparison with the SN~1987A light curve, scaled up to match the simulated magnitudes at 150 days, and the associated Ni masses are shown (purple label and dotted curve). Time is measured in rest-frame days after explosion.\label{fig RDE}}
\end{figure*}
One potential power source of SLSN-I light curves is radioactive decay of $^{56}$Co to stable $^{56}$Fe \citep[e.g.][]{Gal-Yam09}. The half-life time of the $^{56}$Co decay is 77.2 days \citep{Junde11}. As we discussed above, it is rather unlikely that the SLSN peaks are powered by radioactive decay because of the discrepancy between the Ni masses required by the peak luminosities and those required by the late-time decays. One possibility is that the late-time light curves are powered by radioactive decay. Indeed, we showed in Sect. \ref{sec decay} that whenever observable, the SLSN-I light curves tend to slow down, and at late times settle around the 0.01 mag days$^{-1}$ decay rate, which is typical of radioactive $^{56}$Co decay with full trapping. On the other hand, \citet{Inserra17} argued that SLSNe tend to decay faster than the radioactive rate, and therefore could not easily be associated with $^{56}$Co decay. However, the escape of $\gamma$-rays can increase the decay rate.  In this section, we investigate under which conditions $\gamma$-ray escape can efficiently induce a light-curve decline that is faster than the nominal radioactive decay rate.

The radioactive decay energy (RDE) deposition is the heating/excitation/ionization of the SN ejecta because of radioactive emission of $\gamma$-rays (and $e^+$) and the subsequent acceleration of electrons through Compton scattering \citep{Jeffery99}. This phenomenon is important for SNe Ia as well as for core-collapse SNe. After a diffusion phase when the $\gamma$-rays are fully trapped, a transition to a quasi-steady state marks the beginning of a regime where the decay is dominated by RDE deposition (and the diffusion timescale is much larger than the dynamical and radioactive timescales). At the transition point, the SN luminosity is purely determined by the total amount of radioactive material. The quasi-steady state decay is then exponential, starting with a radioactive slope that corresponds to full trapping. In time, the $\gamma$-rays can start to escape, and the decay can appear faster. We investigate here whether $\gamma$-ray escape is important for massive star progenitors.

We simulate the quasi-steady state decay from pure RDE deposition for stars with a density profile that has an inner plateau and decays exponentially \citep[similar to the 's25e12' profile of][]{Dessart11}, where the $^{56}$Ni mass is distributed in the inner ejecta. We consider total ejecta masses between 25 and 100 ${M}_\odot$, $^{56}$Ni masses between 5 and 20 ${M}_\odot$,  and maximum expansion velocity between 10,000 and 20,000 km~s$^{-1}$. The total ejecta mass and maximum expansion velocity determine the absolute value of the density profile at each point. These simulations cannot treat the diffusive phase, but only the light-curve decay beyond maximum light and beyond the transition to the quasi-steady state. As a sanity check, we reproduce the observed  radioactive light-curve phase of SN~1987A given an expansion velocity of 6000 km~s$^{-1}$, the same density profile as we used for SLSNe, total ejecta mass of 10 ${M}_\odot$, and $M(^{56}\rm{Ni})=0.07$ ${M}_\odot$.

Figure~\ref{fig RDE} shows the resulting light curves of our RDE deposition simulations for different initial parameters. In all cases, we observe $\sim$100\% gamma-ray trapping at $\sim$150 days after the explosion. At later times the luminosity can decrease more rapidly because of the reduced trapping due to lower densities. This effect is stronger for high expansion velocities and high $M_{Ni}/M_{ej}$ ejected masses. Within this set of simulations, the deviation from a pure radioactive $^{56}$Co decay ranges from 0.01 (still fully trapped) to a maximum of 0.27 mag ($\sim50$\% escape fraction) at about 450 days after explosion, with a decay rate of 1.33 mag in 100 days.

\begin{figure*}
\epsscale{1.0}
\plotone{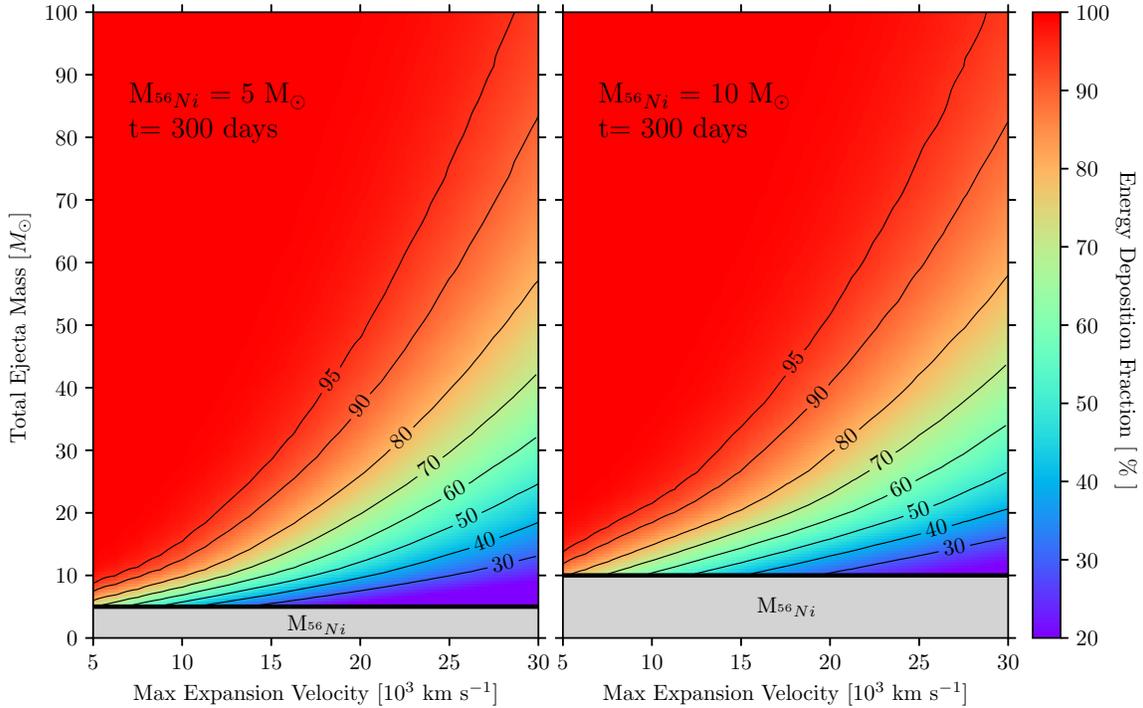}
\caption{Contours show the energy deposition fraction (100\% means full trapping) for different ejecta velocities and total ejecta masses, and in the cases of 5 and $10\,{M}_\odot$ of nickel (left and right panels, respectively). The color scale displays an interpolation of the contours.\label{fig RDE contours}}
\end{figure*}

The contours in Figure~\ref{fig RDE contours} show the energy deposition fraction (where 100\% means full trapping) from our RDE deposition simulations for the cases of 5 and 10 ${M}_\odot$ of nickel. These results confirm that massive star progenitors, with high SN ejecta velocities and high Ni fraction in the ejecta, have limited trapping and therefore can decay faster than the radioactive exponential decline. Figure~\ref{fig RDE contours} can also be used to roughly estimate the total ejecta masses in case the expansion velocity and the trapping are known from the spectra and light curves, respectively. 

An additional factor that can efficiently limit the trapping (and induce faster decays) is the potential mixing of $^{56}$Ni in the outer layers. While this can be a dominant effect, we do not attempt to model it here, because this is highly dependent on the geometry of the mixing, which cannot be constrained. 

\subsection{Magnetar modeling}
\label{sec magnetar}

We investigate whether the SLSN-I light curves could be powered by the spin-down of a magnetar. We consider an analytic magnetar model sourced from \citet{Arnett82} and \citet{Kasen10}. The fitting technique is described in detail in Rubin et al. (2018, in preparation). The main model parameters are the initial pulsar spin $P$, the magnetic field $B$, the diffusion timescale $t_m \approxeq M_{\rm ej}^{3/4} E_k^{−1/4}$ (where $M_{\rm ej}$ and $E_k$ are the ejected mass and kinetic energy, respectively), and the explosion time $t_{\rm exp}$. This is a basic modeling, that includes neither photon escape nor multiple components. The uncertainties are derived with a Monte Carlo treatment and shown in Figs. \ref{fig magnetar errors 09cnd} to \ref{fig magnetar errors 13ajg}. 
The treatment of the opacity is the same as in \citet{Inserra13}.

\begin{figure}
\epsscale{1.2}
\plotone{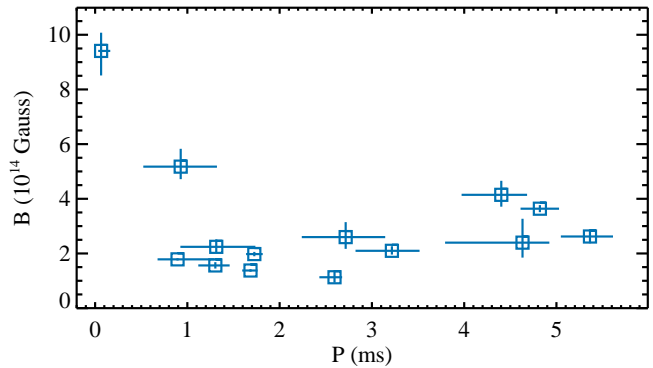}
\caption{Magnetic field strength and magnetar initial spin, as listed in Table~\ref{table magnetar}.\label{fig B14 spin}}
\end{figure}

Figures \ref{fig magnetar 1} and \ref{fig magnetar 2} show a fit of the magnetar model described above to the bolometric SLSN-I light curves. The best-fit parameters are reported in Table~\ref{table magnetar}, and displayed in Fig.~\ref{fig B14 spin}. In most cases, we obtain a satisfactory overall description of the light curves, with the exception of PTF~10hgi, PTF~10vwg, and PTF~11rks, for which we observe a different decay than predicted from our magnetar fit. In addition, the magnetar model does not describe well the light curve of PTF~11dij, both for the late-time decay and the early rise.\footnote{Forcing the explosion date to be before $-28$ days improves the fit at late times, but cannot well explain the data around peak and at $\sim50$ days after peak.} The confidence levels of the best-fit parameters are shown in Figs. \ref{fig magnetar errors 09cnd} to \ref{fig magnetar errors 13ajg}.

Recently, \citet{Nicholl17} has modeled a large literature SLSN-I sample with a magnetar model, including several published (i)PTF objects. The spin values $P$ that we obtain are mostly consistent with the values of \citet{Nicholl17}. For PTF~12dam, we find a spin value $P=1.87^{+0.07}_{-0.08}$~ms, which is a bit lower than the $\sim2.3$~ms values that were derived by \citet{Nicholl17} and \citet{Vreeswijk17}, and also lower than that by \citet{Chen15}. The magnetic fields $B$ that we obtain are in all cases higher than what was derived by \citet{Nicholl17}.
 For PTF~13ajg, our $B$ value is more similar to that found by \citet{Vreeswijk14}. For PTF~12dam, our $B$ value is more similar to those found by \citet{Chen15} and \citet{Vreeswijk17}. 

\begin{figure}
\epsscale{1.2}
\plotone{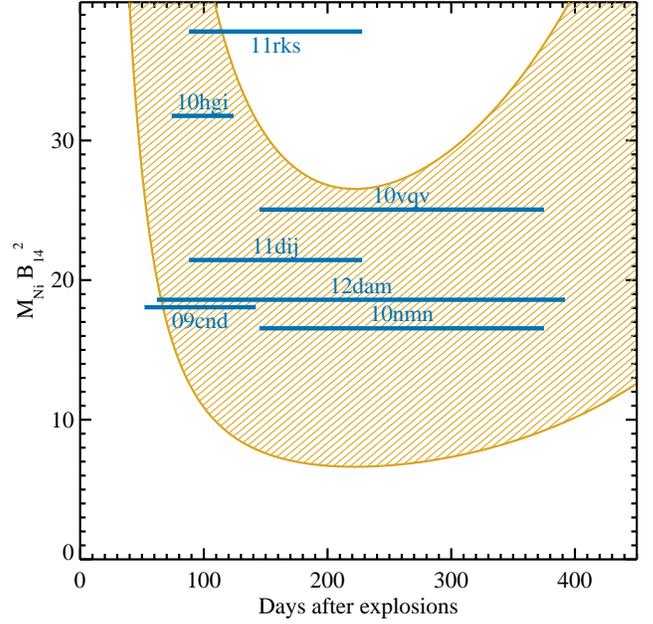}
\caption{Parameter space where a magnetar (in the dipole case) can mimic the radioactive $^{56}$Co decay \citep[shaded area;][]{Moriya17}. The horizontal lines mark the epochs when SLSNe are observed to decay consistently with the radioactive decay.\label{fig NiB2}}
\end{figure}

The late-time luminosities expected from the spin-down of a magnetar decline as $t^{-2}$ \citep[e.g.][]{Woosley10}. In principle, such model may also be able to mimic the $^{56}$Co decay of about 1 mag per 100 days, which we observe in SLSN-I late-time light curves. In the case of pure dipole radiation, the magnetar light curves start to have the same decay rate around 200 days postpeak. At later times, e.g. 400 days postpeak, the magnetar model is expected to have a decay rate that is noticeably slower than from the 1 mag/100 days \citep[e.g.][]{Inserra13}. However, the capability of a magnetar to mimic a radioactive decay would require a pure dipole radiation and a narrow set of fine-tuned parameters, in particular for the magnetic field and Ni masses \citep{Moriya17}. In Fig.~\ref{fig NiB2} we display the space parameter where a magnetar (in the dipole case) can mimic the radioactive $^{56}$Co decay from \citet{Moriya17}, and compare it with the results from our magnetar fit on the SLSN light curves. The reference time intervals are derived from the observed times after peak where the SLSN decay follows the radioactive rate (Fig.~\ref{fig MNi from decay}), and using the explosion times from the magnetar fits to the data (Figs. \ref{fig magnetar 1} and \ref{fig magnetar 2}). The magnetic field $B$ is taken from the magnetar fit, while the Ni masses are taken from scaling the late-time light-curve decays (Sect. \ref{sec Ni masses}). The SLSNe for which these measurements are available are lying in the parameter space where the magnetar decay mimics the radioactive decay of $^{56}$Co, with the marginal exception of PTF~11rks, for which a magnetar model does not describe the data well. Given that we do observe radiative-like light-curve decays at late times in SLSNe, it is perhaps not surprising that most of the derived magnetar parameters that we derive fulfill the radiative-mimicking criteria. These results indicate that we cannot disentangle between the magnetar and the radioactive decay models at these epochs, up to 400 days after explosion, but that observations at later times can be extremely powerful in disentangling between the two models.     

\begin{figure}
\epsscale{1.2}
\plotone{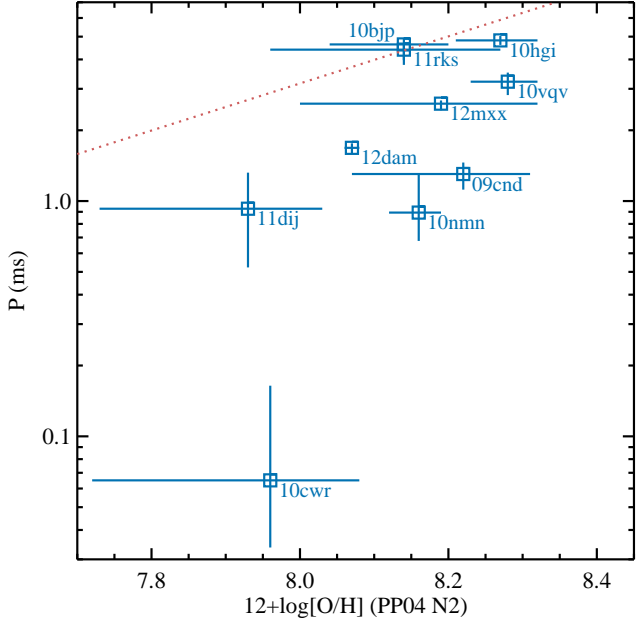}
\caption{Magnetar spin and host-galaxy metallicity. The red dotted line shows the spin-metallicity correlation found by \citet{Chen17a}.\label{fig spin met}}
\end{figure}

Intriguingly, a correlation between the magnetar initial spin and the host galaxy metallicity was found by \citet{Chen17a}. In Fig.~\ref{fig spin met} we show these quantities for our SLSN data using the host metallicities derived in \citet[][; PP04 N2 scale]{Perley16} and compare them to the correlation of \citet{Chen17a}. We cannot confirm the existence of such a correlation in the PTF sample, or at least we find a large scatter (more than a factor of 10) in the derived spin periods, for a similar metallicity range.

\begin{table}
\begin{center}
\caption{Resulting parameter from a magnetar fit to the bolometric light curves. Note. The last column lists the host-galaxy metallicity from \citet{Perley16}.\label{table magnetar}}
\begin{tabular}{@{}l@{\hspace{2mm}} c@{\hspace{2mm}} c@{\hspace{2mm}} c@{\hspace{2mm}} c@{\hspace{0mm}} c@{}} 
\tableline\tableline

ID   & $B$         & $P$ & $\tau_m$ & $t_{\rm exp}$  &  12 + log[O/H] \\

        &($10^{14}$ G) & (ms) & (days) & (days) &  (PP04 N2)\\
\tableline
 09cnd  & $1.56^{+0.12}_{-0.11}$ & $ 1.30^{+ 0.16}_{- 0.18}$ & $ 63.0^{+  3.2}_{-  3.0}$ & $ -45.0^{+   1.0}_{-   1.0}$ & $8.22 ^{+0.09 }_{-0.15}$\\
 09cwl  & $2.24^{+0.27}_{-0.20}$ & $ 1.31^{+ 0.42}_{- 0.39}$ & $ 57.7^{+  6.2}_{-  8.1}$ & $ -37.1^{+   4.4}_{-   2.2}$ & --                     \\
 10bjp  & $2.40^{+0.87}_{-0.55}$ & $ 4.63^{+ 0.29}_{- 0.84}$ & $ 38.3^{+  8.8}_{-  7.0}$ & $ -46.6^{+   3.0}_{-   2.6}$ & $8.14 ^{+0.06 }_{-0.10}$\\
 10cwr  & $9.41^{+0.67}_{-0.90}$ & $ 0.06^{+ 0.10}_{- 0.03}$ & $ 34.8^{+  0.9}_{-  1.0}$ & $ -12.6^{+   0.6}_{-   0.8}$ & $7.96 ^{+0.12 }_{-0.24}$\\
 10hgi  & $3.64^{+0.13}_{-0.13}$ & $ 4.82^{+ 0.21}_{- 0.21}$ & $ 39.0^{+  1.9}_{-  2.1}$ & $ -42.0^{+   1.1}_{-   1.3}$ & $8.27 ^{+0.05 }_{-0.06}$\\
 10nmn  & $1.78^{+0.04}_{-0.04}$ & $ 0.89^{+ 0.40}_{- 0.22}$ & $ 48.5^{+  2.4}_{-  1.9}$ & $-104.7^{+   0.4}_{-   1.0}$ & $8.16 ^{+0.03 }_{-0.04}$\\
 10vqv  & $2.10^{+0.17}_{-0.14}$ & $ 3.22^{+ 0.30}_{- 0.39}$ & $ 29.9^{+  6.1}_{-  4.9}$ & $ -28.5^{+   2.3}_{-   2.3}$ & $8.28 ^{+0.04 }_{-0.05}$\\
 10vwg  & $2.60^{+0.55}_{-0.43}$ & $ 2.72^{+ 0.43}_{- 0.48}$ & $ 29.1^{+  6.5}_{-  4.8}$ & $ -34.1^{+   2.8}_{-   7.1}$ & --                     \\
 11dij  & $5.18^{+0.65}_{-0.46}$ & $ 0.93^{+ 0.39}_{- 0.41}$ & $ 39.2^{+  1.4}_{-  1.3}$ & $ -27.6^{+   0.4}_{-   0.8}$ & $7.93 ^{+0.10 }_{-0.20}$\\
 11rks  & $4.15^{+0.51}_{-0.43}$ & $ 4.40^{+ 0.28}_{- 0.43}$ & $ 26.7^{+  1.7}_{-  1.7}$ & $ -22.8^{+   1.1}_{-   1.3}$ & $8.14 ^{+0.13 }_{-0.18}$\\
 12dam  & $1.38^{+0.05}_{-0.05}$ & $ 1.68^{+ 0.08}_{- 0.09}$ & $ 76.8^{+  2.1}_{-  2.2}$ & $ -56.0^{+   0.8}_{-   0.8}$ & $8.07 ^{+0.01 }_{-0.01}$\\
 12gty  & $2.62^{+0.27}_{-0.25}$ & $ 5.36^{+ 0.25}_{- 0.32}$ & $ 54.0^{+  4.1}_{-  4.2}$ & $ -58.5^{+   2.0}_{-   2.3}$ & --                     \\
 12mxx  & $1.13^{+0.26}_{-0.19}$ & $ 2.59^{+ 0.08}_{- 0.16}$ & $ 44.3^{+  5.0}_{-  4.2}$ & $ -48.4^{+   2.1}_{-   1.9}$ & $8.19 ^{+0.13 }_{-0.19}$\\
 13ajg  & $1.98^{+0.08}_{-0.08}$ & $ 1.73^{+ 0.09}_{- 0.09}$ & $ 33.8^{+  1.3}_{-  1.3}$ & $ -28.7^{+   1.4}_{-   0.8}$ & --                     \\
\tableline
\end{tabular}
\end{center}
\end{table}

\subsection{Cosmology tests}
\label{sec cosmo}
Because SLSNe can be observed out to large distances, as far out to $z\sim 4$ \citep{Cooke12} and likely well beyond with future facilities, the prospects of using SLSNe for cosmological distance determinations is of primary interest. Indeed, \cite{Inserra14} suggested that SLSNe-I may be standardizable, based on \textit{i)} the narrowness of the peak magnitude distribution, \textit{ii)} a weak correlation between the peak magnitude (at rest-frame 400~nm) and its decay after a certain time, and \textit{iii)} the dependence of the peak magnitude (at rest-frame 400~nm) on the SLSN-I color (rest-frame 400--520~nm). We test similar correlations in the current sample.

\textit{i)} We use the rest-frame $g$ band as a proxy for the rest-frame 400 nm band of \citet{Inserra14}. The peak magnitudes of the SLSNe in our sample are widely distributed around their mean value ($<M_g> = -21.14$ mag), with a standard deviation of 0.75 mag, which is almost twice as in the sample of \cite{Inserra14}.

\begin{figure*}
\epsscale{1.0}
\plotone{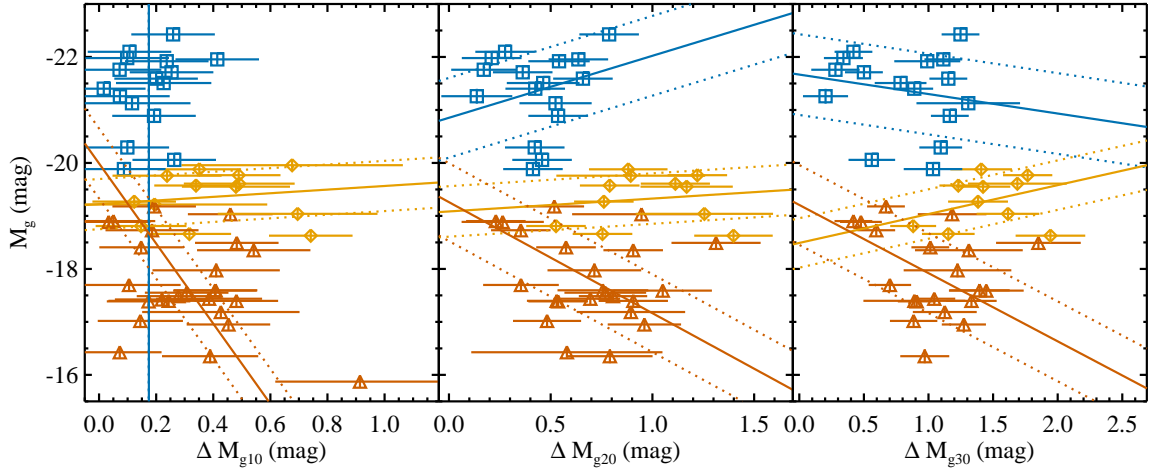}
\caption{Rest-frame $g$ magnitude at peak and its decay from the peak after 10 (left), 20 (middle), and 30 days (right panel). Blue squares show SLSNe-I, gold diamonds show SNe Ic-BL, and orange triangles SNe Ic, all from PTF. The solid and dotted lines show linear fits through the data and the intrinsic scatter, respectively. \label{fig Mg vs DeltaMg}}
\end{figure*}

\textit{ii)} Figure~\ref{fig Mg vs DeltaMg} shows the distribution of the rest-frame $g$ peak magnitude, $M_{g,\rm{peak}}$, with its decay, $\Delta M_g$, at 10, 20, and 30 days after the peak. These were all calculated from the smoothed light curves (Sect. \ref{sec smooth}) and can therefore be slightly different from the $M_{g,\rm{peak}}$ calculated with the second-order polynomial around the peak (Sect. \ref{sec peak mag}). As a comparison, SLSNe-I and also Type Ic and Ic-BL SNe from PTF are shown in Fig.~\ref{fig Mg vs DeltaMg}. There are very weak correlations, highlighted by the linear fits to the data. However, none of these trends are significant correlations. In every case, the Pearson correlation coefficient is $\mid r \mid <0.3$, and the intrinsic scatter is up to $\sim0.8$~mag for SLSNe-I; see Fig.~\ref{fig Mg vs DeltaMg}. The intrinsic scatter of the correlation is the scatter required for the correlation to have $\chi^2\sim1$ \citep{Bedregal06,Williams10}, and it is a way of discriminating the observational scatter from what is intrinsic to the correlation. The data and fit results for SLSNe-I are reported in Tables \ref{table Mg vs DeltaMg} and \ref{table fit Mg vs DeltaMg}, respectively.\\

\begin{table}
\begin{center}
\caption{Rest-frame $g$ magnitude at peak and its decay from the peak after 10, 20, and 30 days, calculated from the smoothed light curves (Sect. \ref{sec smooth}).\label{table Mg vs DeltaMg}}
\begin{tabular}{@{}l l l l l@{} }
\tableline\tableline
PTF ID &  $M_{g,\rm{peak}}$ & $\Delta M_{g10}$ & $\Delta M_{g20}$ &$\Delta M_{g30}$ \\
\tableline
 09atu  & $-21.8\pm  0.1$ & $  0.1\pm  0.1$ & $  0.2\pm  0.2$ & $  0.3\pm  0.2 $ \\
 09cnd  & $-22.1\pm  0.1$ & $  0.1\pm  0.1$ & $  0.3\pm  0.1$ & $  0.4\pm  0.1 $ \\
 09cwl  & $-22.0\pm  0.1$ & $  0.1\pm  0.1$ & $  0.2\pm  0.1$ & $  0.3\pm  0.2 $ \\
10aagc  & $-20.1\pm  0.1$ & $  0.7\pm  0.5$ & $  1.2\pm  0.5$ & $  2.0\pm  0.6 $ \\
 10cwr  & $-21.6\pm  0.1$ & $  0.2\pm  0.1$ & $  0.7\pm  0.1$ & $  1.2\pm  0.1 $ \\
 10hgi  & $-20.3\pm  0.1$ & $  0.1\pm  0.1$ & $  0.4\pm  0.1$ & $  1.1\pm  0.2 $ \\
 10vqv  & $-21.5\pm  0.1$ & $  0.2\pm  0.2$ & $  0.5\pm  0.2$ & $  0.8\pm  0.2 $ \\
 10vwg  & $-21.9\pm  0.1$ & $  0.2\pm  0.1$ & $  0.5\pm  0.2$ & $  1.0\pm  0.3 $ \\
 11dij  & $-21.4\pm  0.1$ & $  0.0\pm  0.1$ & $  0.4\pm  0.1$ & $  0.9\pm  0.1 $ \\
 11rks  & $-20.9\pm  0.1$ & $  0.2\pm  0.1$ & $  0.5\pm  0.1$ & $  1.2\pm  0.1 $ \\
 12dam  & $-21.7\pm  0.1$ & $  0.3\pm  0.1$ & $  0.4\pm  0.1$ & $  0.5\pm  0.1 $ \\
 12gty  & $-20.1\pm  0.1$ & $  0.3\pm  0.1$ & $  0.5\pm  0.1$ & $  0.6\pm  0.2 $ \\
 12hni  & $-19.9\pm  0.1$ & $  0.1\pm  0.1$ & $  0.4\pm  0.1$ & $  1.0\pm  0.2 $ \\
 13ajg  & $-22.4\pm  0.1$ & $  0.3\pm  0.1$ & $  0.8\pm  0.1$ & $  1.2\pm  0.1 $ \\
 13cjq  & $-21.1\pm  0.1$ & $  0.1\pm  0.2$ & $  0.5\pm  0.2$ & $  1.3\pm  0.4 $ \\
 13dcc  & $-22.0\pm  0.1$ & $  0.4\pm  0.1$ & $  0.6\pm  0.1$ & $  1.1\pm  0.1 $ \\
 13ehe  & $-21.3\pm  0.1$ & $  0.1\pm  0.2$ & $  0.1\pm  0.2$ & $  0.2\pm  0.2 $ \\

 \tableline
\end{tabular}
\end{center}
\end{table}

\begin{table}
\begin{center}
\caption{Fit parameters of the $M_{g,\rm{peak}}= A + B \times \Delta M_{g}$ relation for SLSNe at different epochs (days after peak, see Fig.~\ref{fig Mg vs DeltaMg} and Table~\ref{table Mg vs DeltaMg}). Note. $\sigma_{\rm int}$ is the intrinsic scatter (see text). $r$ and $\rho$ are the Pearson and Spearman correlation coefficients, respectively, and are listed with their respective null probability ($p_r$ and $p_\rho$). \label{table fit Mg vs DeltaMg}}
\begin{tabular}{@{}l l l l l l l l@{}}
\tableline\tableline
$t$ &  $A$   &  $B$  &  $\sigma_{\rm int}$ & $r$ & $p_r$ & $\rho$ & $p_\rho$ \\
\tableline
  10 & --            & --            & --    &$-0.24$&$ 0.37$&$-0.23$&$ 0.39$\\
  20 &$-20.9\pm  0.5$&$ -1.2\pm  1.1$&$ 0.75$&$-0.10$&$ 0.71$&$-0.11$&$ 0.69$\\
  30 &$-21.7\pm  0.5$&$  0.4\pm  0.5$&$ 0.76$&$ 0.14$&$ 0.60$&$ 0.15$&$ 0.57$\\

  \tableline
\end{tabular}
\end{center}
\end{table}

\begin{figure}
\epsscale{1.2}
\plotone{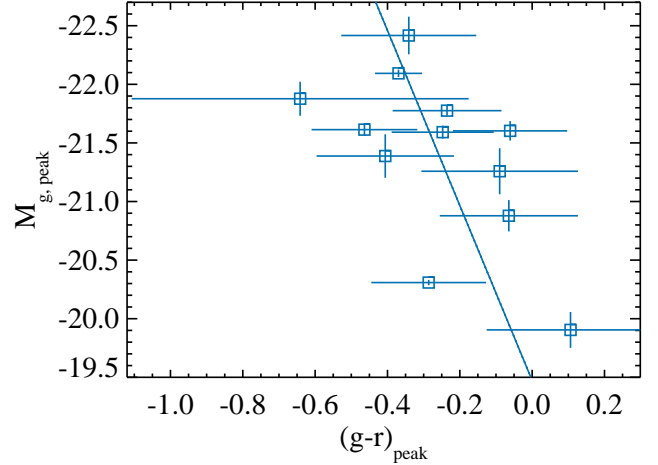}
\caption{Rest-frame $g$ peak magnitude versus the rest-frame $g-r$ color at peak, for PTF SLSNe. The solid line shows a linear fit to the data. The Pearson correlation coefficient is $r=0.6$. \label{fig Mg vs g-r}}
\end{figure}

\textit{iii)} In Fig.~\ref{fig Mg vs g-r} and Table~\ref{table Mg vs gr}, we compare the rest-frame $g$ peak magnitude with the rest-frame $g-r$ color at peak. The rest-frame $r$ was derived from the observed $i$-band photometry using the same techniques as described for the $r$ to rest-frame $g$ conversion. The $k$-correction values are listed in Table~\ref{table k-corr ir}. Both the rest-frame $g$- and $r$-band peaks were estimated by fitting a second-order polynomial to the data around peak. We could then constrain the rest-frame $g-r$ for a subsample of SLSNe-I, as listed in Table~\ref{table Mg vs gr}.
\begin{table}
\begin{center}
\caption{Rest-frame $g$ and rest-frame $g-r$ magnitudes at peak for the SLSNe in our sample (see Fig.~\ref{fig Mg vs g-r}).\label{table Mg vs gr}}
\begin{tabular}{@{}l  l l@{}}
\tableline\tableline
PTF ID &  $M_{g,\rm{peak}}$ & $g-r(\rm peak)$ \\
\tableline
 09atu & $-21.78\pm 0.04$ & $-0.24\pm 0.15 $ \\
 09cnd & $-22.09\pm 0.03$ & $-0.37\pm 0.07 $ \\
 10cwr & $-21.59\pm 0.06$ & $-0.25\pm 0.14 $ \\
 10hgi & $-20.31\pm 0.02$ & $-0.29\pm 0.16 $ \\
 11dij & $-21.39\pm 0.19$ & $-0.41\pm 0.19 $ \\
 11rks & $-20.88\pm 0.13$ & $-0.06\pm 0.19 $ \\
 12dam & $-21.61\pm 0.05$ & $-0.46\pm 0.15 $ \\
 12gty & $-19.90\pm 0.15$ & $ 0.11\pm 0.23 $ \\
 12mxx & $-21.60\pm 0.08$ & $-0.06\pm 0.16 $ \\
 13ajg & $-22.42\pm 0.16$ & $-0.34\pm 0.19 $ \\
 13dcc & $-21.88\pm 0.15$ & $-0.64\pm 0.47 $ \\
 13ehe & $-21.26\pm 0.20$ & $-0.09\pm 0.22 $ \\

  \tableline
\end{tabular}
\end{center}
\end{table}

The correlation $M_{g,\rm{peak}} = A + B \times (g-r)_{\rm peak}$ is weak, with a Pearson correlation coefficient of $0.59$ (null probability $p_r=0.04$) and a Spearman correlation coefficient of $0.57$ (null probability $p_\rho=0.04$). The normalization and slope are $A=-19.5\pm1.2$ and $B=7.5\pm3.8$. In this case, the intrinsic scatter is consistent with zero. While this may suggest a potentially real correlation, it may simply be the consequence of the large error bars that we measure for $g-r$. A larger statistical sample is needed to further investigate the solidity of this relation. 

The possible dust reddening $E(B-V)$ of the SLSN-I host galaxies is not taken into account here. While this should we expect the reddening to be small \citep[e.g.][]{Perley16}, there may be regions that are locally more dusty \citep[e.g.][]{Cikota17}. However, a case-to-case characterization of the host-galaxies' $E(B-V)$ is not possible here and is beyond the scope of this paper.

While we find the same overall trends as \citet{Inserra14}, the diagnostics above show mostly weaker correlations than reported by their work. One exception is the relation between the peak magnitude and $g-r$ at peak, for which we find a similar Pearson correlation coefficient to \citet{Inserra14}. 

We further investigate possible correlations of $M_g$ with other variables, such as the early decay slope or host metallicity, with no convincing results. The weak trends of $M_g$ with the rise and decay times are shown in Fig.~\ref{fig rise decay 1mag}. We therefore cannot strengthen the claim that SLSNe might be standardizable candles with the current data. Future transient surveys may clarify this issue with much improved statistics \citep[e.g.][]{Scovacricchi16}.

\section{Conclusions}
\label{sec conclusions}

We study a sample of 26 SLSNe-I, all discovered by the PTF survey with light-curve coverage out to late times, well beyond 100 days after peak for half of the sample. Based on our analysis, we conclude the following.

\begin{enumerate}

\item The absolute peak magnitudes of PTF spectroscopically classified SLSNe-I are $-22.5\lesssim M_g\lesssim-20$~mag (Sect. \ref{sec peak mag} and \ref{sec disc peak}). The mean SLSN-I peak magnitude is $-21.14$~mag, which is brighter than the mean magnitudes of SNe Ic-BL and SNe Ib/c by about 2 and 4 mag, respectively. When including volumetric corrections, the peak-magnitude distribution evolves smoothly from SNe Ib/c to SNe Ic-BL, and to SLSNe-I. There is only very little overlap between the faintest SLSNe-I and the brightest SNe Ic-BL. 

\item At early times ($< 60$ days after peak) SLSNe-I tend to decay faster (0.04 mag day$^{-1}$ on average) and with a wider range of decay rates than at late times ($>60$ days); see Sect. \ref{sec decay} and \ref{sec disc slopes}. 

\item At late times, all SLSN-I light curves for which sufficient data are available cluster around the decay rate of $\sim1$~mag per 100~days, which is consistent with the radioactive decay of $^{56}$Co to stable $^{56}$Fe (Sect. \ref{sec decay} and \ref{sec disc slopes}). 

\item We observe no gap between fast- and slow-declining SLSNe-I. Nevertheless, virtually all slow-declining events (SLSN-R) show early- and late-time bumps/plateau which are not as common in classical SLSNe-I. Thus, the possibility is still open that SLSN-I/R represent a subclass of SLSN-I (Sect. \ref{sec decay} and \ref{sec disc slopes}).

\item SLSNe-I rise more slowly than SNe Ib/c and Ic-BL (i.e. SLSNe have longer rise timescales $t^{\rm \Delta1mag}_{\rm rise}$ by roughly 10 days), even for similar decay times. Indeed, the rise times correlate differently with the decay times for SLSNe-I and SNe Ib/c (Sect. \ref{sec rise fall 1mag}, \ref{sec rise fall t12}, and \ref{sect disc rise decay times}).

\item This implies that the light curves of SLSNe-I are different from SNe Ib/c and Ic-BL, and therefore SLSNe-I can be considered as a separate population photometrically, as well as spectroscopically.

\item The peaks of SLSNe-I are not powered by the production of radioactive nickel, unless there are strong asymmetries in the ejecta (Sect. \ref{sec Ni masses}). 

\item Late-time light curves can be explained with the radioactive decay of Ni masses ranging from 1 to 10 ${M}_\odot$. Radioactive decay might be an important powering source at these stages (Sect. \ref{sec Ni masses}). 

\item The slope of the late-time decay is in a few SLSNe-I faster than the radioactive decay. This can be explained by the escape of $\gamma$-rays from the massive ejecta. Our simulations of the radiative decay energy deposition for massive progenitors shows that the trapping is reduced for higher expansion velocities and higher Ni fractions of the ejected masses (Sect. \ref{sec RDE}). 

\item The majority of the SLSN-I light curves can reasonably well be reproduced also with a spinning-down magnetar model, with the exceptions of PTF~10hgi, PTF~10vwg, PTF~11dij, and PTF~11rks (Sect. \ref{sec magnetar}). The derived magnetic fields lie in the parameter space where a magnetar model can mimic the radioactive decay of $^{56}$Co. 

\item We cannot distinguish between a radioactively powered and magnetar light curves at this stage. Very late-time observations are needed to disentangle between the magnetar and radioactive models.

\item We find no correlation between the magnetar spin and the host metallicity (Sect. \ref{sec magnetar}). 

\item We find similar correlations to those claimed to make SLSN-I standardizable candles \citep{Inserra14}; see Sect. \ref{sec cosmo}. These correlations are significantly weaker, except for the correlation between the rest-frame $g$-band peak magnitude with the rest-frame $g-r$ at peak. With the current data, we cannot strengthen the potential of exploiting SLSNe for cosmology.

\end{enumerate}

\acknowledgments
We thank the referee for a useful and constructive report, which helped make the paper more robust. We thank L. Dessart, A. Jerkstrand, A. Kozyreva, and the SLSN MIAPP 2017 workshop participants for insightful discussions. A.D.C. acknowledges support by the Weizmann Institute of Science Koshland Center for Basic Research. Support for I.A. was provided by NASA through the Einstein Fellowship Program, grant PF6-170148. M.S. acknowledges support from EU/FP7-ERC grant No. [615929]. K.M. acknowledges support from the STFC through an Ernest Rutherford Fellowship. E.O.O. is grateful for support by grants from the Israel Science Foundation, Minerva, Israeli ministry of Science, the US-Israel Binational Science Foundation and the I-CORE Program of the Planning and Budgeting Committee and The Israel Science Foundation. A.C. acknowledges support from the NSF CAREER award 1455090. M.M.K. acknowledges support from the GROWTH project funded by the National Science Foundation under Grant No. 1545949. The intermediate Palomar Transient Factory project is a scientific collaboration among the California Institute of Technology, Los Alamos National Laboratory, the University of Wisconsin, Milwaukee, the Oskar Klein Center, the Weizmann Institute of Science, the TANGO Program of the University System of Taiwan, and the Kavli Institute for the Physics and Mathematics of the Universe. LANL participation in iPTF is supported by the US Department of Energy as a part of the Laboratory Directed Research and Development program. A portion of this work was carried out at the Jet Propulsion Laboratory under a Research and Technology Development Grant, under contract with the National Aeronautics and Space Administration. This research has made use of the NASA/ IPAC Infrared Science Archive, which is operated by the Jet Propulsion Laboratory, California Institute of Technology, under contract with the National Aeronautics and Space Administration. This research has made use of NASA's Astrophysics Data System. This paper made use of Lowell Observatory's Discovery Channel Telescope (DCT). Lowell operates the DCT in partnership with Boston University, Northern Arizona University, the University of Maryland, and the University of Toledo. Partial support of the DCT was provided by Discovery Communications. Large Monolithic Imager (LMI) on DCT was built by Lowell Observatory using funds from the National Science Foundation (AST-1005313). This work makes use of observations taken with the LCO network. This research has made use of the NASA/IPAC Extragalactic Database (NED) which is operated by the Jet Propulsion Laboratory, California Institute of Technology, under contract with the National Aeronautics and Space Administration. Some of the data presented herein were obtained at the W.M. Keck Observatory, which is operated as a scientific partnership among the California Institute of Technology, the University of California and the National Aeronautics and Space Administration. The Observatory was made possible by the generous financial support of the W.M. Keck Foundation. The authors wish to recognize and acknowledge the very significant cultural role and reverence that the summit of Maunakea has always had within the indigenous Hawaiian community. We are most fortunate to have the opportunity to conduct observations from this mountain.

\bibliographystyle{apj} 

\bibliography{biblio}

\newpage

\appendix

\section{SLSN-I light-curve properties}

The light-curve properties of the (i)PTF SLSN-I sample that we derived above (i.e. peak magnitude, early- and late-time decay, rise and decay timescales) are reported in Table \ref{table derived}.

\vspace{3cm}
\floattable
\begin{deluxetable}{@{}l|@{\hspace{0.0mm}} c@{\hspace{0.0mm}} c@{\hspace{0.0mm}} c@{\hspace{0.0mm}} c@{\hspace{0.0mm}} c@{\hspace{0.0mm}} c@{\hspace{0.0mm}} c@{\hspace{0.0mm}} c@{\hspace{0.0mm}} c@{\hspace{0.0mm}} c@{\hspace{0.0mm}} c@{\hspace{0.0mm}} c@{}}
\rotate
\tablecaption{Light-curve properties of the H-poor SLSN sample \label{table derived}}
\tablewidth{0pt}
\tablehead{
\colhead{PTF ID}  &   \colhead{$MJD_{\rm peak}$}  &   \colhead{$M{g, \rm{peak}}$}  &   \colhead{$\chi^2_{\nu, \rm{peak}}/\nu$}  &   \colhead{Slope1$^a$}  &   \colhead{$\chi^2_{\nu, \rm{Slope1}}/\nu$}  &   \colhead{Slope2$^a$}  &   \colhead{$\chi^2_{\nu, \rm{Slope2}}/\nu$}  &   \colhead{Int$^b$}  & \colhead{$t^{\rm 1mag}_{\rm rise}$}  & \colhead{$t^{\rm 1mag}_{\rm fall}$} & \colhead{$t_{\rm rise, 1/2}$}  & \colhead{$t_{\rm fall, 1/2}$} \\
 &  &   \colhead{(mag)} &  & \colhead{(mag day$^{-1}$)} & & \colhead{(mag day$^{-1}$)} & & \colhead{(days)} & \colhead{(days)} & \colhead{(days)} & \colhead{(days)} & \colhead{(days)} \\    
\tabletypesize{\scriptsize}
}
\startdata 
               09as & $  54918.20$ &  $(-19.51\pm0.13)$ & $  0.30/   8$ & $  0.0830\pm0.0121$ & $  0.27/   9$ &             --           &    --    &   --     	   &               --     &          $ 11\pm  3$ &               --     &          $  9\pm  3$   \\
              09atu & $  55062.32$ &  $ -21.78\pm0.04 $ & $  0.20/  22$ & $  0.0137\pm0.0052$ & $  0.27/  13$ &             --           &    --    &   --     	   &          $ 37\pm  2$ &          $ 70\pm  8$ &          $ 32\pm  3$ &          $ 61\pm 11$   \\
              09cnd & $  55086.35$ &  $ -22.09\pm0.03 $ & $  0.17/  28$ & $  0.0226\pm0.0018$ & $  0.61/   8$ &             --           &    --    &   --     	   &          $ 33\pm  2$ &          $ 54\pm  6$ &          $ 29\pm  2$ &          $ 45\pm  4$   \\
              09cwl & $  55067.25$ &  $ -22.03\pm0.13 $ & $  0.44/   3$ & $  0.0320\pm0.0012$ & $  0.98/   2$ &             --           &    --    &   --     	   &          $ 25\pm  2$ &               --     &          $ 21\pm  2$ &          $ 54\pm  7$   \\
             10aagc & $  55499.48$ &  $ -20.12\pm0.25 $ & $  1.69/  28$ & $  0.0745\pm0.0042$ & $  1.16/  26$ &             --           &    --    &   --     	   &               --     &          $ 17\pm  9$ &               --     &          $ 11\pm  5$   \\
              10bfz & $  55227.46$ &  $(-20.86\pm0.03)$ & $  0.15/   5$ & $  0.0658\pm0.0026$ & $  0.85/  30$ &             --           &    --    &   --     	   &               --     &          $ 33\pm  4$ &               --     &          $ 28\pm  2$   \\
              10bjp & $  55252.52$ &  $ -20.52\pm0.16 $ & $  0.41/   9$ & $  0.0241\pm0.0154$ & $  1.73/   4$ &             --           &    --    &   --     	   &          $ 42\pm  2$ &               --     &               --     &               --       \\
              10cwr & $  55281.23$ &  $ -21.59\pm0.06 $ & $  0.46/  10$ & $  0.0675\pm0.0038$ & $  3.76/   2$ &             --           &    --    &   --     	   &               --     &          $ 28\pm  2$ &               --     &          $ 23\pm  2$   \\
     \textbf{10hgi} & $  55367.43$ &  $ -20.31\pm0.02 $ & $  0.09/  22$ & $  0.0519\pm0.0033$ & $  0.92/  12$ & $  0.0211\pm0.0011$ & $  0.93/  12$ & $   37.7$	   &          $ 32\pm  2$ &          $ 28\pm  2$ &          $ 29\pm  2$ &          $ 25\pm  2$   \\
              10nmn & $  55384.20$ &  $ -20.53\pm0.04 $ & $  0.02/   2$ &             --           &    --    & $  0.0080\pm0.0003$ & $  0.26/  16$ &   --     	   &               --     &          $ 61\pm 18$ &               --     &               --       \\
              10uhf & $  55452.25$ &  $ -20.60\pm0.22 $ & $  1.50/  12$ & $  0.0356\pm0.0072$ & $  1.48/   6$ &             --           &    --    &   --     	   &               --     &               --     &               --     &               --       \\
              10vqv & $  55470.52$ &  $ -21.58\pm0.11 $ & $  0.40/  14$ & $  0.0338\pm0.0043$ & $  0.12/   6$ & $  0.0186\pm0.0084$ & $  0.27/   5$ & $   55.6$	   &               --     &          $ 36\pm  5$ &               --     &          $ 29\pm  5$   \\
              10vwg & $  55455.29$ &  $ -21.94\pm0.20 $ & $  2.89/  19$ & $  0.0451\pm0.0019$ & $  9.48/   8$ & $  0.0119\pm0.0029$ & $  0.67/   5$ & $   43.0$	   &          $ 39\pm  2$ &               --     &          $ 33\pm  3$ &          $ 25\pm  4$   \\
     \textbf{11dij} & $  55684.37$ &  $ -21.39\pm0.19 $ & $  5.44/  22$ & $  0.0550\pm0.0005$ & $  4.92/  30$ & $  0.0141\pm0.0018$ & $  0.26/   3$ & $   74.5$	   &          $ 15\pm  2$ &          $ 31\pm  2$ &          $ 13\pm  2$ &          $ 27\pm  2$   \\
              11hrq & $  55753.48$ &  $(-19.80\pm0.04)$ & $  0.32/  23$ & $  0.0156\pm0.0006$ & $  0.45/  45$ & $  0.0163\pm0.0006$ & $  1.25/  58$ & $    7.4$	   &               --     &          $ 55\pm  6$ &               --     &          $ 37\pm  7$   \\
     \textbf{11rks} & $  55935.14$ &  $ -20.88\pm0.13 $ & $  1.84/  50$ & $  0.0501\pm0.0021$ & $  1.62/  30$ & $  0.0085\pm0.0072$ & $  1.39/   2$ & $   60.0$	   &          $ 17\pm  2$ &          $ 27\pm  2$ &          $ 15\pm  2$ &          $ 24\pm  2$   \\
     \textbf{12dam} & $  56092.33$ &  $ -21.61\pm0.05 $ & $  4.72/  45$ & $  0.0137\pm0.0002$ & $  4.84/  17$ & $  0.0178\pm0.0002$ & $  5.35/  52$ & $  113.4$	   &          $ 34\pm  2$ &          $ 62\pm  5$ &          $ 29\pm  2$ &          $ 49\pm  6$   \\
     \textbf{12gty} & $  56143.36$ &  $ -19.90\pm0.15 $ & $  0.80/  20$ & $  0.0277\pm0.0031$ & $  2.41/   9$ & $  0.0115\pm0.0028$ & $  0.88/   7$ & $   54.9$	   &          $ 36\pm  3$ &          $ 41\pm 11$ &          $ 28\pm  3$ &          $ 36\pm  5$   \\
              12hni & $  56154.25$ &  $ -19.92\pm0.10 $ & $  1.78/  11$ & $  0.0470\pm0.0016$ & $  1.21/  11$ & $ -0.0217\pm0.0071$ & $  0.52/   4$ & $   74.1$	   &               --     &          $ 29\pm  4$ &               --     &          $ 25\pm  2$   \\
              12mxx & $  56292.14$ &  $ -21.60\pm0.08 $ & $  0.93/  30$ &             --           &    --    &             --           &    --    &   --     	   &          $ 29\pm  2$ &               --     &          $ 26\pm  2$ &               --       \\
     \textbf{13ajg} & $  56410.35$ &  $ -22.42\pm0.16 $ & $  1.00/  69$ & $  0.0376\pm0.0033$ & $  1.86/  50$ & $  0.0096\pm0.0008$ & $ 10.32/   1$ & $   87.6$	   &          $ 24\pm  2$ &          $ 25\pm  2$ &          $ 21\pm  2$ &          $ 19\pm  2$   \\
              13bdl & $  56493.22$ &  $ -20.36\pm0.22 $ & $  0.66/  28$ &             --           &    --    &             --           &    --    &   --     	   &               --     &               --     &               --     &               --       \\
              13bjz & $  56438.17$ &  $ -20.81\pm0.18 $ & $  2.53/  15$ &             --           &    --    &             --           &    --    &   --     	   &               --     &               --     &          $  8\pm  2$ &               --       \\
              13cjq & $  56506.28$ &  $ -21.13\pm0.09 $ & $  0.66/  45$ & $  0.0409\pm0.0056$ & $  2.71/   7$ & $  0.0046\pm0.0052$ & $  0.79/   2$ & $   49.3$	   &               --     &          $ 25\pm  5$ &               --     &          $ 23\pm  3$   \\
              13dcc & $  56612.35$ &  $ -21.88\pm0.15 $ & $  0.59/  49$ & $  0.0476\pm0.0044$ & $  1.96/  15$ & $  0.0122\pm0.0007$ & $  0.88/   3$ & $   62.3$	   &               --     &          $ 28\pm  2$ &               --     &          $ 24\pm  3$   \\
              13ehe & $  56669.54$ &  $ -21.26\pm0.20 $ & $  0.92/  31$ & $  0.0062\pm0.0065$ & $  1.29/   4$ & $  0.0158\pm0.0007$ & $  0.14/   1$ & $   37.9$	   &               --     &          $114\pm  7$ &               --     &          $ 97\pm 10$   \\
\enddata
\tablecomments{SLSNe with best sampled data, where the light curves can be fully characterized in their rise and fall times, peaks, early- and late-time declines, are highlighted in bold. Peak magnitudes are in parenthesis for the cases where the peak is not sufficiently covered, so these represent magnitude upper limits. The peak magnitude for PTF~10nmn is taken from Yaron et al. (2018 in preparation), which includes a more complete coverage of the peak. In the case of PTF~12hni the late-time decay slope is negative because of the light curve rebrightening. The $\chi^2_\nu$ are sometimes large because the formal errors in the observed magnitudes within the fitted time interval are very small (e.g. between 0.006 and 0.024 mag around peak for PTF~12dam). $^a$ Slopes of the postpeak early- and late-time linear fit to the data (Sect. \ref{sec decay}). $^b$ Intersection between the postpeak early- and late-times linear fits to the data (in days after peak, Sect. \ref{sec decay}).}
\end{deluxetable}

\newpage

\section{Appendix Figures}

The figures below show the light curves of (i)PTF SLSNe-I in our sample, their rise and decay timescales in flux and magnitudes, the magnetar fits to the data, and their confidence levels.

\begin{figure*}[!ht]
\epsscale{1.1}
\plotone{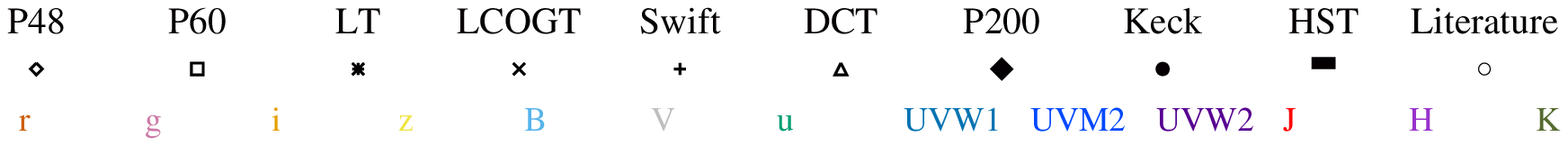}
\end{figure*}
\begin{figure*}[!hb]
\epsscale{1.0}
\plottwo{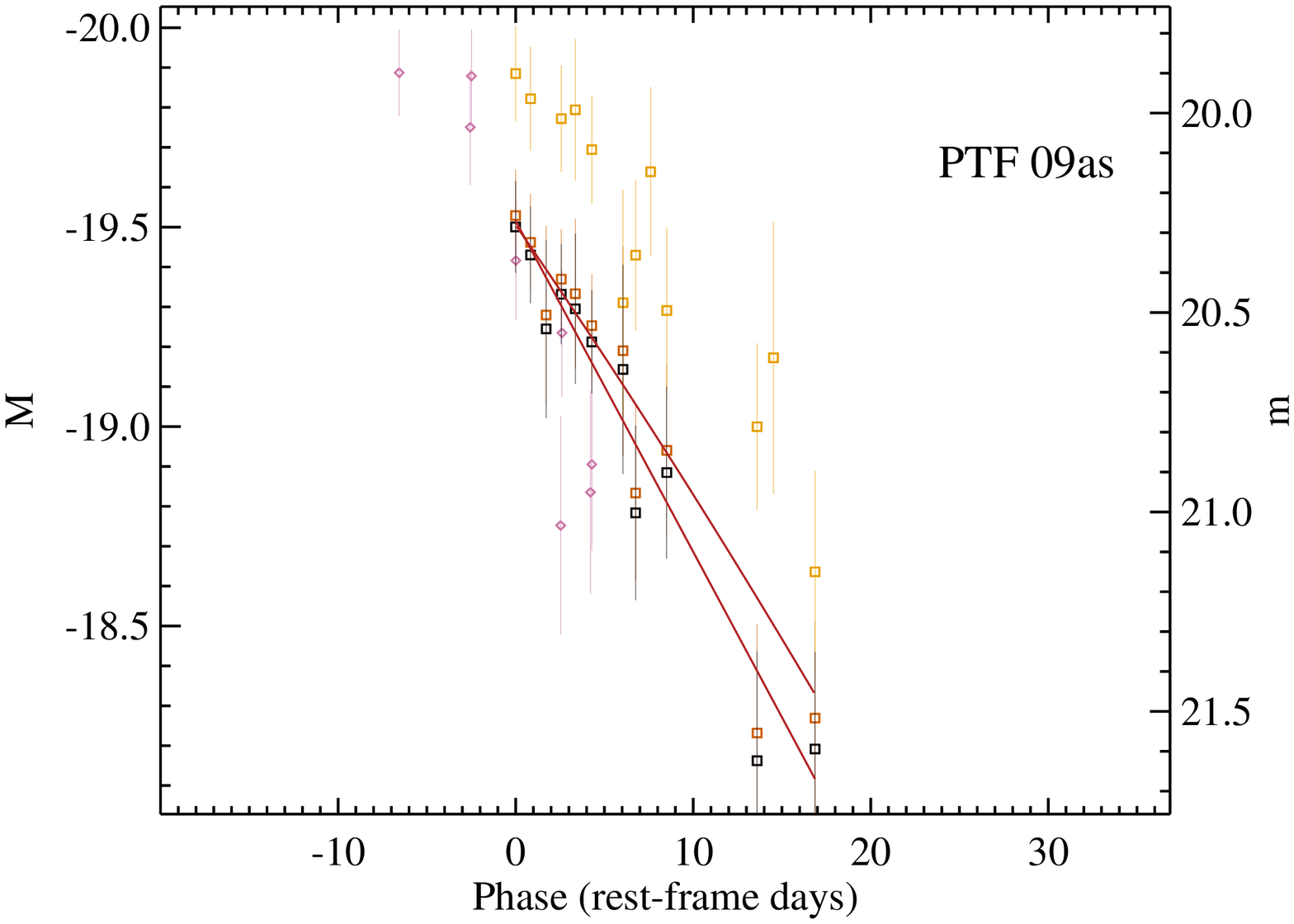}{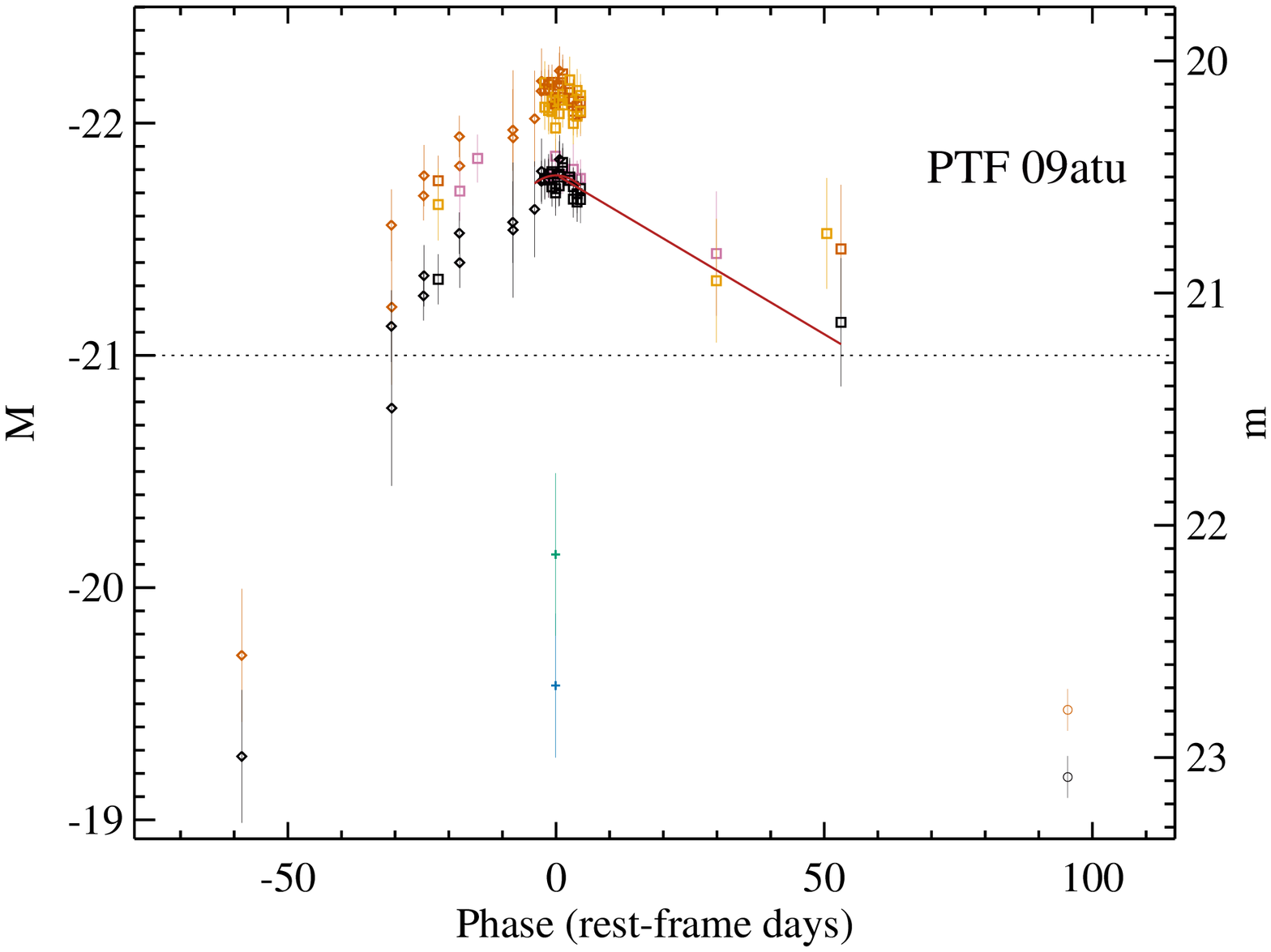}
\plottwo{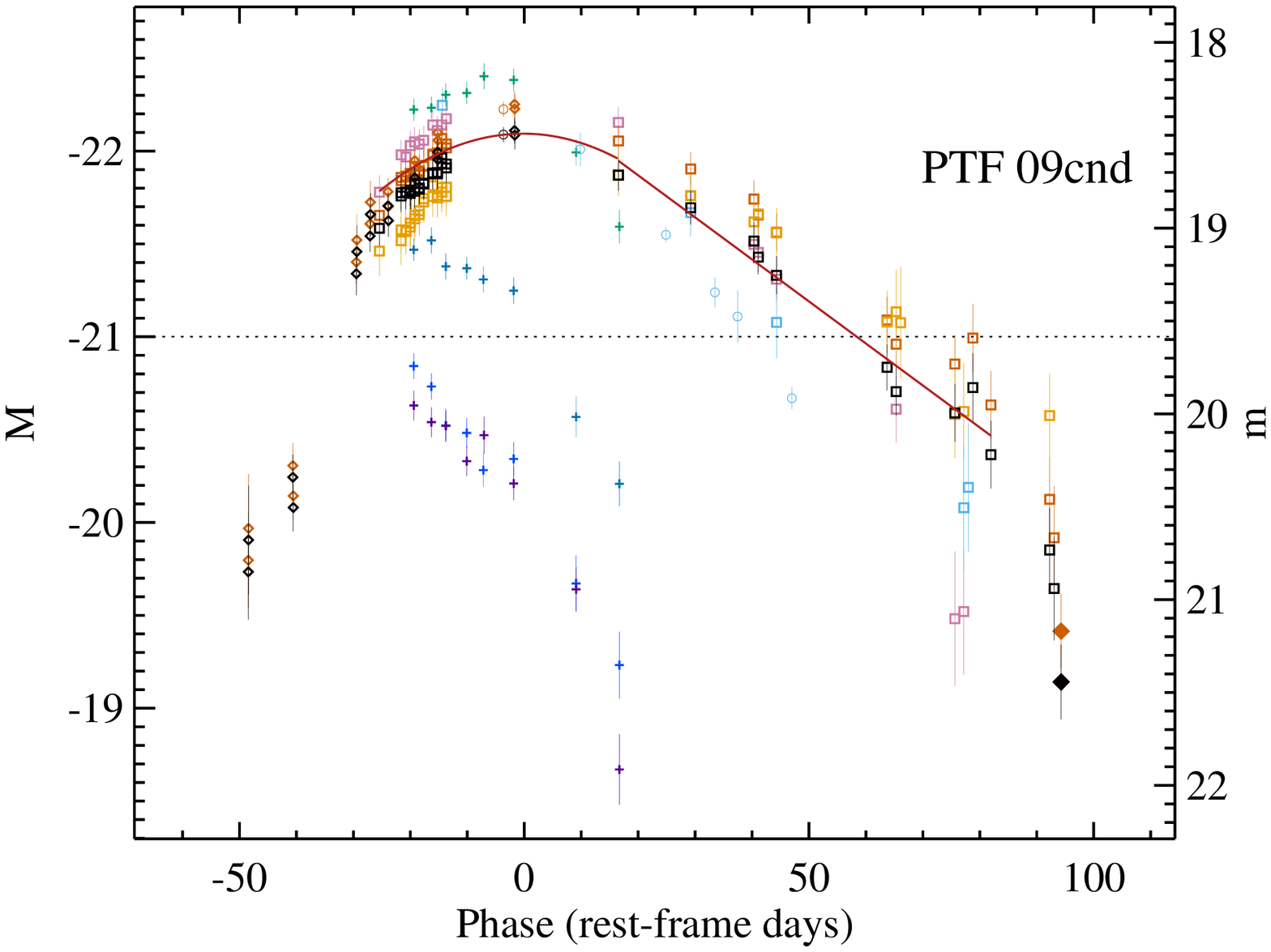}{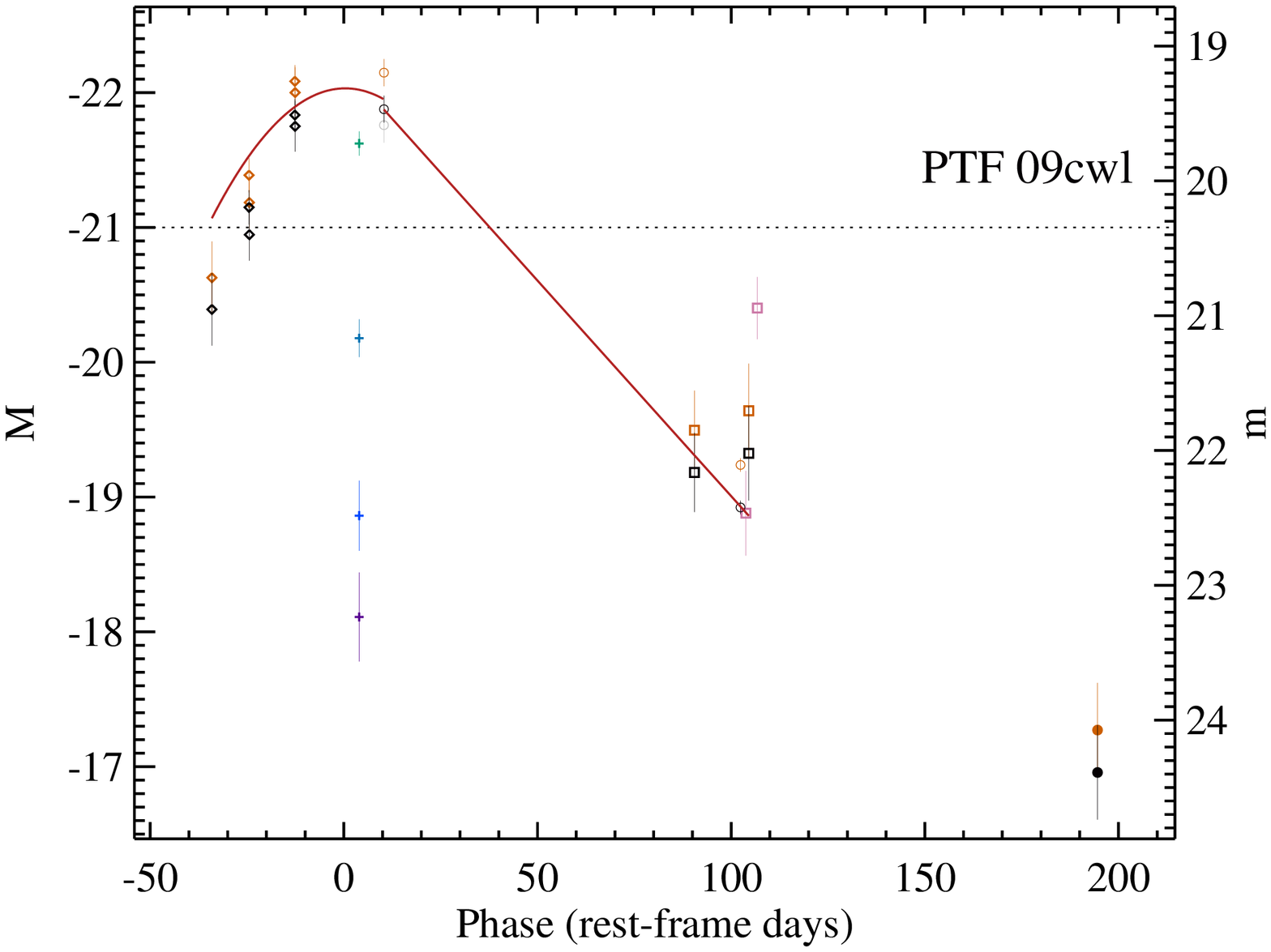}
\plottwo{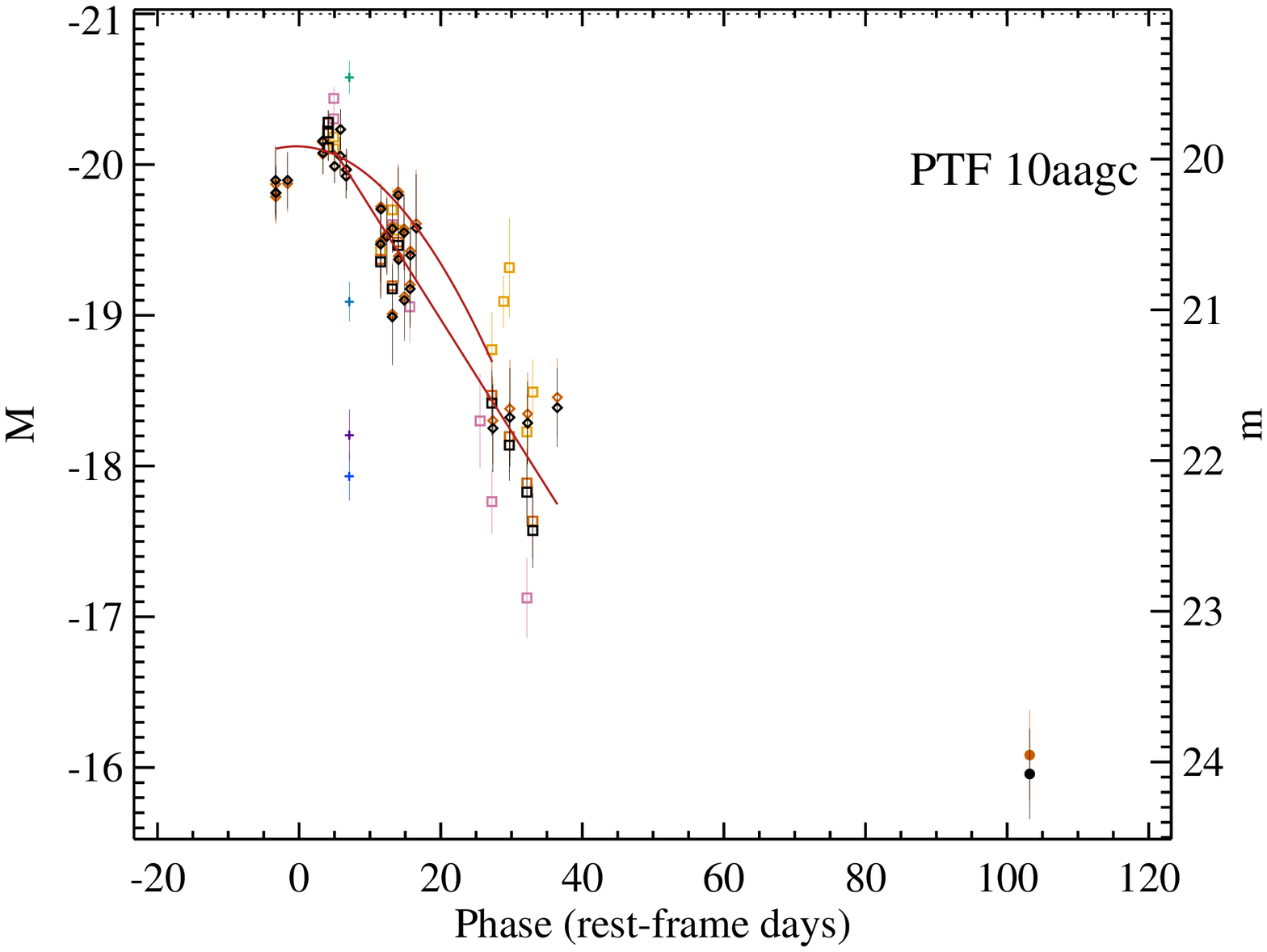}{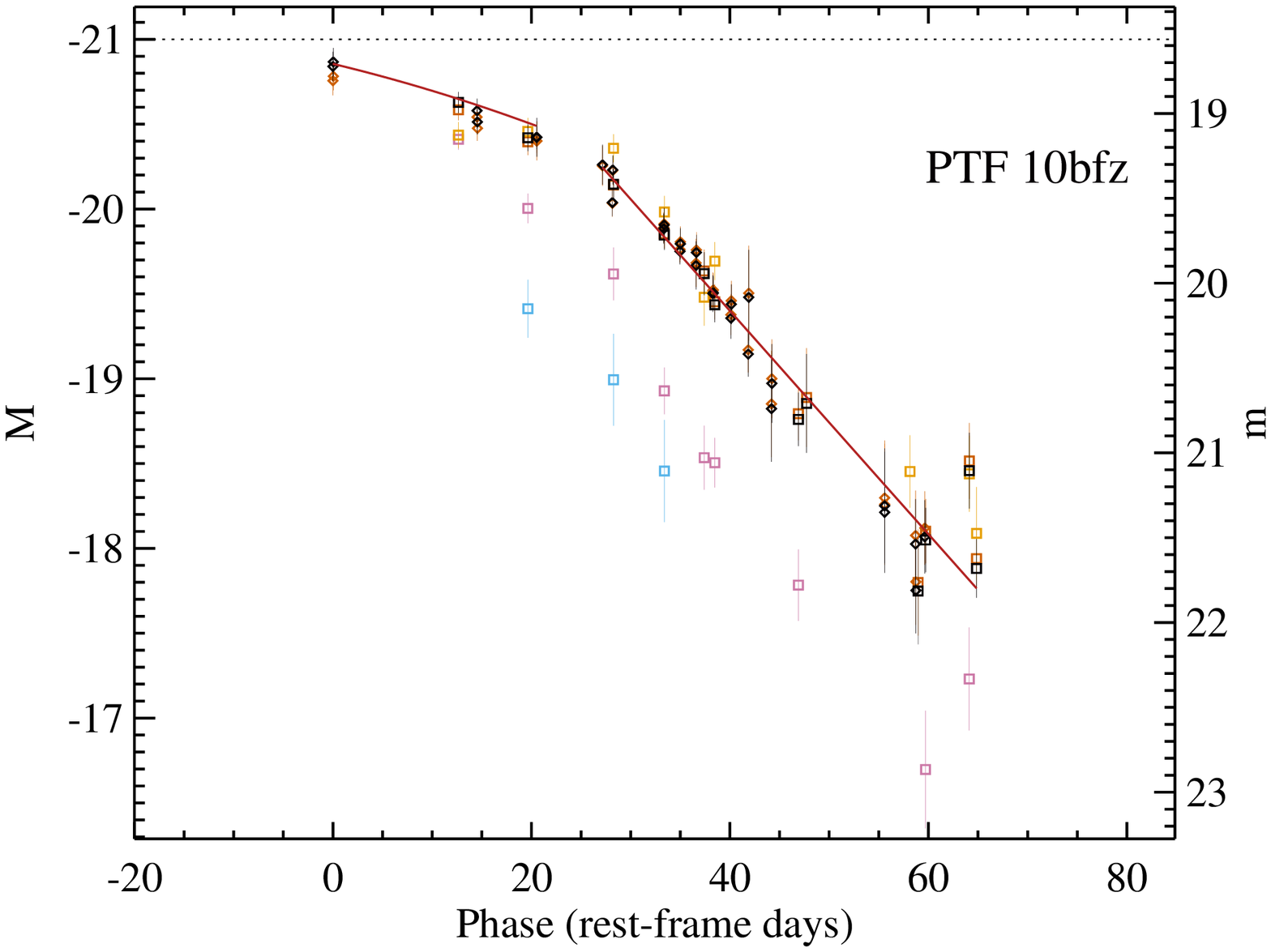}
\caption{Light curves of (i)PTF SLSNe-I in our sample for individual observed filters and rest-frame $M_g$ (black). Both apparent and absolute magnitudes are corrected for Galactic foreground dust extinction (see Sect. \ref{sec k-corr}). The solid lines show the second-order polynomial fit around peak (Sect. \ref{sec peak mag}) and the postpeak early- and late-time decay linear fits to the data (Sect. \ref{sec decay}). The horizontal dotted line marks the $M=-21$ mag ``hystorical'' threshold for SLSNe, for comparison.\label{fig ind abs g mag 1}}
\end{figure*}

\newpage
\begin{figure*}[!ht]
\epsscale{1.1}
\plotone{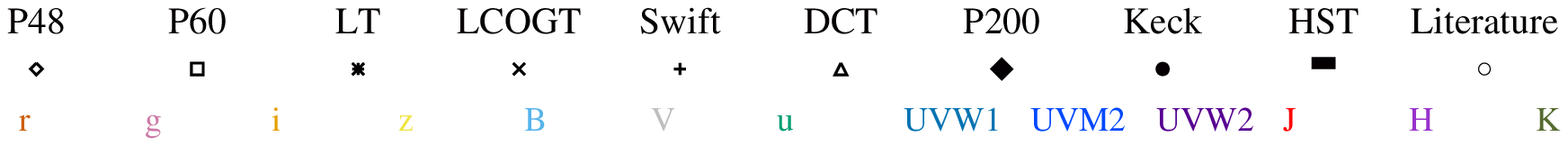}
\end{figure*}
\begin{figure*}[!hb]
\epsscale{1.15}
\plottwo{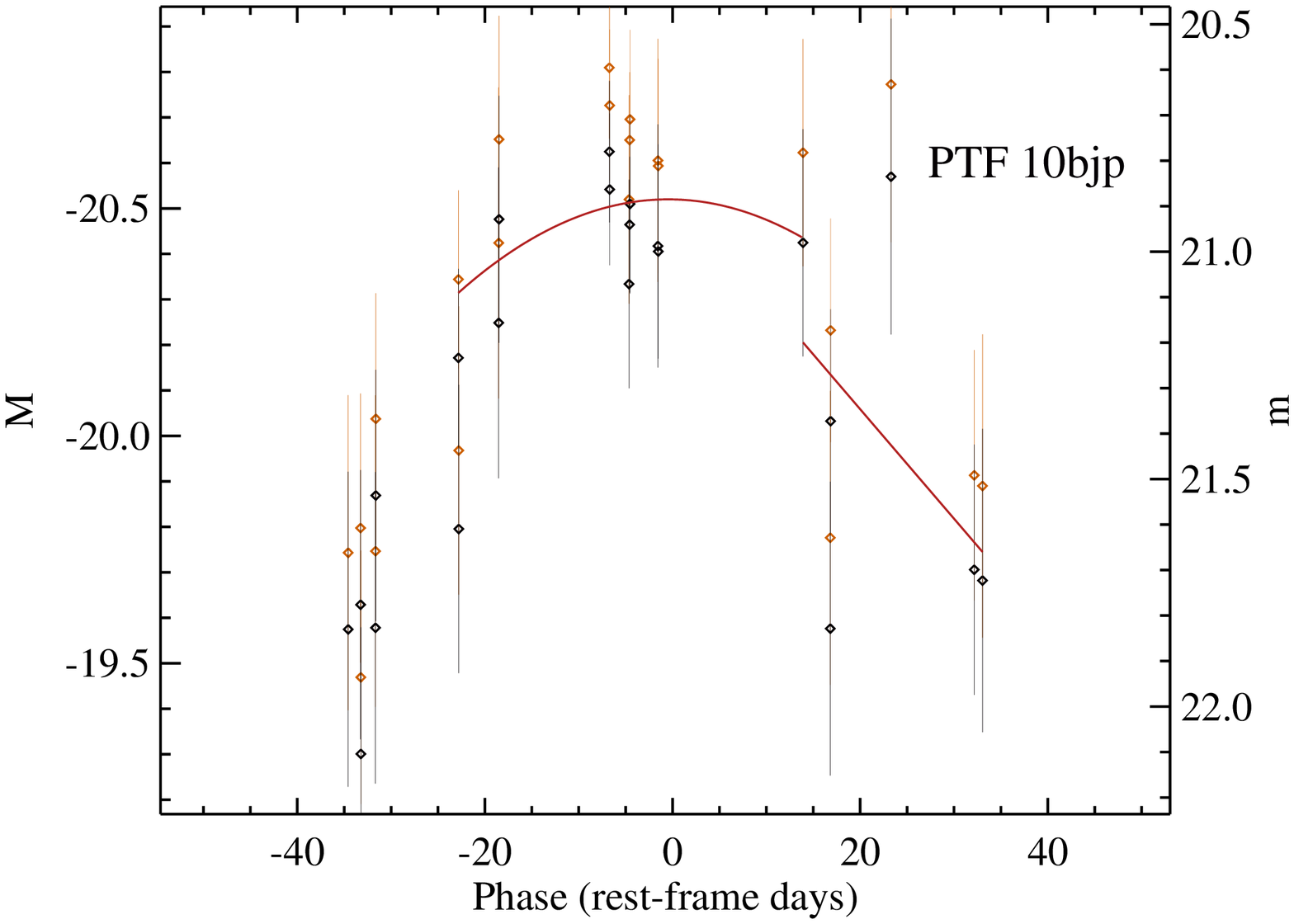}{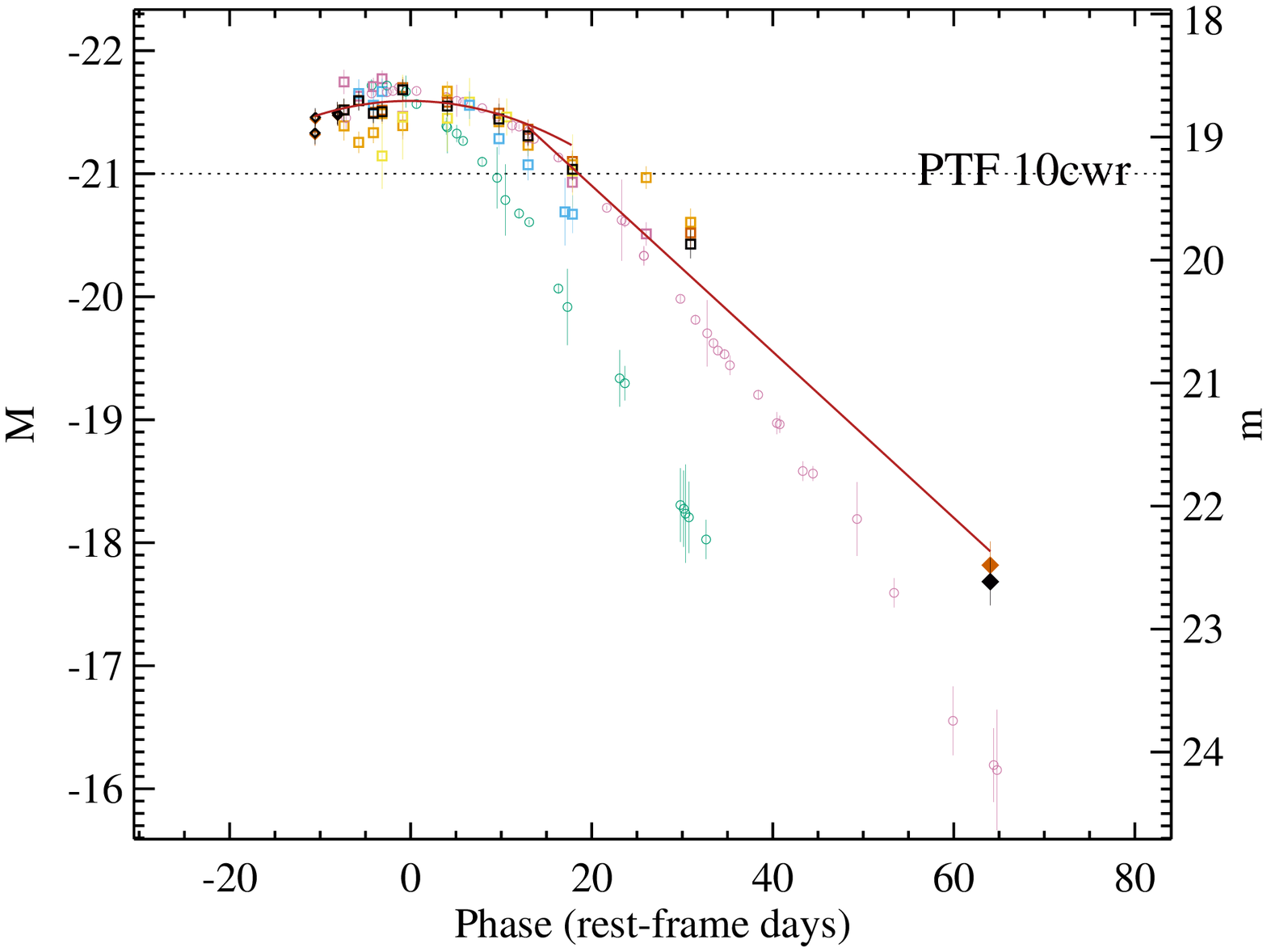}
\plottwo{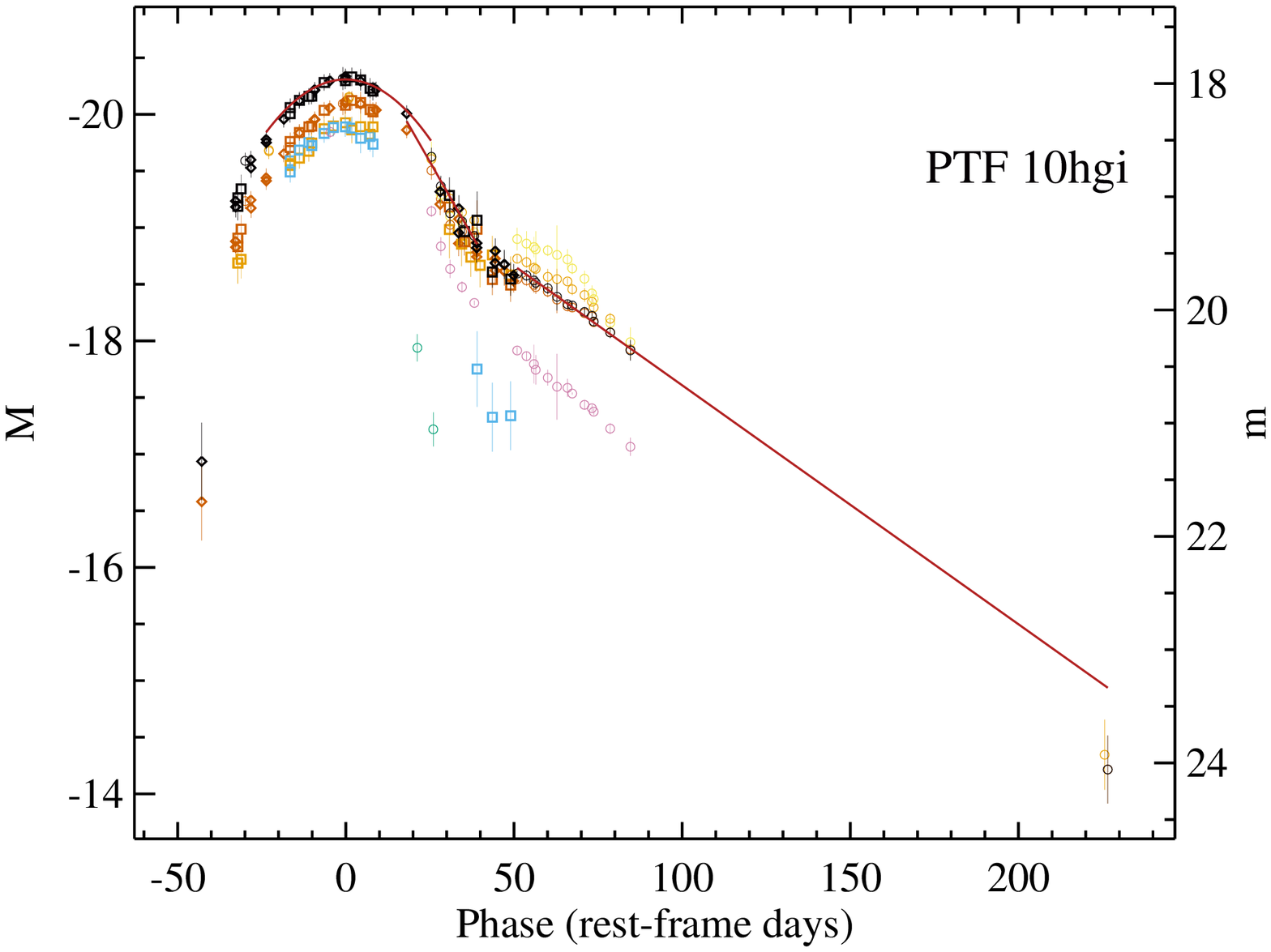}{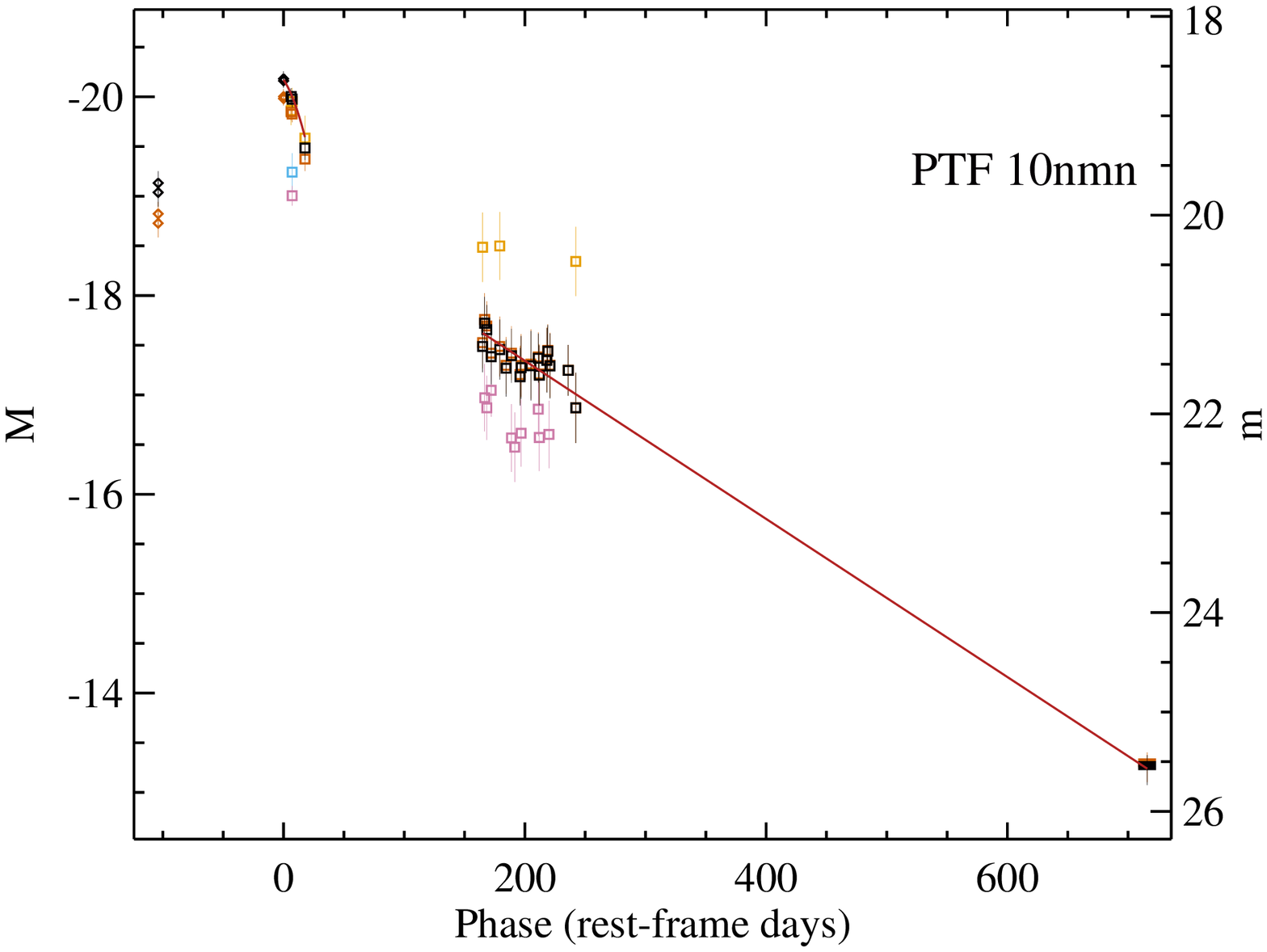}
\plottwo{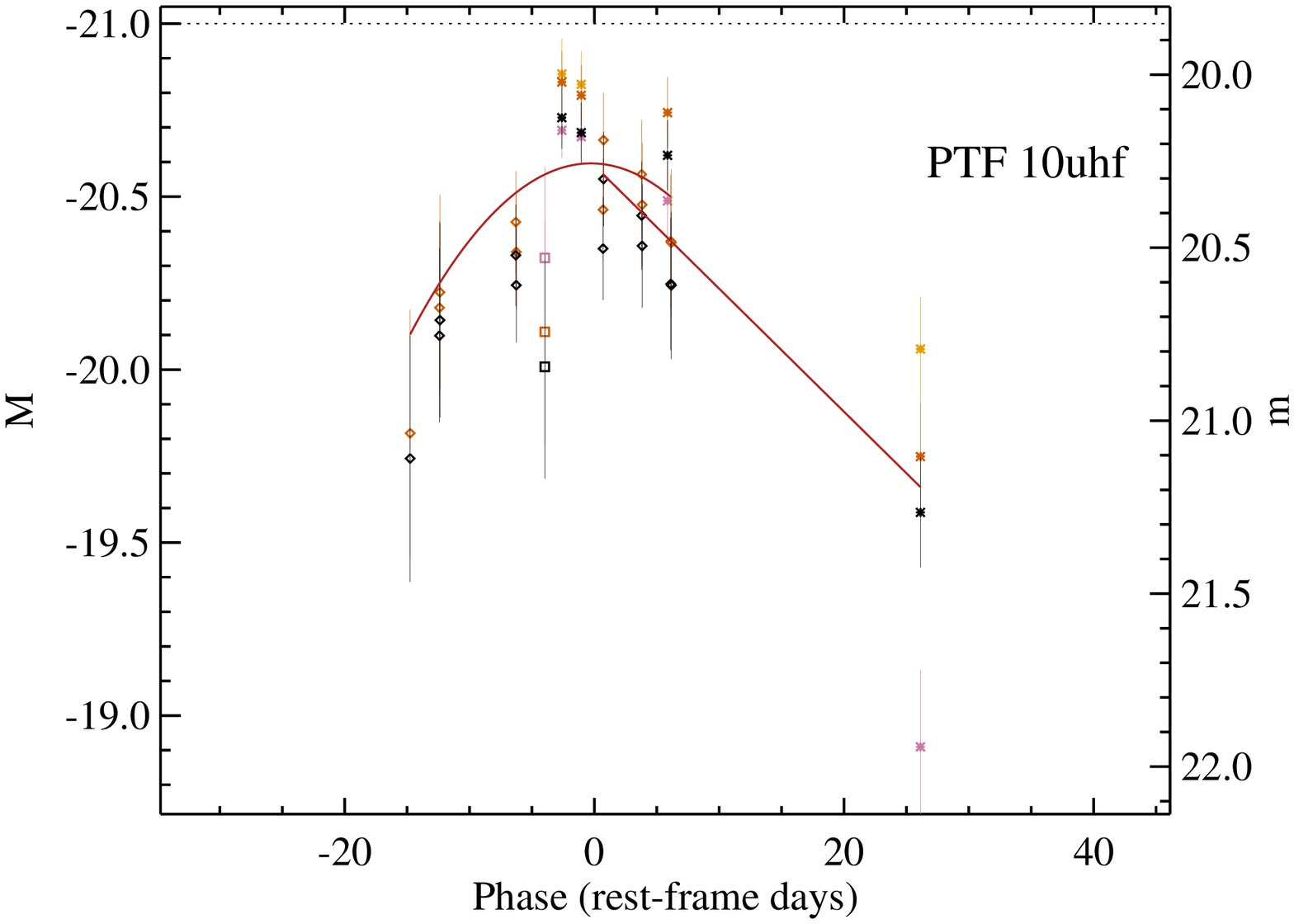}{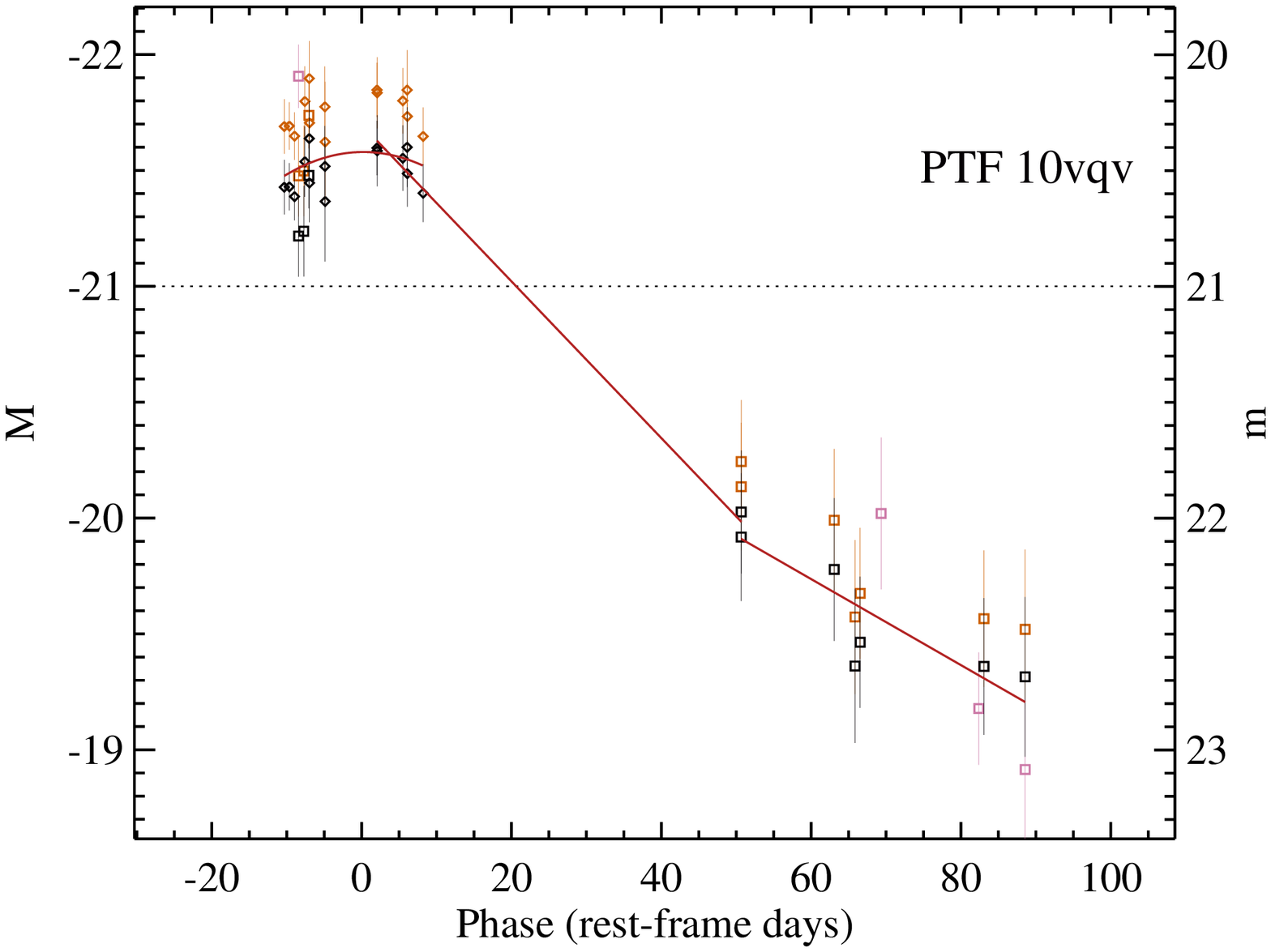}
\caption{Continuation of Fig.~\ref{fig ind abs g mag 1}.\label{fig ind abs g mag 2}}
\end{figure*}

\newpage
\begin{figure*}[!ht]
\epsscale{1.1}
\plotone{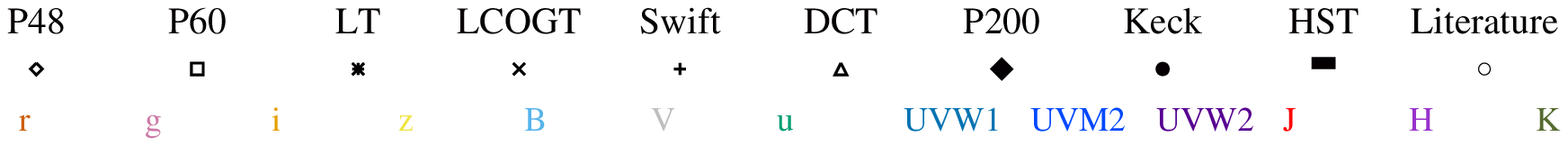}
\end{figure*}
\begin{figure*}[!hb]
\epsscale{1.15}
\plottwo{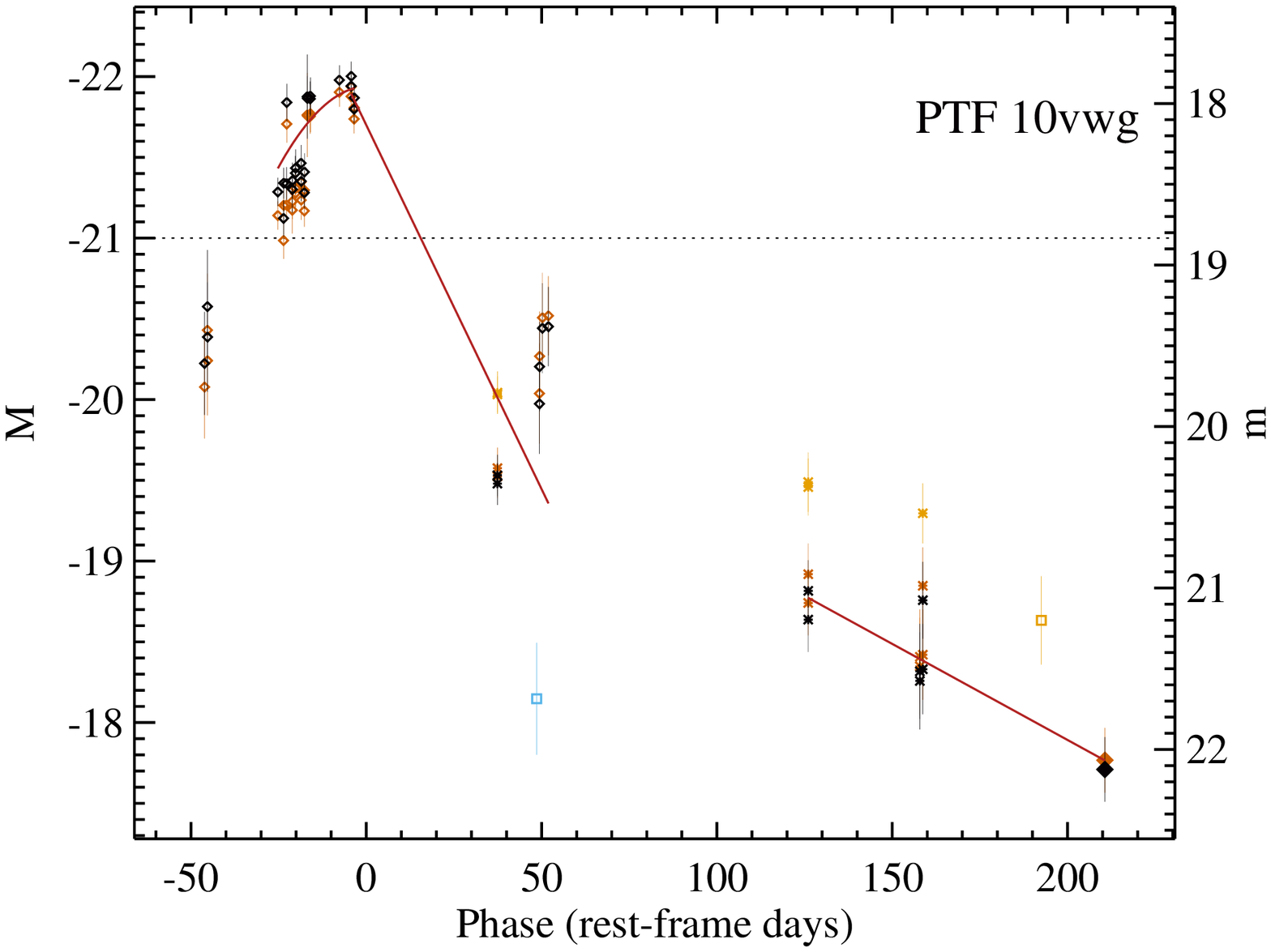}{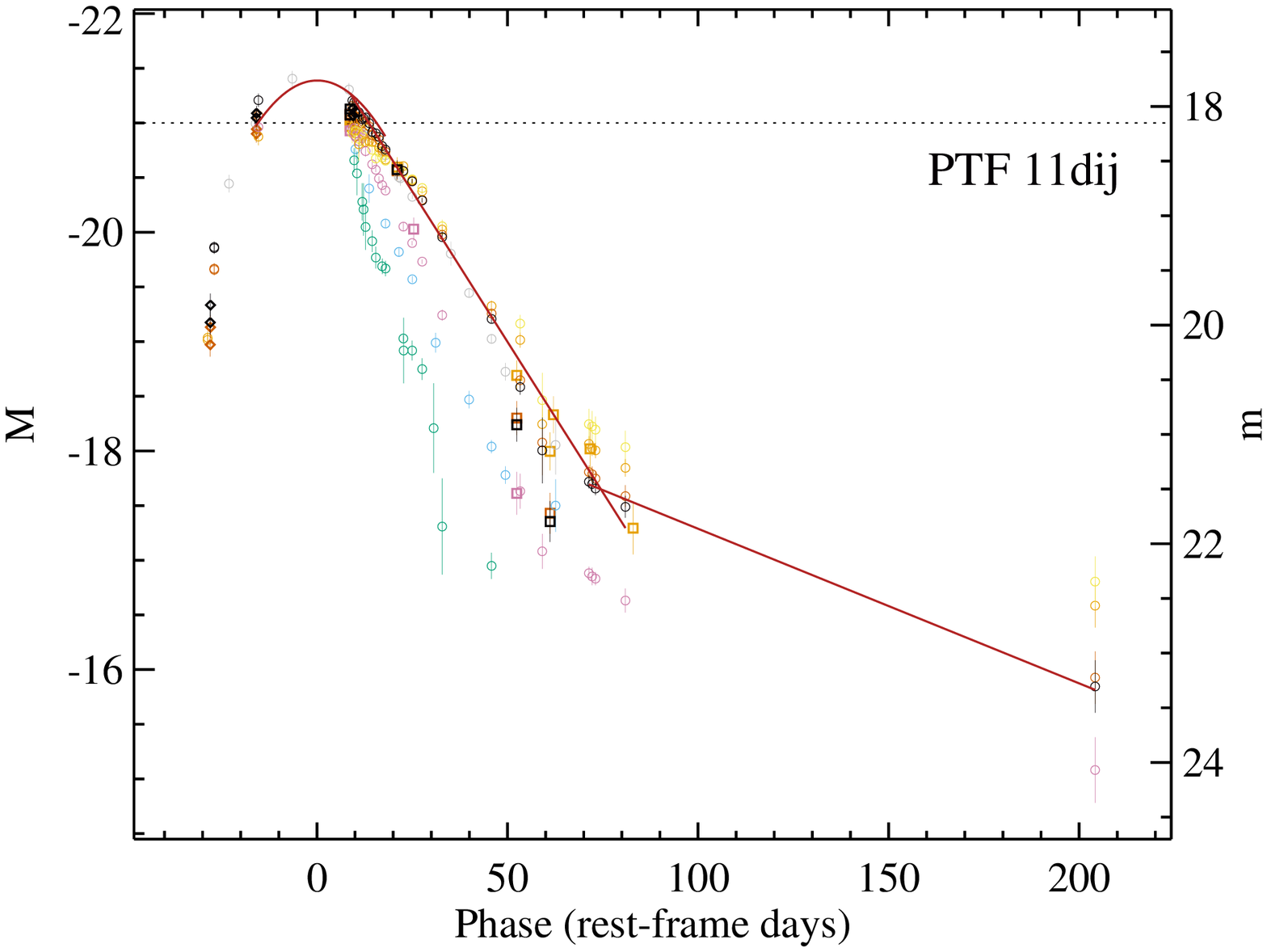}
\plottwo{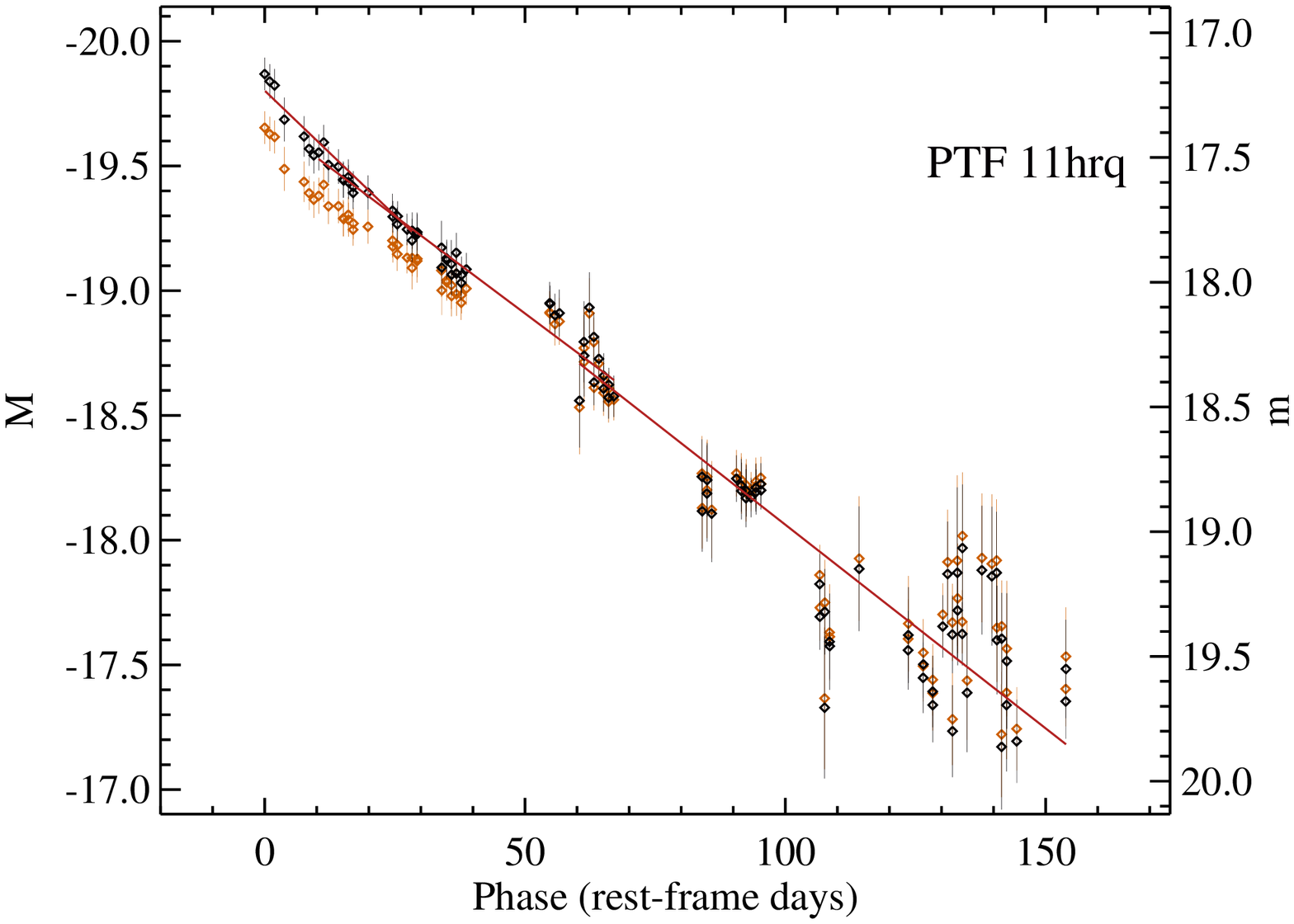}{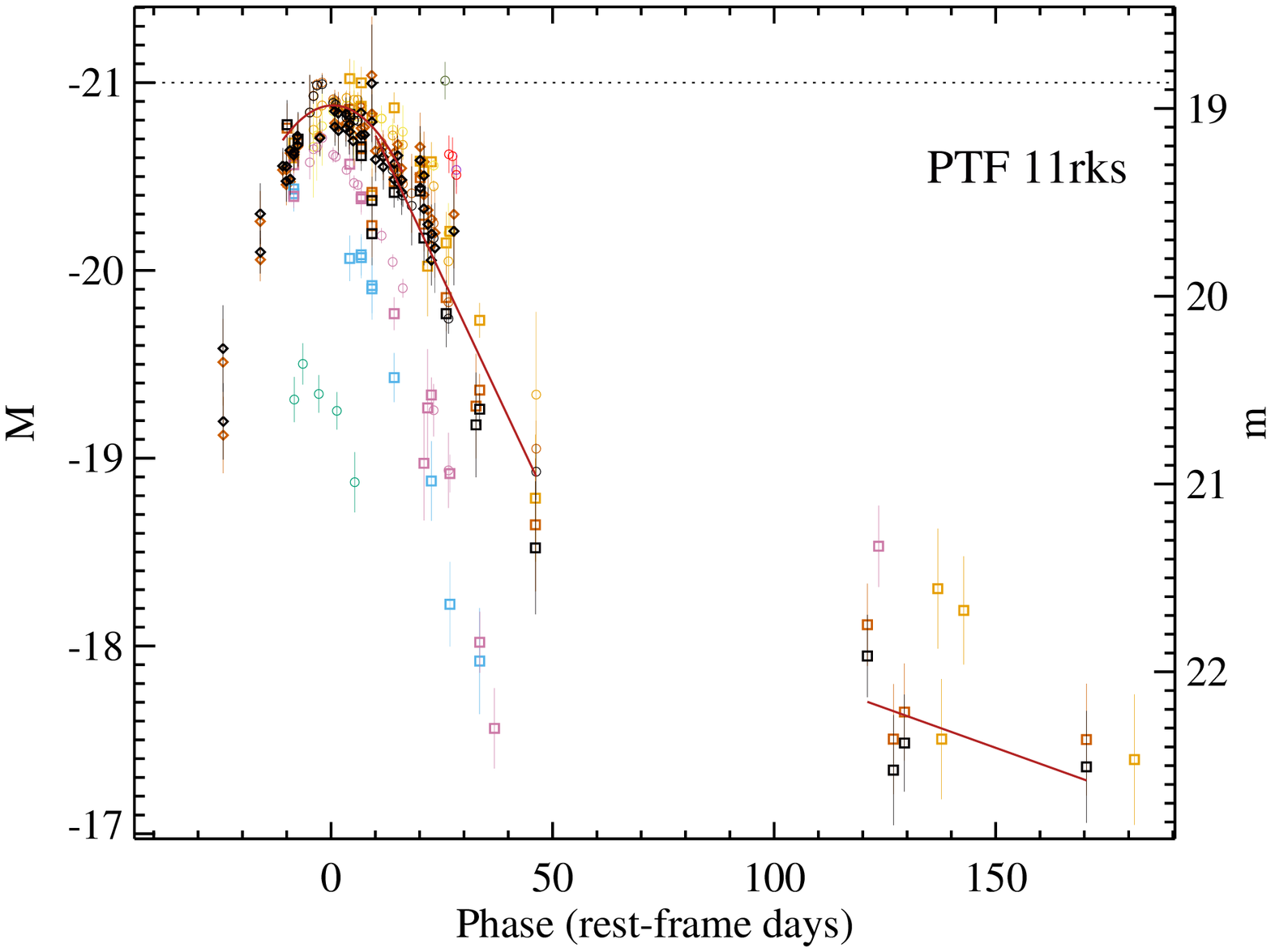}
\plottwo{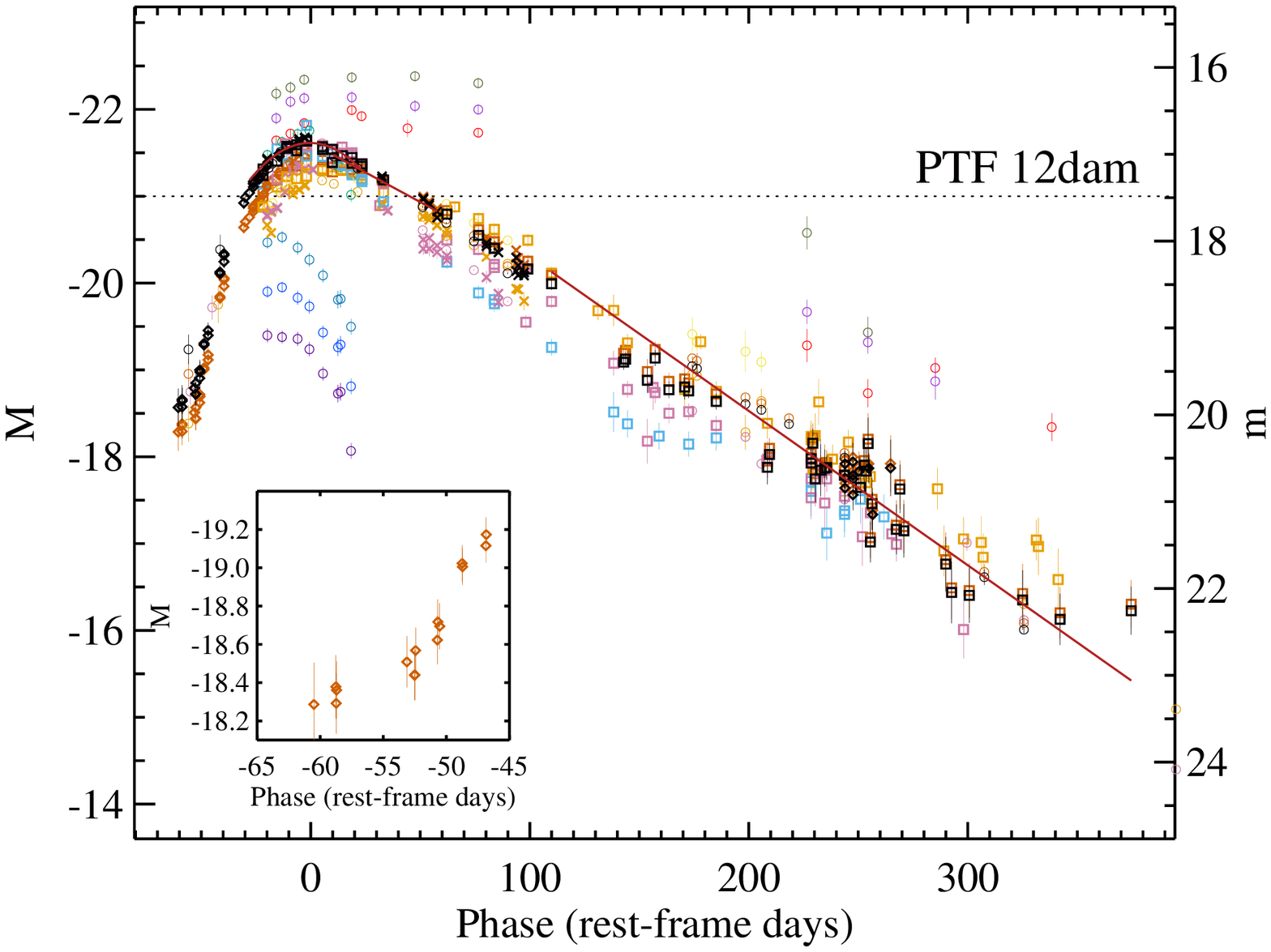}{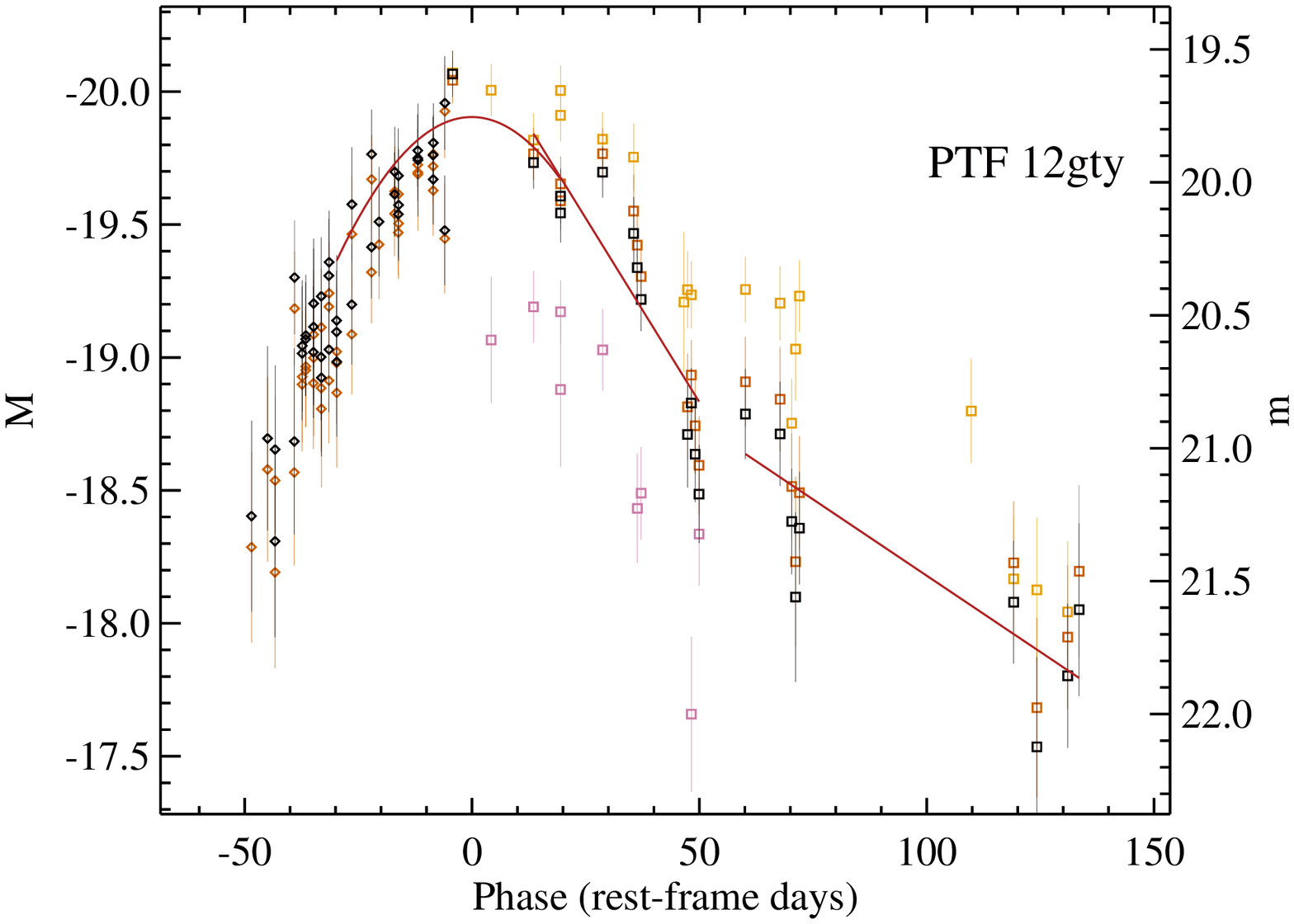}
\caption{Continuation of Fig.~\ref{fig ind abs g mag 1}.\label{fig ind abs g mag 3}}
\end{figure*}

\newpage
\begin{figure*}[!ht]
\epsscale{1.1}
\plotone{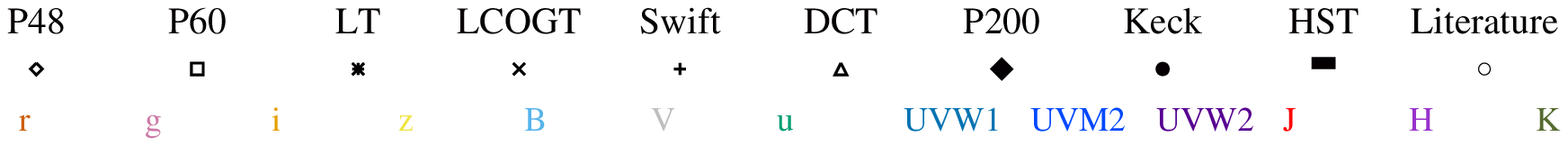}
\end{figure*}
\begin{figure*}[!hb]
\epsscale{1.15}
\plottwo{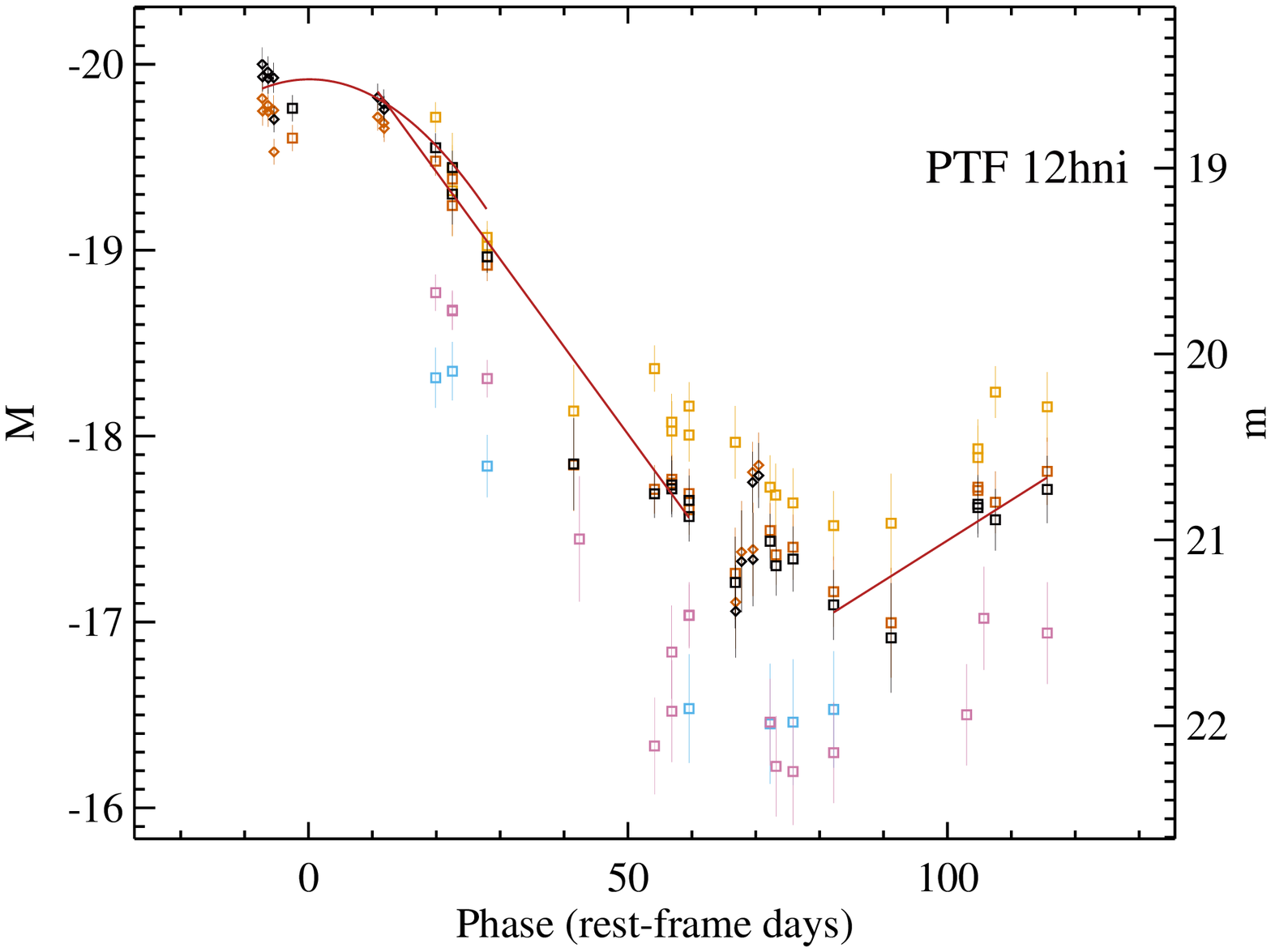}{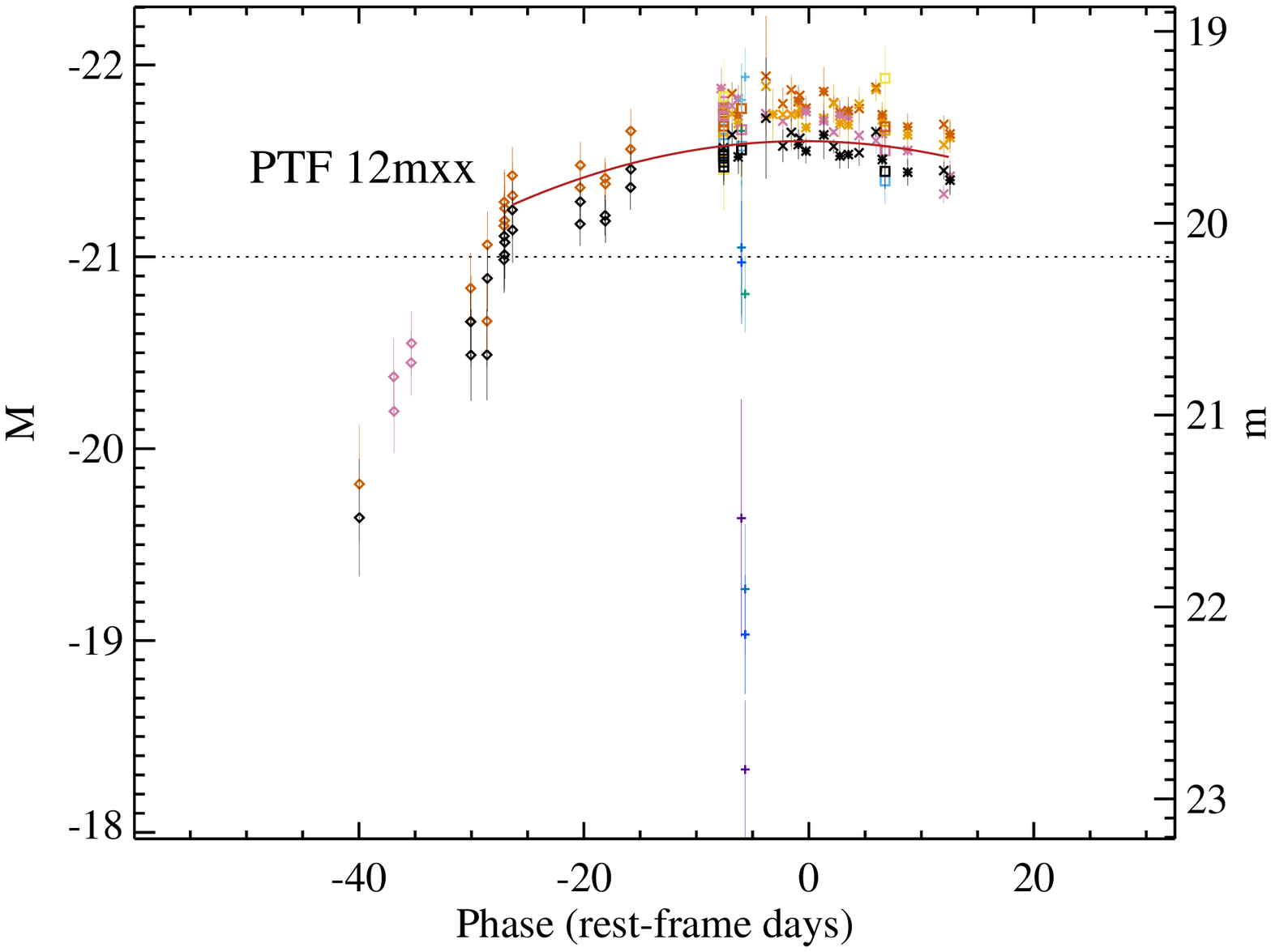}
\plottwo{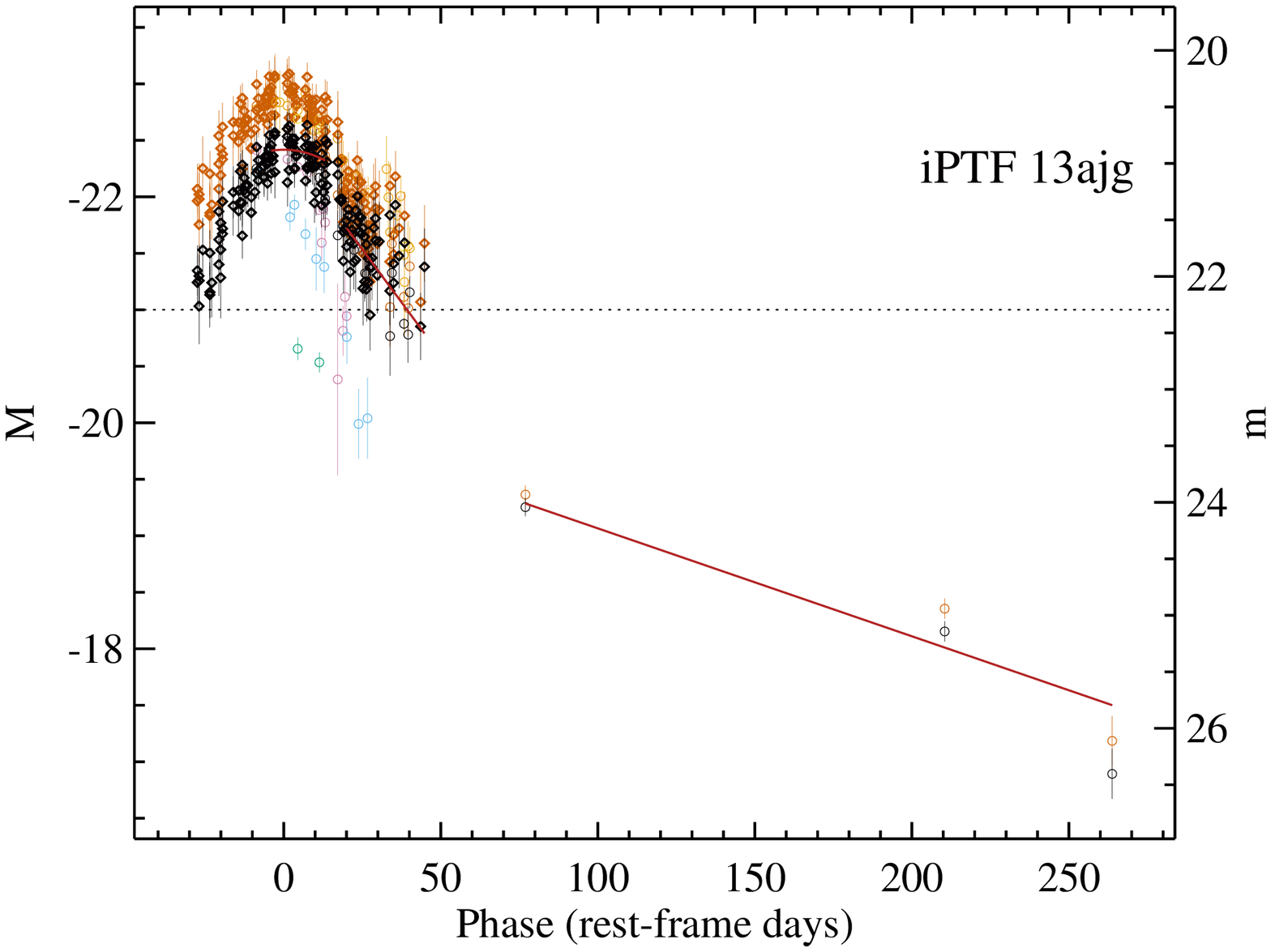}{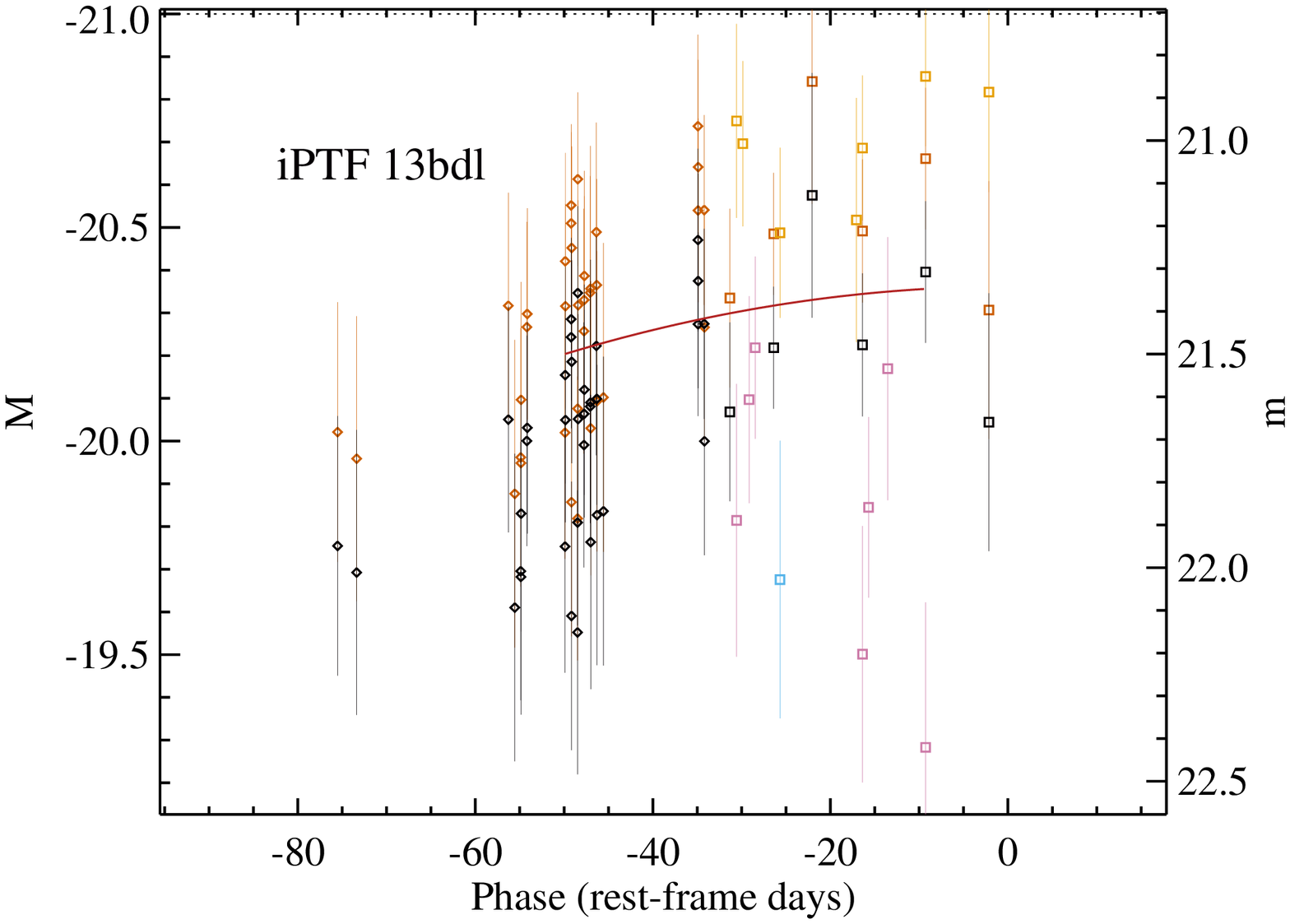}
\plottwo{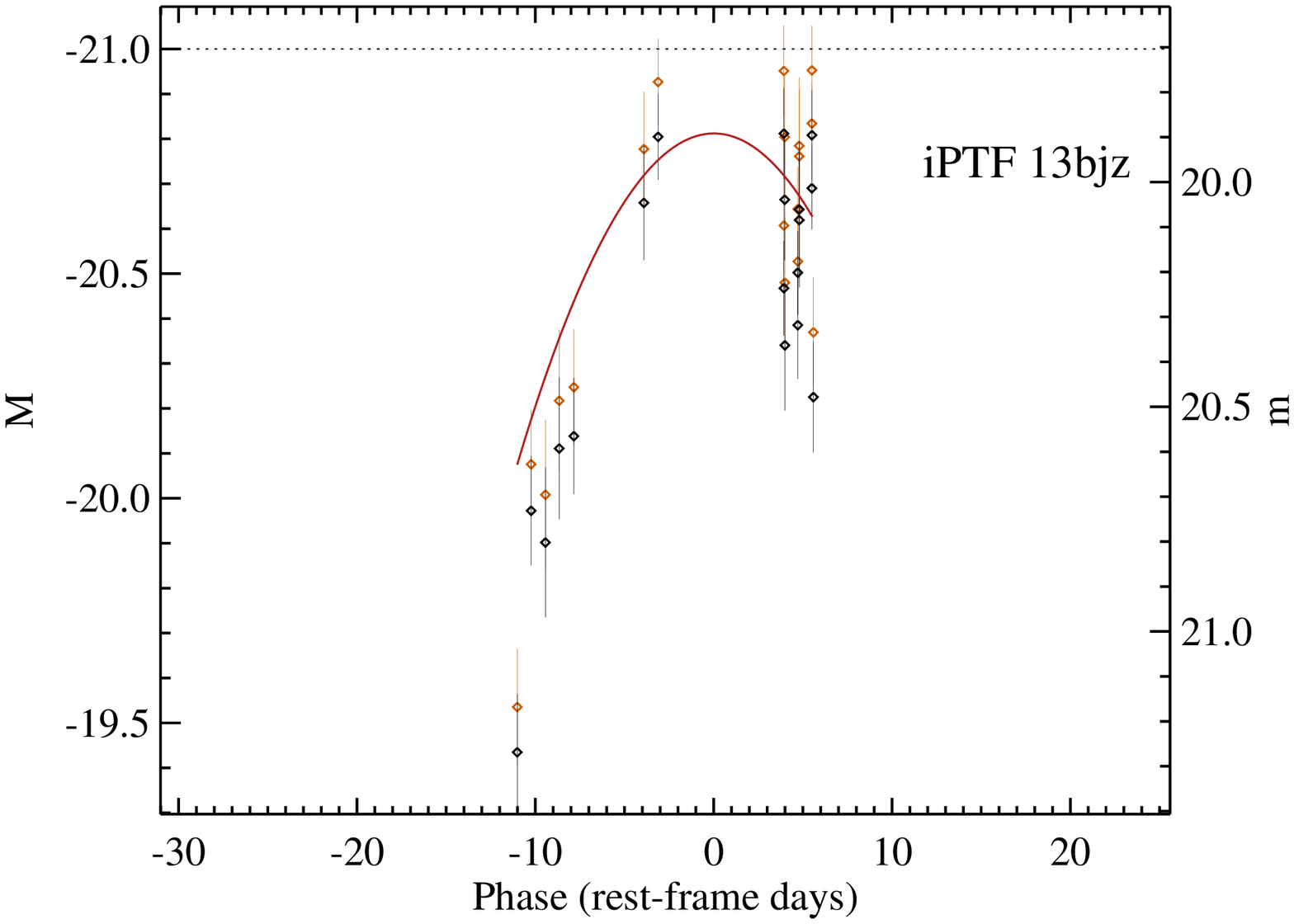}{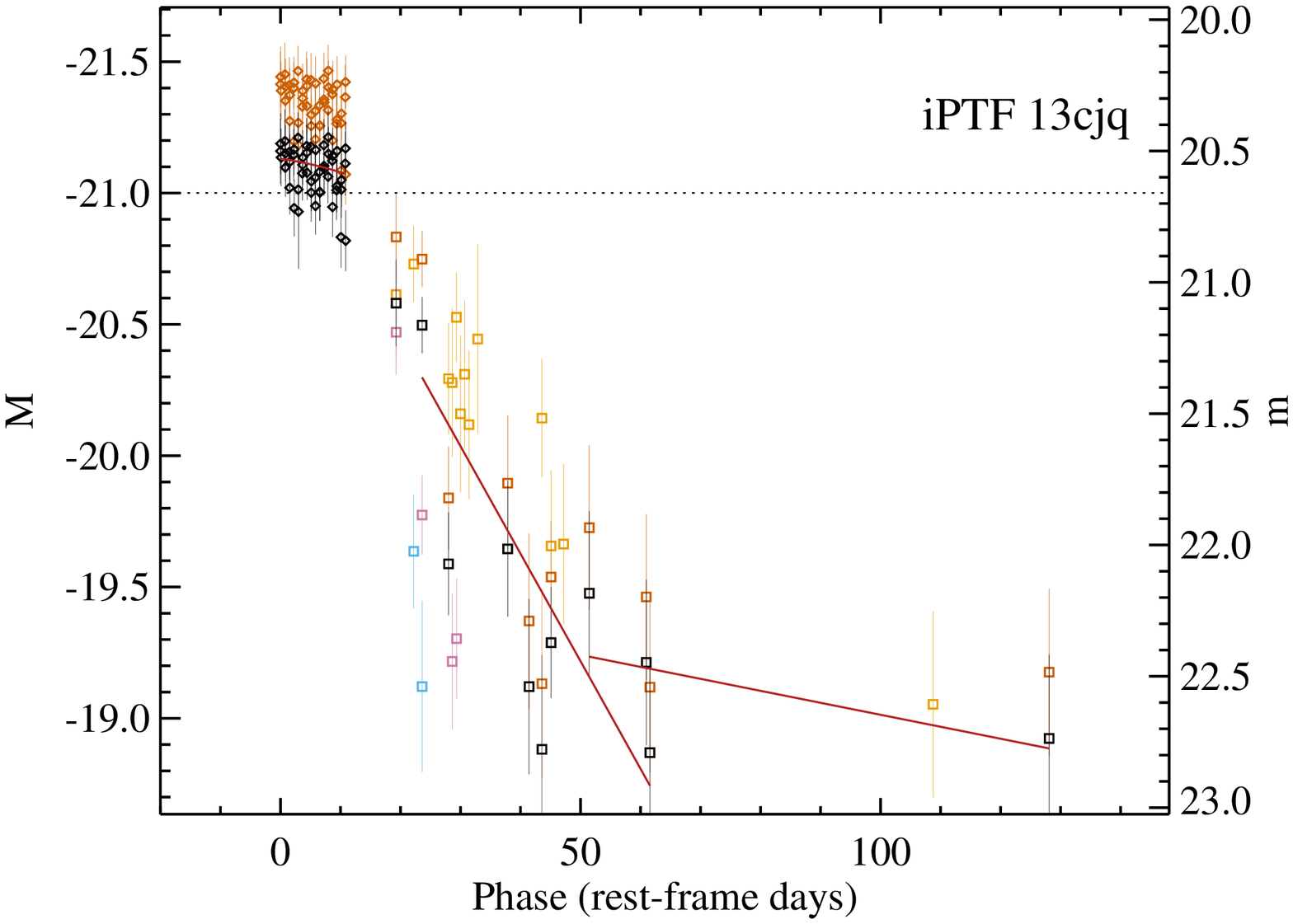}
\caption{Continuation of Fig.~\ref{fig ind abs g mag 1}.\label{fig ind abs g mag 4}}
\end{figure*}

\begin{figure*}
\gridline{\fig{fig/legend.eps}{\textwidth}{}}
\gridline{\fig{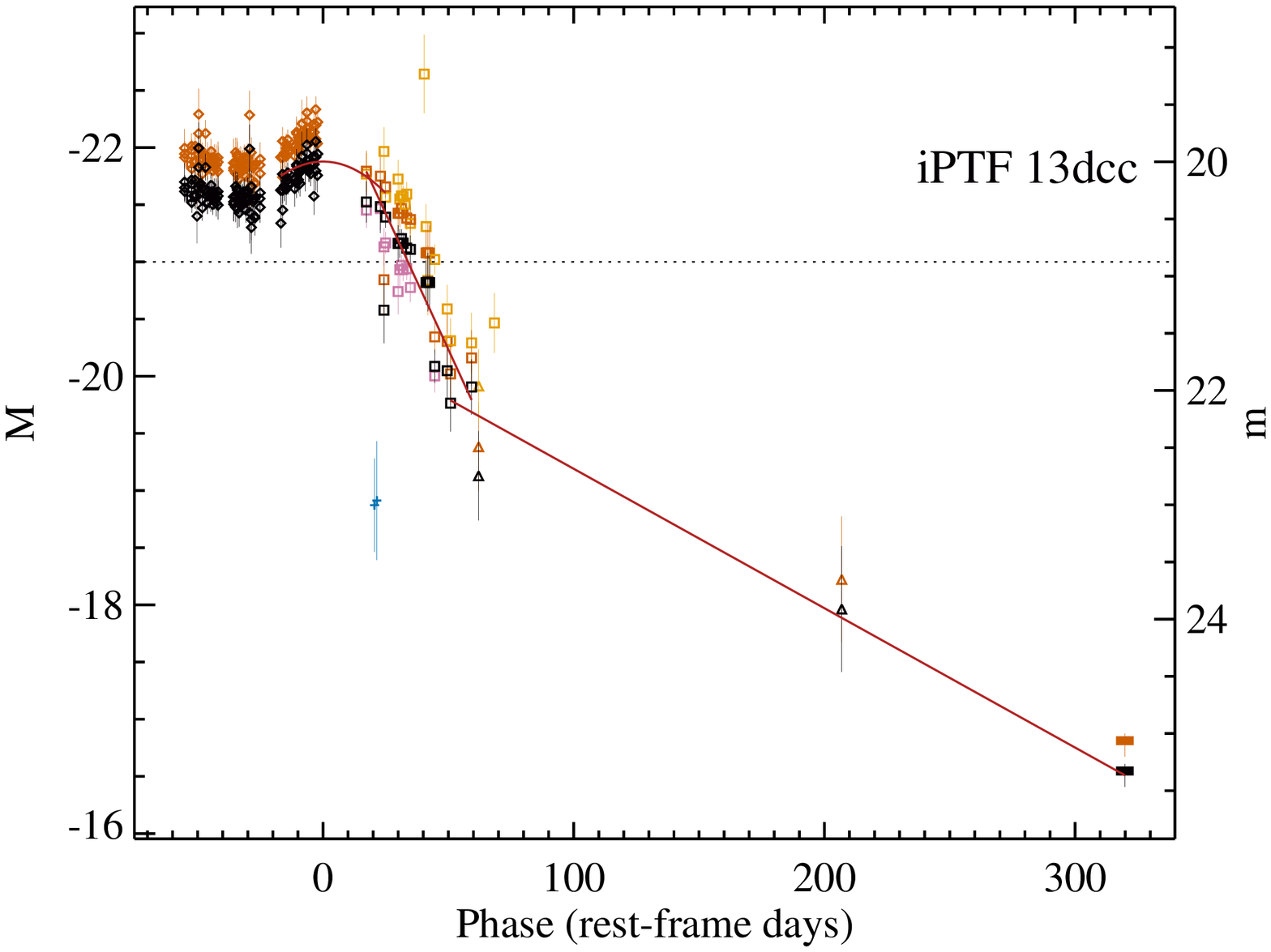}{0.5\textwidth}{}
          \fig{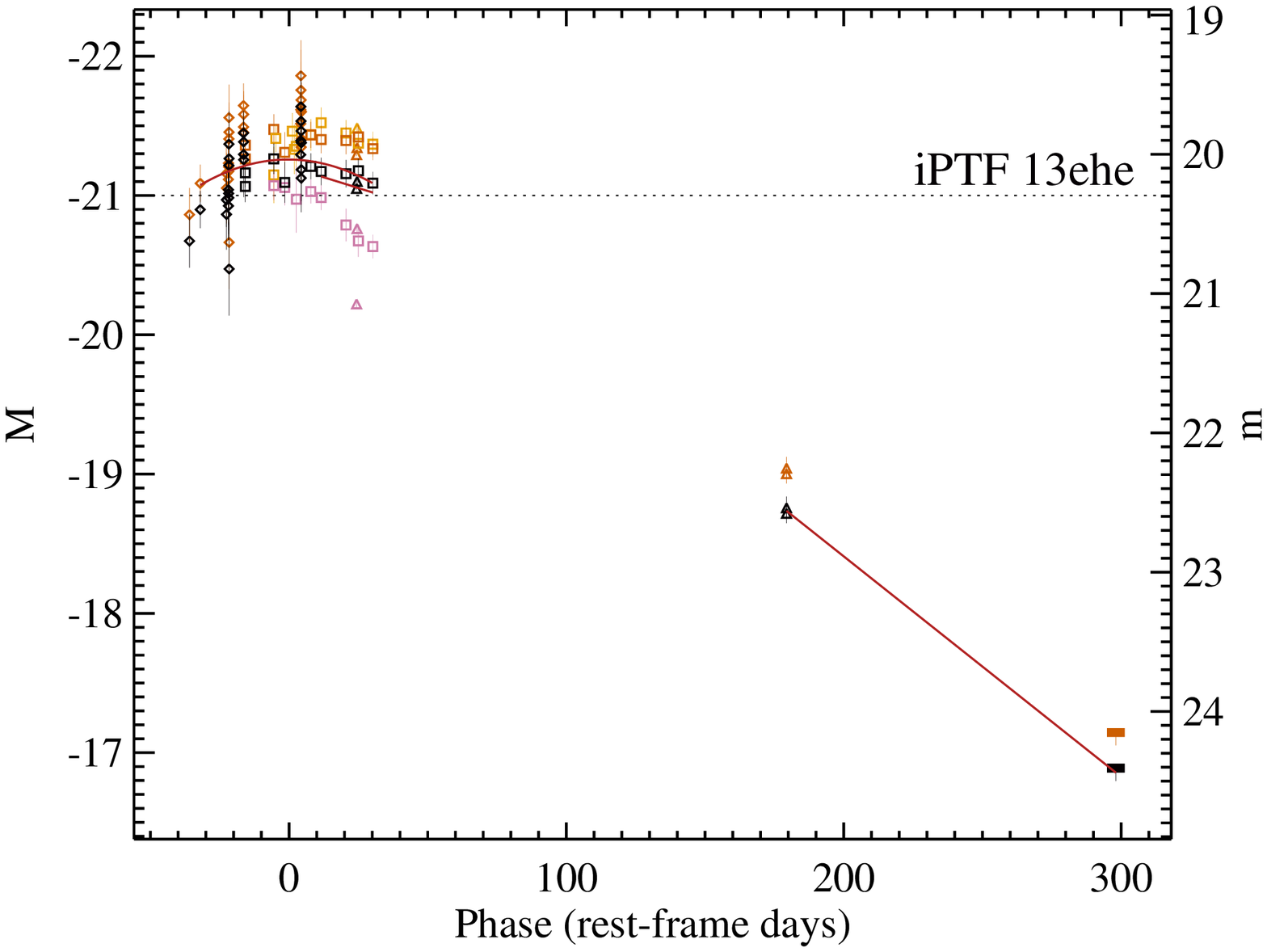}{0.5\textwidth}{}}
\caption{Continuation of Fig.~\ref{fig ind abs g mag 1}.\label{fig ind abs g mag 5}}
\end{figure*}


\begin{figure*}[!h]
\epsscale{1.1}
\plottwo{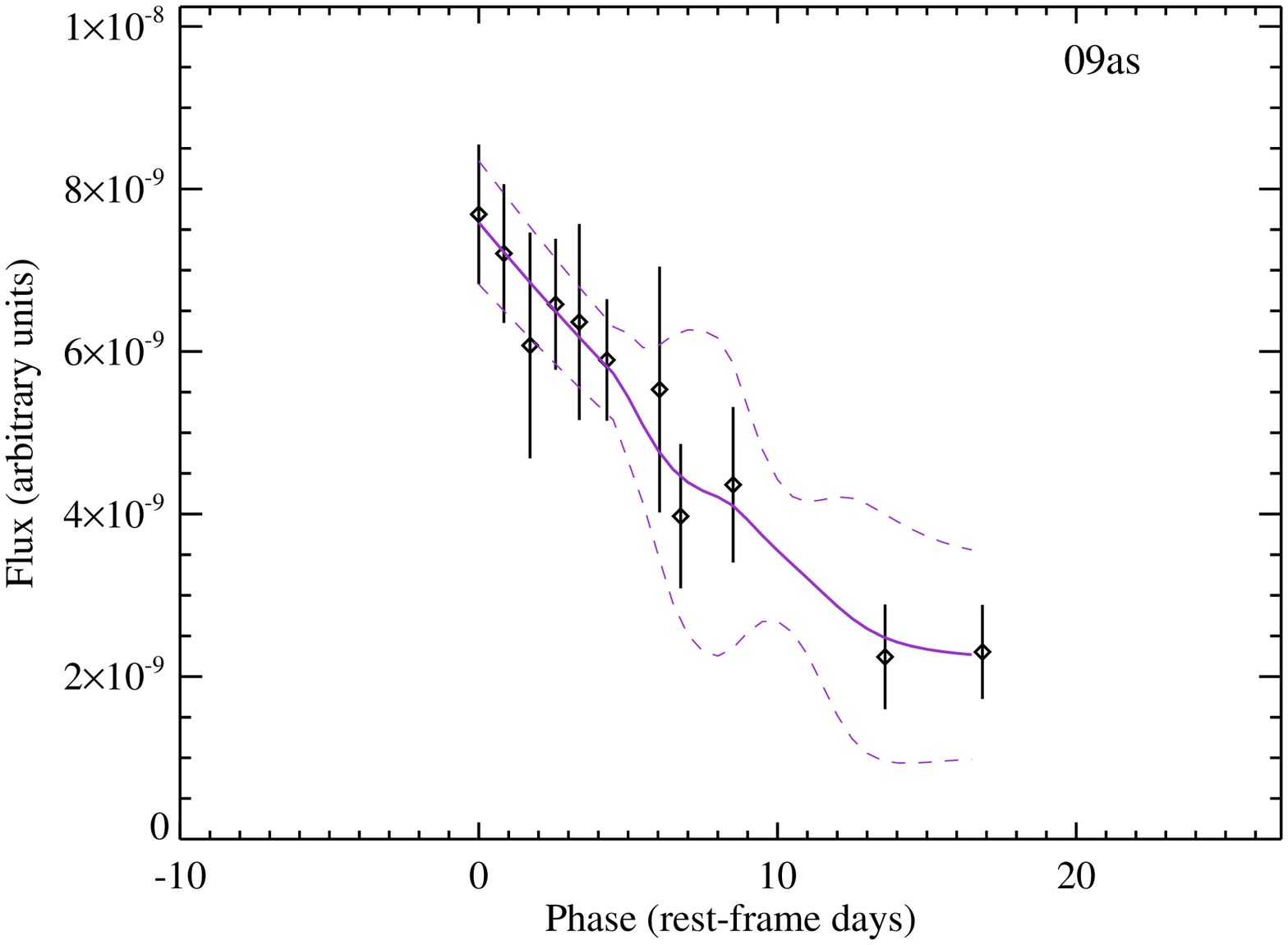}{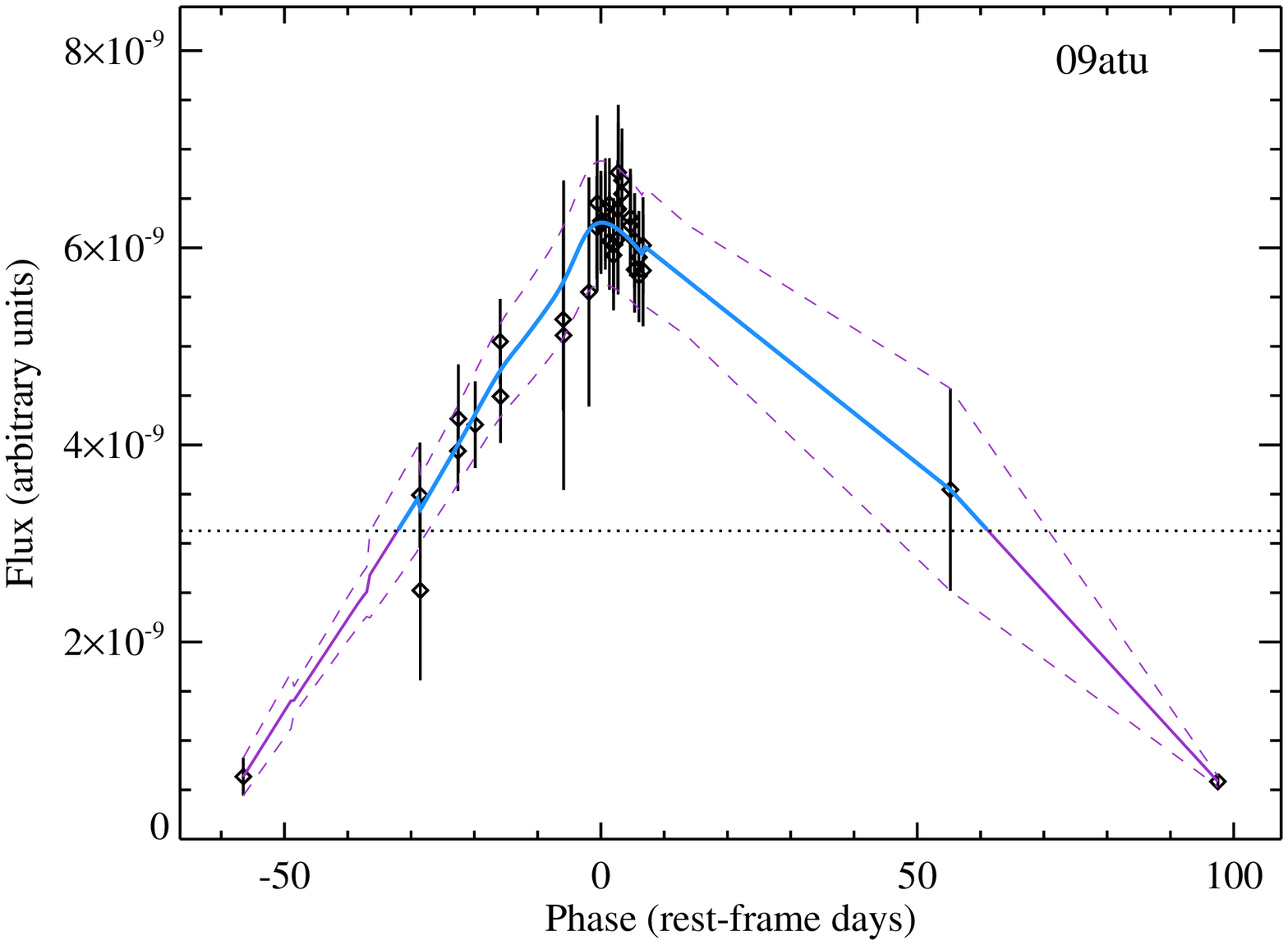}
\plottwo{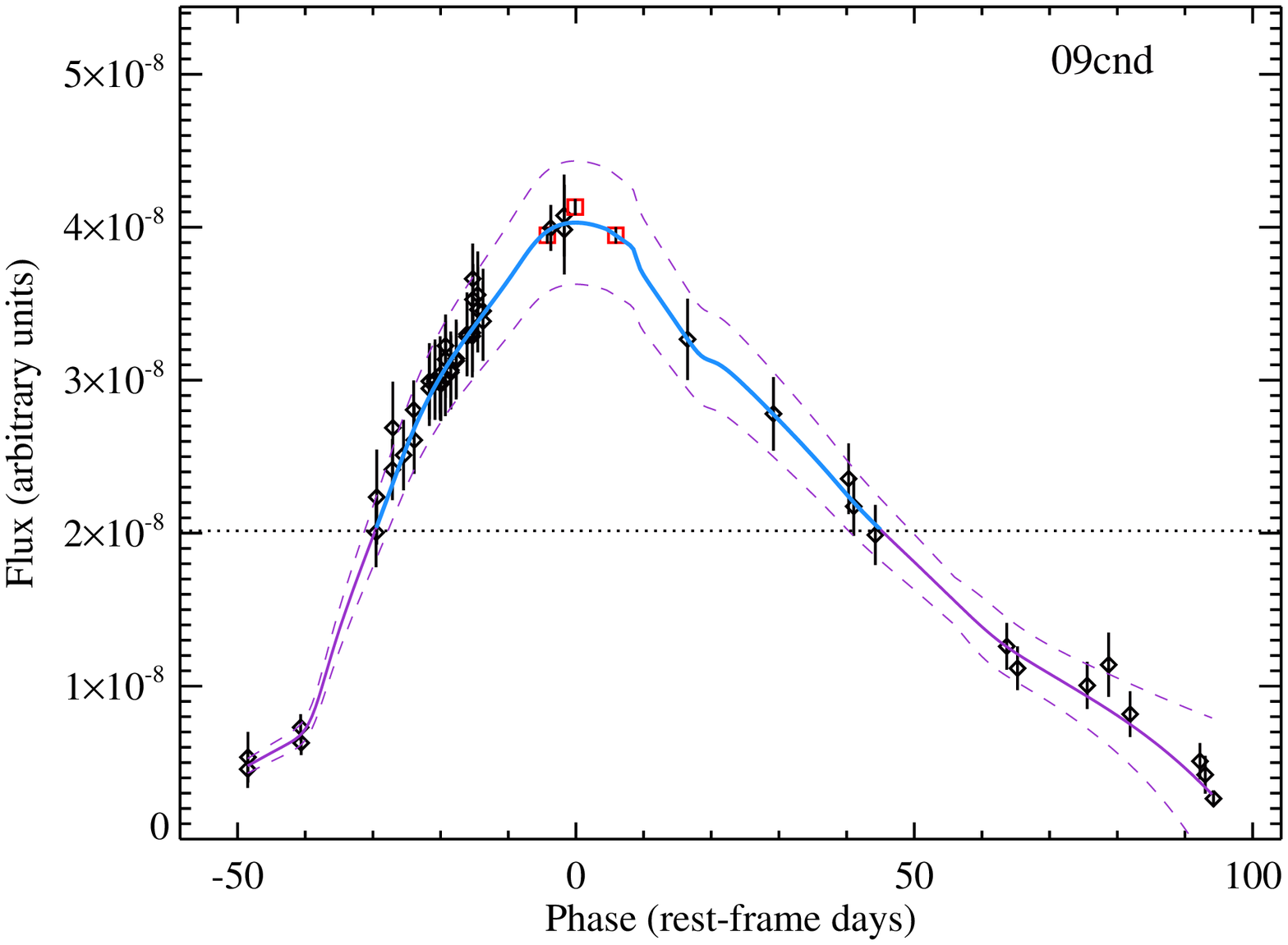}{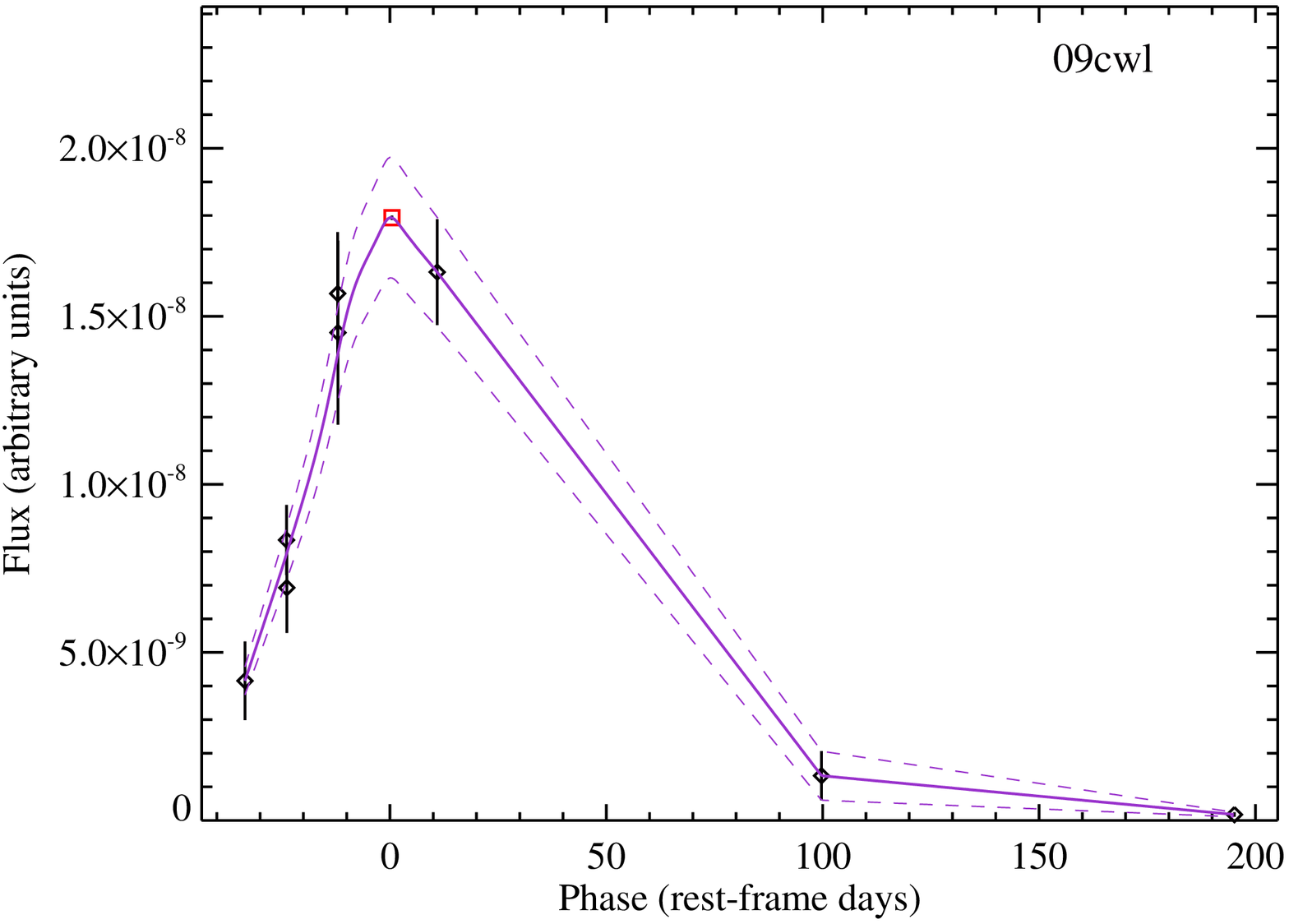}
\plottwo{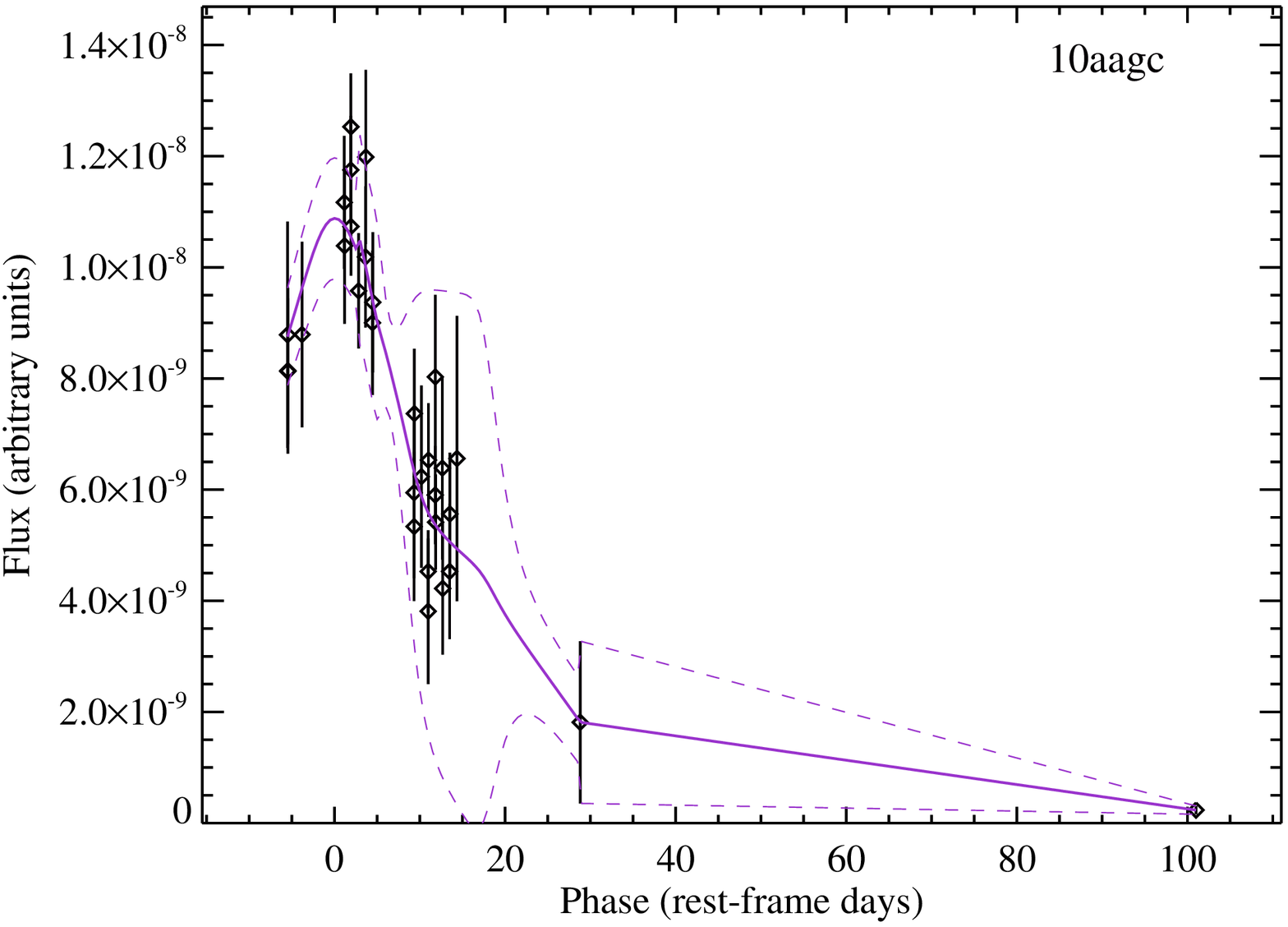}{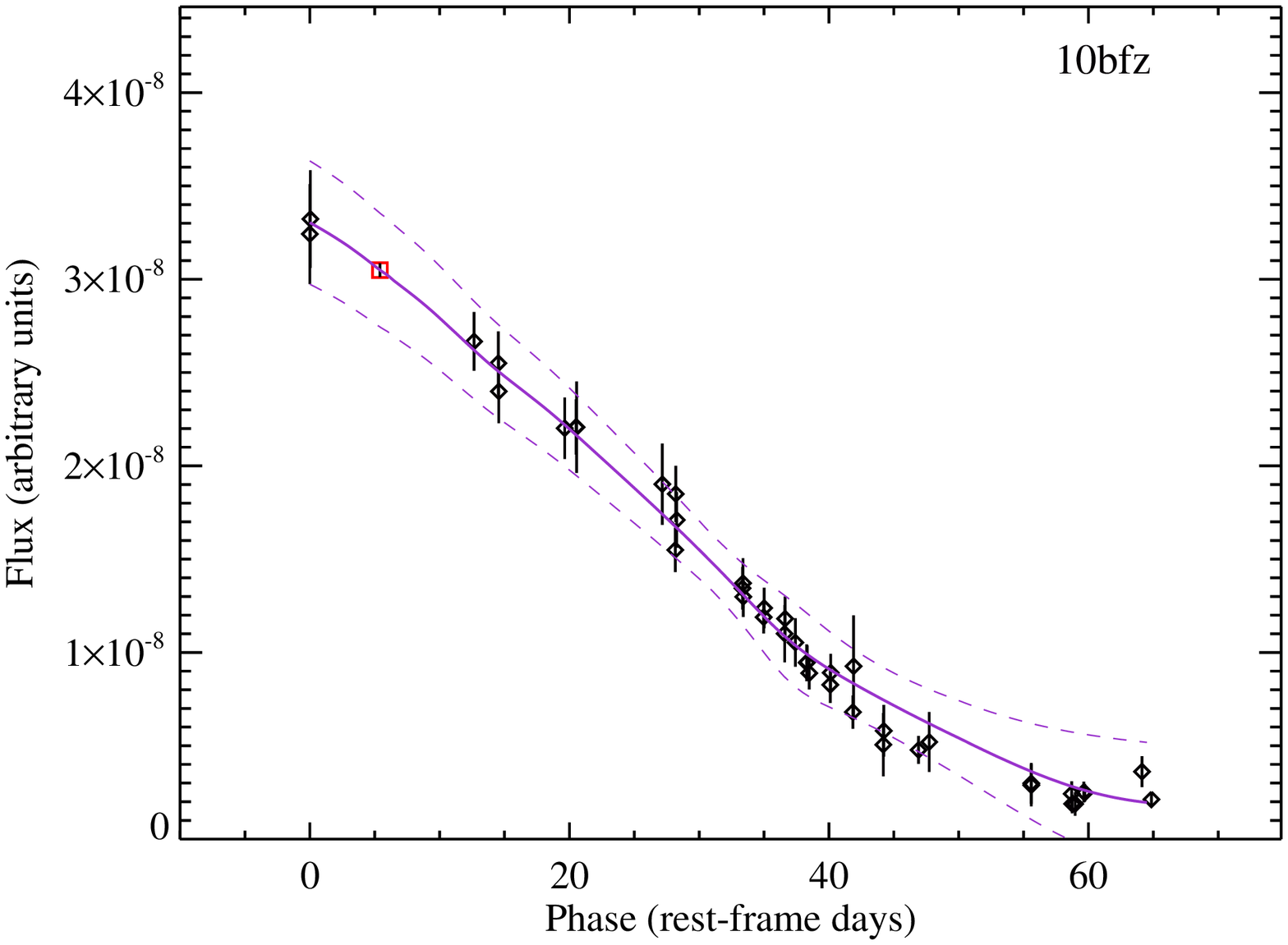}
\caption{Rest-frame $g$-band light curves of the SLSNe-I in flux (black diamonds). The smoothed light curves (described in Sect. \ref{sec smooth}) are shown by the solid curves. The red squares indicate the auxiliary points introduced for the light-curve smoothing. The horizontal dotted line marks the half-flux limit, which is used to calculate the $t_{\rm rise, 1/2}$ and $t_{\rm fall, 1/2}$ timescales (Sect. \ref{sect disc rise decay times}), when the light curve is well-characterized between the half-flux limit and the peak (highlighted blue solid curves).\label{fig ind smooth flux 1}}
\end{figure*}

\begin{figure*}[!h]
\epsscale{1.1}
\plottwo{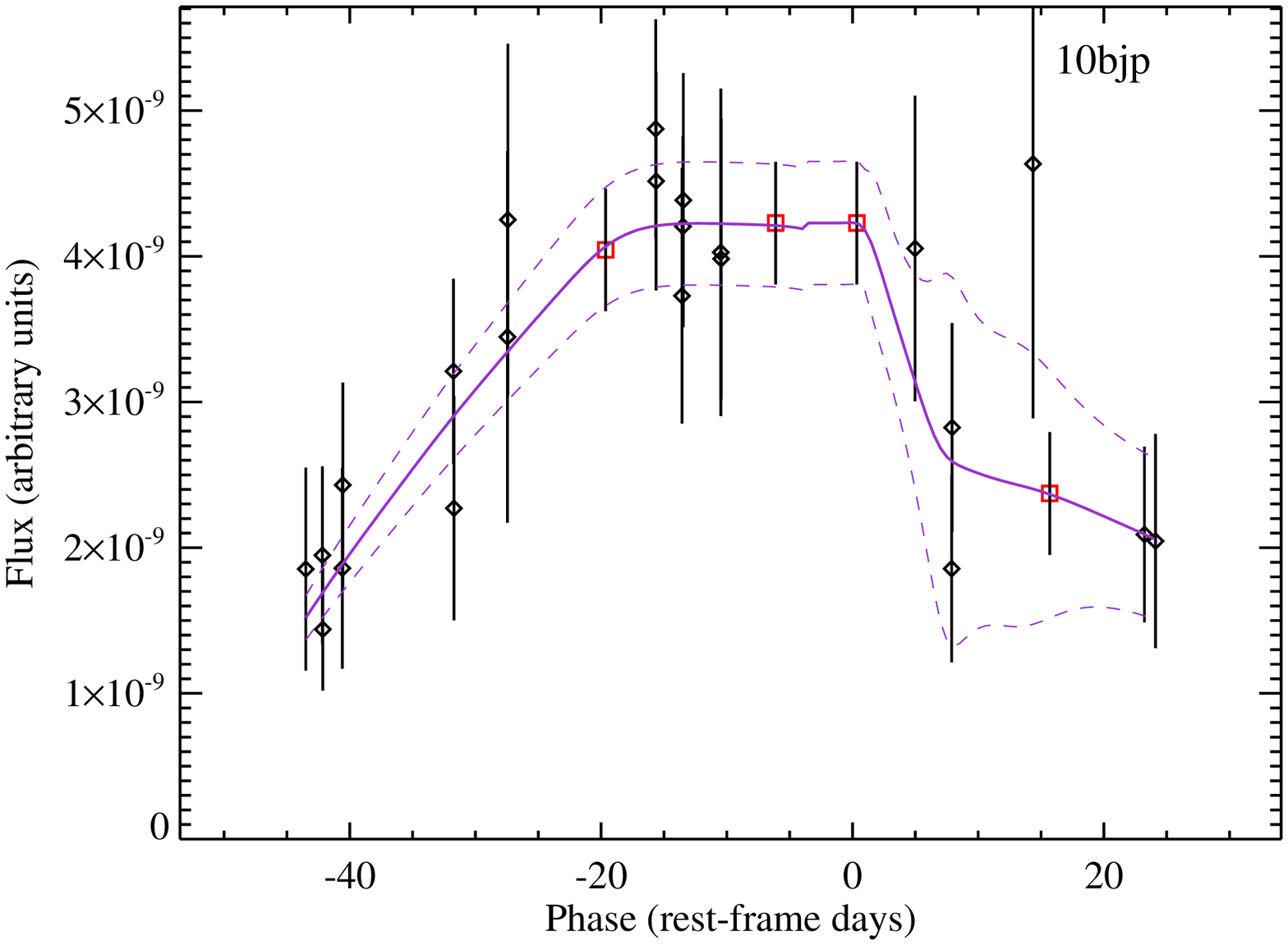}{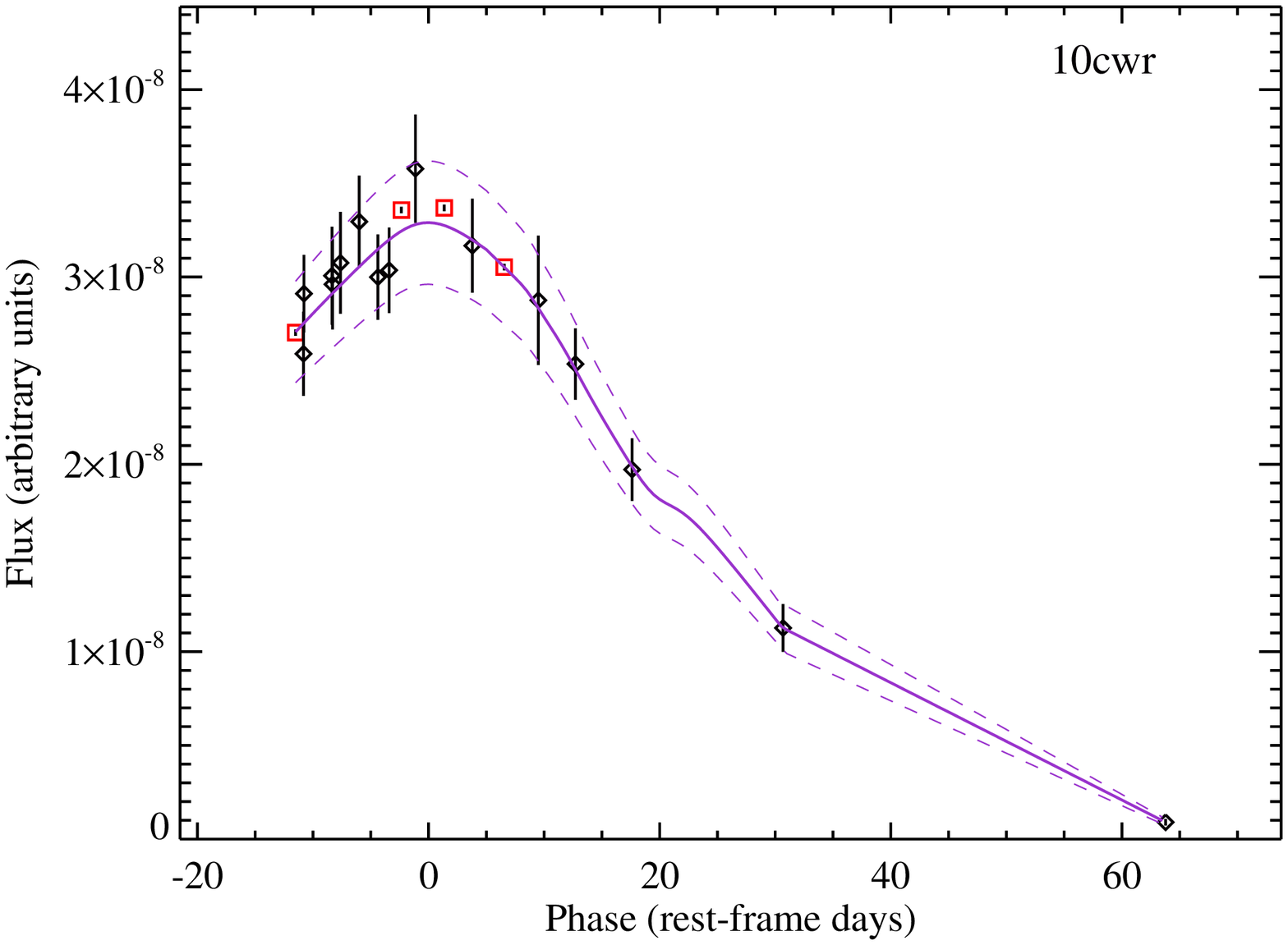}
\plottwo{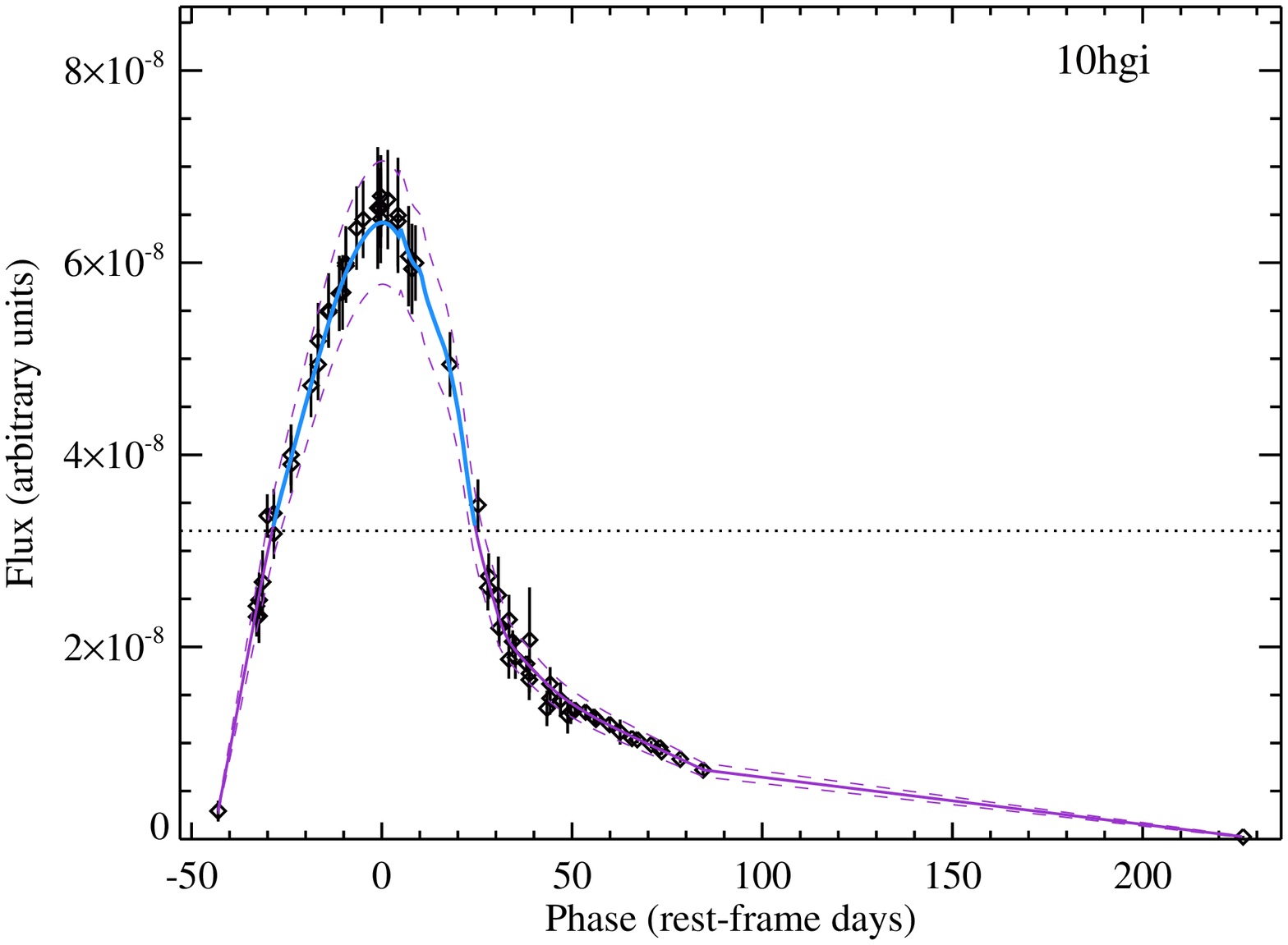}{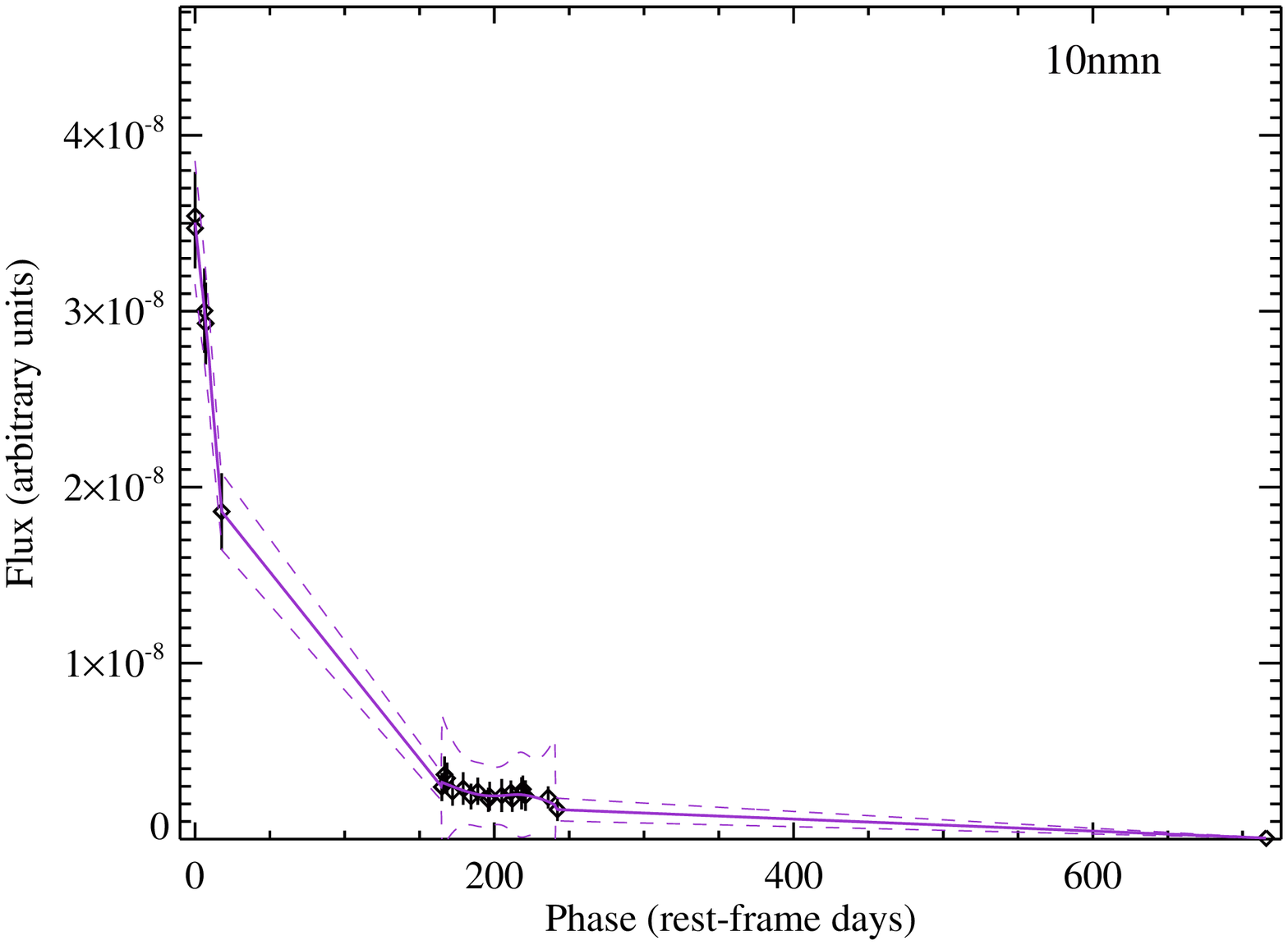}
\plottwo{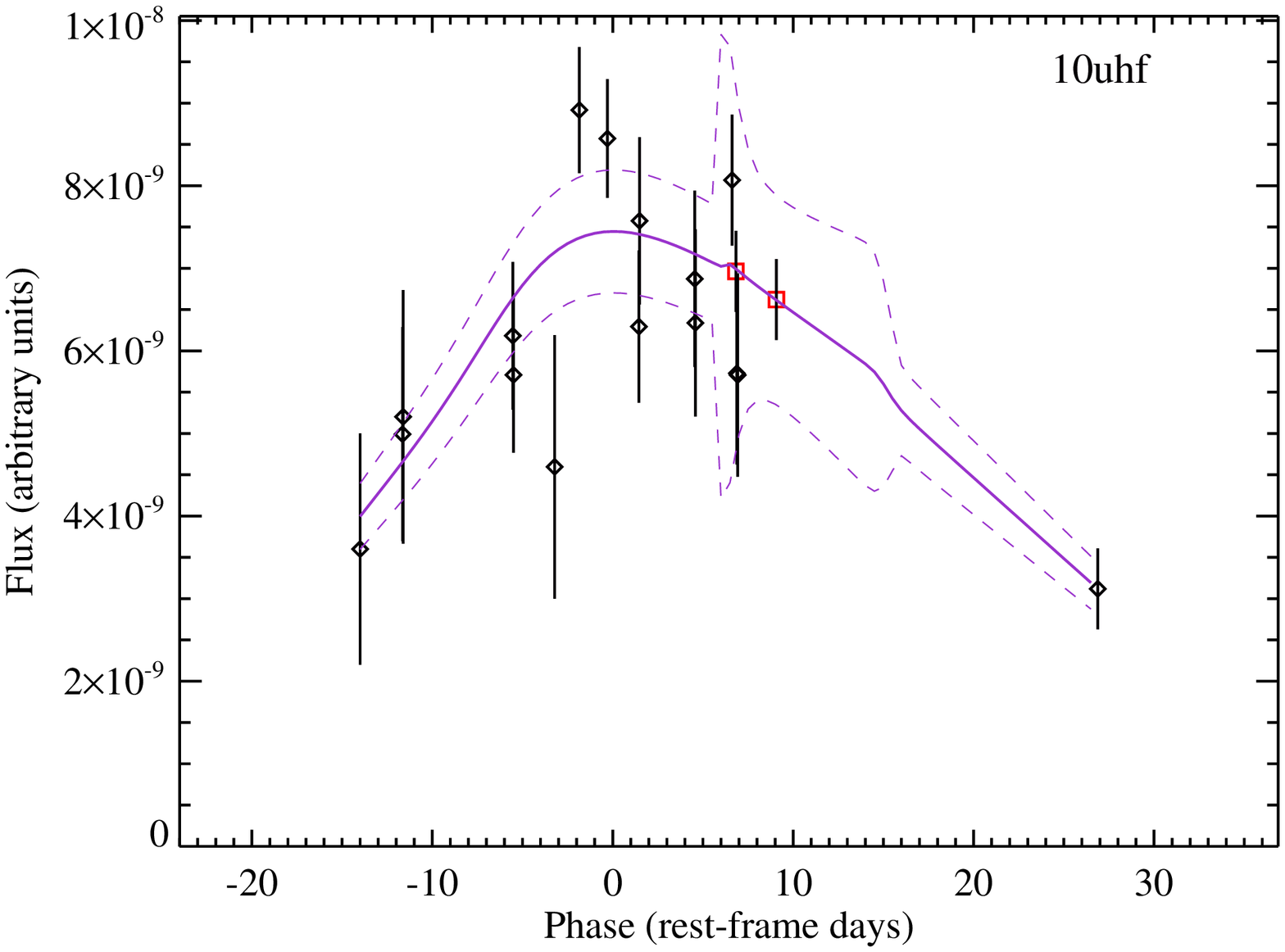}{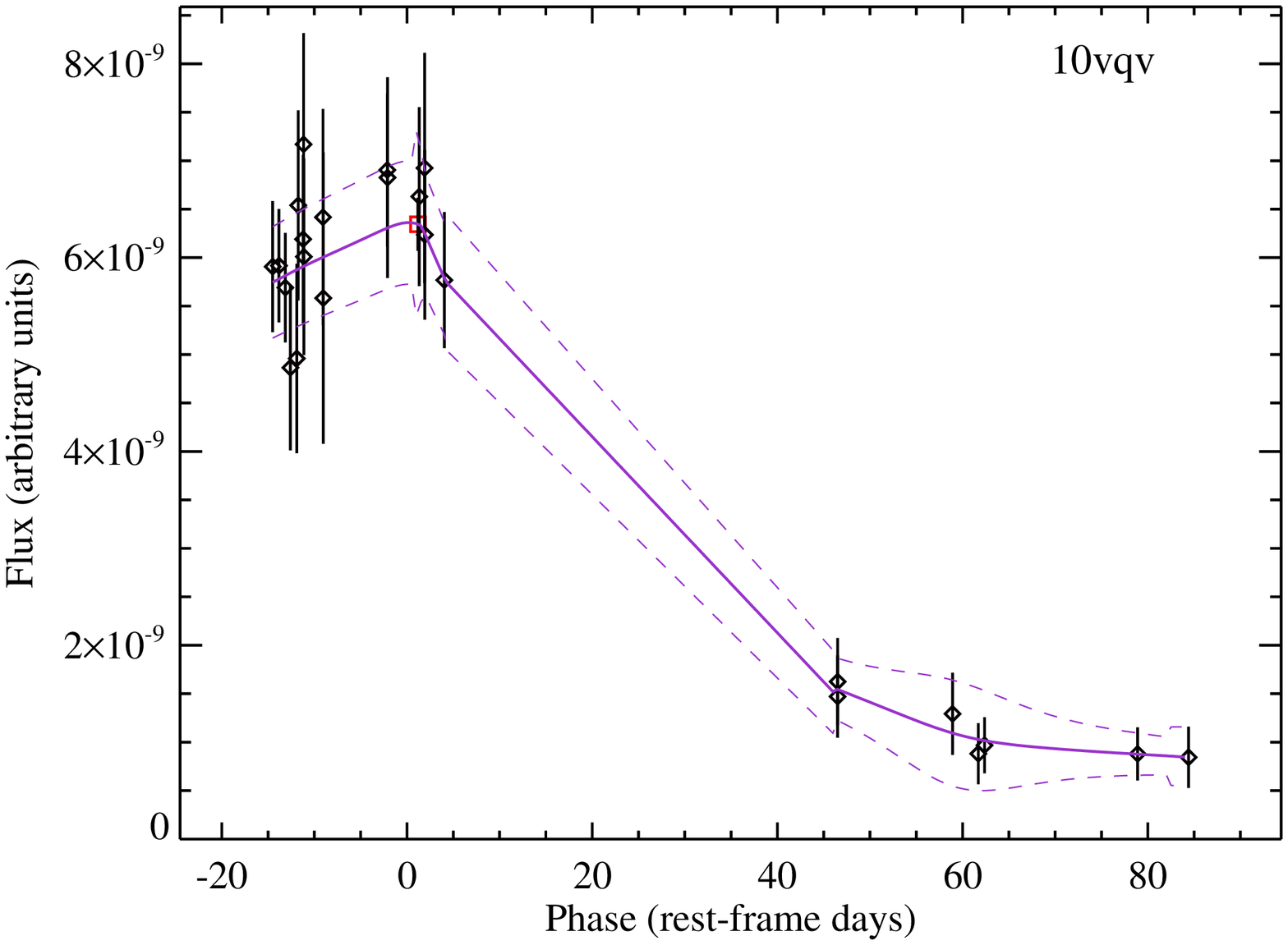}
\caption{Continuation of Fig.~\ref{fig ind smooth flux 1}. \label{fig ind smooth flux 2}}
\end{figure*}

\begin{figure*}[!h]
\epsscale{1.1}
\plottwo{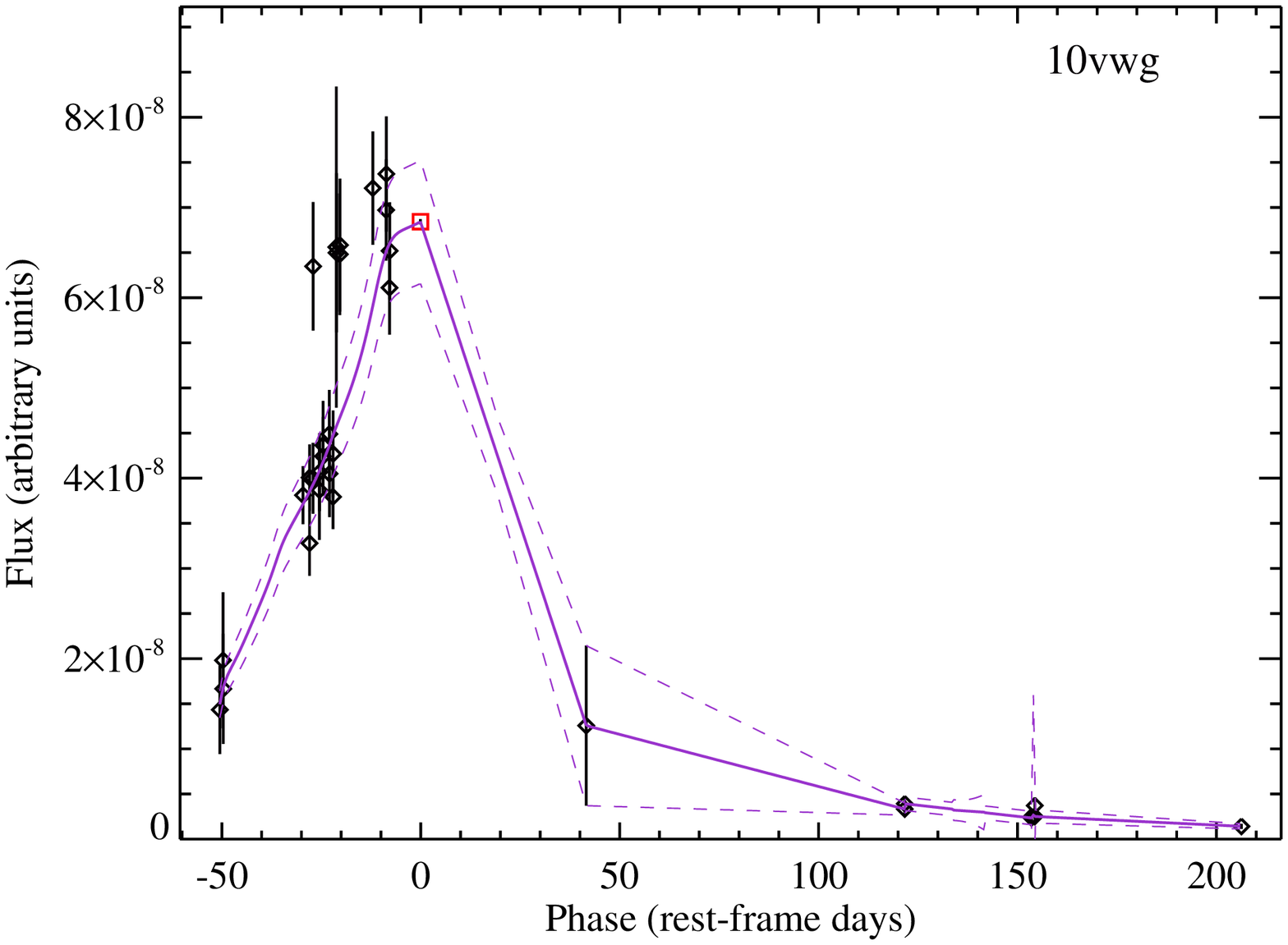}{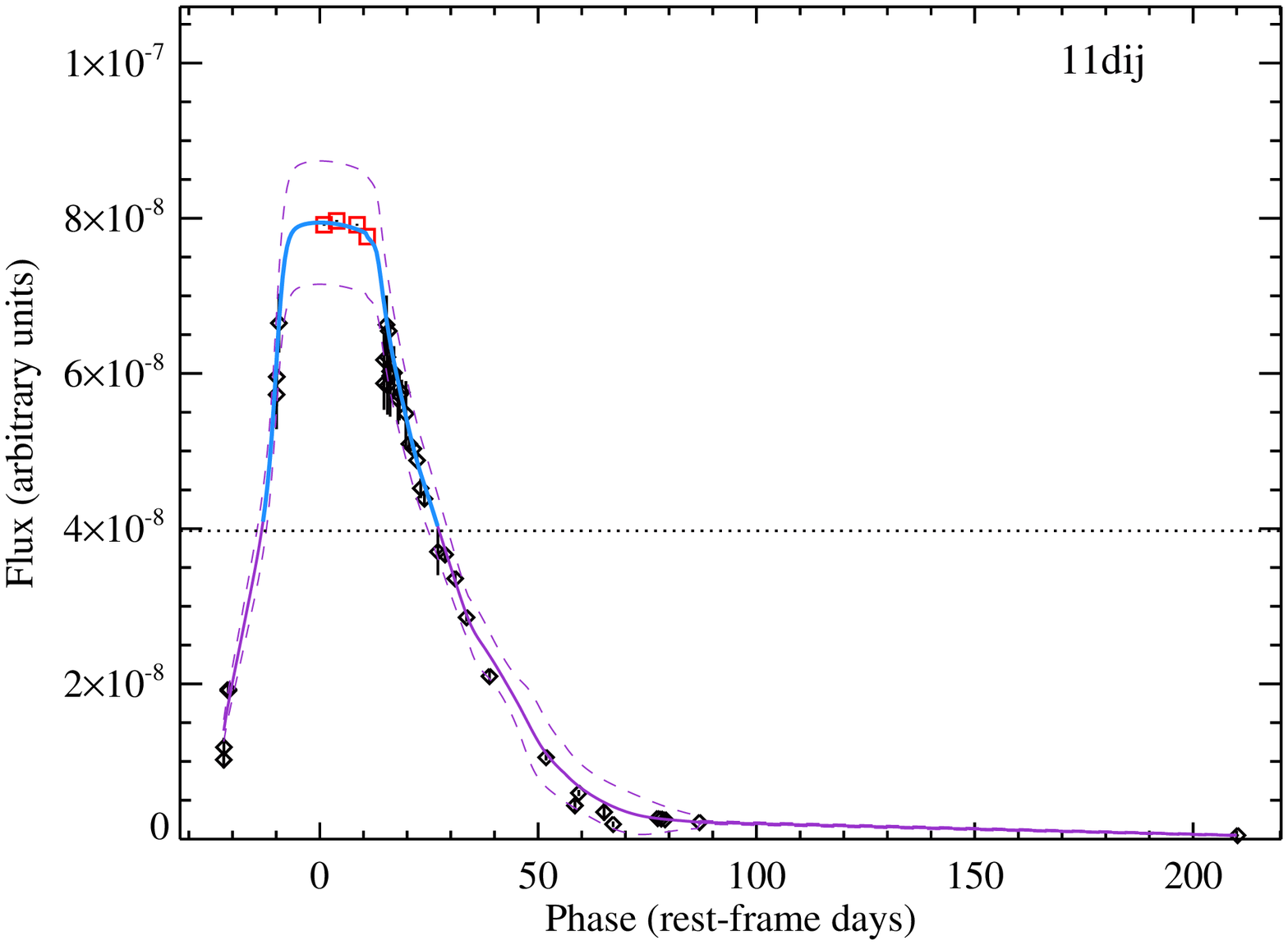}
\plottwo{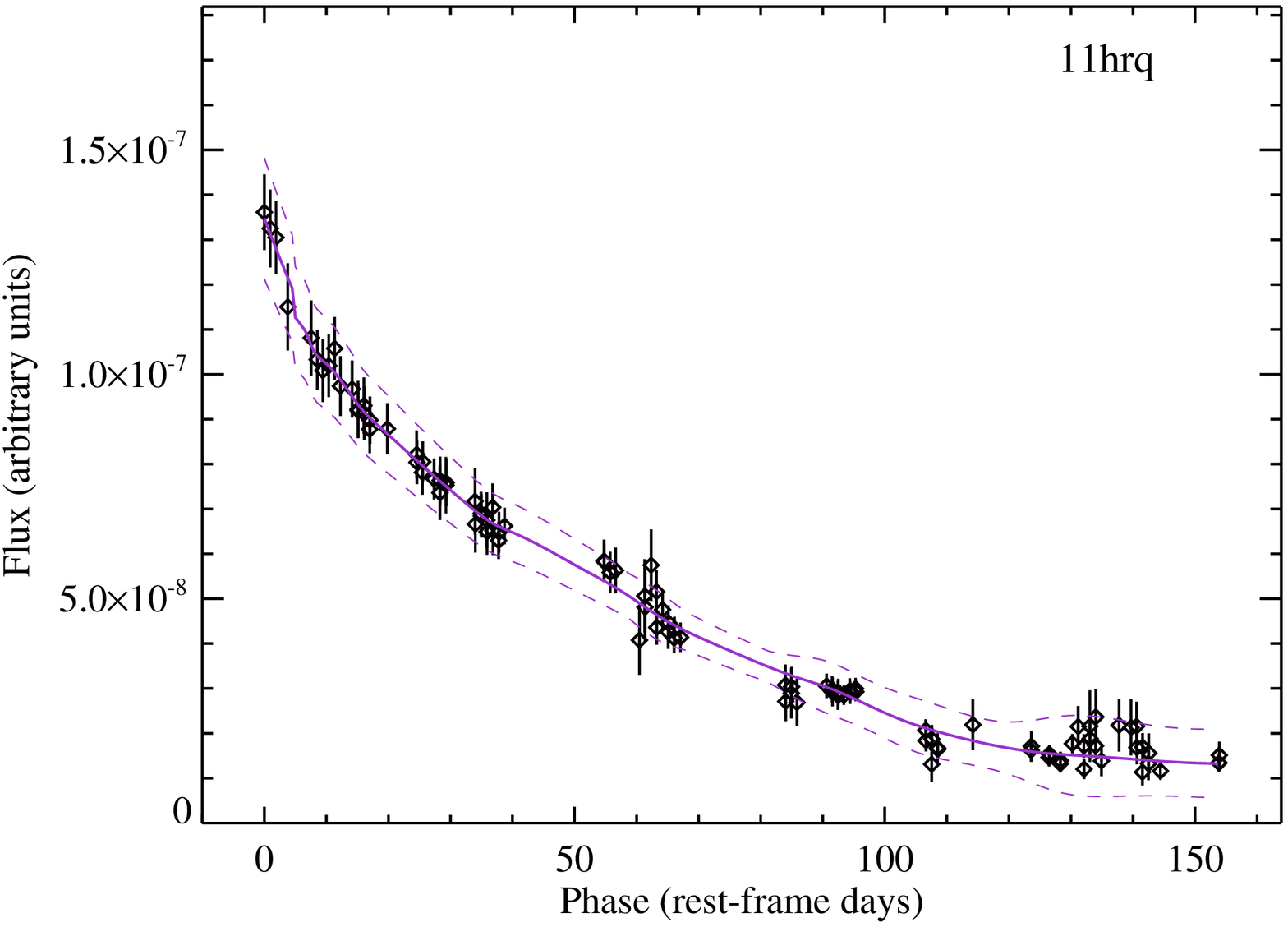}{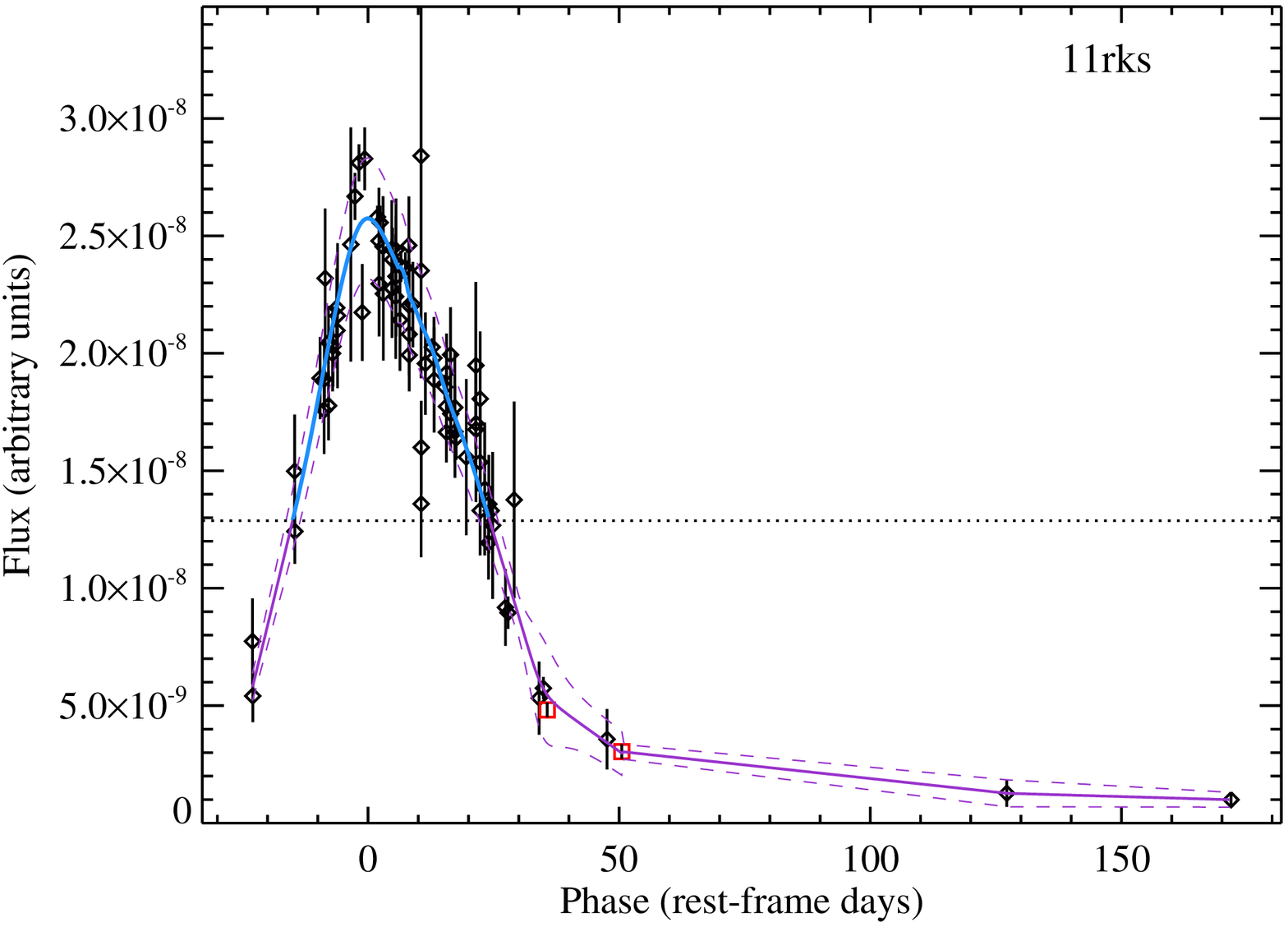}
\plottwo{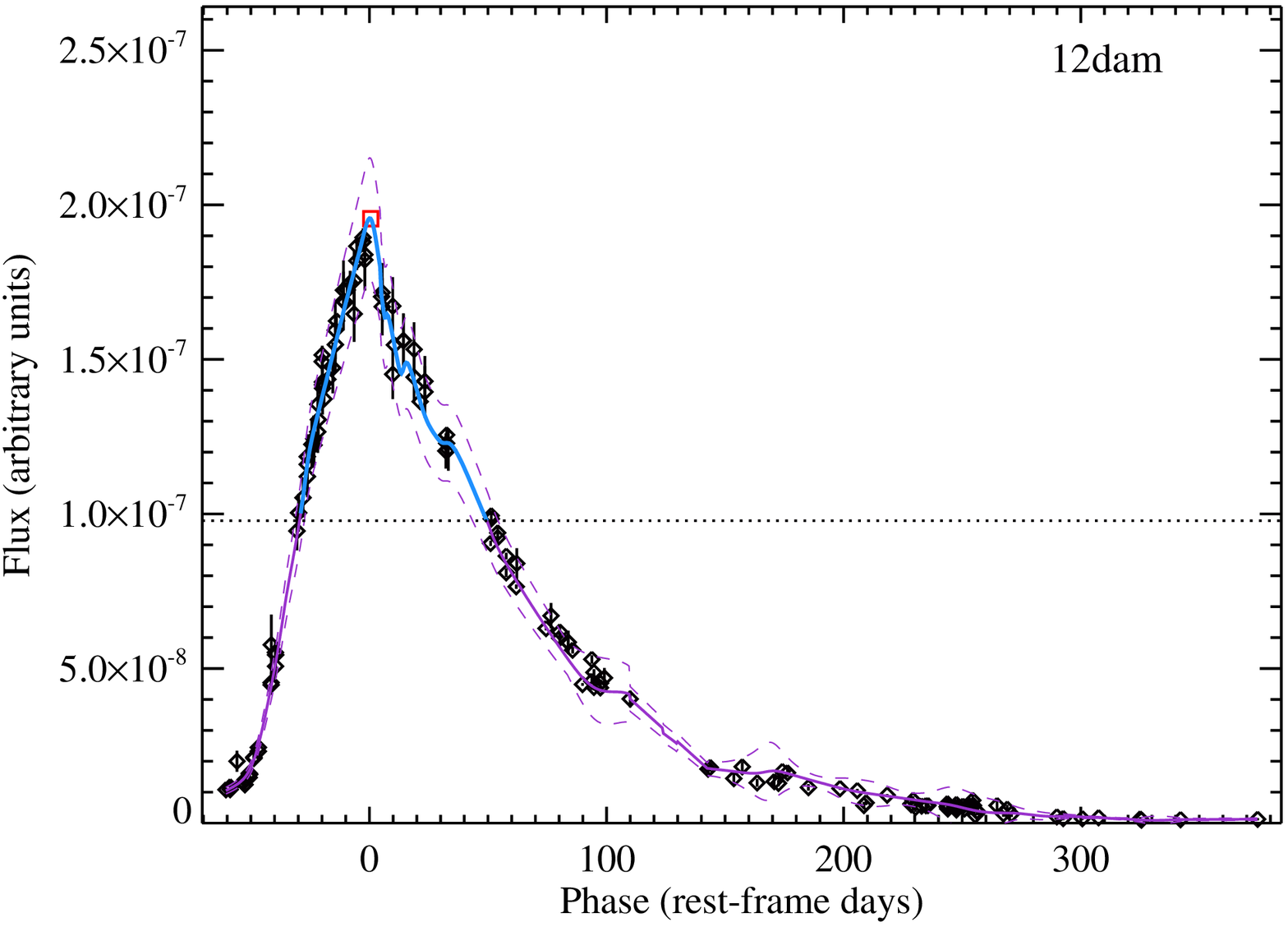}{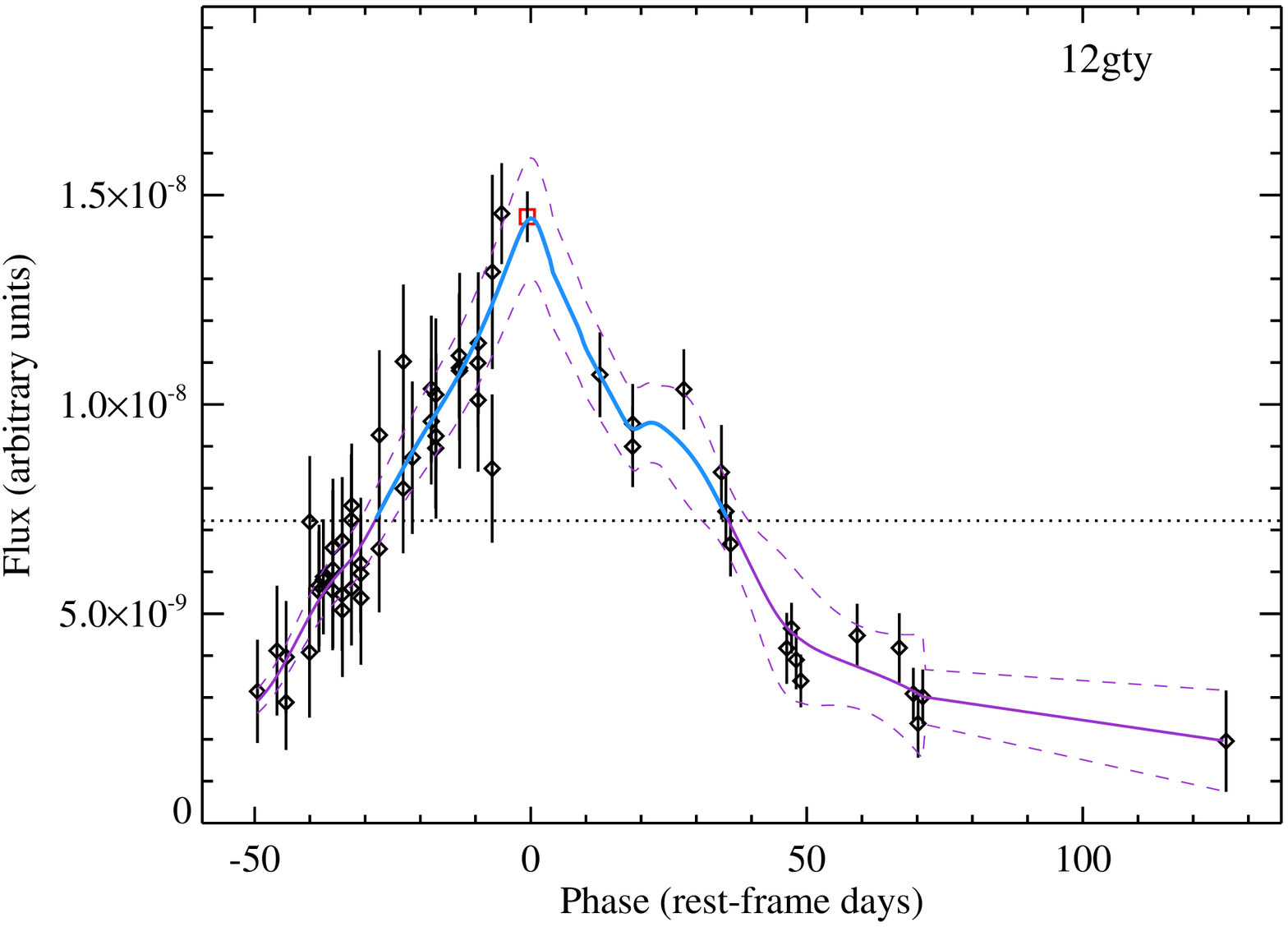}
\caption{Continuation of Fig.~\ref{fig ind smooth flux 1}. \label{fig ind smooth flux 3}}
\end{figure*}

\clearpage

\begin{figure}[!h]
\epsscale{1.1}
\plottwo{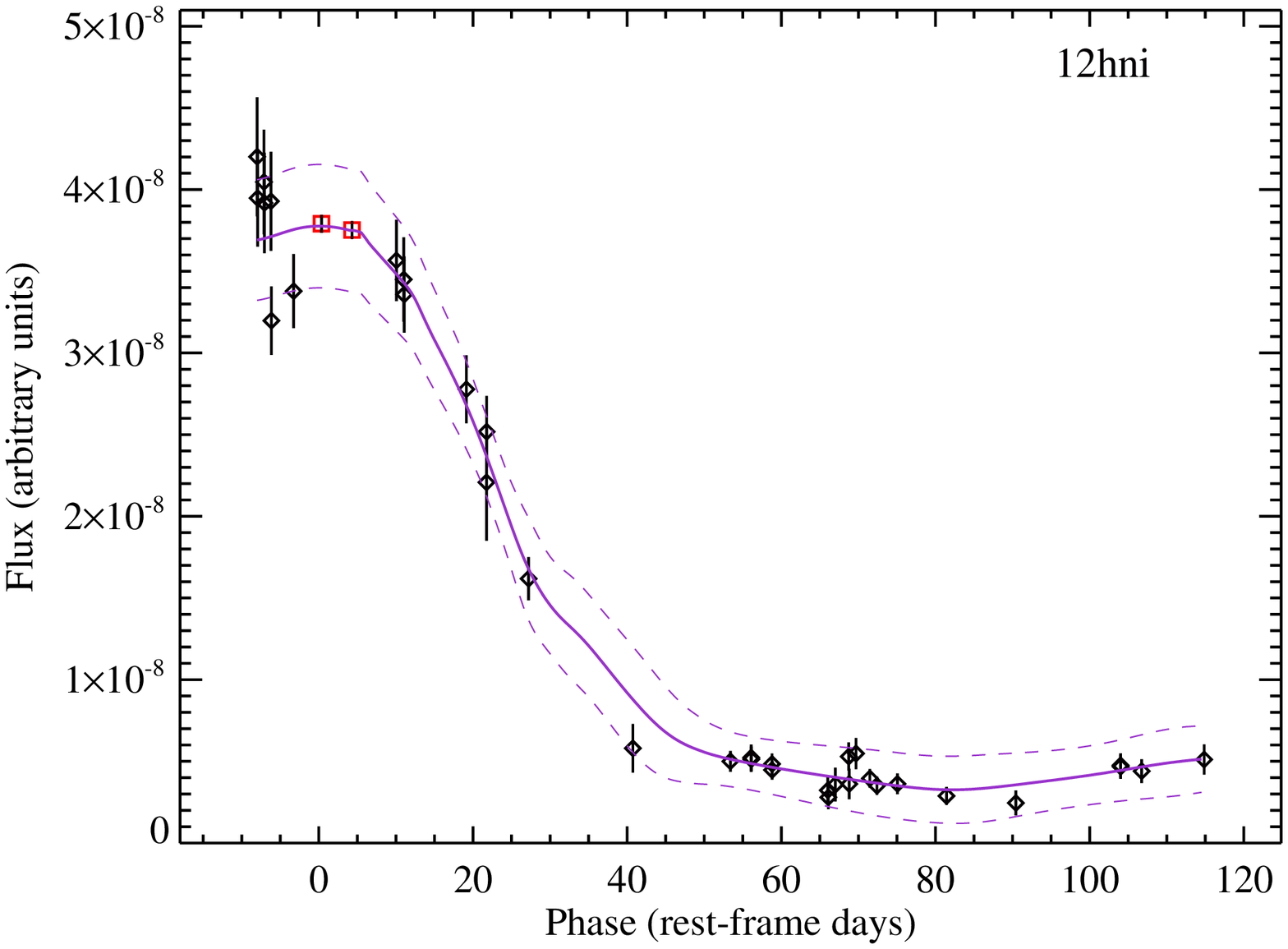}{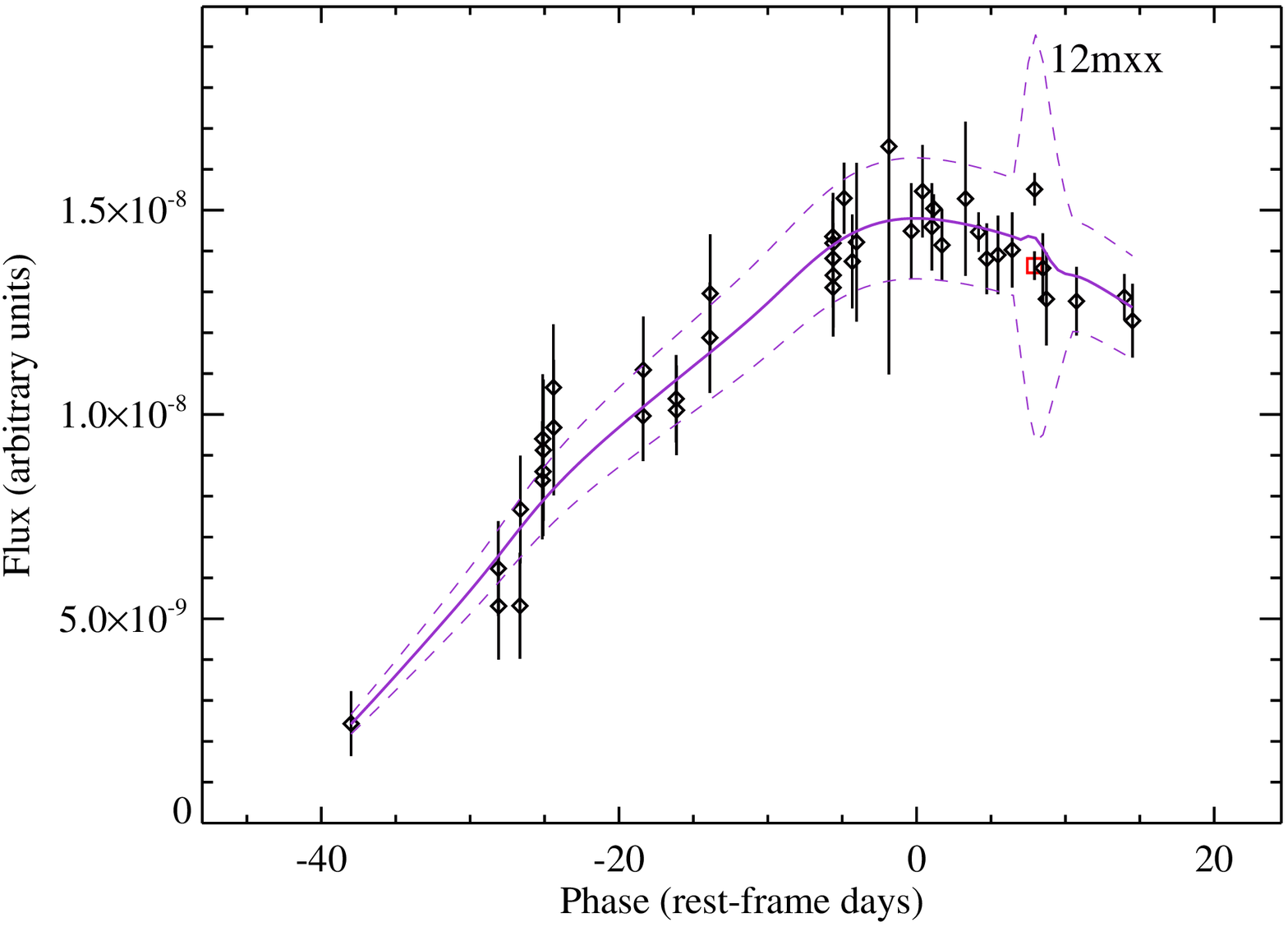}
\plottwo{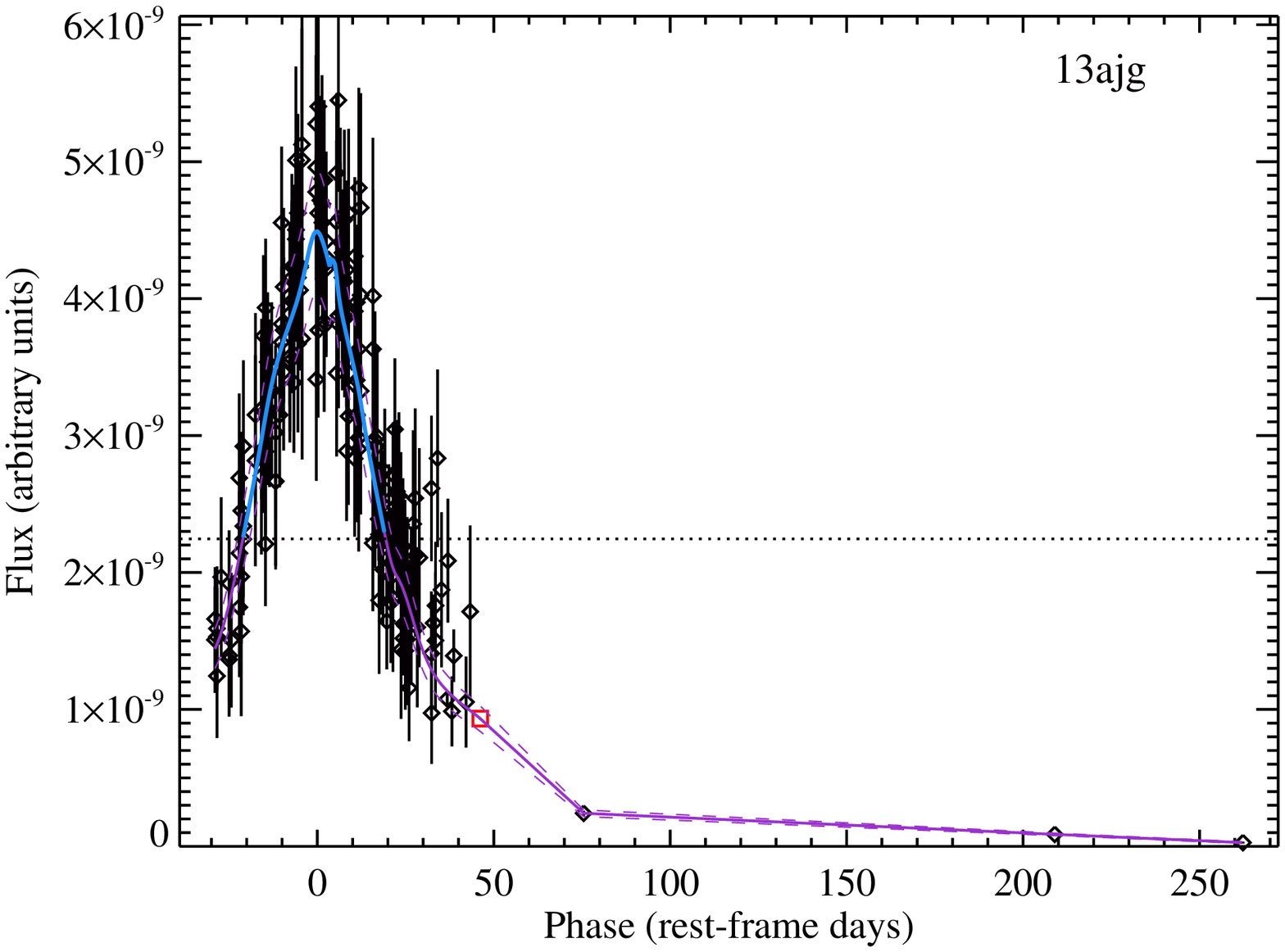}{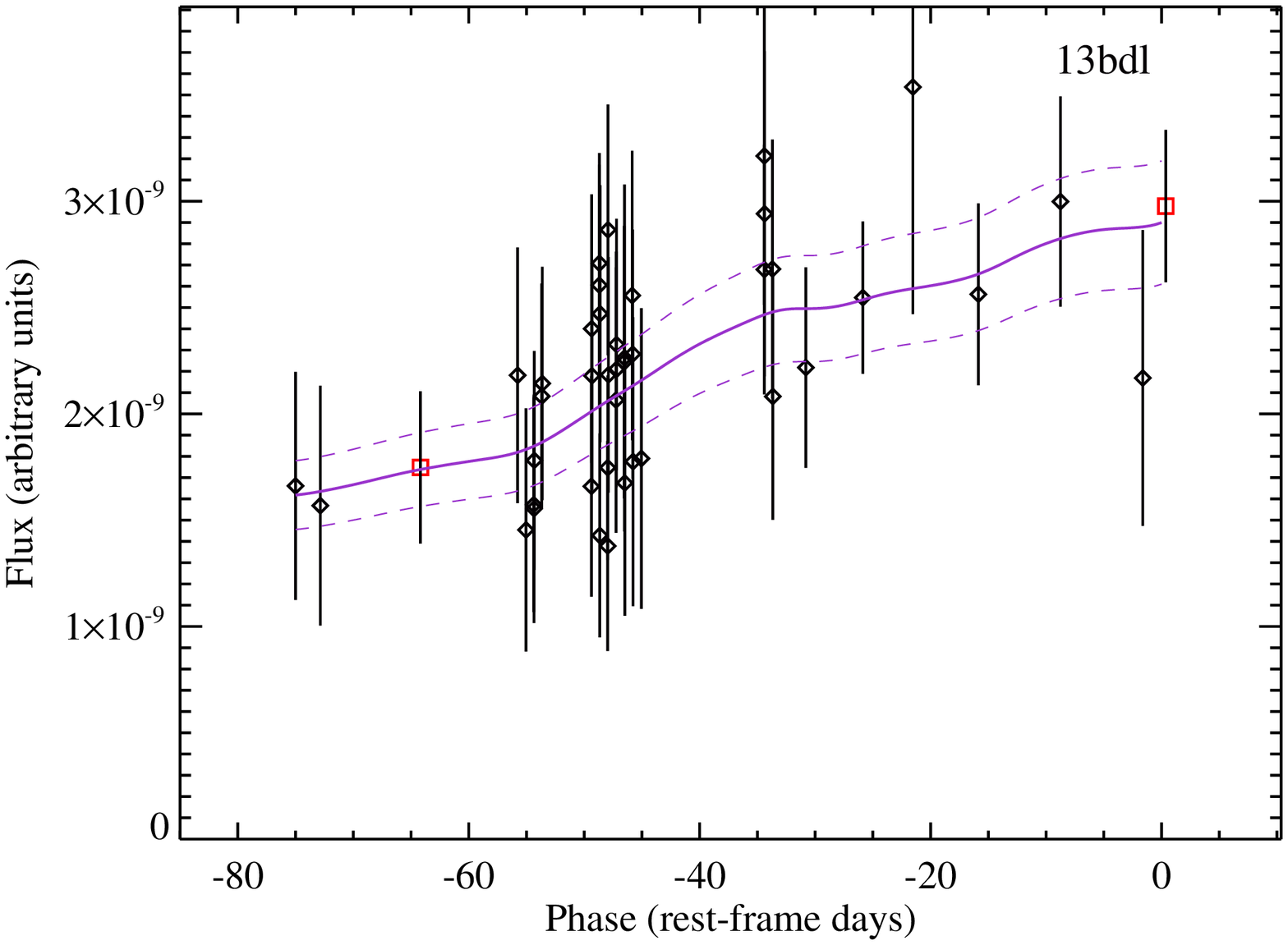}
\plottwo{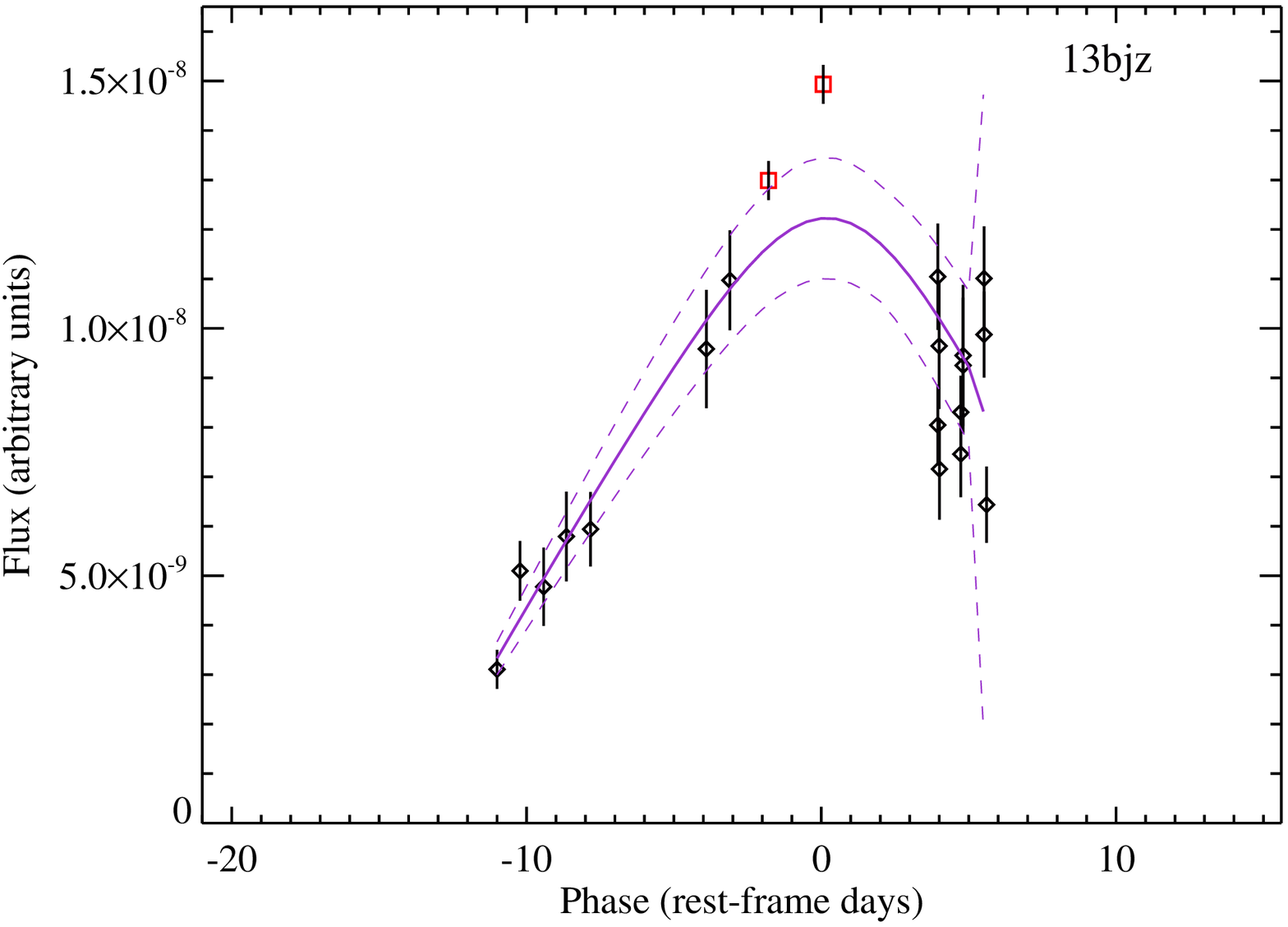}{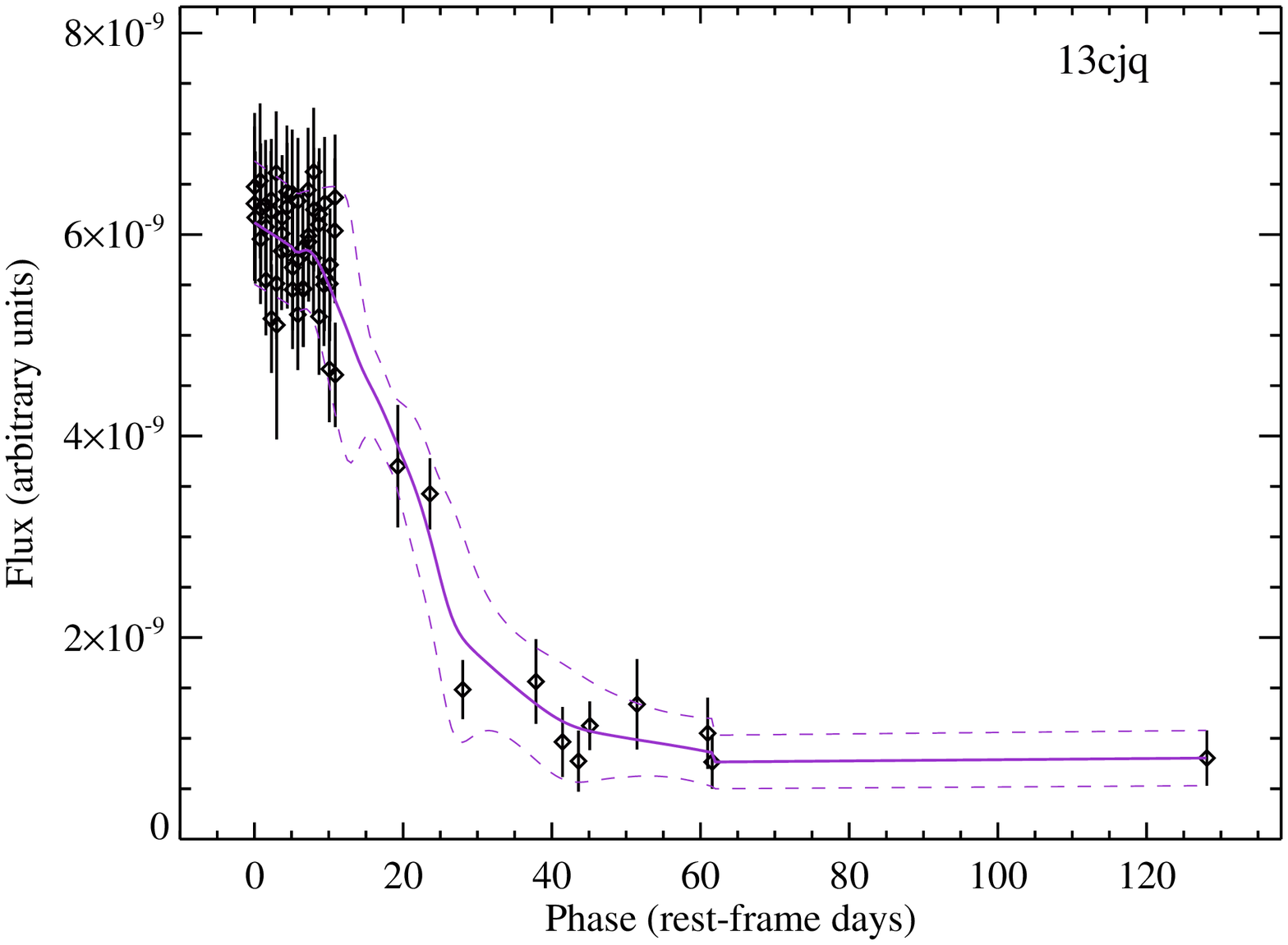}
\caption{Continuation of Fig.~\ref{fig ind smooth flux 1}. \label{fig ind smooth flux 4}}
\end{figure}

\clearpage

\begin{figure}[!h]
\epsscale{1.1}
\plottwo{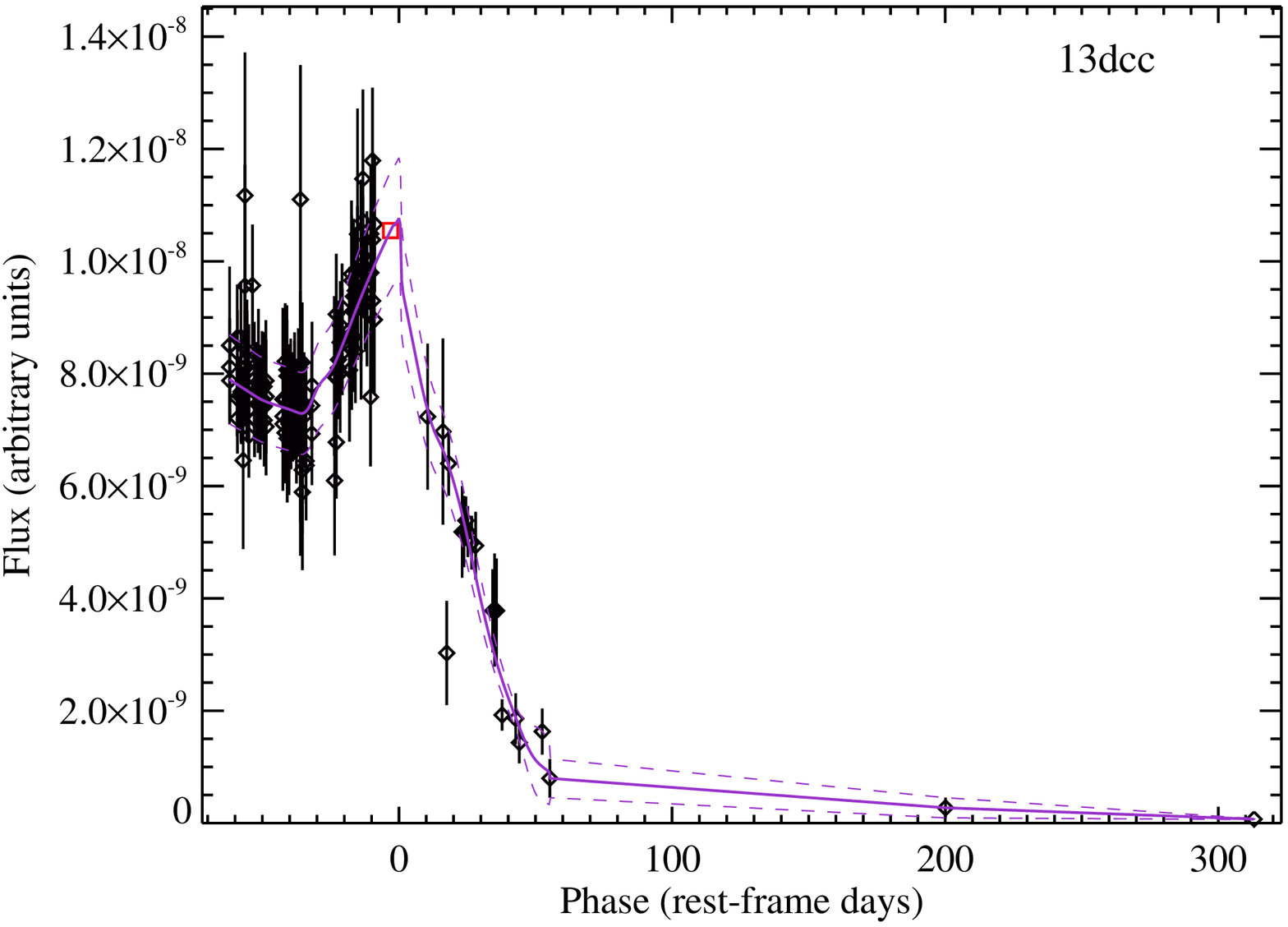}{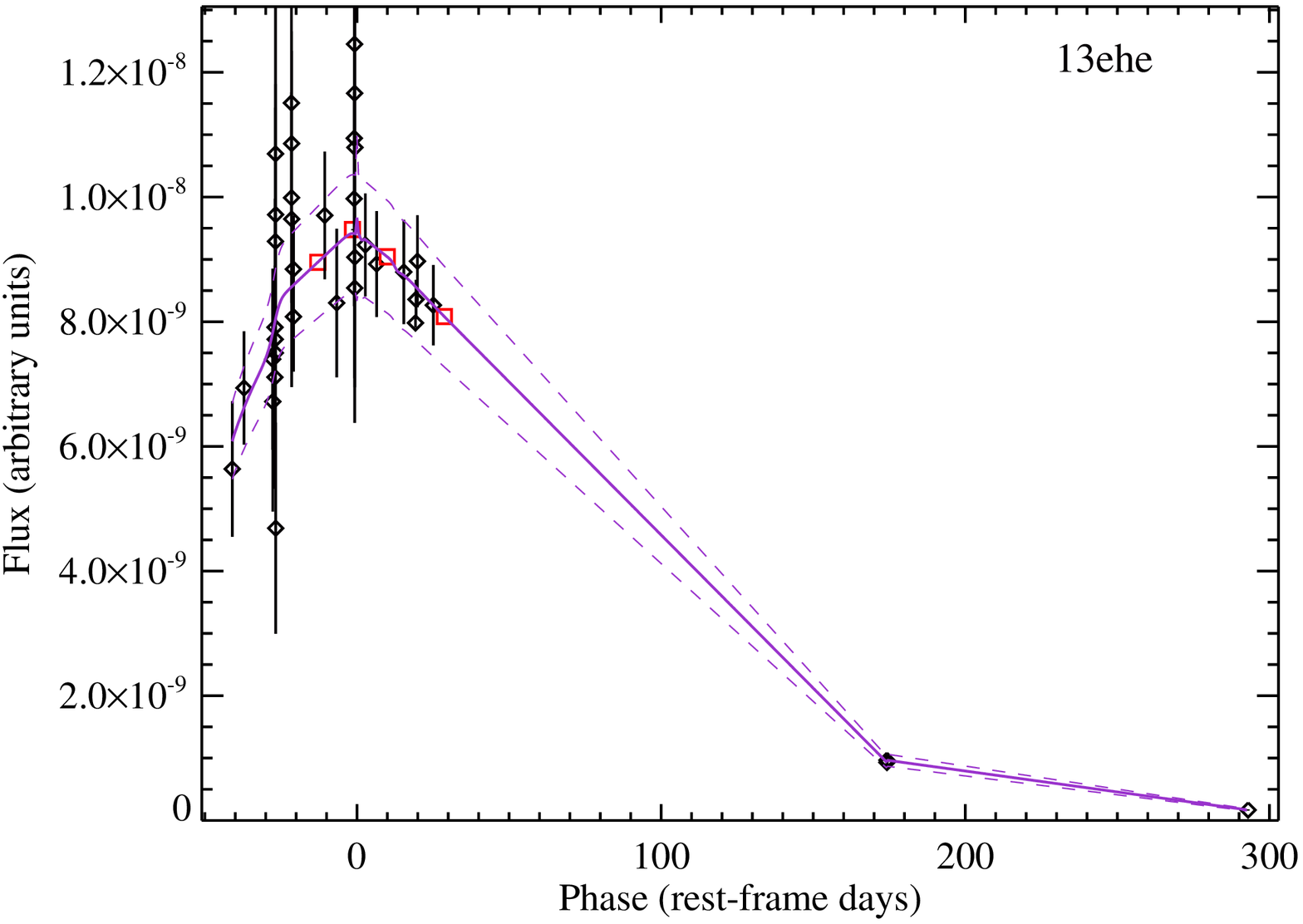}
\caption{Continuation of Fig.~\ref{fig ind smooth flux 1}. \label{fig ind smooth flux 5}}
\end{figure}


\begin{figure*}[!h]
\epsscale{1.1}
\plottwo{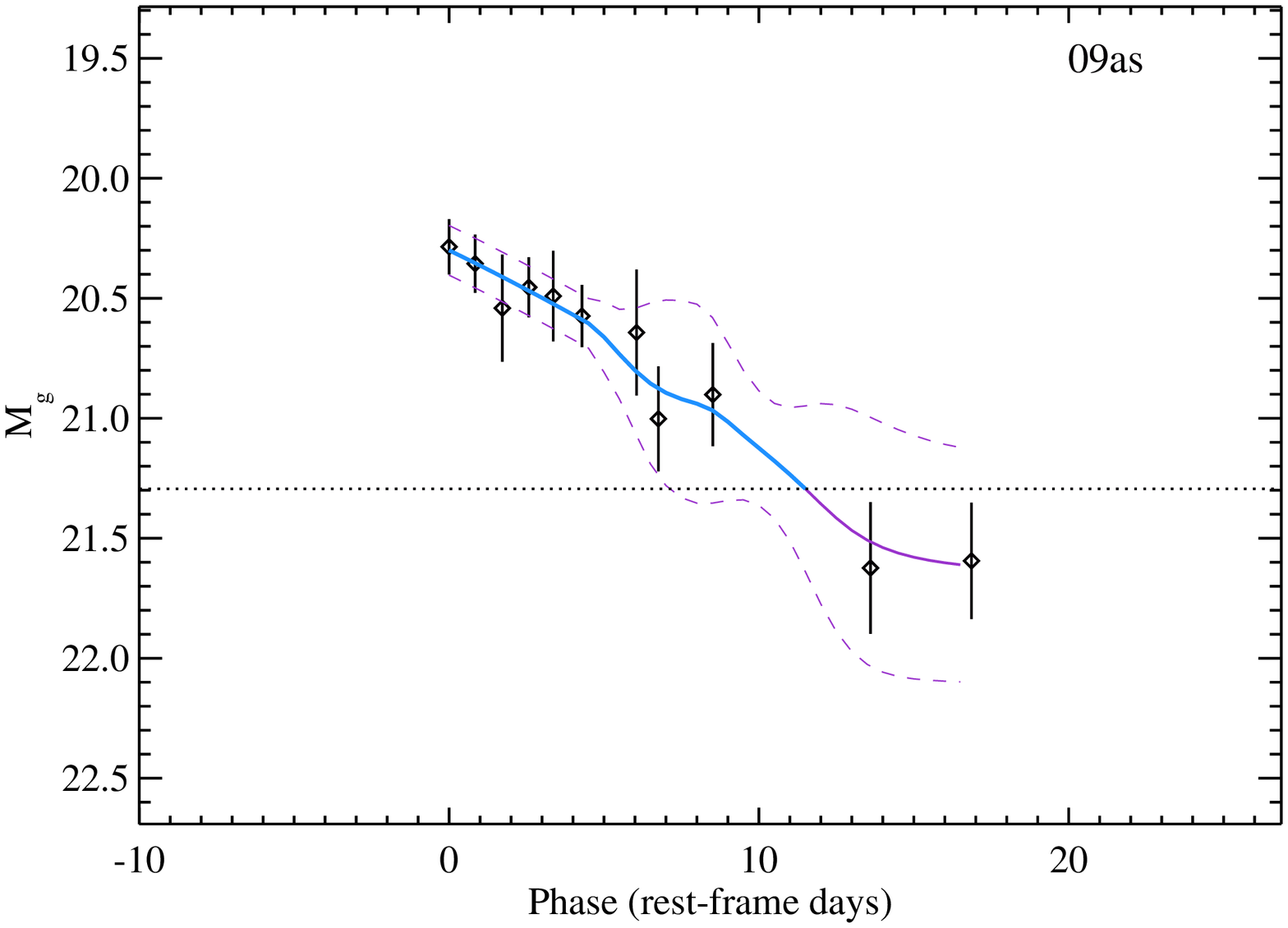}{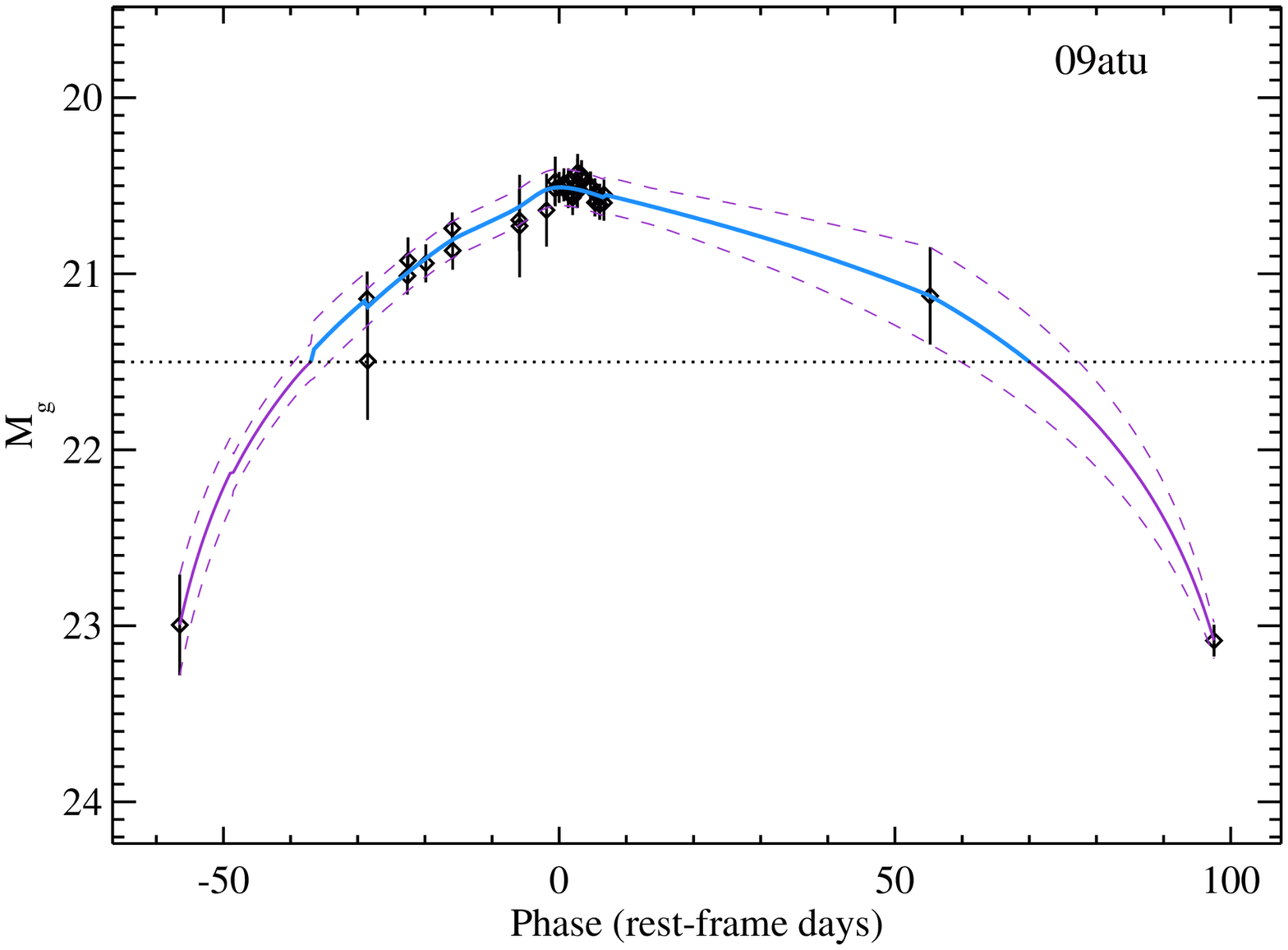}
\plottwo{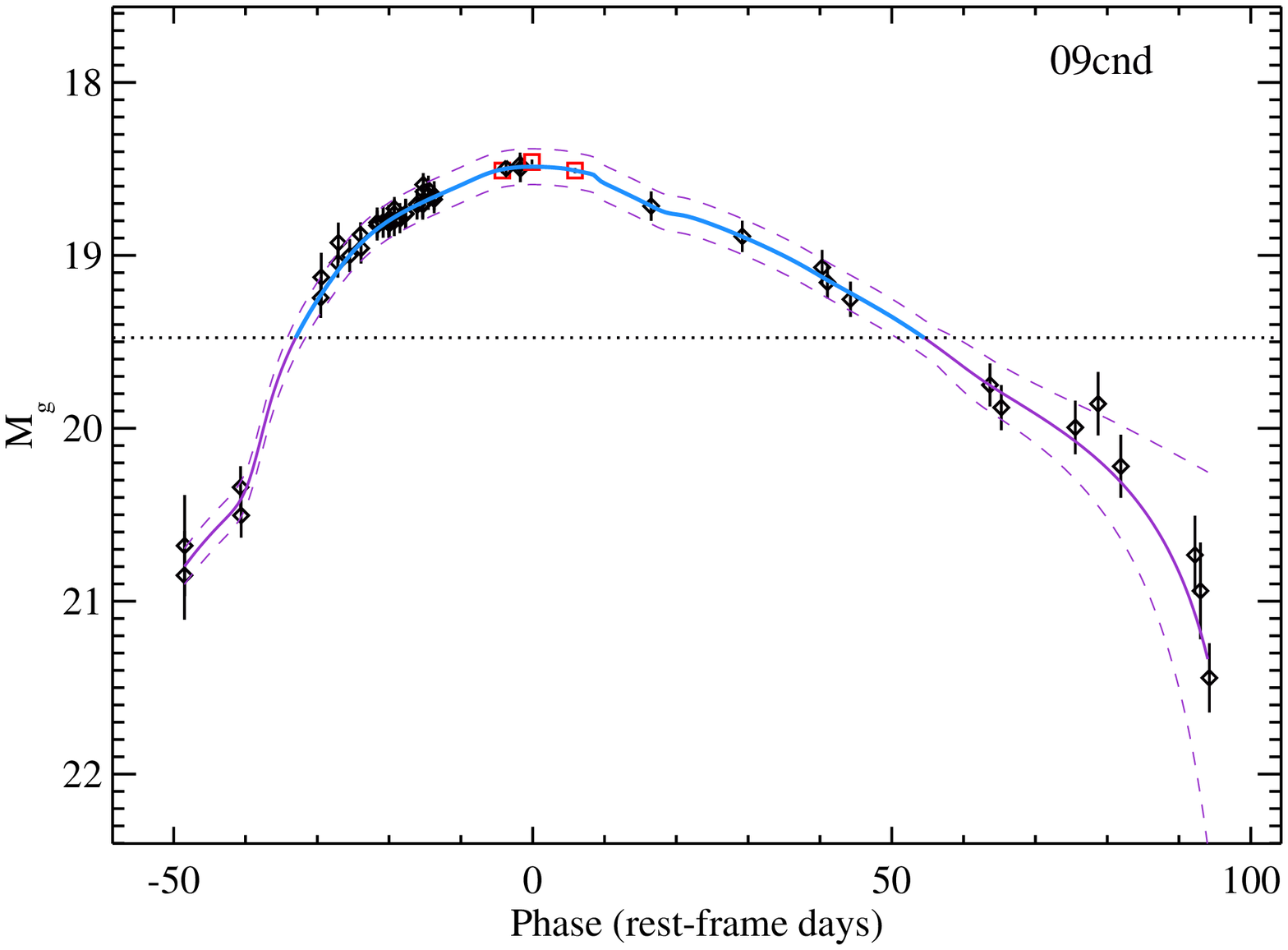}{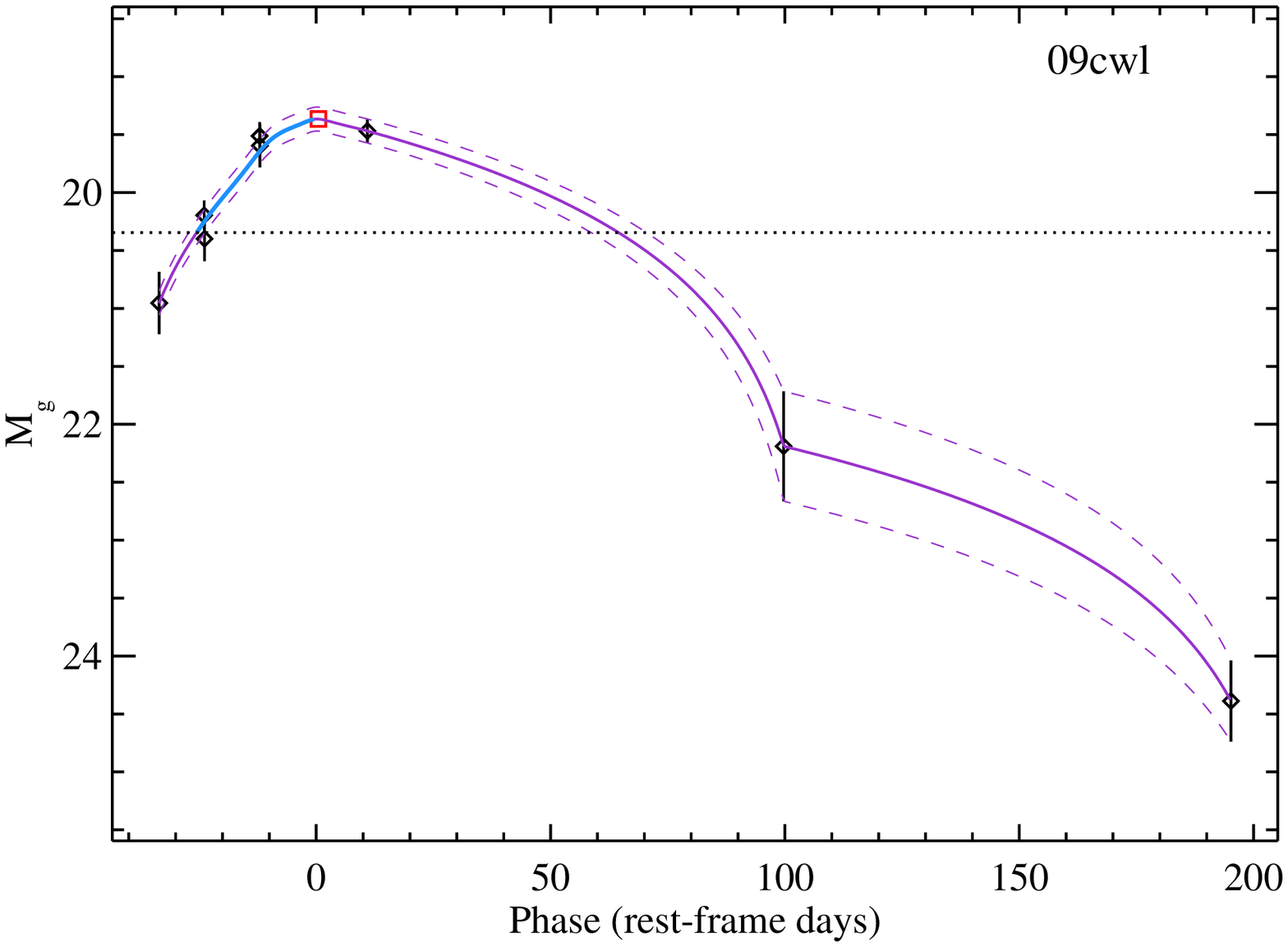}
\plottwo{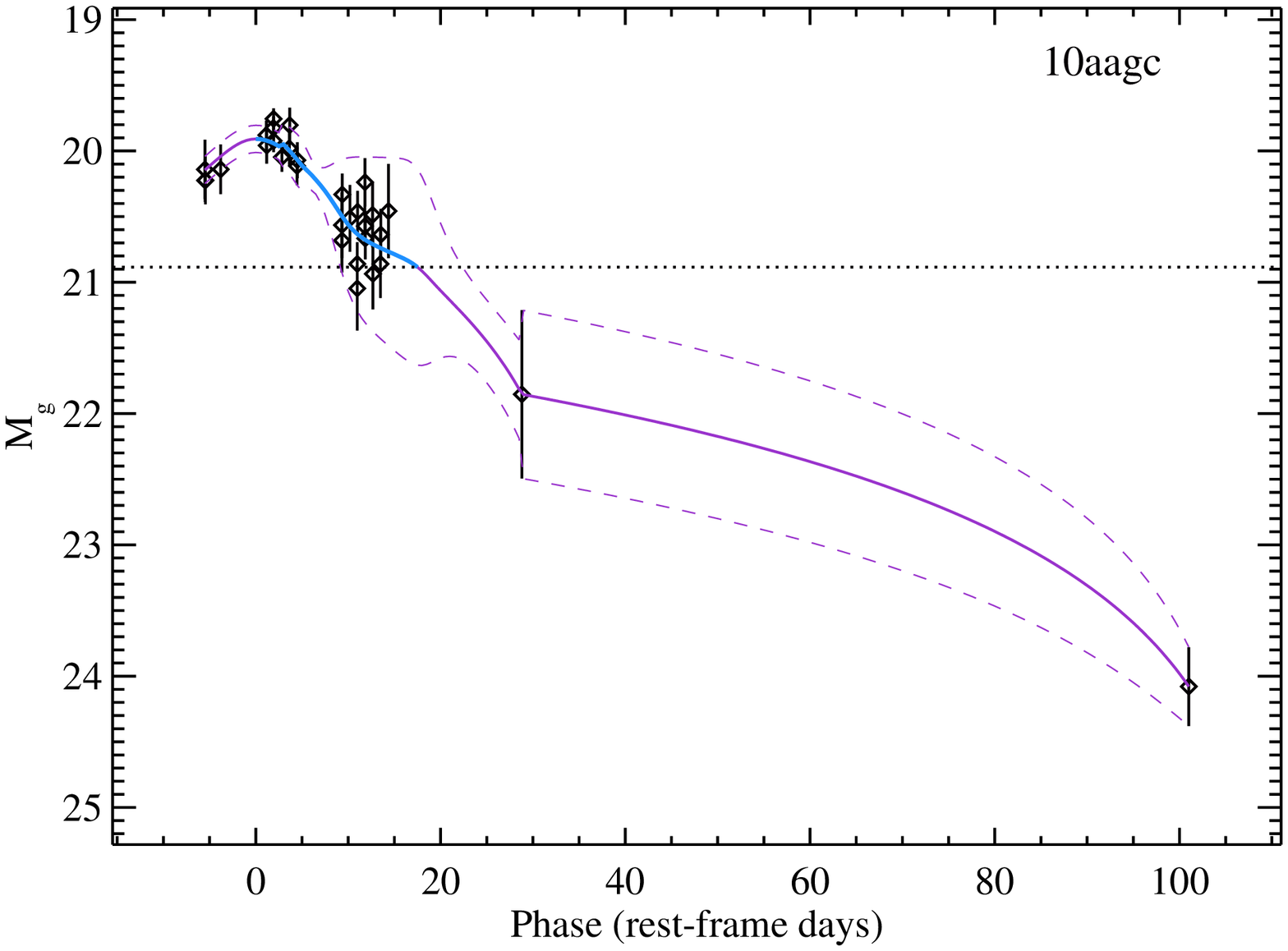}{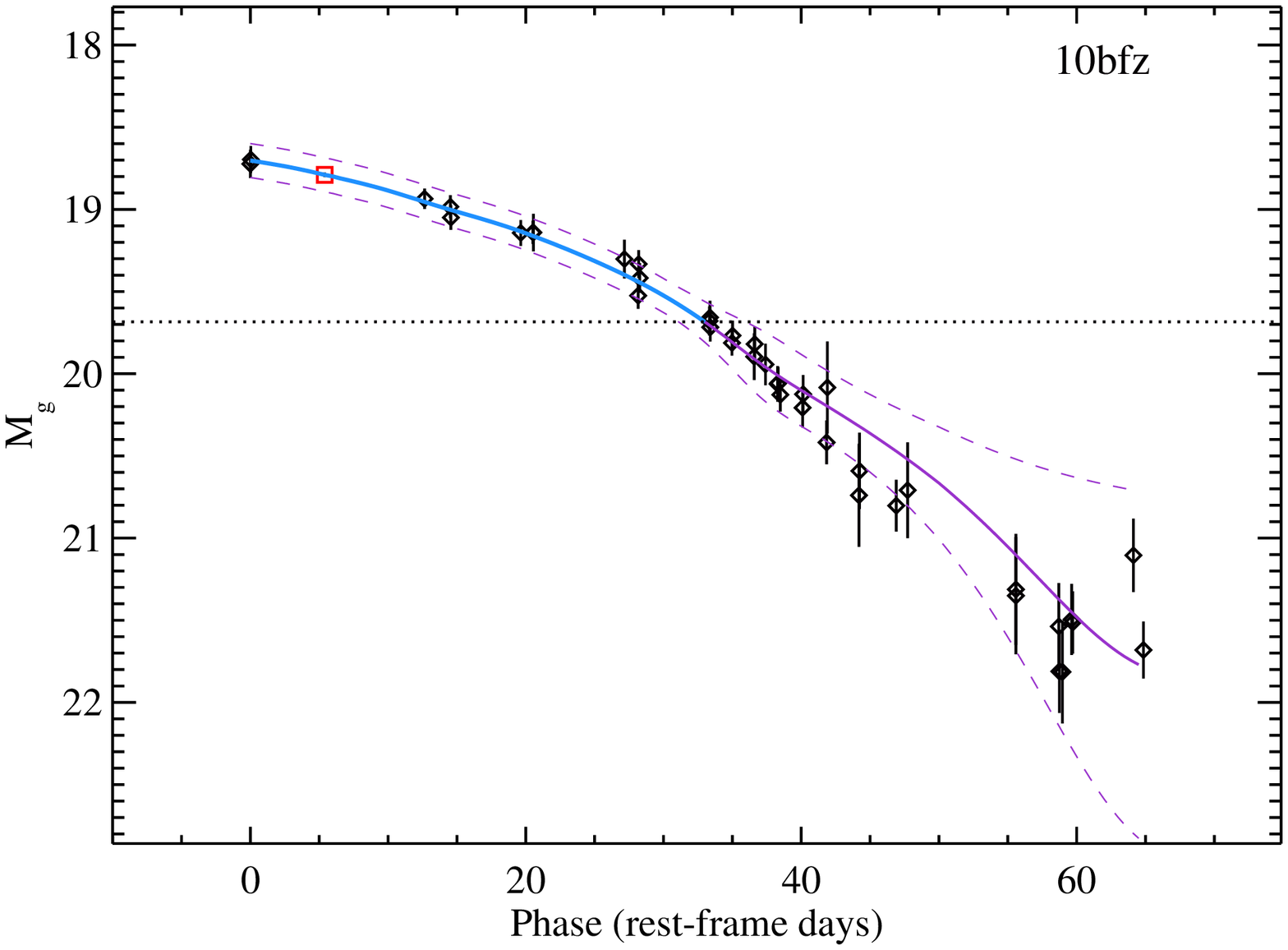}
\caption{Same as Fig.\ref{fig ind smooth flux 1}, but for magnitudes. The horizontal dotted lines mark here the distance of 1 mag from peak, which is used to calculate the $t^{\rm \Delta1mag}_{\rm rise}$ and $t^{\rm \Delta1mag}_{\rm fall}$ timescales, when the light curves are well-characterized above this threshold (highlighted blue solid curves).  \label{fig ind smooth mag 1}}
\end{figure*}

\begin{figure*}[!h]
\epsscale{1.1}
\plottwo{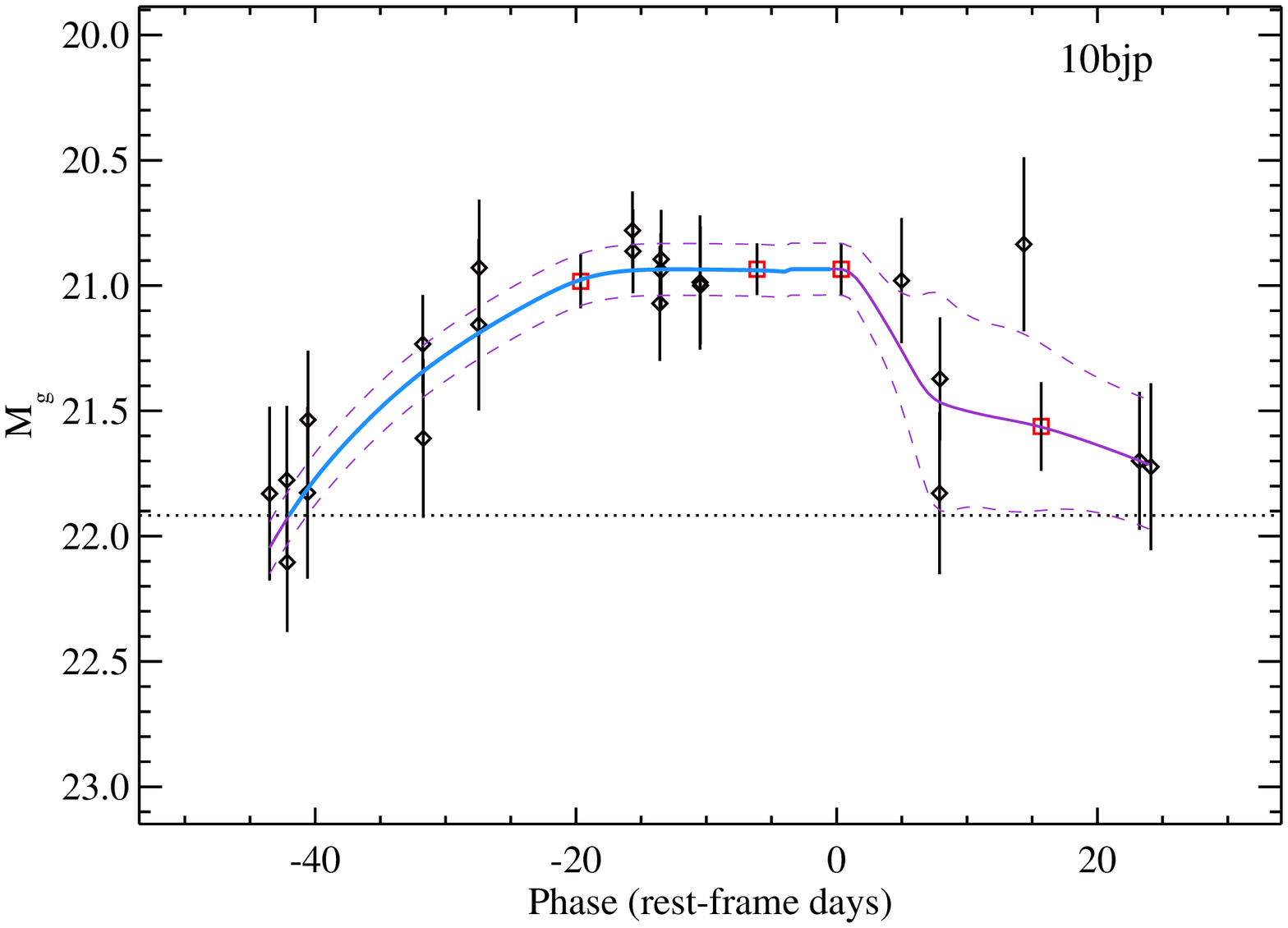}{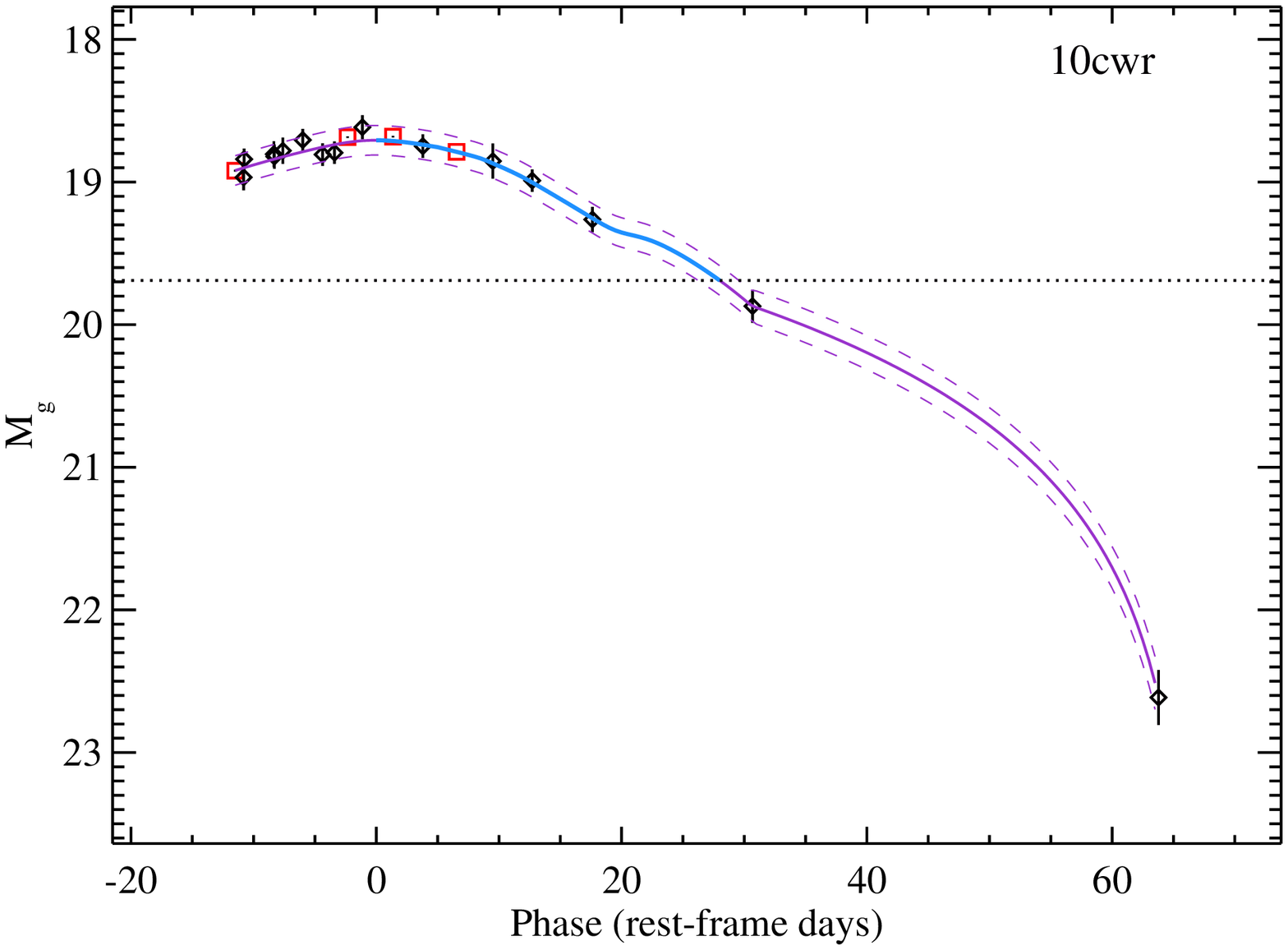}
\plottwo{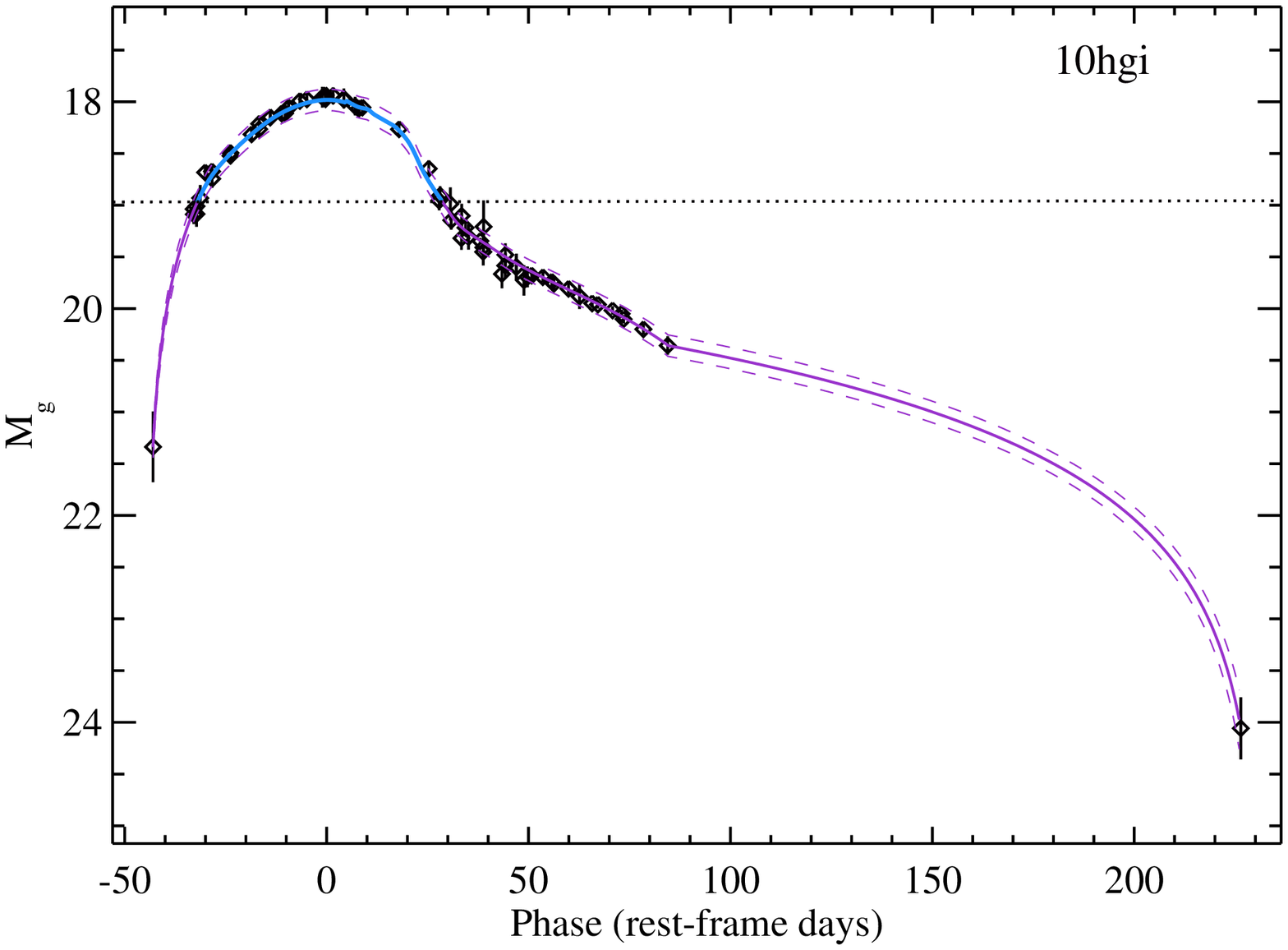}{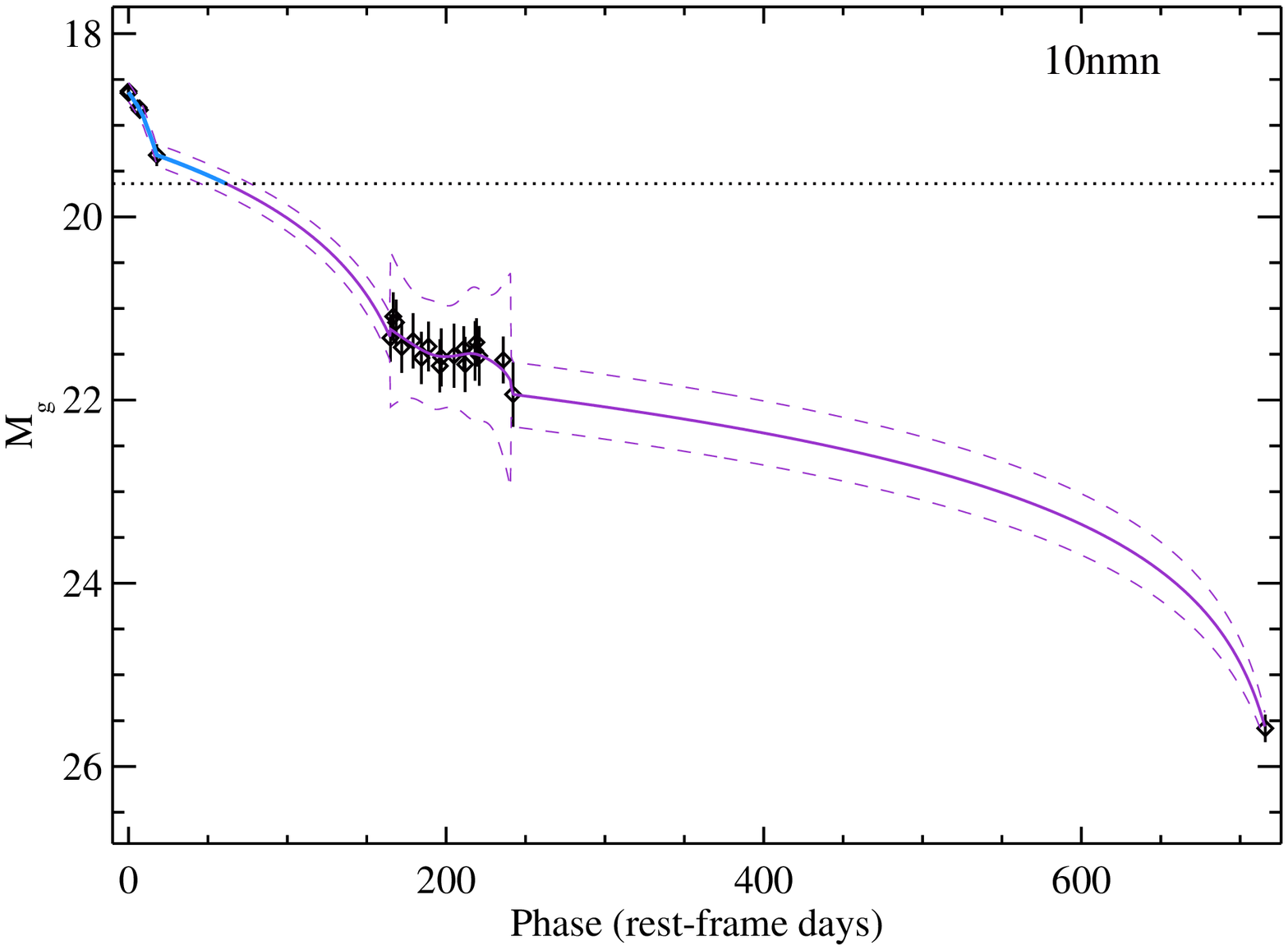}
\plottwo{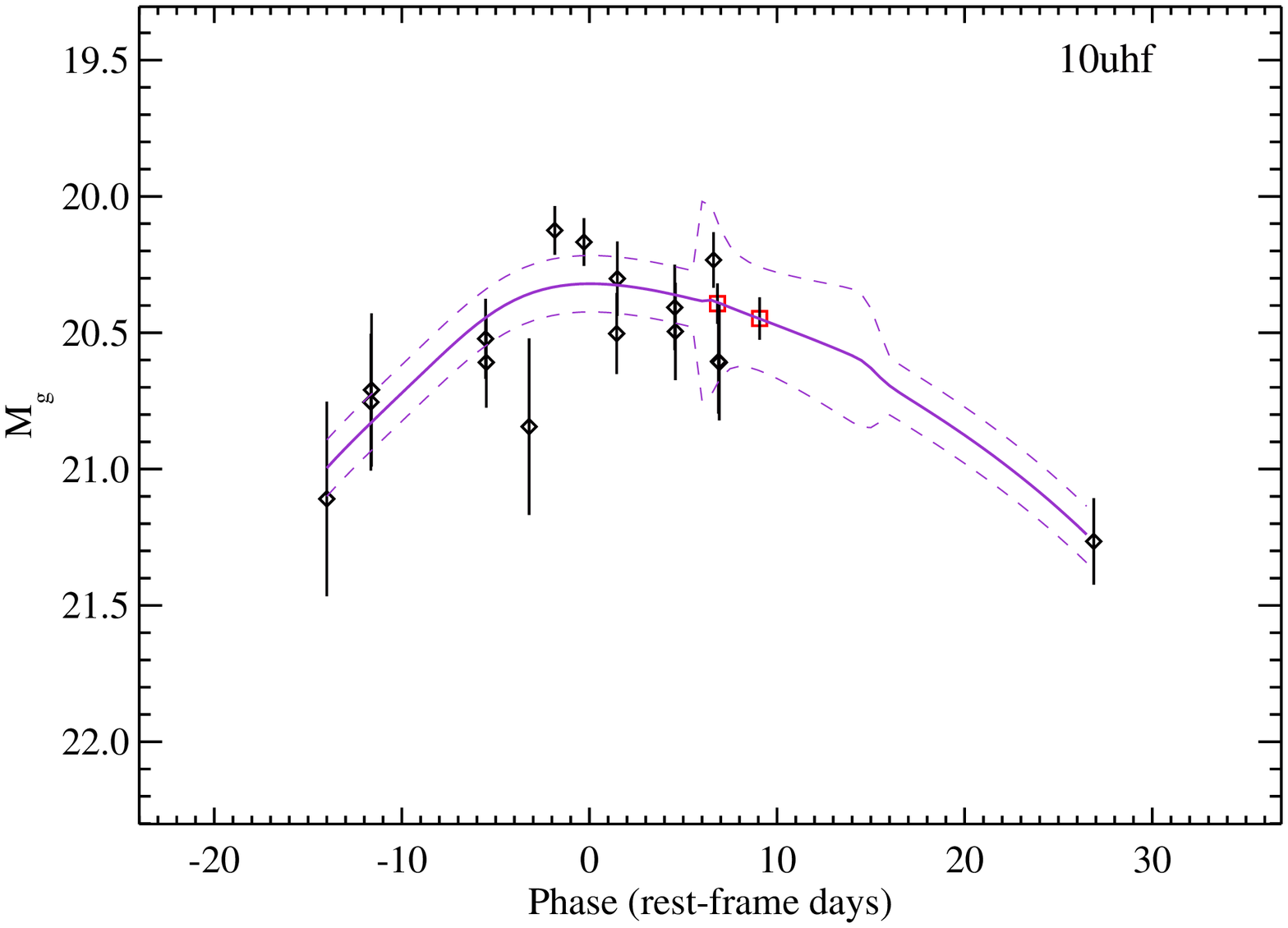}{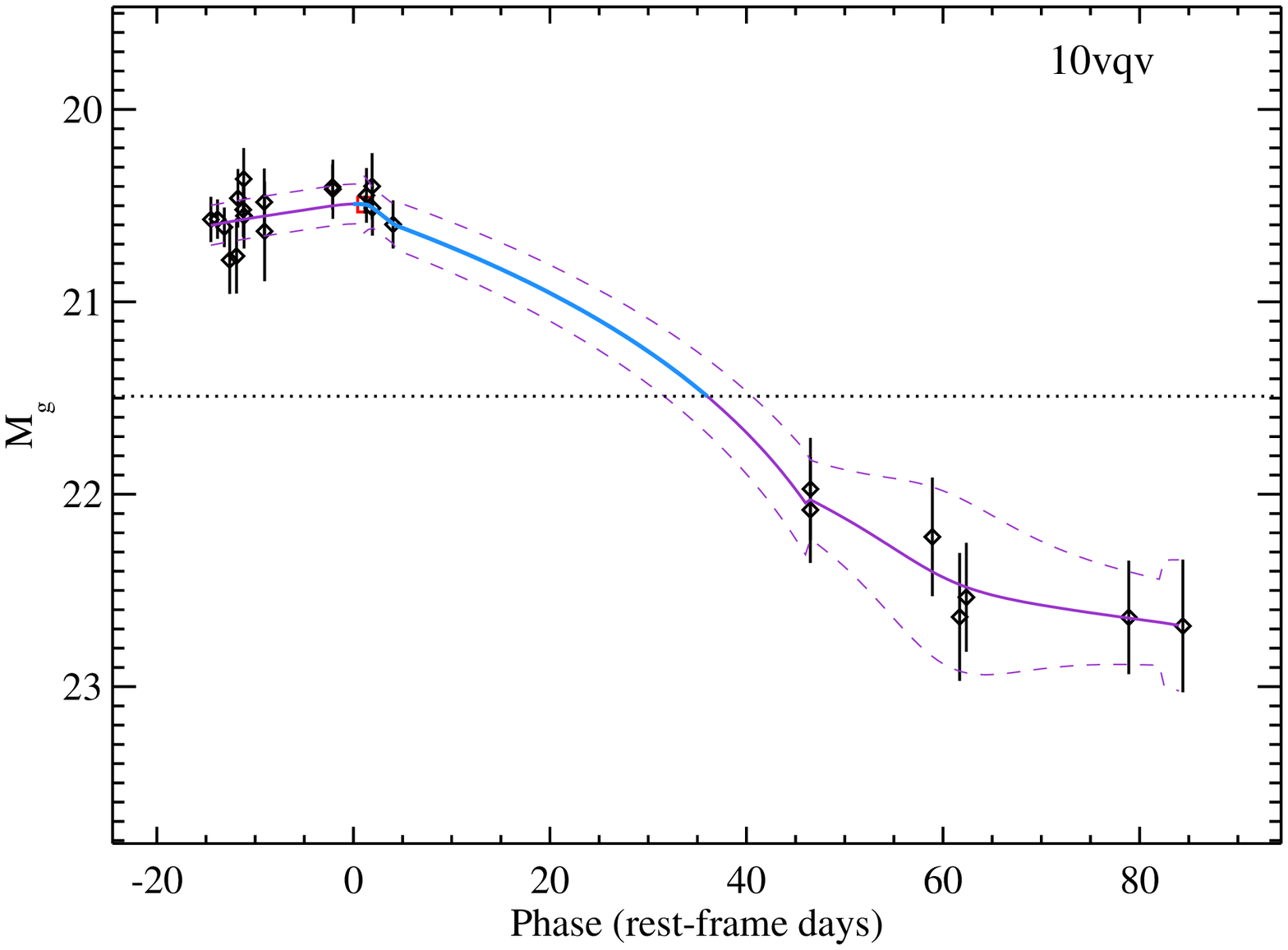}
\caption{Continuation of Fig.~\ref{fig ind smooth mag 1}. \label{fig ind smooth mag 2}}
\end{figure*}

\begin{figure*}[!h]
\epsscale{1.1}
\plottwo{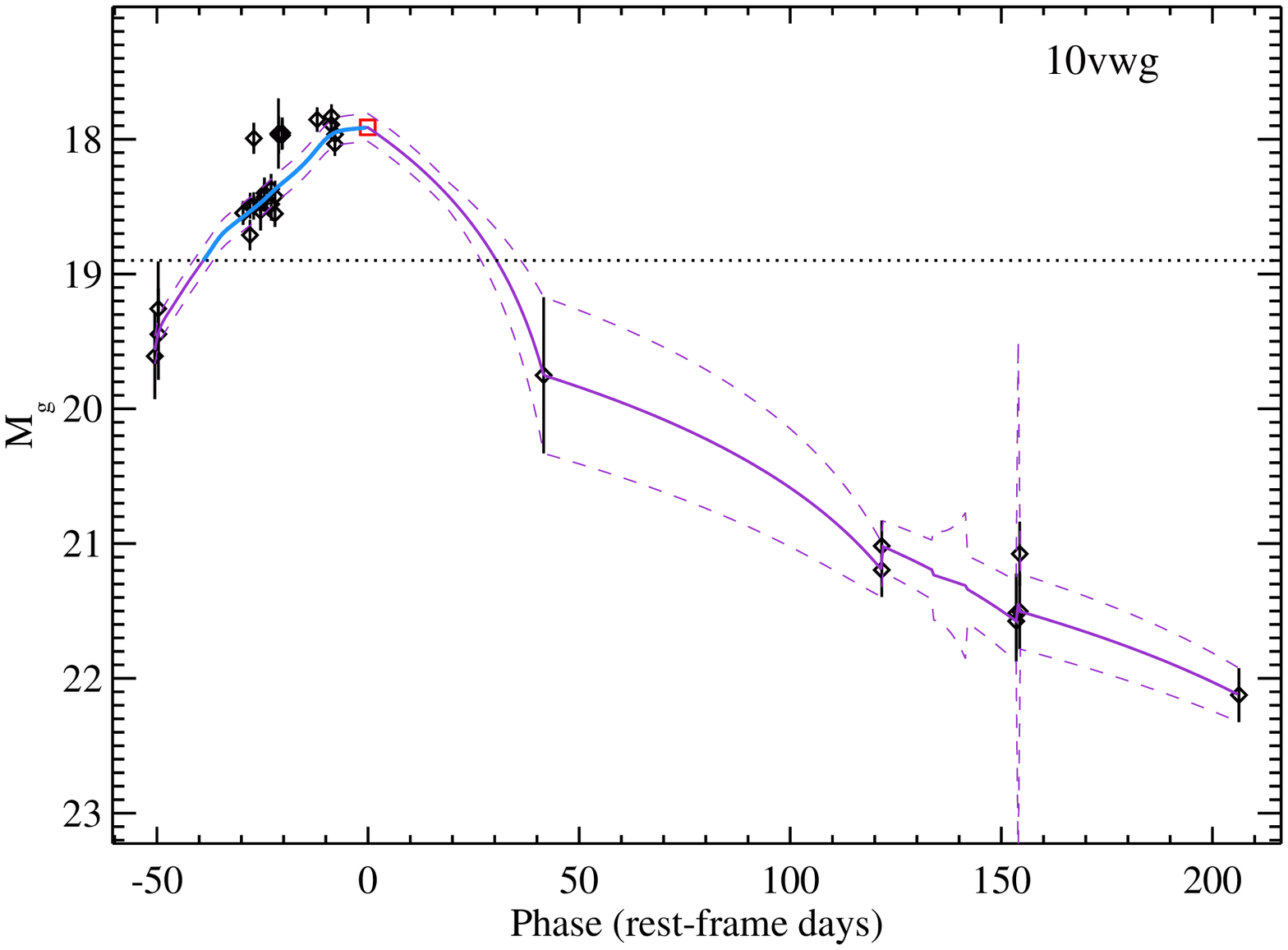}{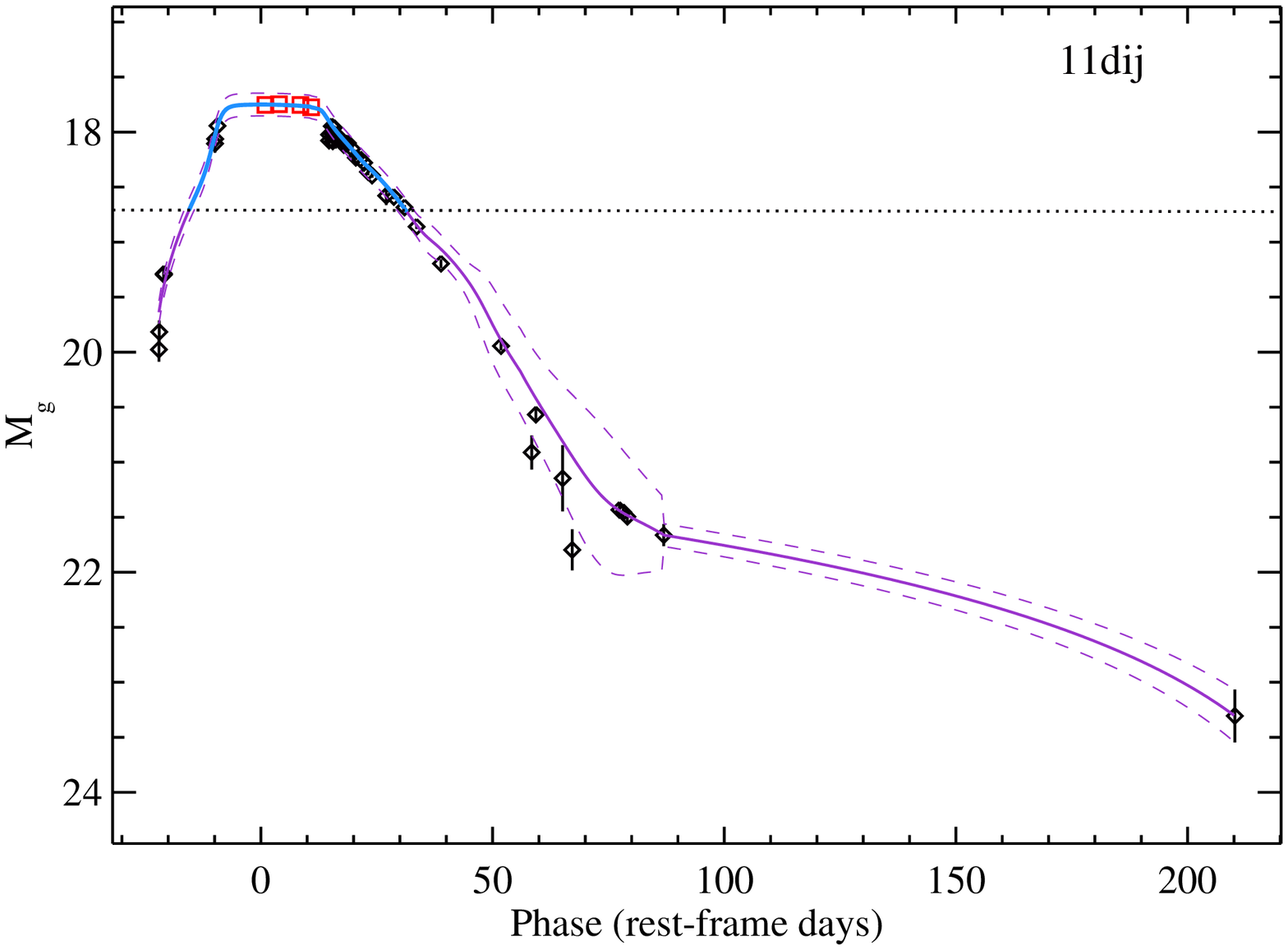}
\plottwo{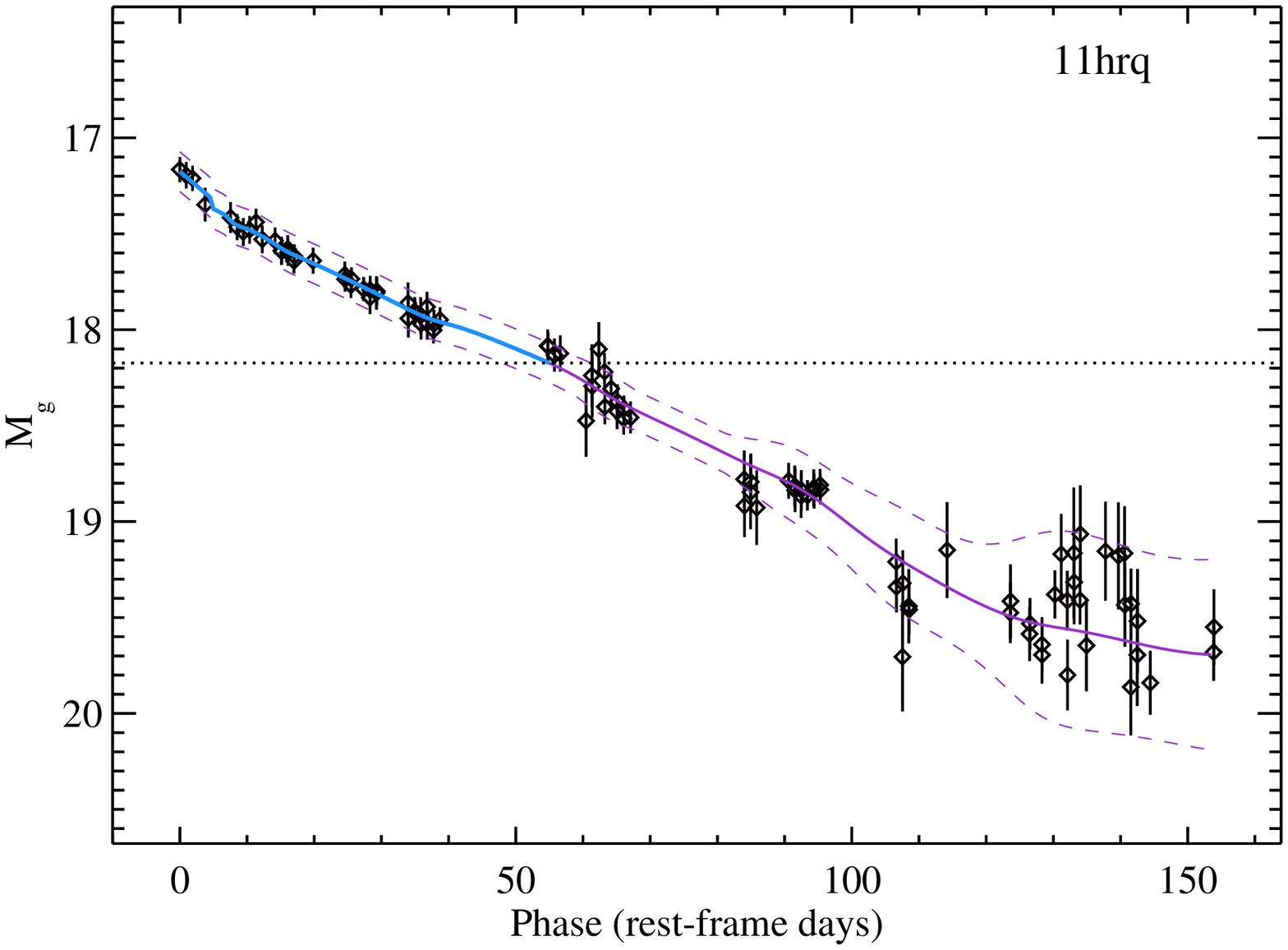}{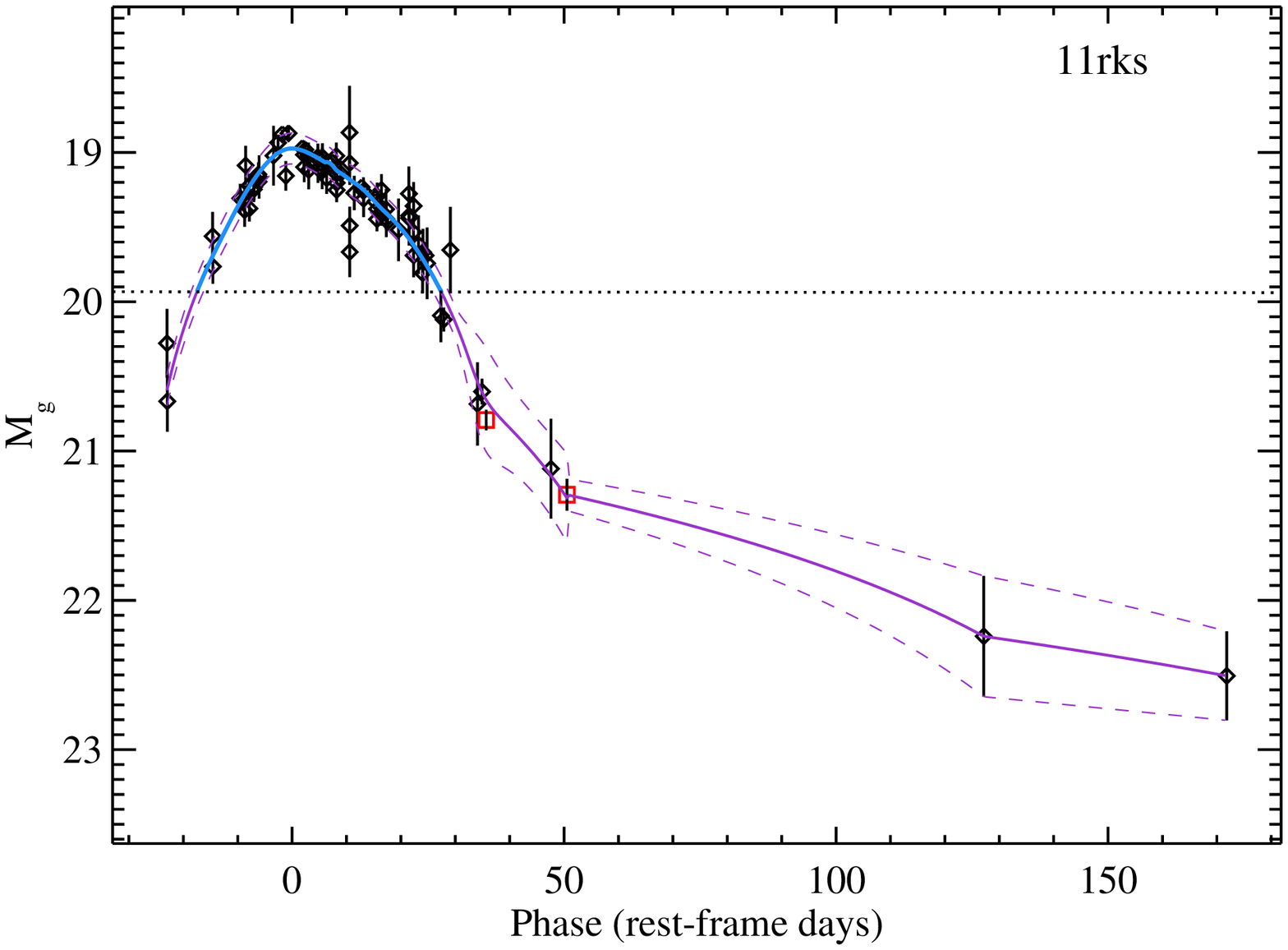}
\plottwo{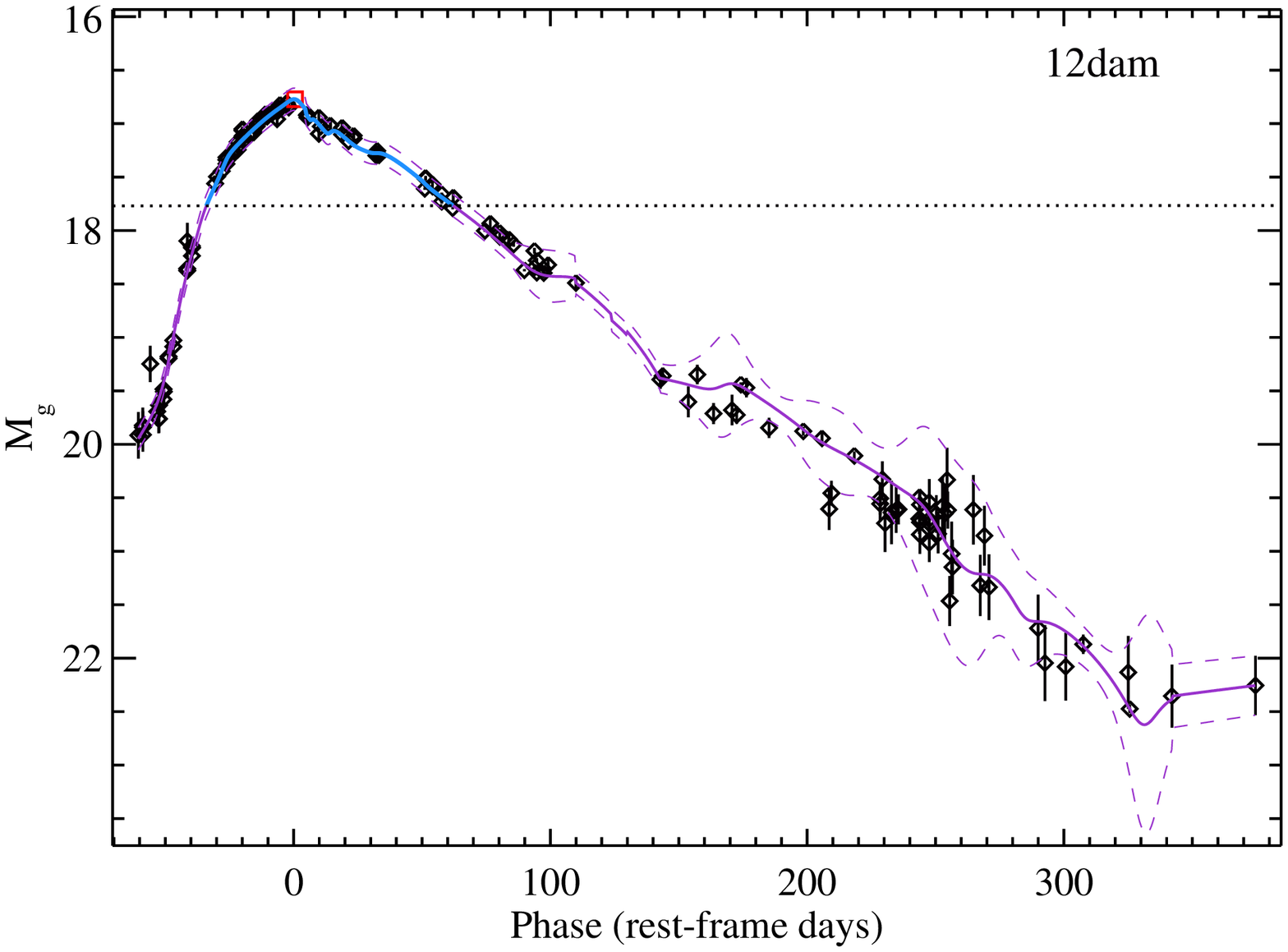}{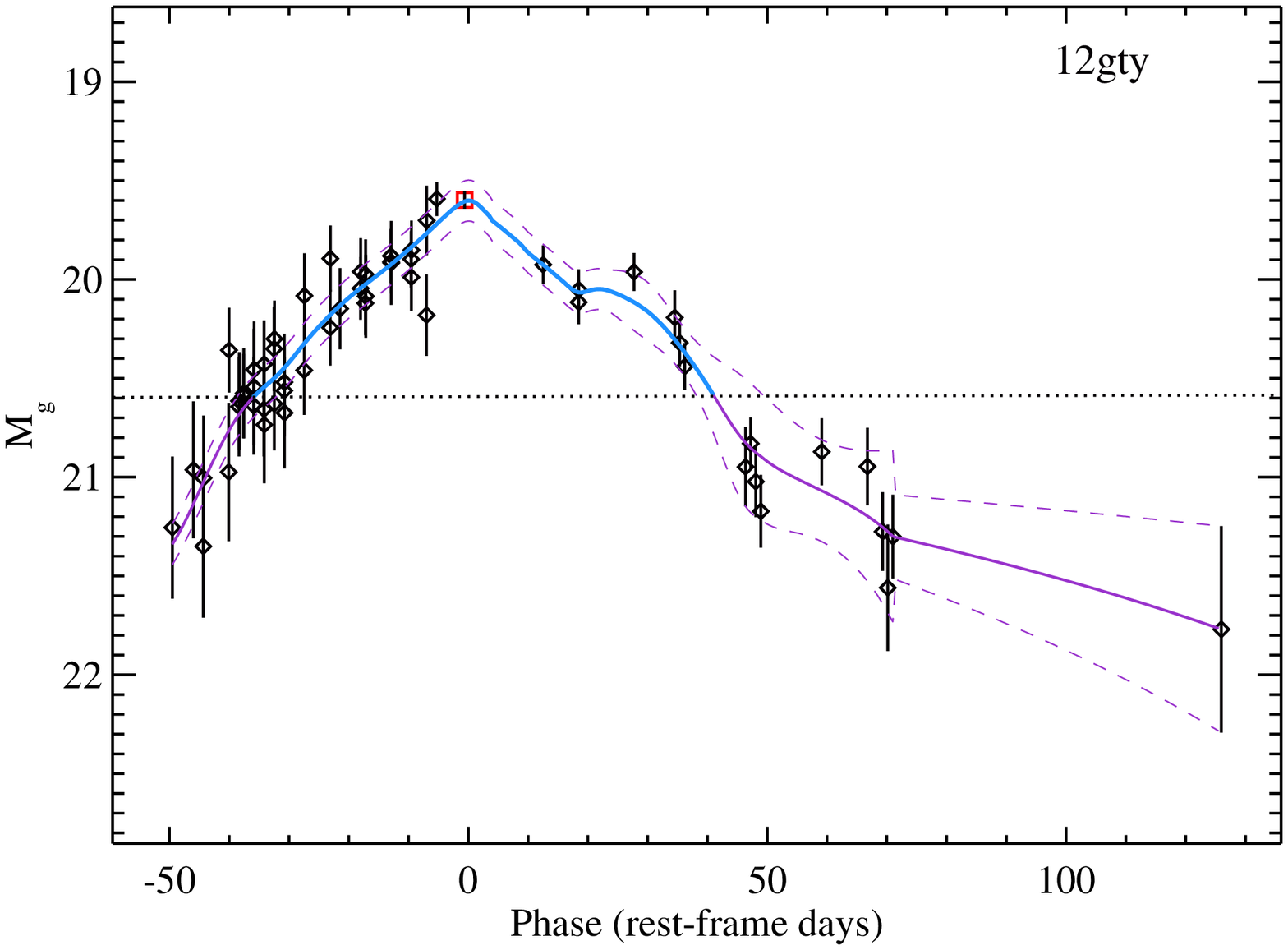}
\caption{Continuation of Fig.~\ref{fig ind smooth mag 1}. \label{fig ind smooth mag 3}}
\end{figure*}

\clearpage
\begin{figure*}[!h]
\epsscale{1.1}
\plottwo{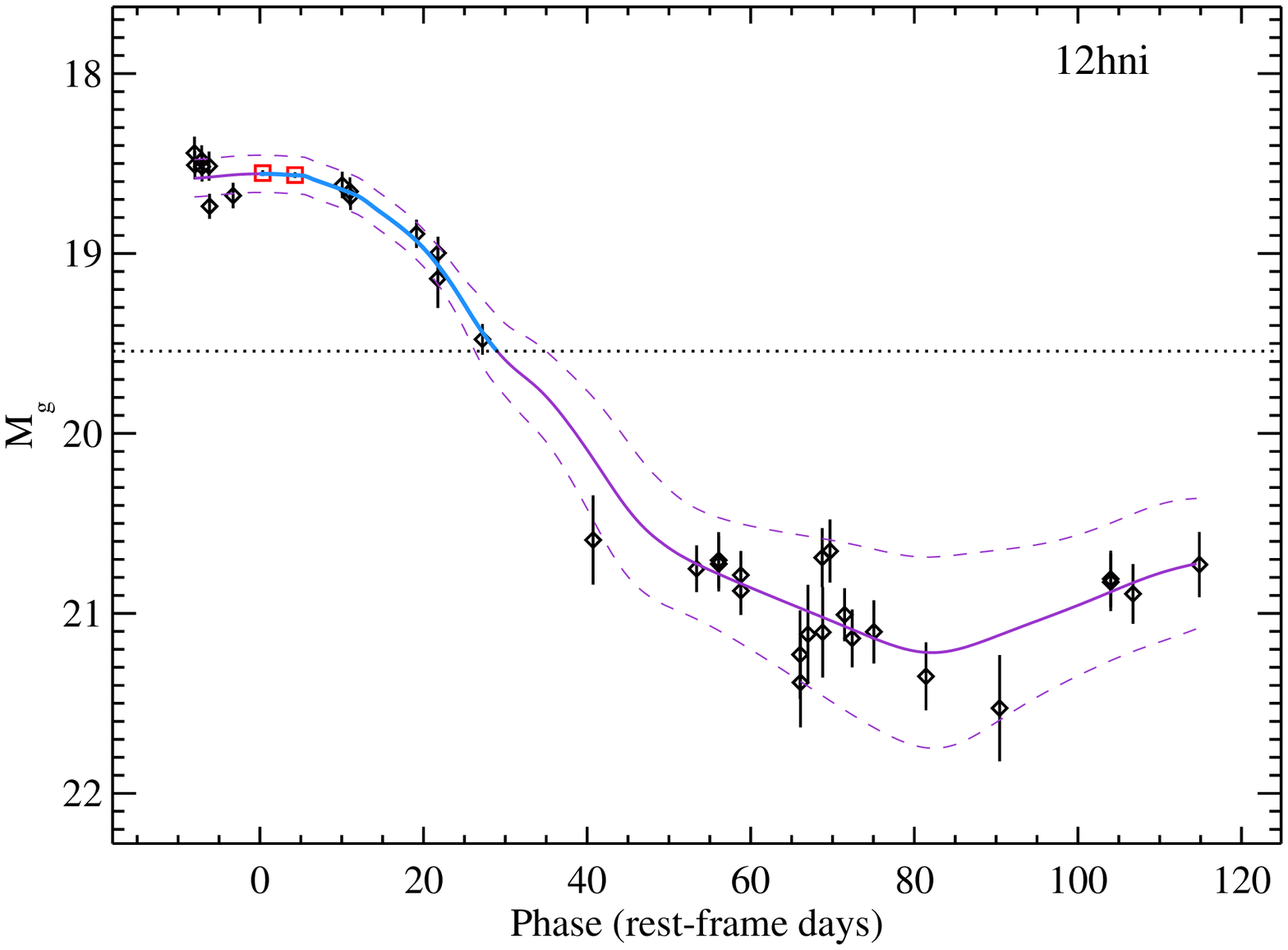}{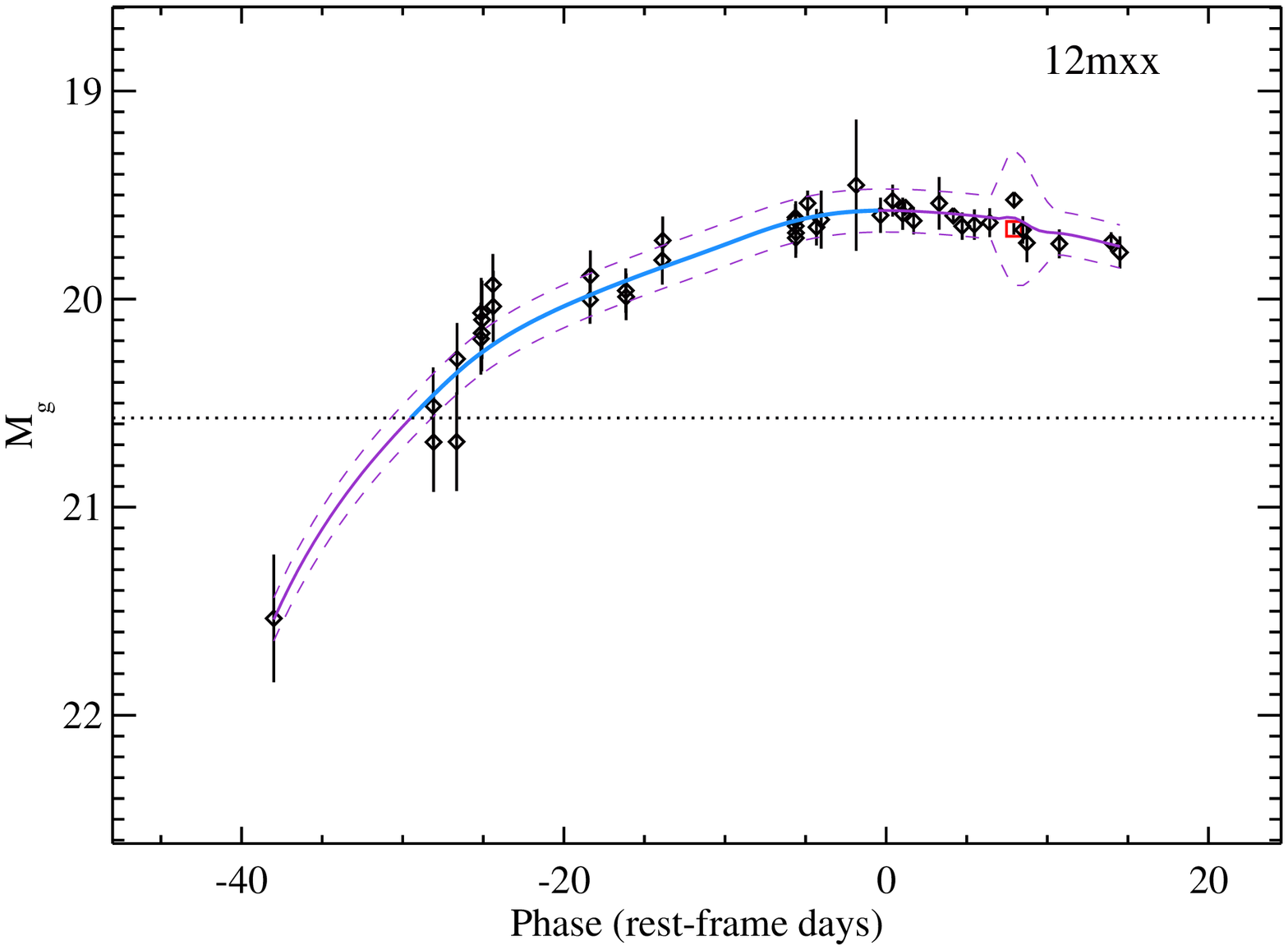}
\plottwo{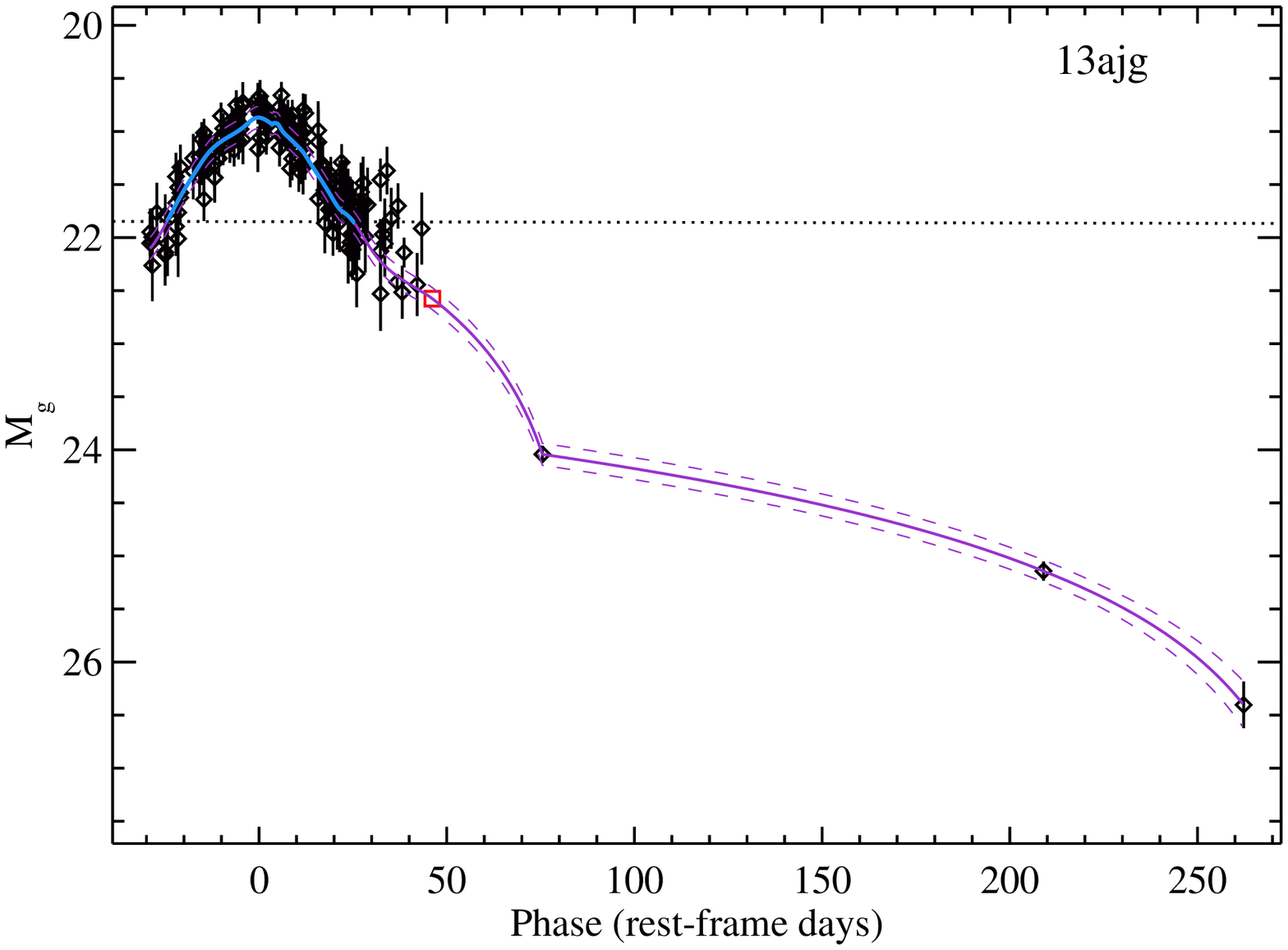}{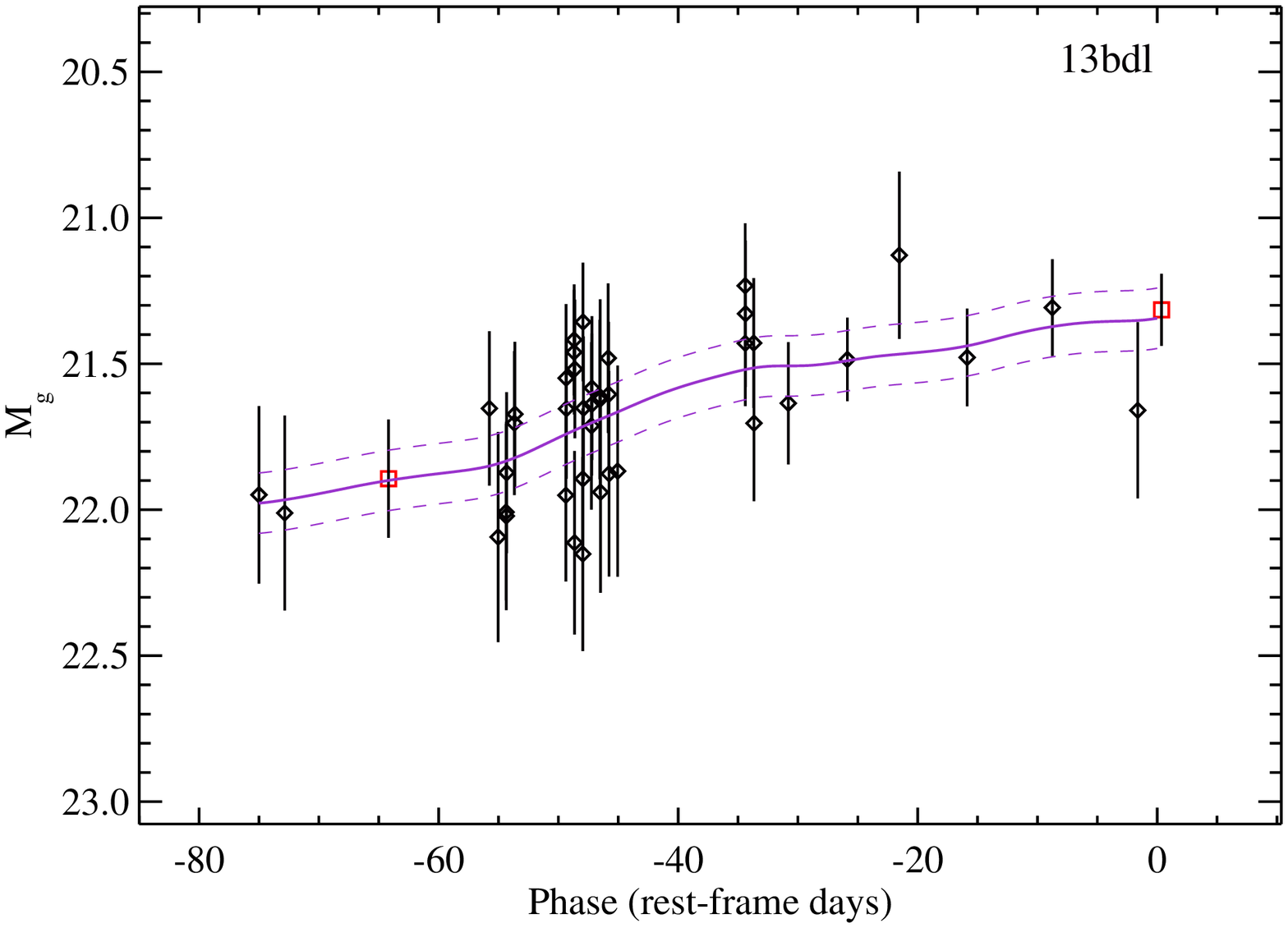}
\plottwo{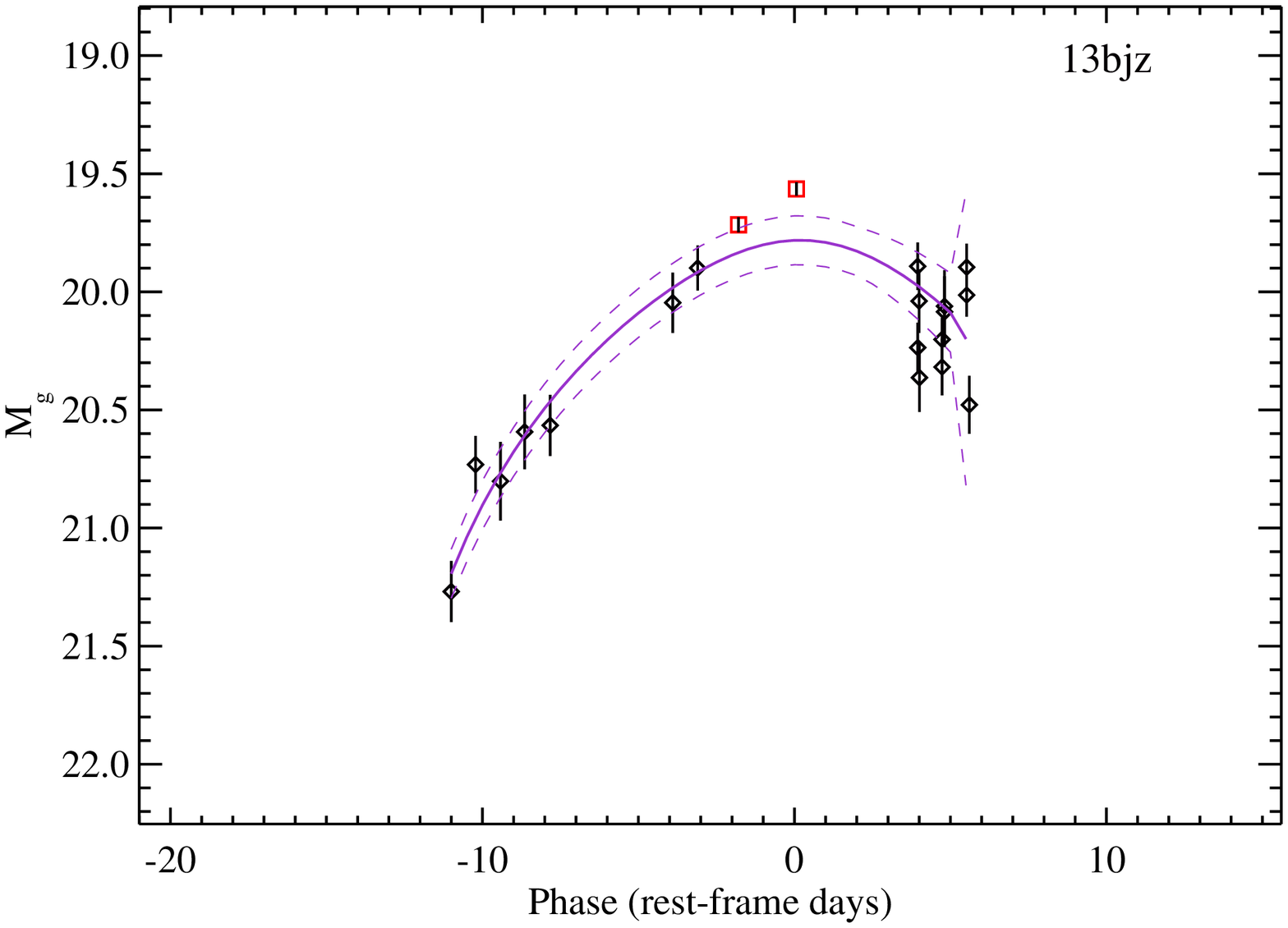}{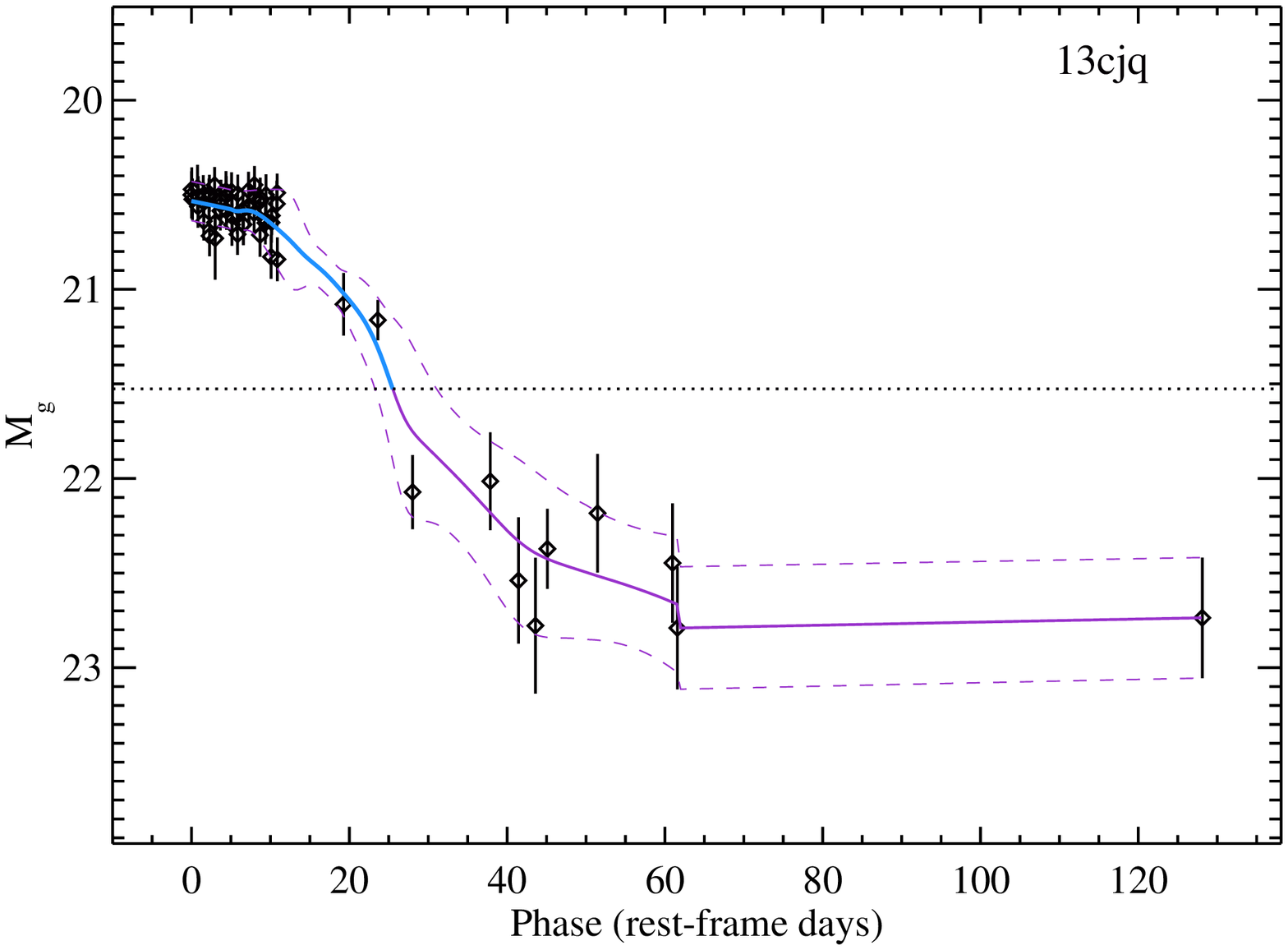}
\caption{Continuation of Fig.~\ref{fig ind smooth mag 1}. \label{fig ind smooth mag 4}}
\end{figure*}

\clearpage

\begin{figure}[!h]
\epsscale{1.1}
\plottwo{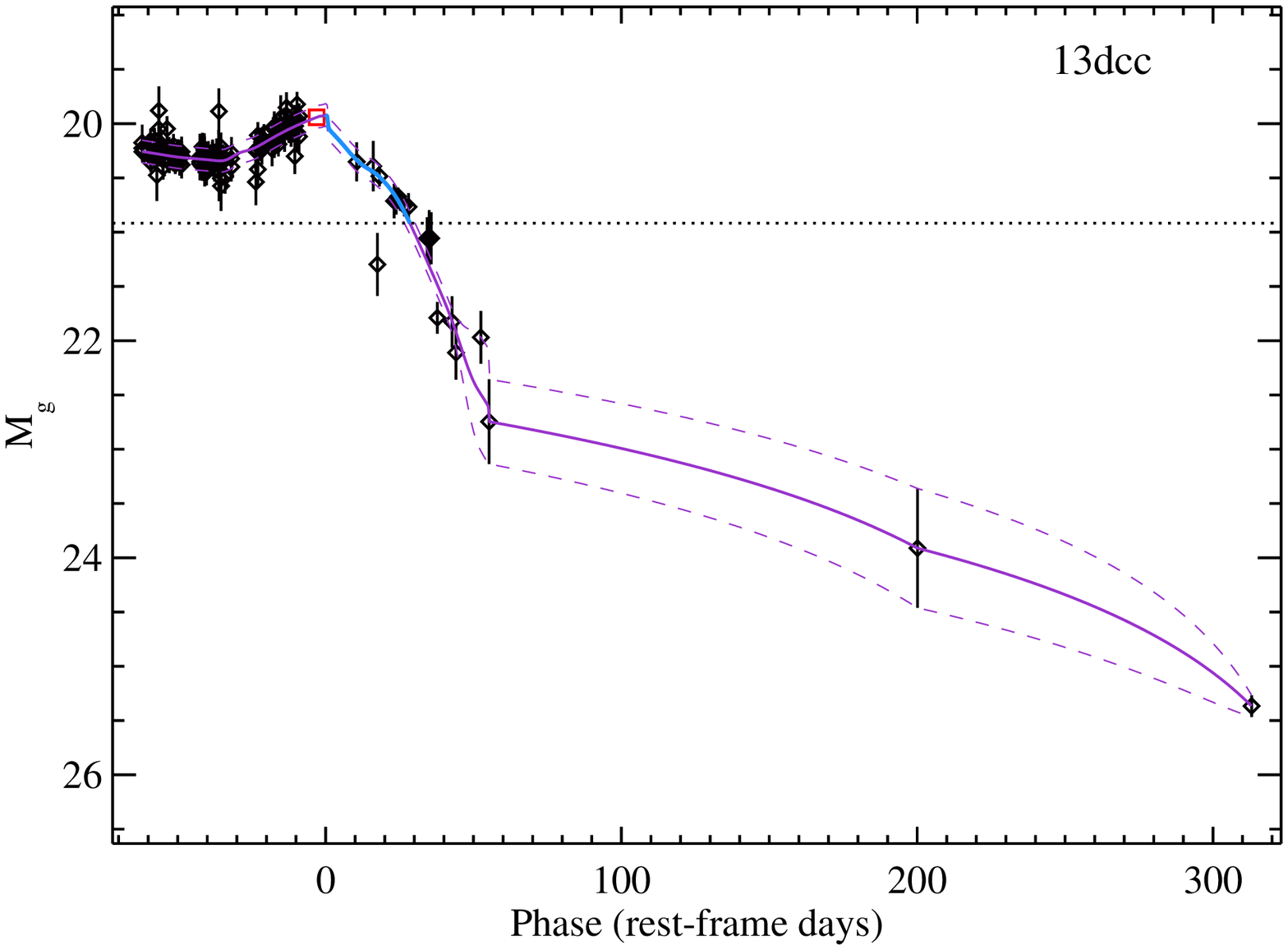}{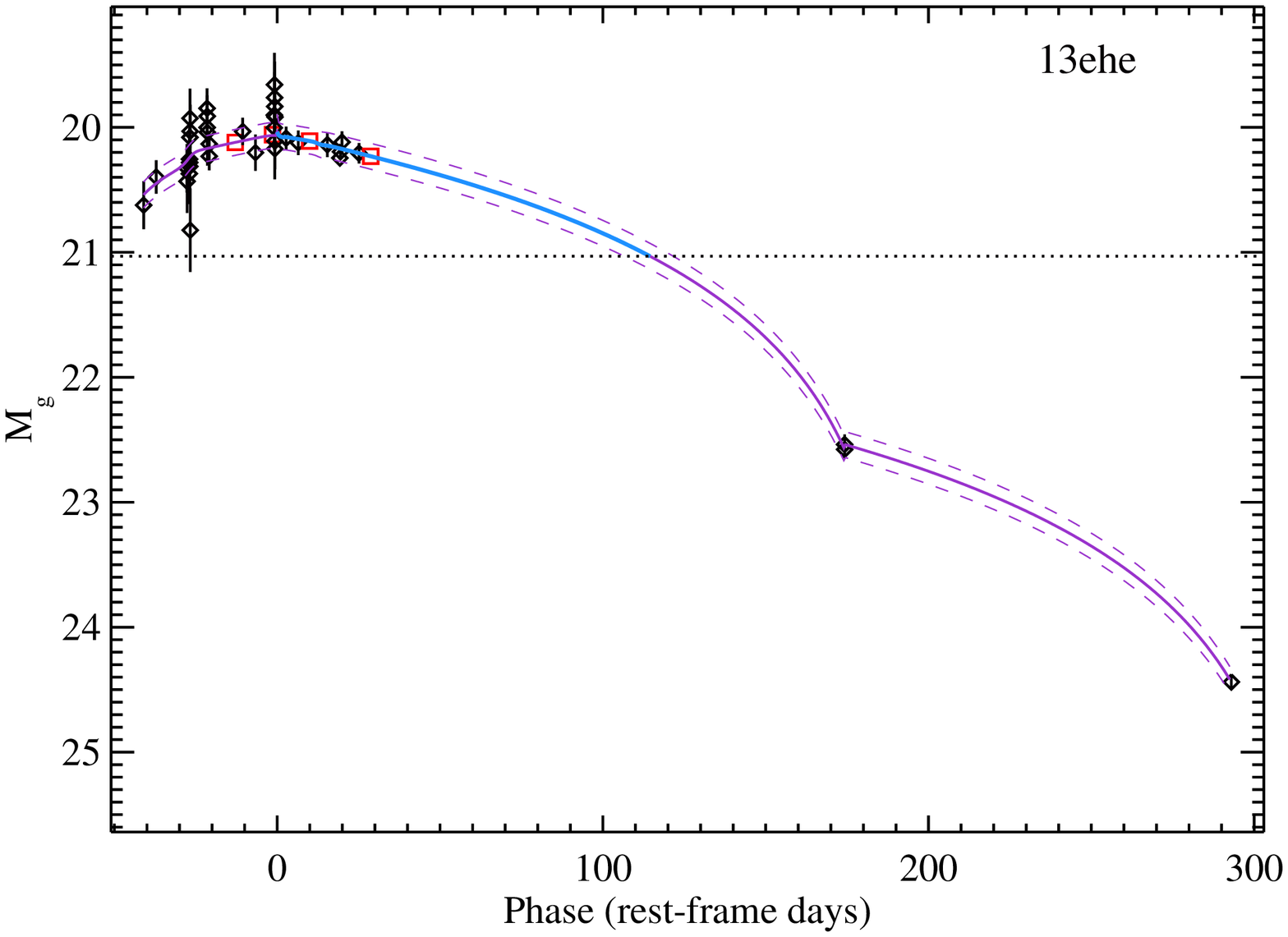}
\caption{Continuation of Fig.~\ref{fig ind smooth mag 1}. \label{fig ind smooth mag 5}}
\end{figure}


\newpage
\begin{figure*}[!hb]
\epsscale{1.15}
\plottwo{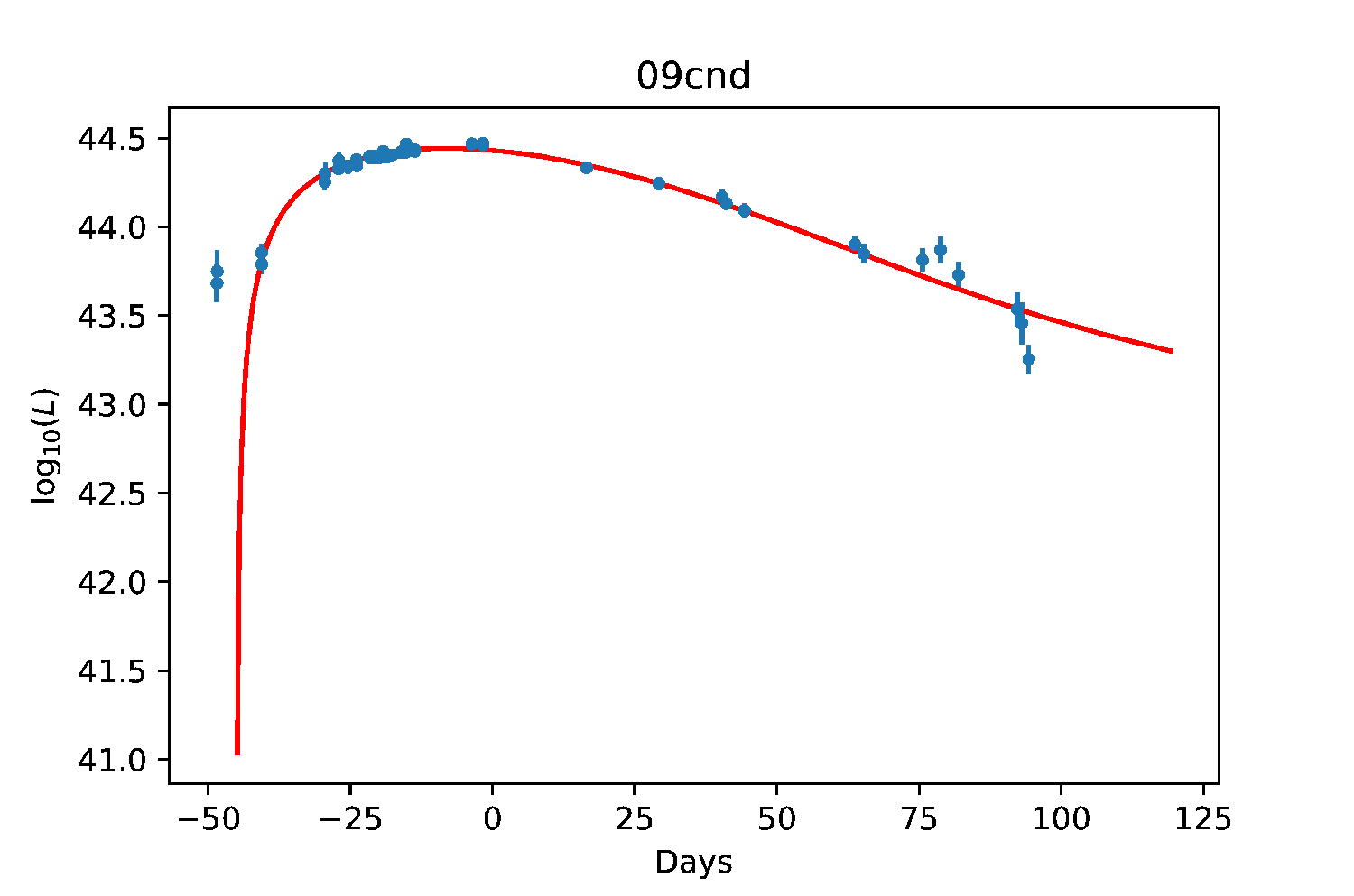}{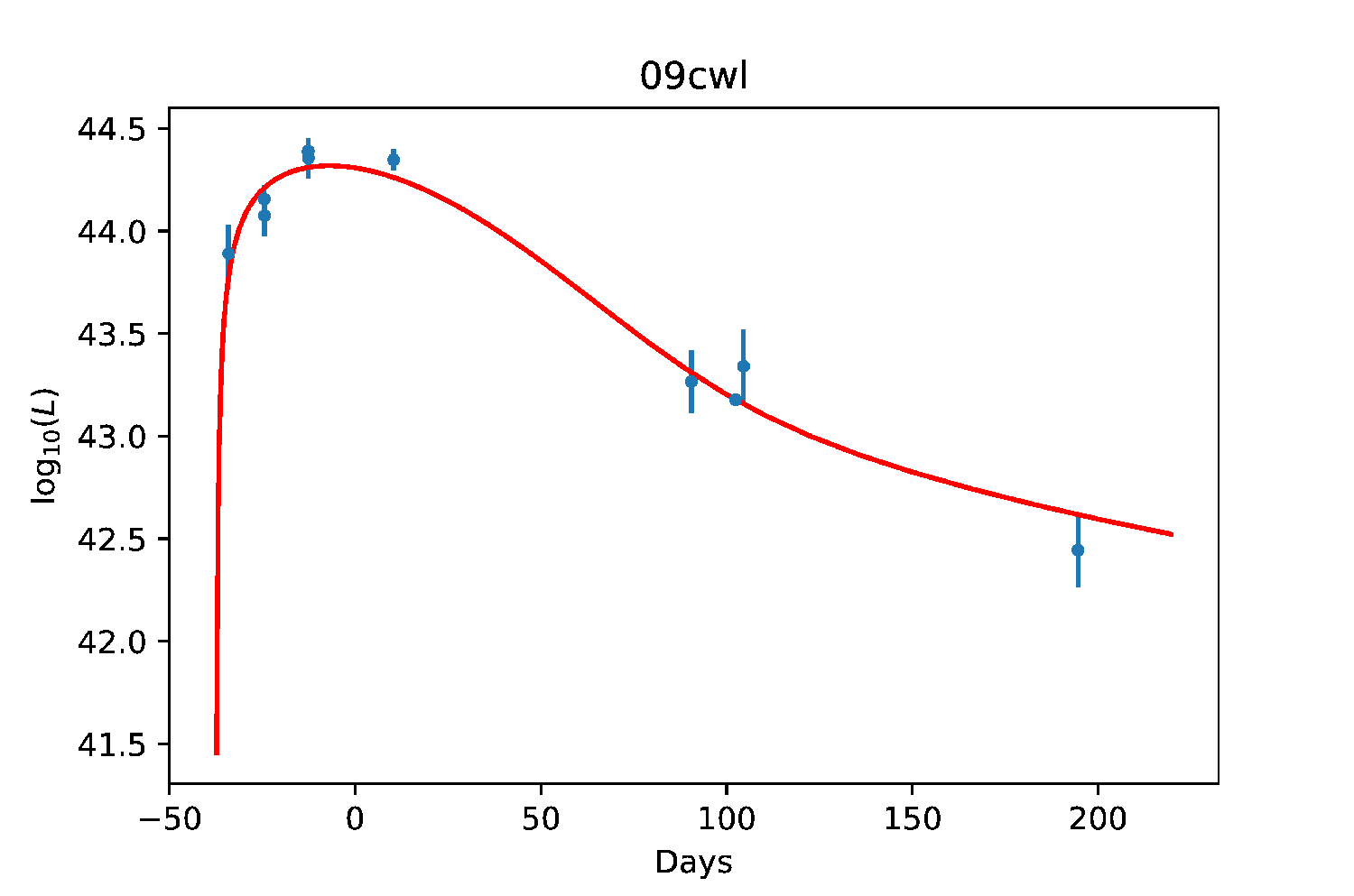}
\plottwo{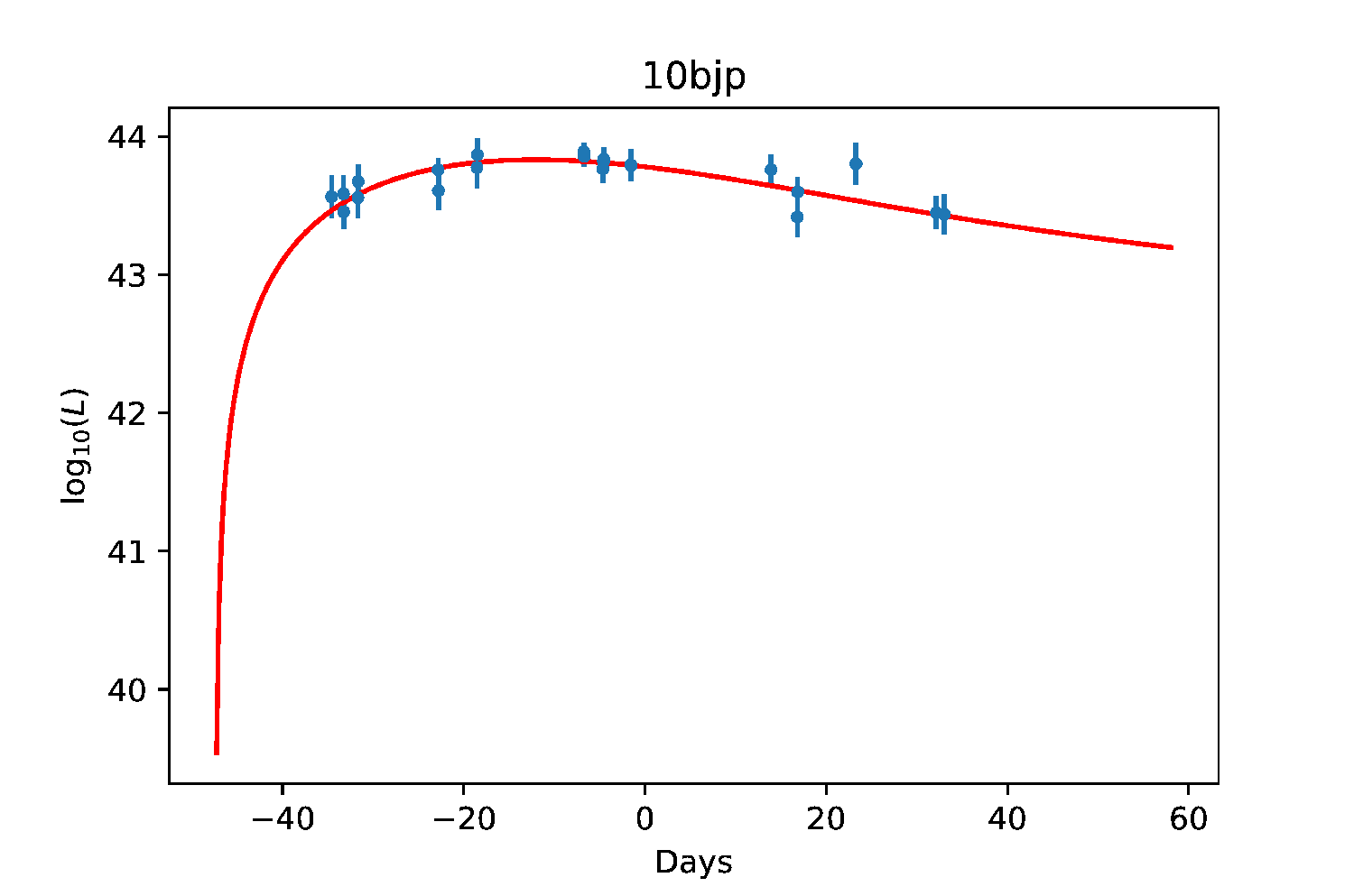}{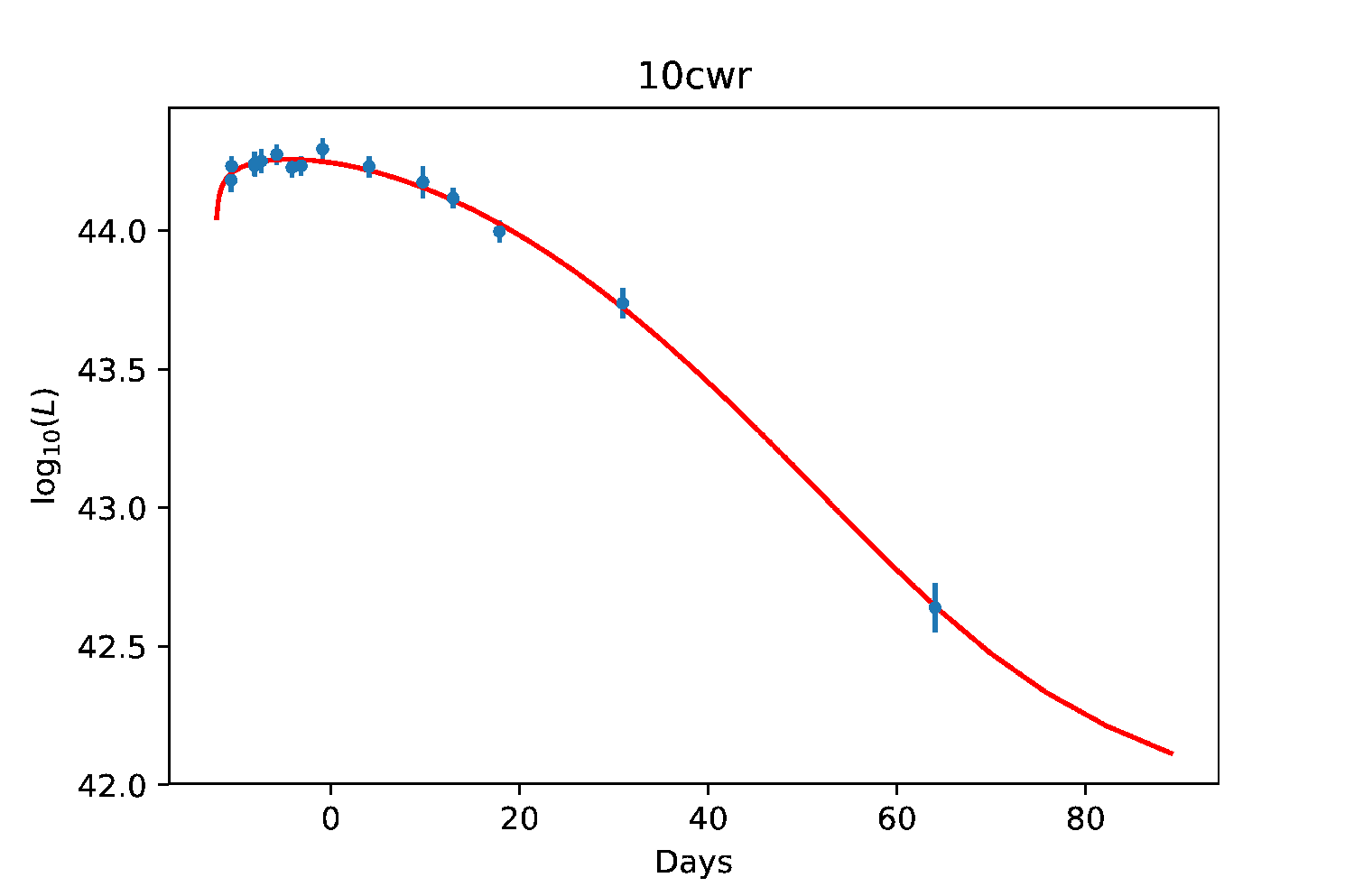}
\plottwo{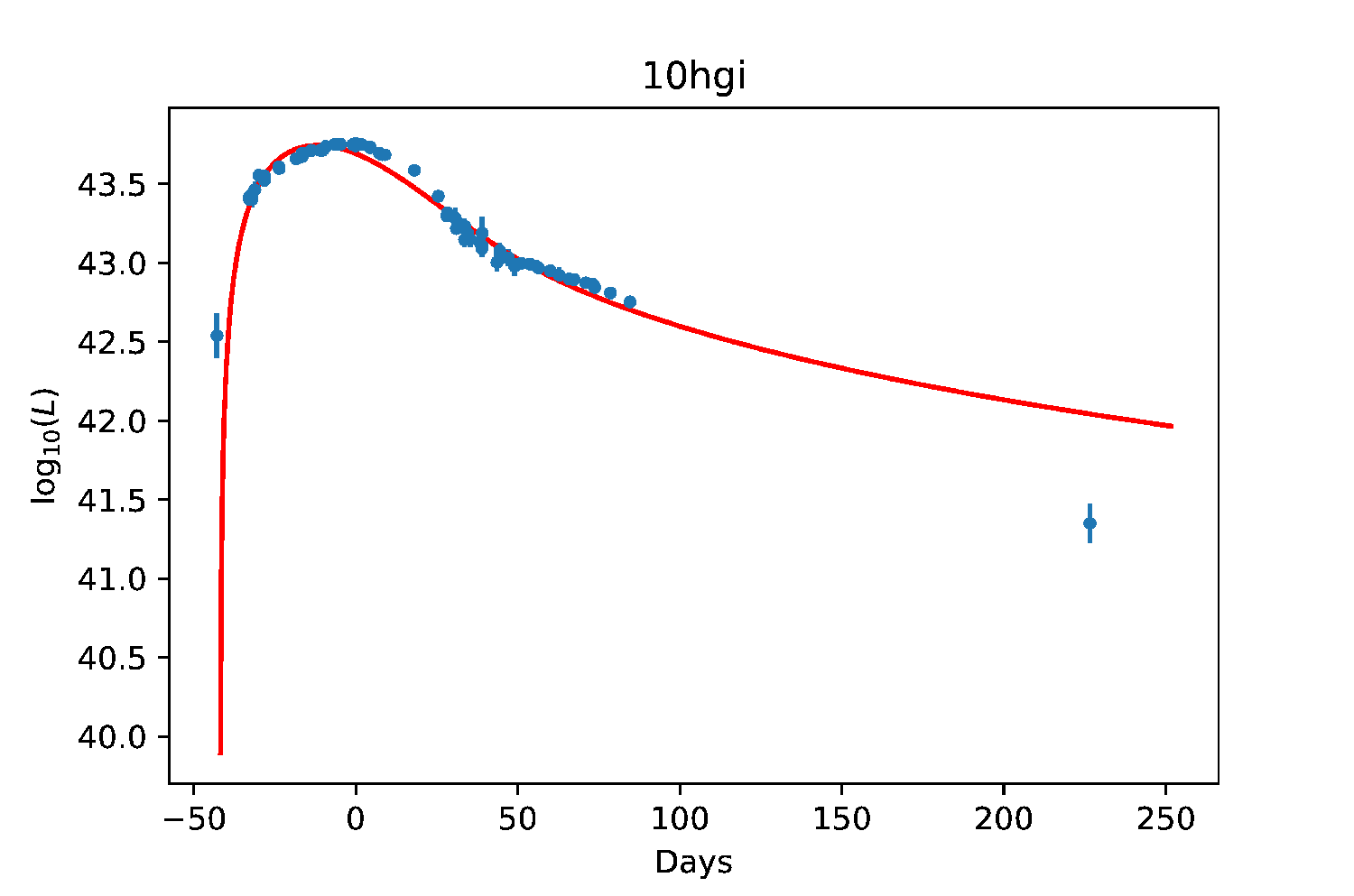}{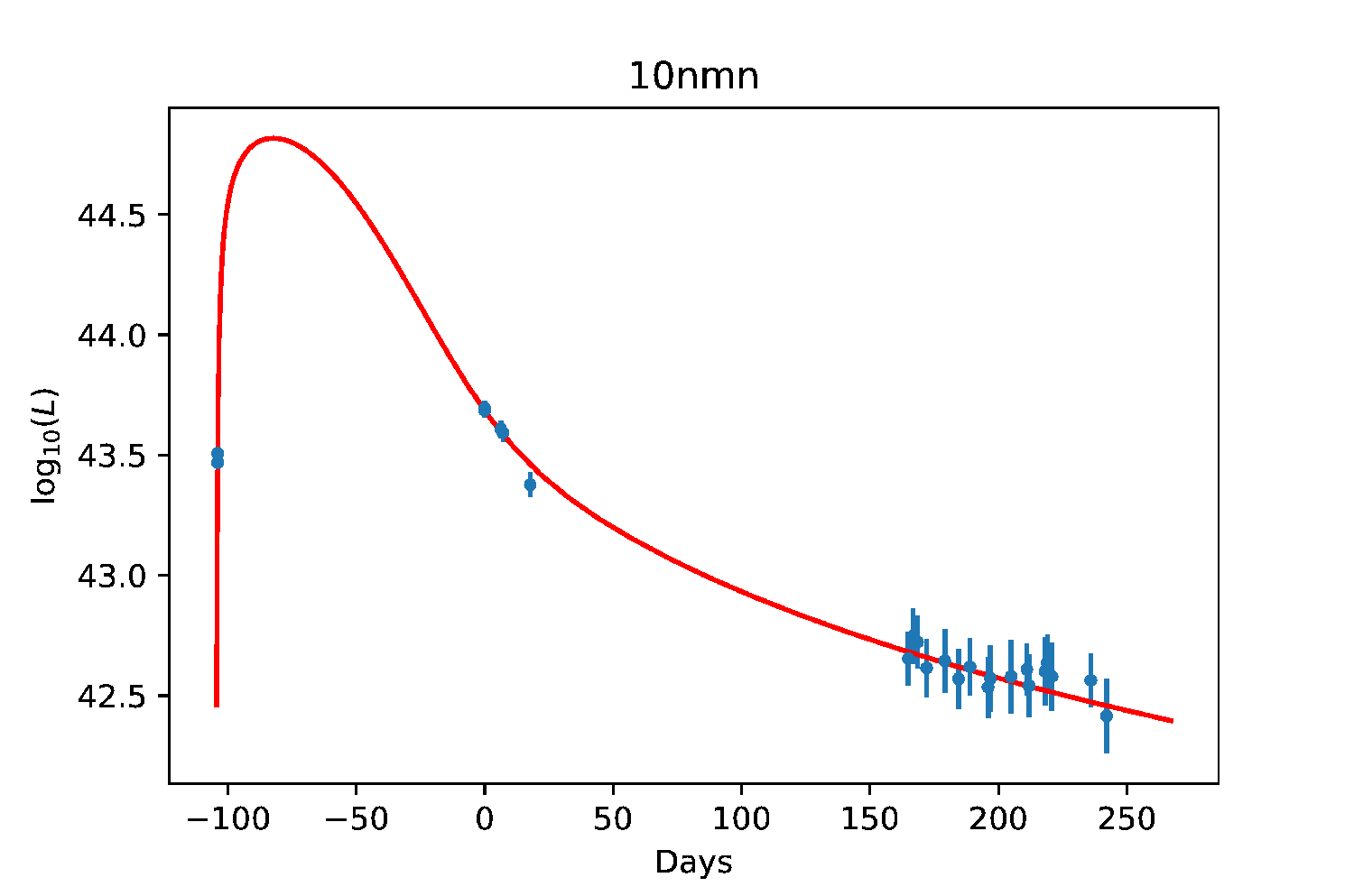}
\caption{Magnetar fit to the bolometric light curves.\label{fig magnetar 1}}
\end{figure*}
\newpage
\begin{figure*}[!hb]
\epsscale{1.15}
\plottwo{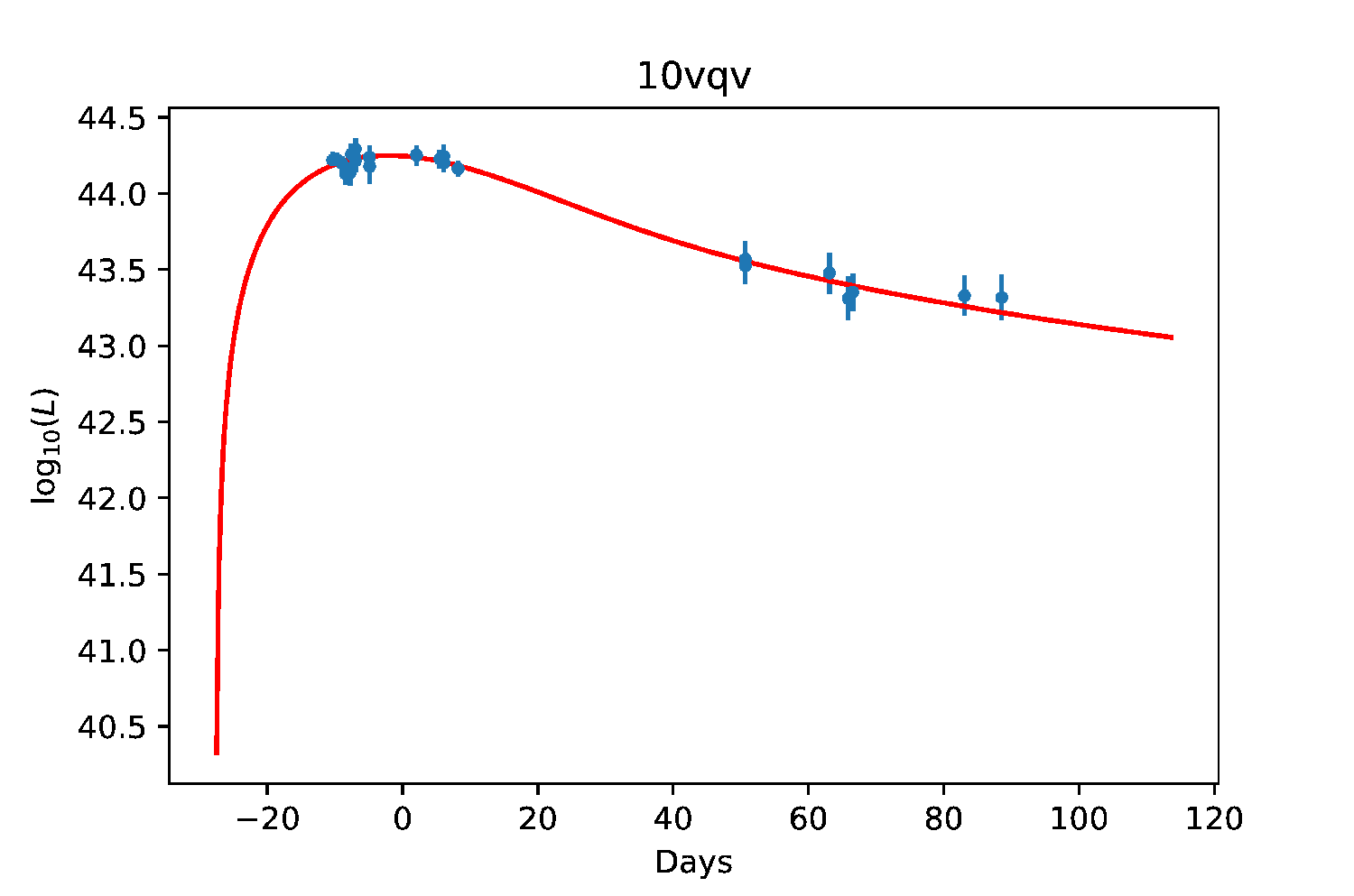}{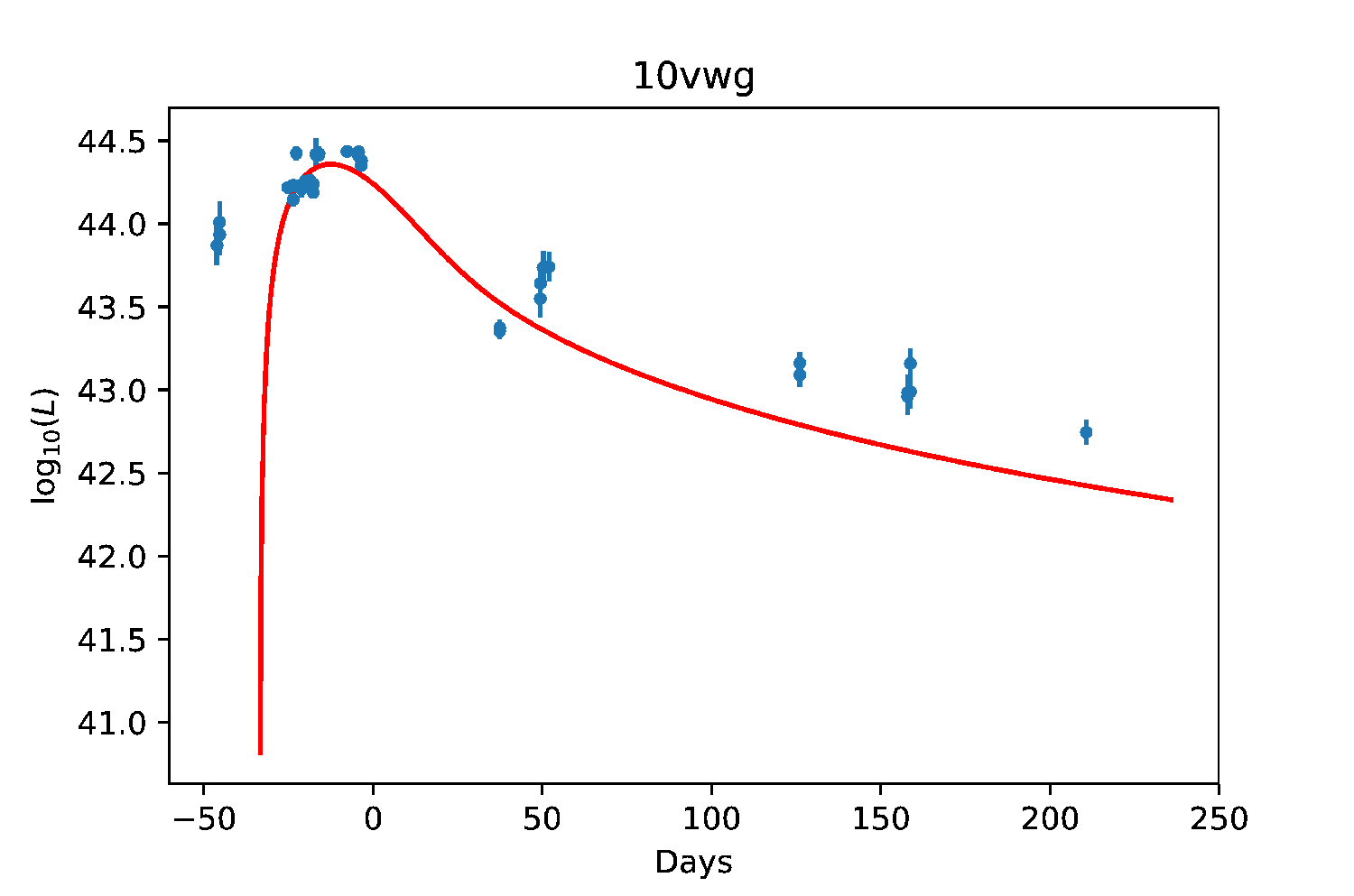}
\plottwo{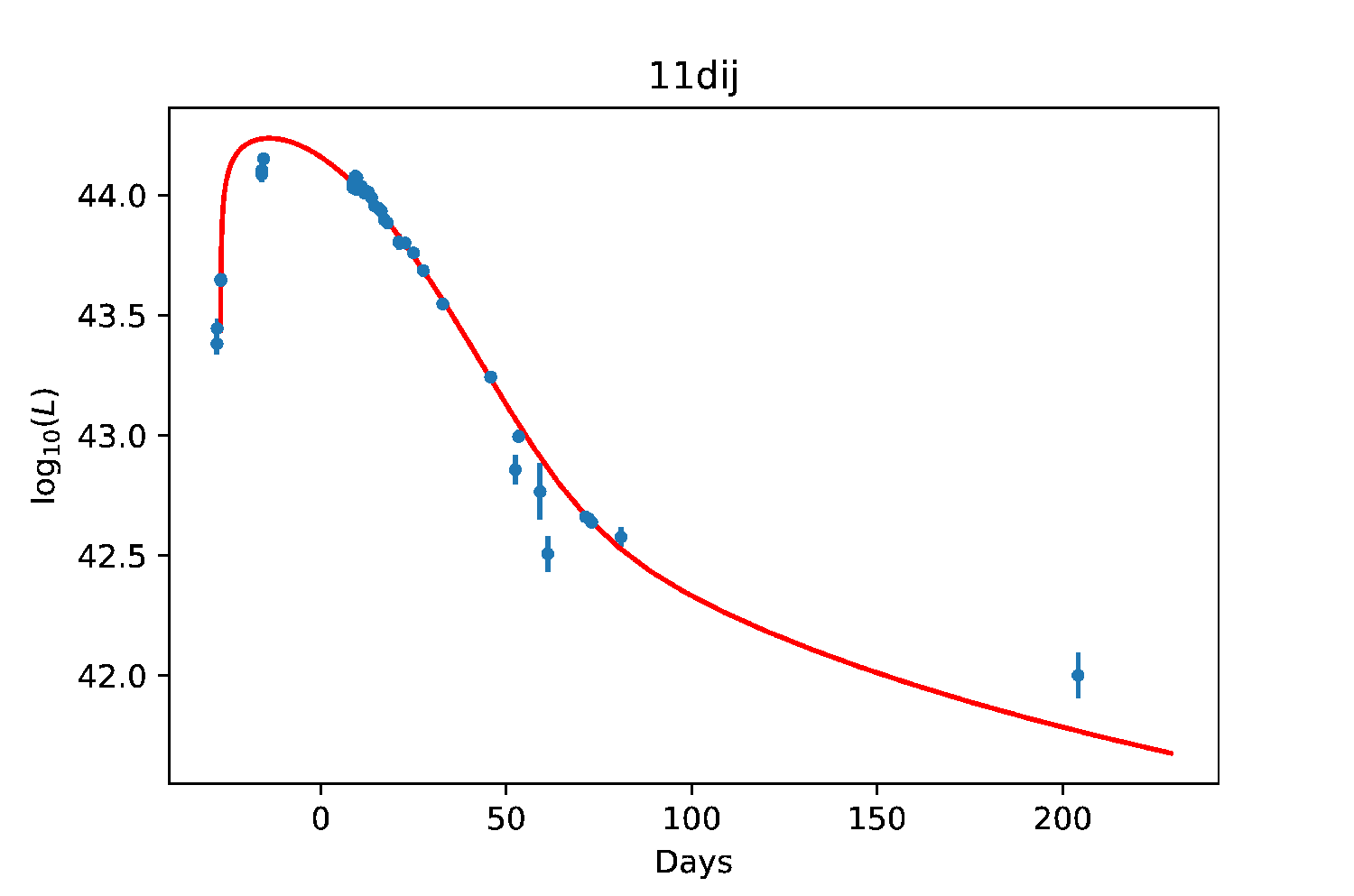}{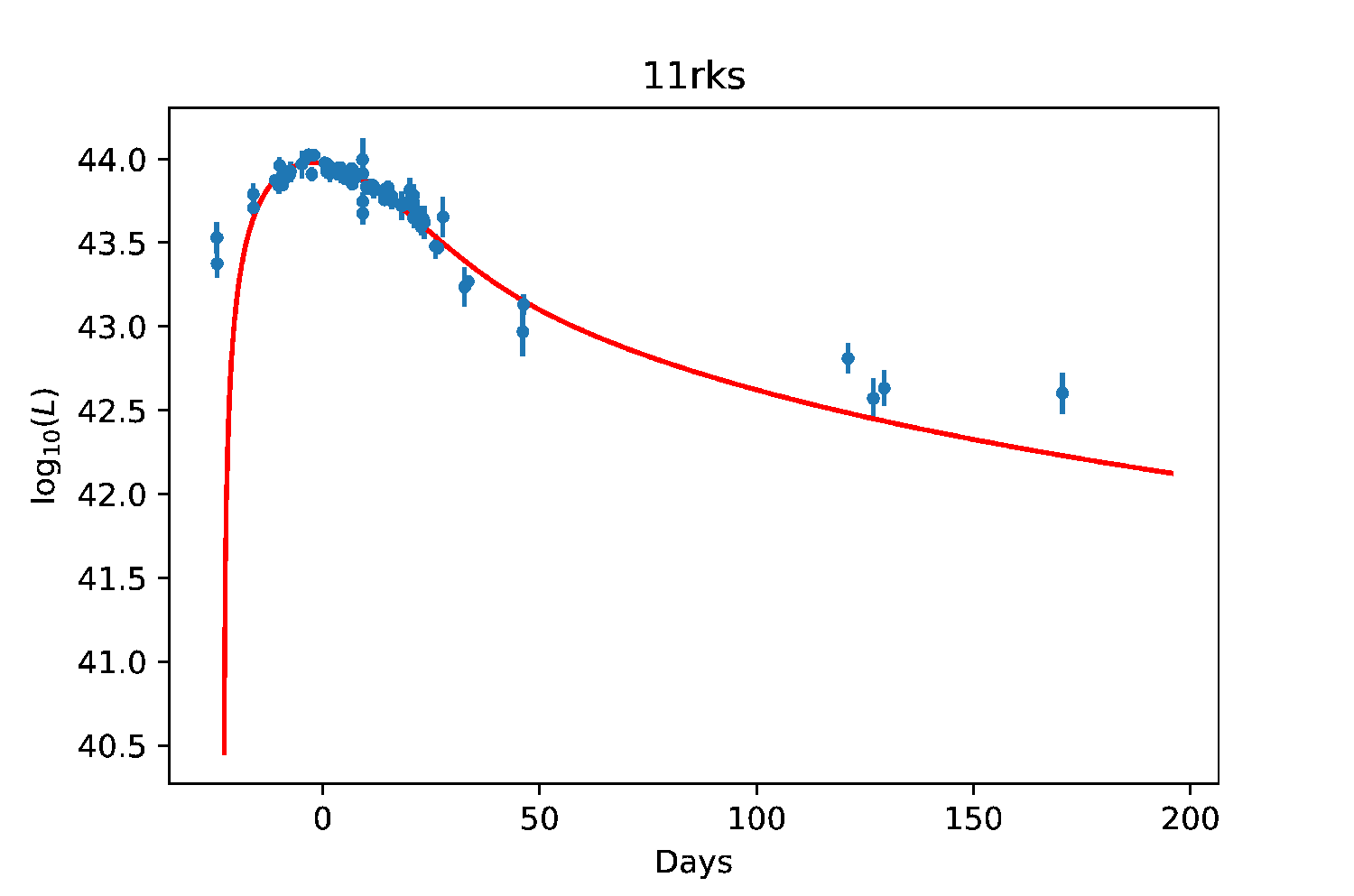}
\plottwo{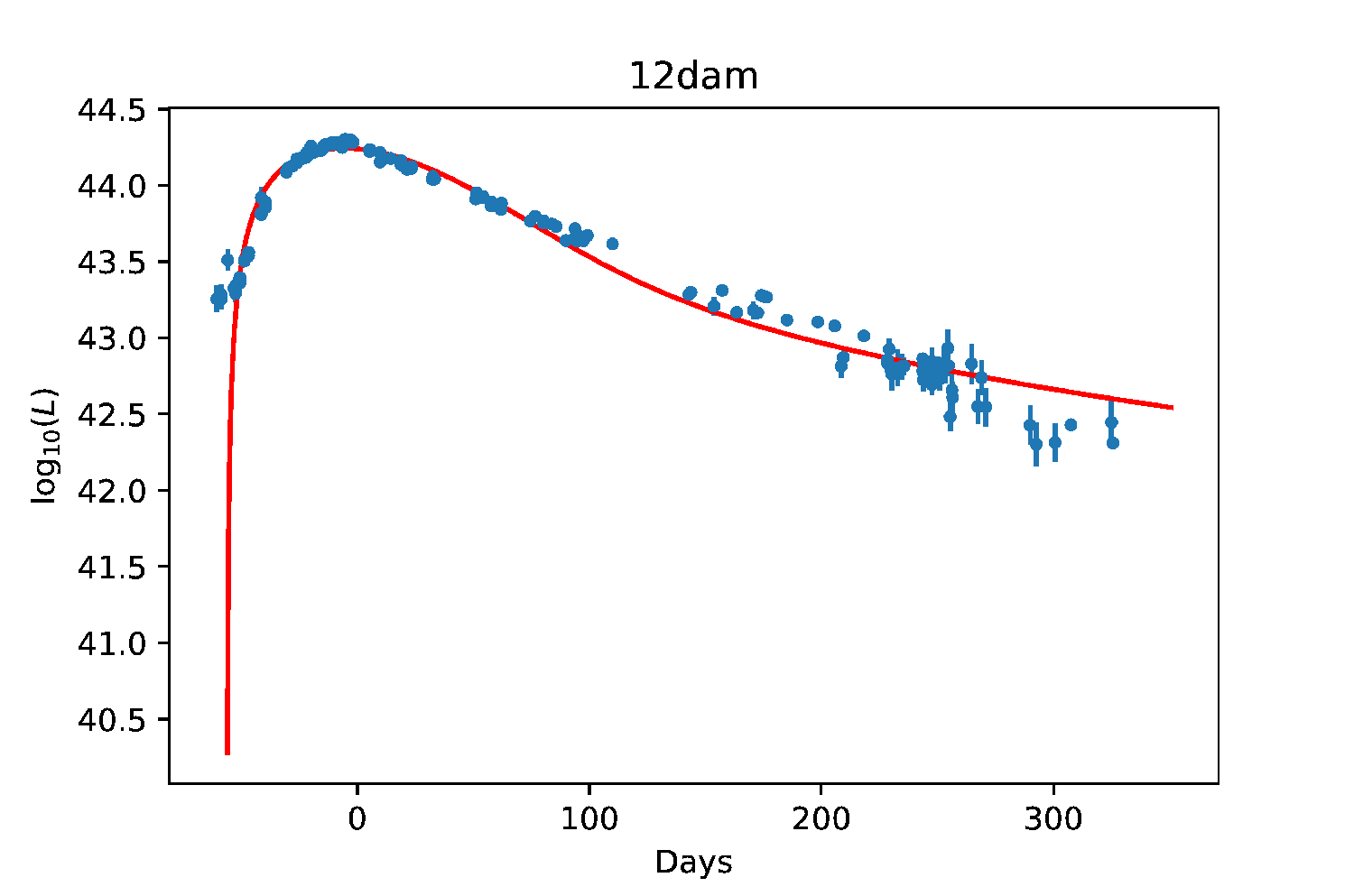}{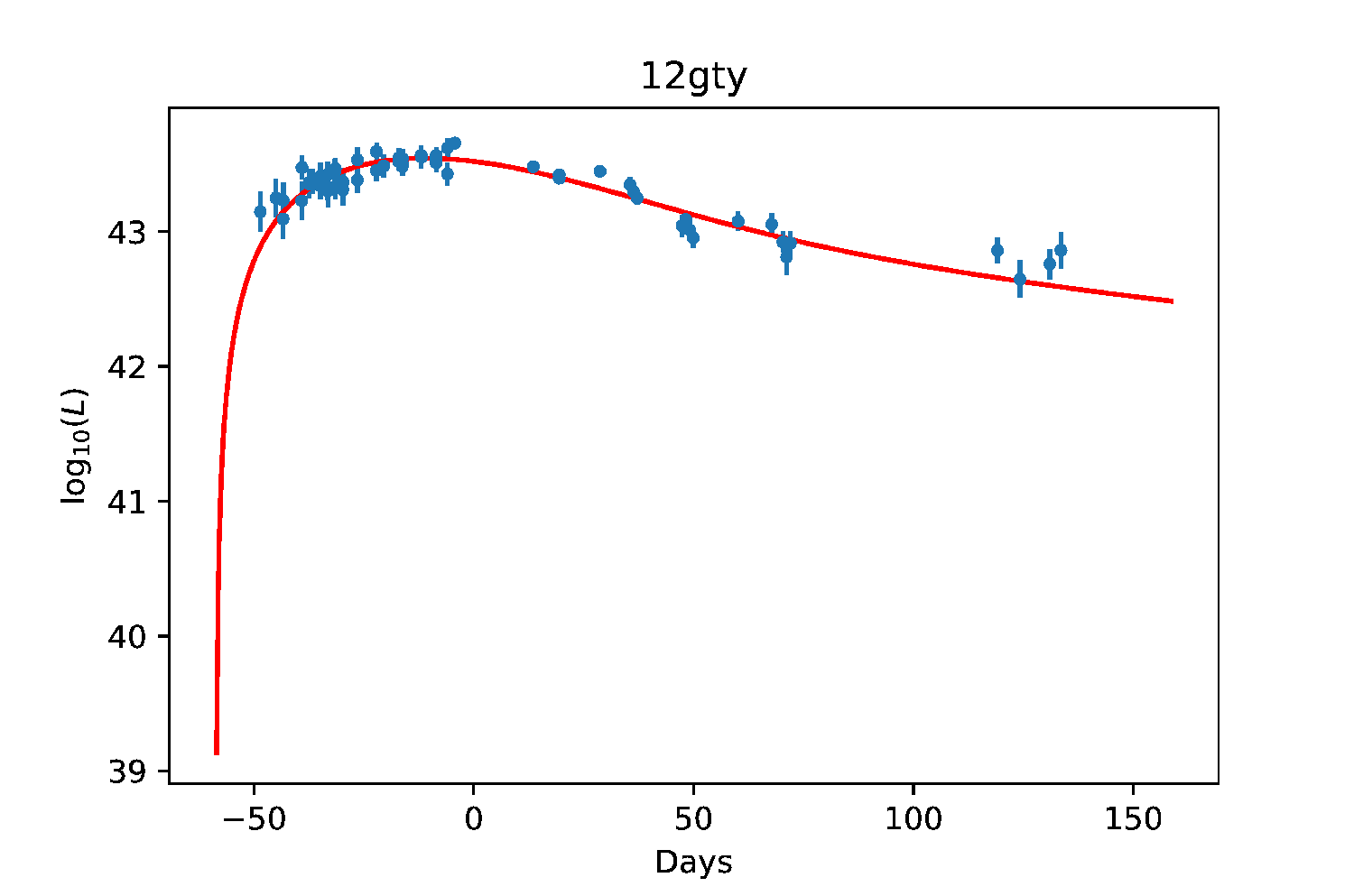}
\plottwo{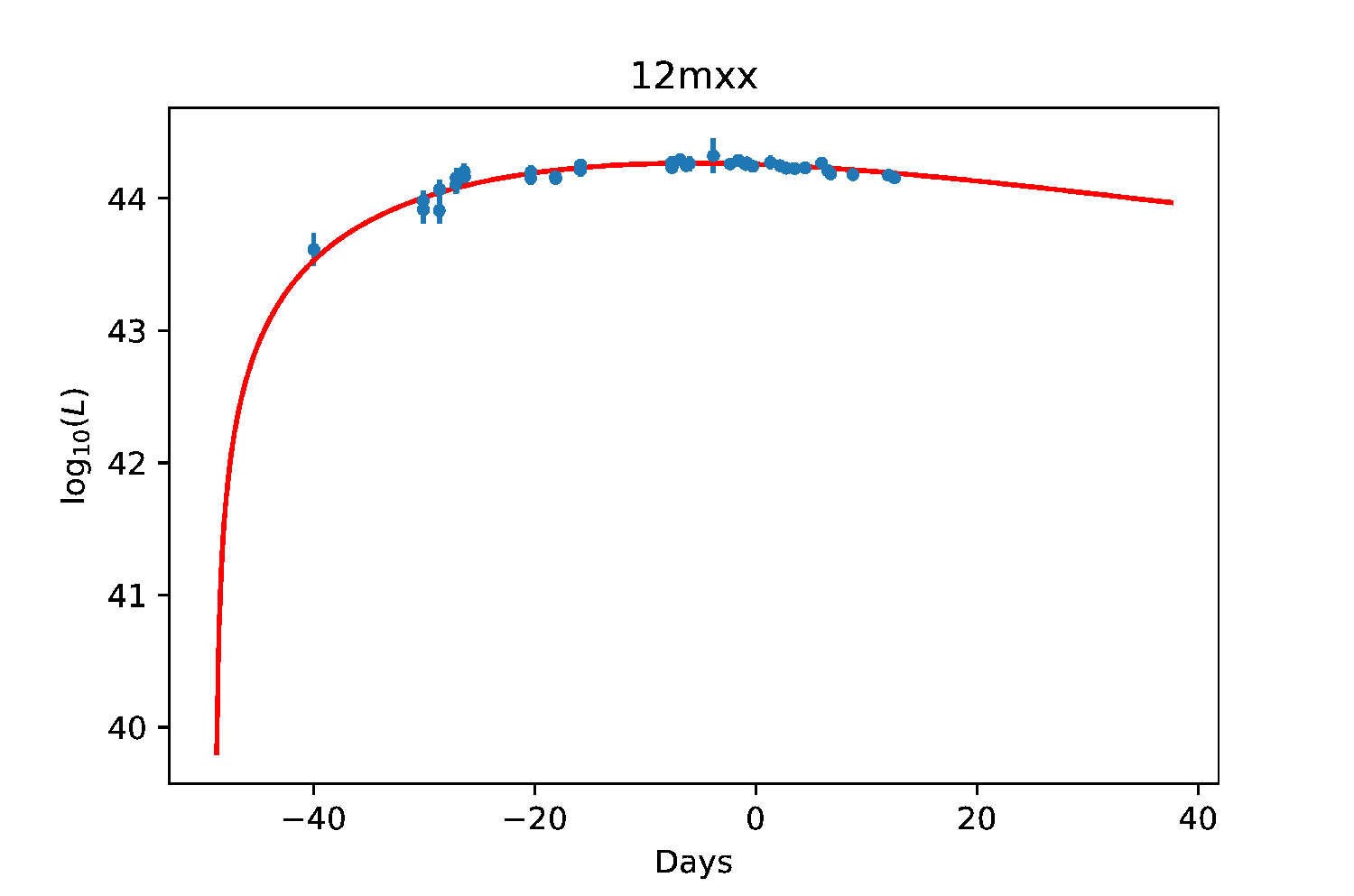}{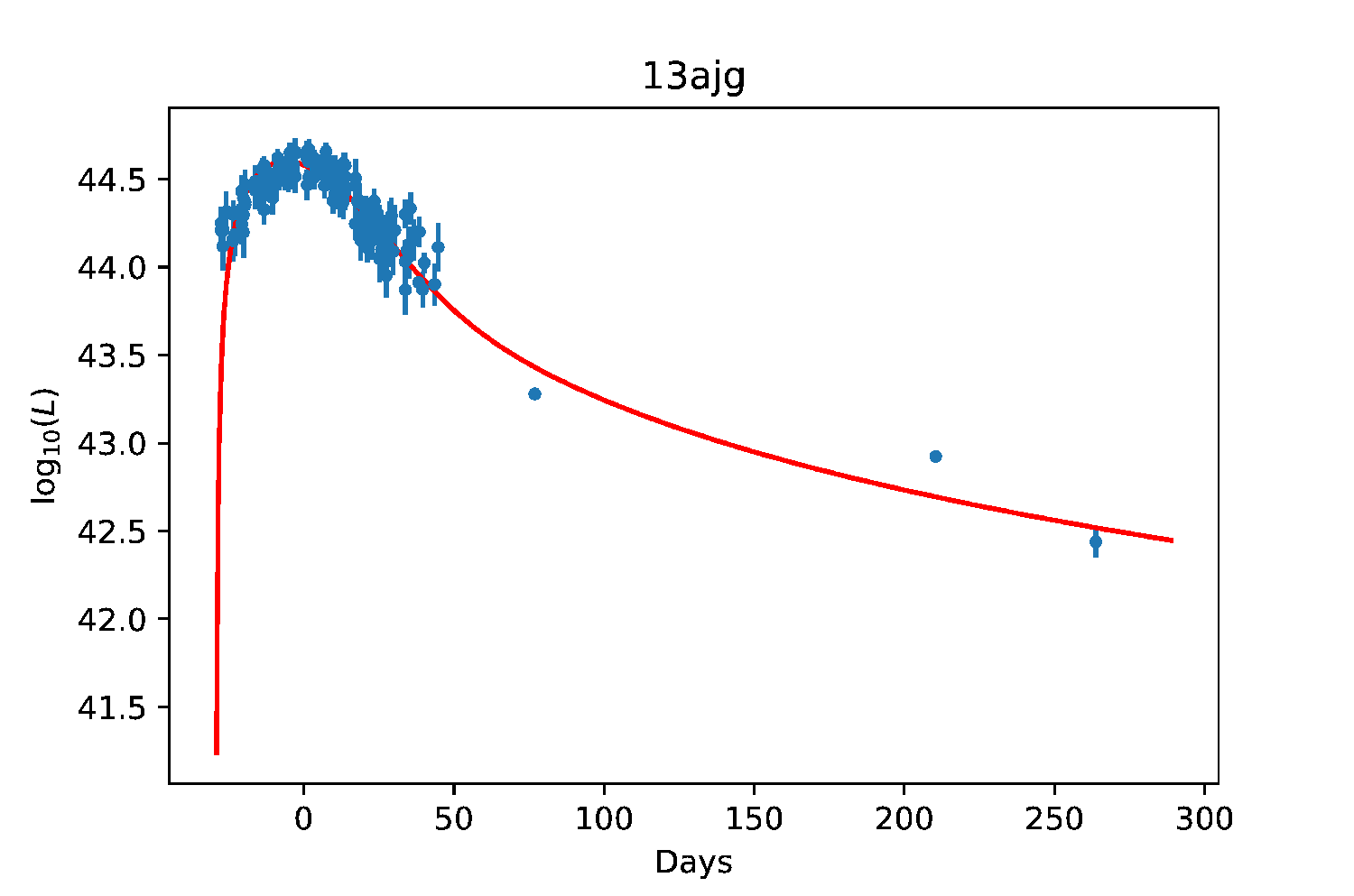}
\caption{Continuation of Fig.~\ref{fig magnetar 1}. \label{fig magnetar 2}}
\end{figure*}

\begin{figure*}[!ht]
\epsscale{0.7}
\plotone{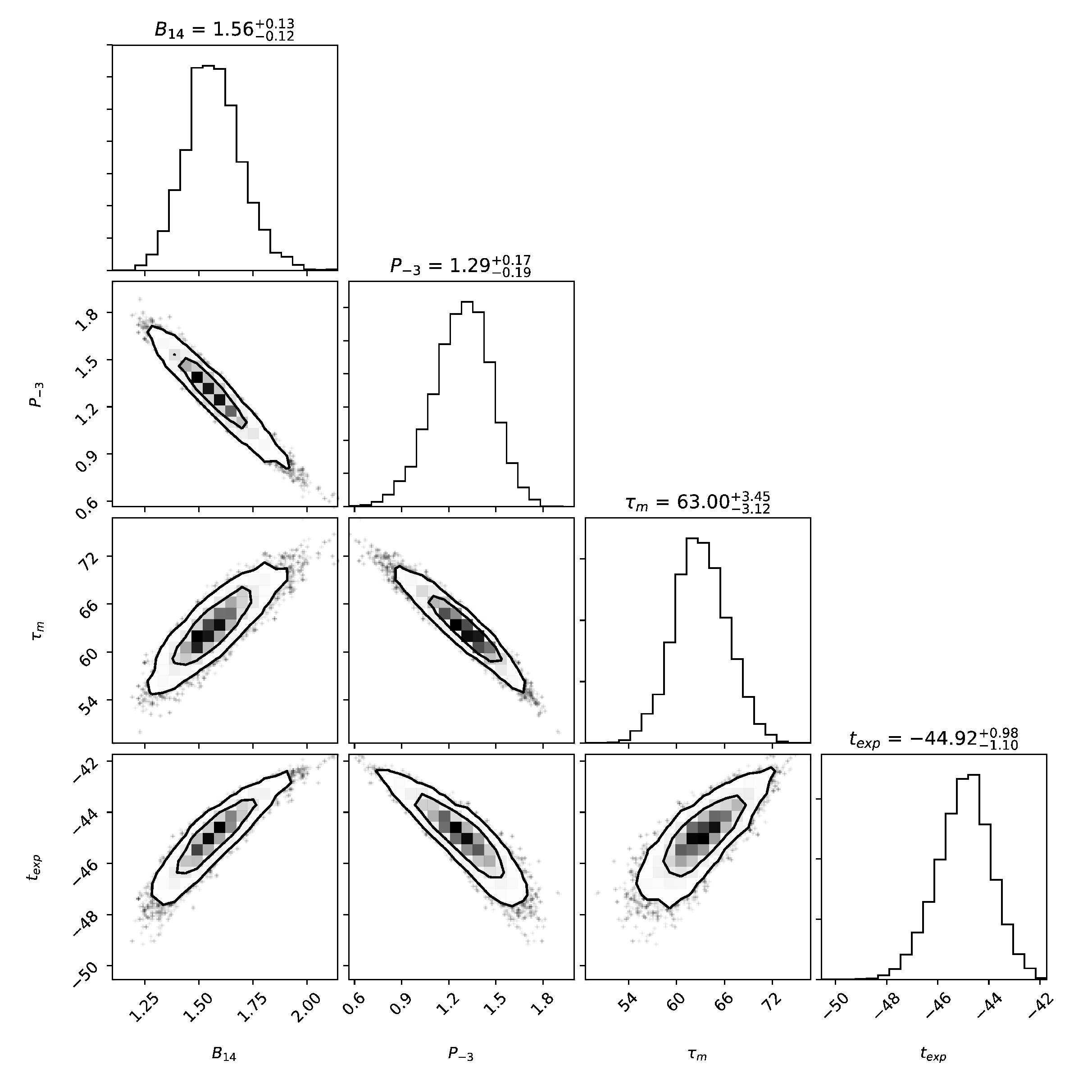}
\caption{Confidence levels of the best-fit parameters for the magnetar model of PTF~09cnd.\label{fig magnetar errors 09cnd}}
\end{figure*}
\begin{figure*}[!hb]
\epsscale{0.7}
\plotone{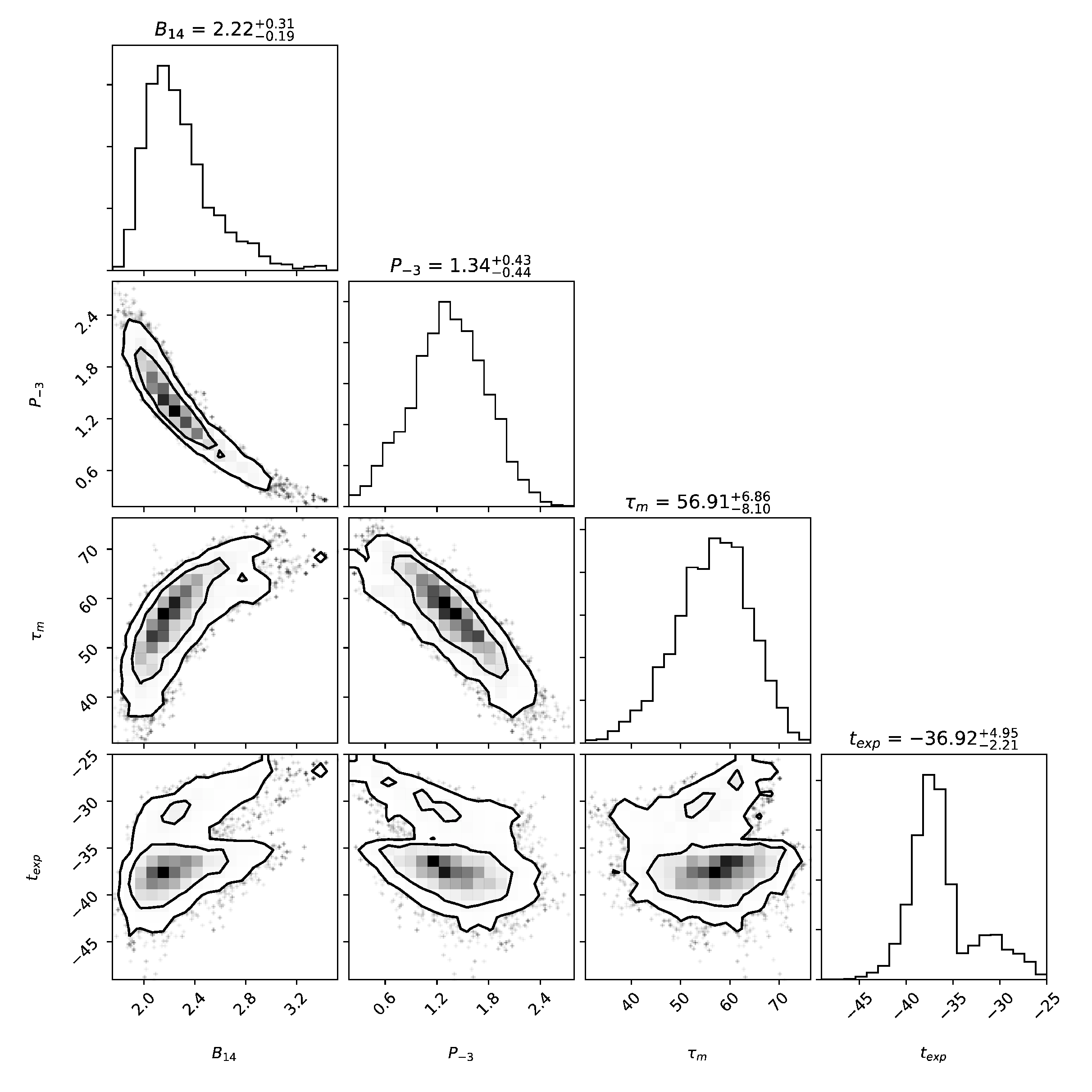}
\caption{Confidence levels of the best-fit parameters for the magnetar model of PTF~09cwl.\label{fig magnetar errors 09cwl}}
\end{figure*}
\newpage
\begin{figure*}[!ht]
\epsscale{0.7}
\plotone{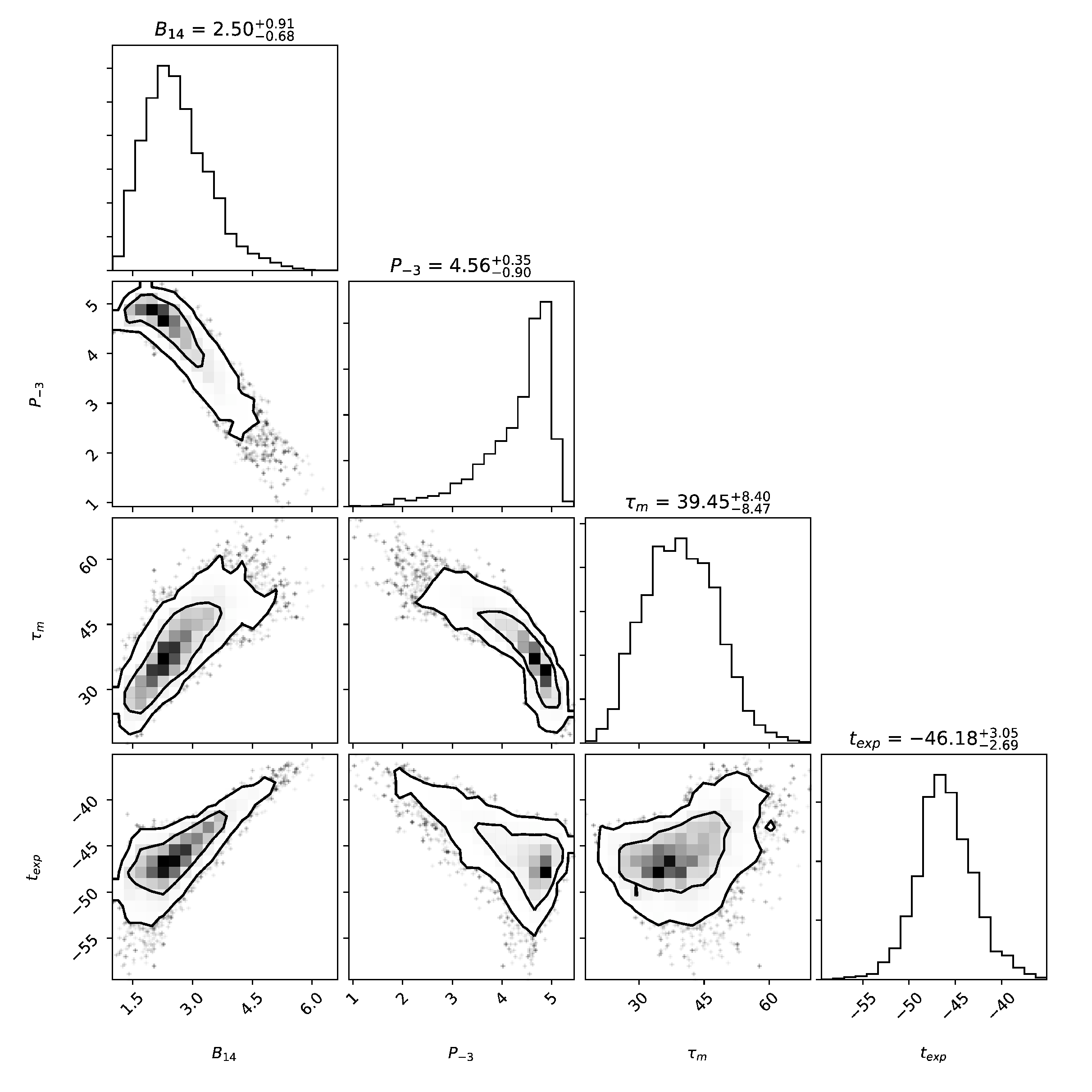}
\caption{Confidence levels of the best-fit parameters for the magnetar model of PTF~10bjp.\label{fig magnetar errors 10bjp}}
\end{figure*}
\begin{figure*}[!hb]
\epsscale{0.7}
\plotone{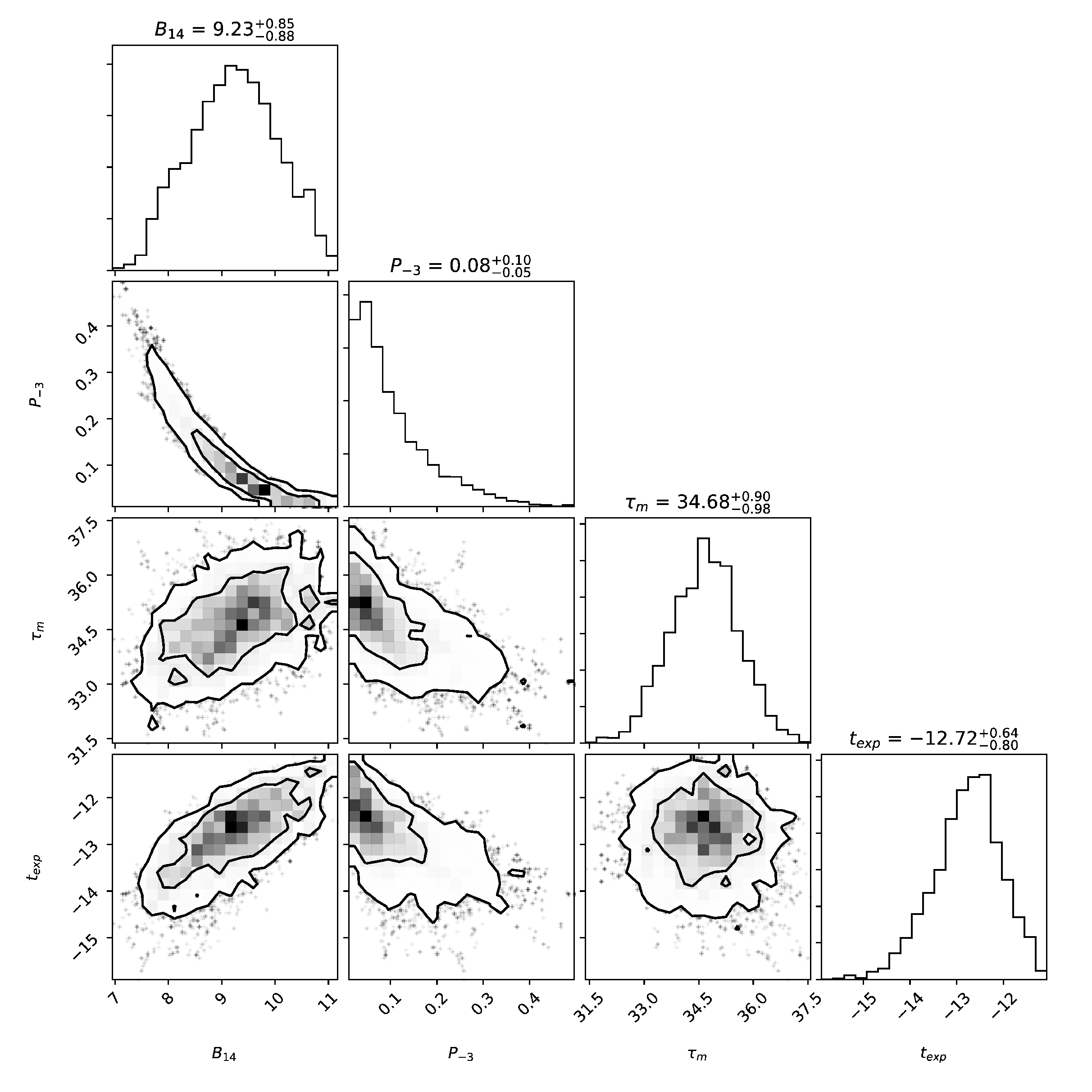}
\caption{Confidence levels of the best-fit parameters for the magnetar model of PTF~10cwr.\label{fig magnetar errors 10cwr}}
\end{figure*}
\newpage
\begin{figure*}[!ht]
\epsscale{0.7}
\plotone{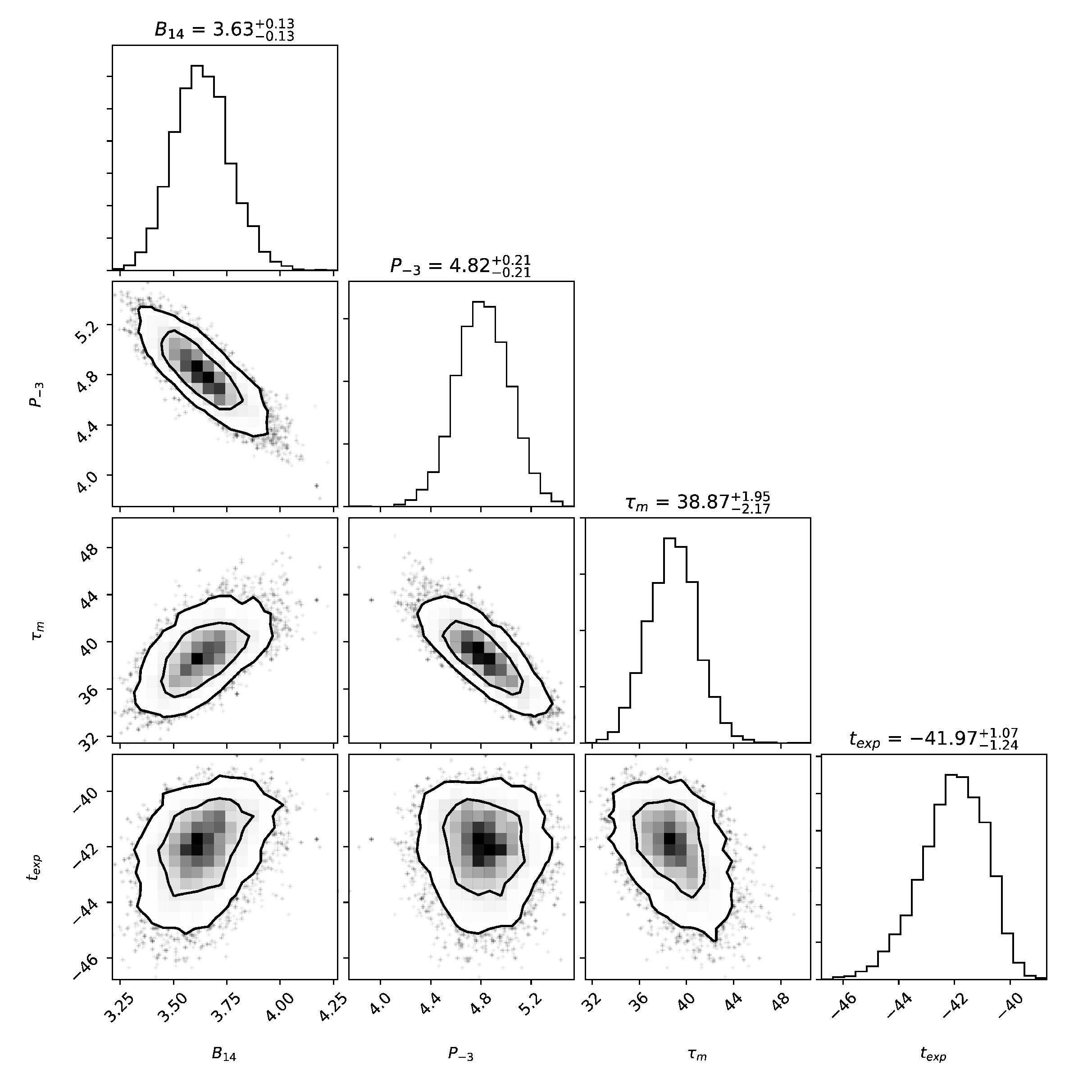}
\caption{Confidence levels of the best-fit parameters for the magnetar model of PTF~10hgi.\label{fig magnetar errors 10hgi}}
\end{figure*}
\begin{figure*}[!hb]
\epsscale{0.7}
\plotone{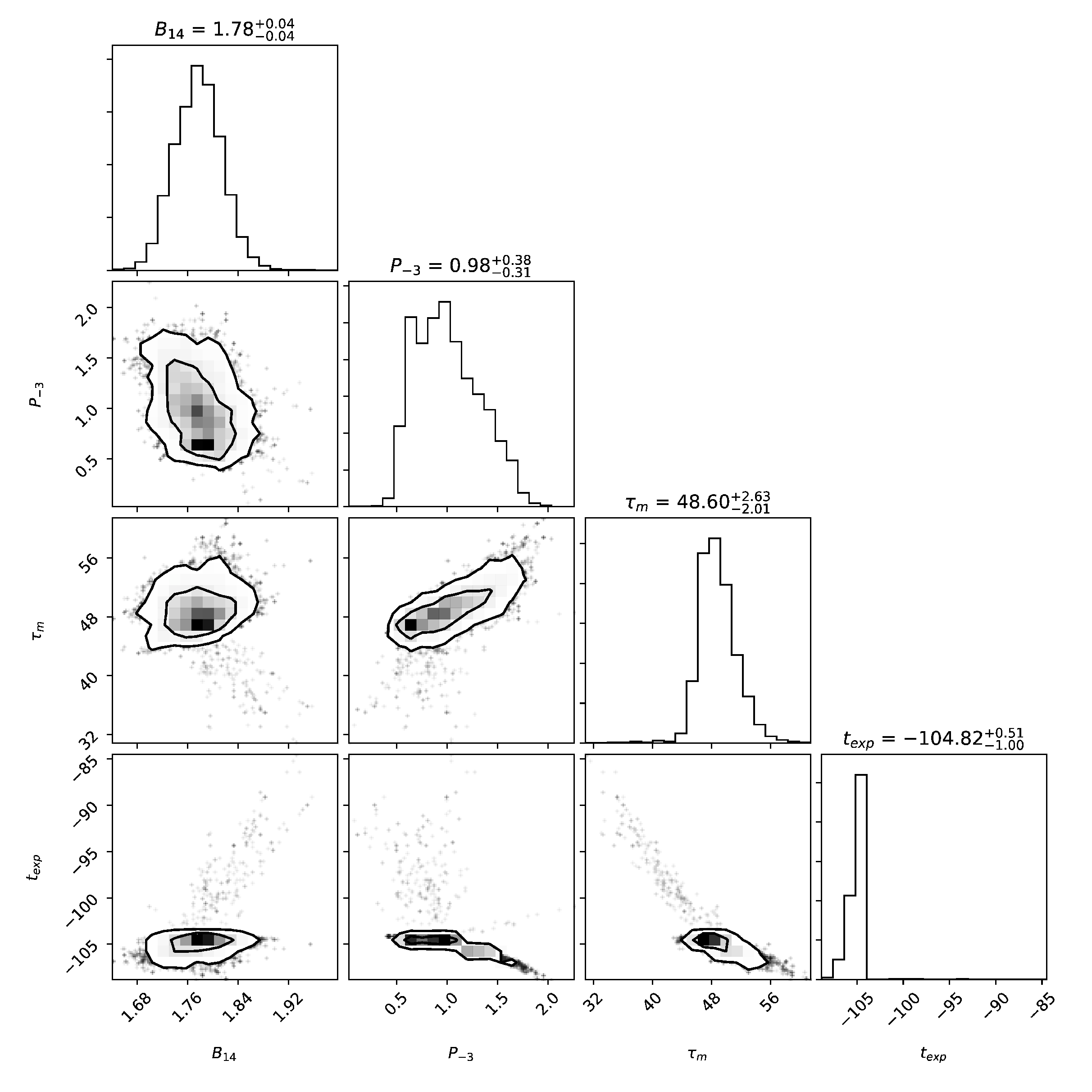}
\caption{Confidence levels of the best-fit parameters for the magnetar model of PTF~10nmn.\label{fig magnetar errors 10nmn}}
\end{figure*}
\newpage
\begin{figure*}[!ht]
\epsscale{0.7}
\plotone{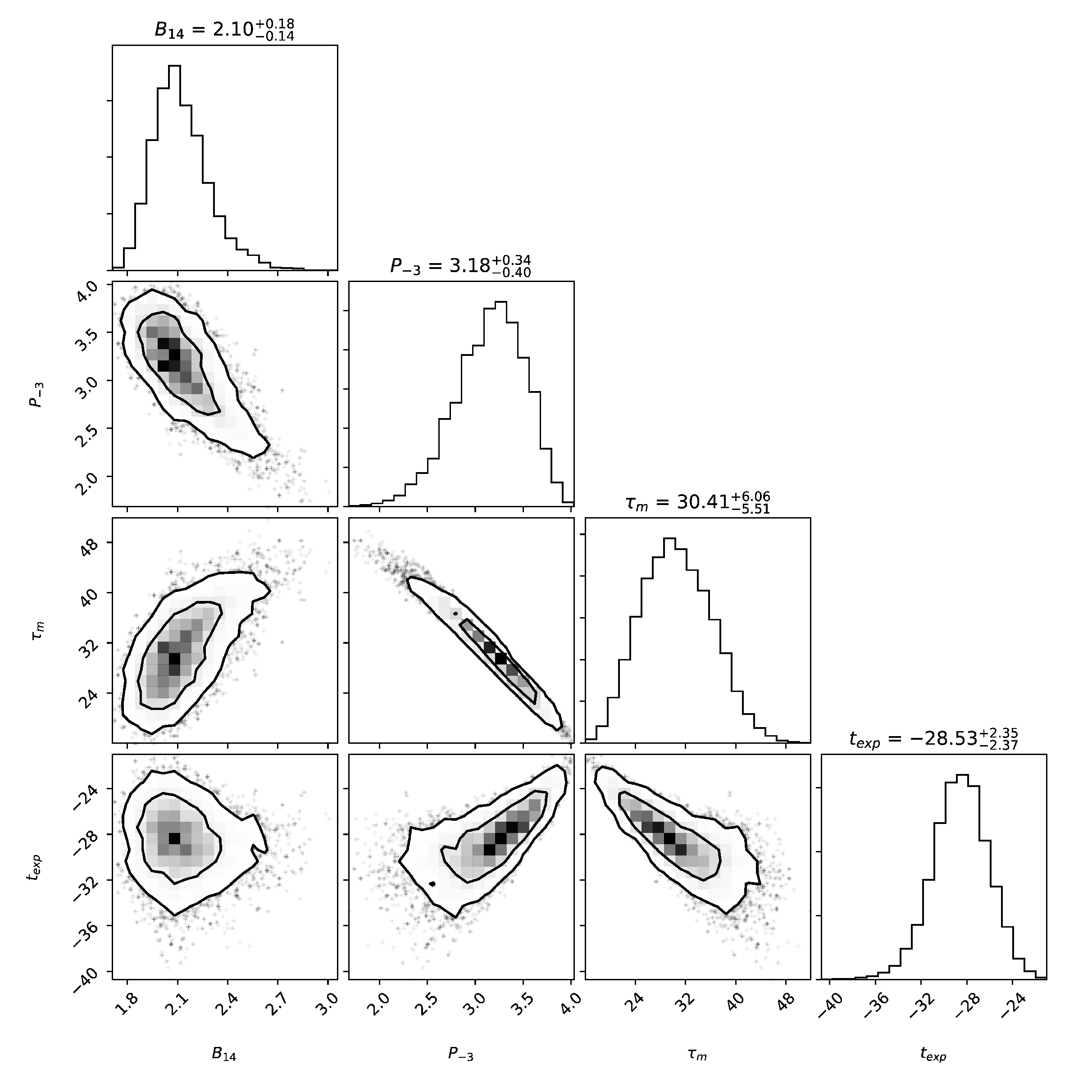}
\caption{Confidence levels of the best-fit parameters for the magnetar model of PTF~10vqv.\label{fig magnetar errors 10vqv}}
\end{figure*}
\begin{figure*}[!hb]
\epsscale{0.7}
\plotone{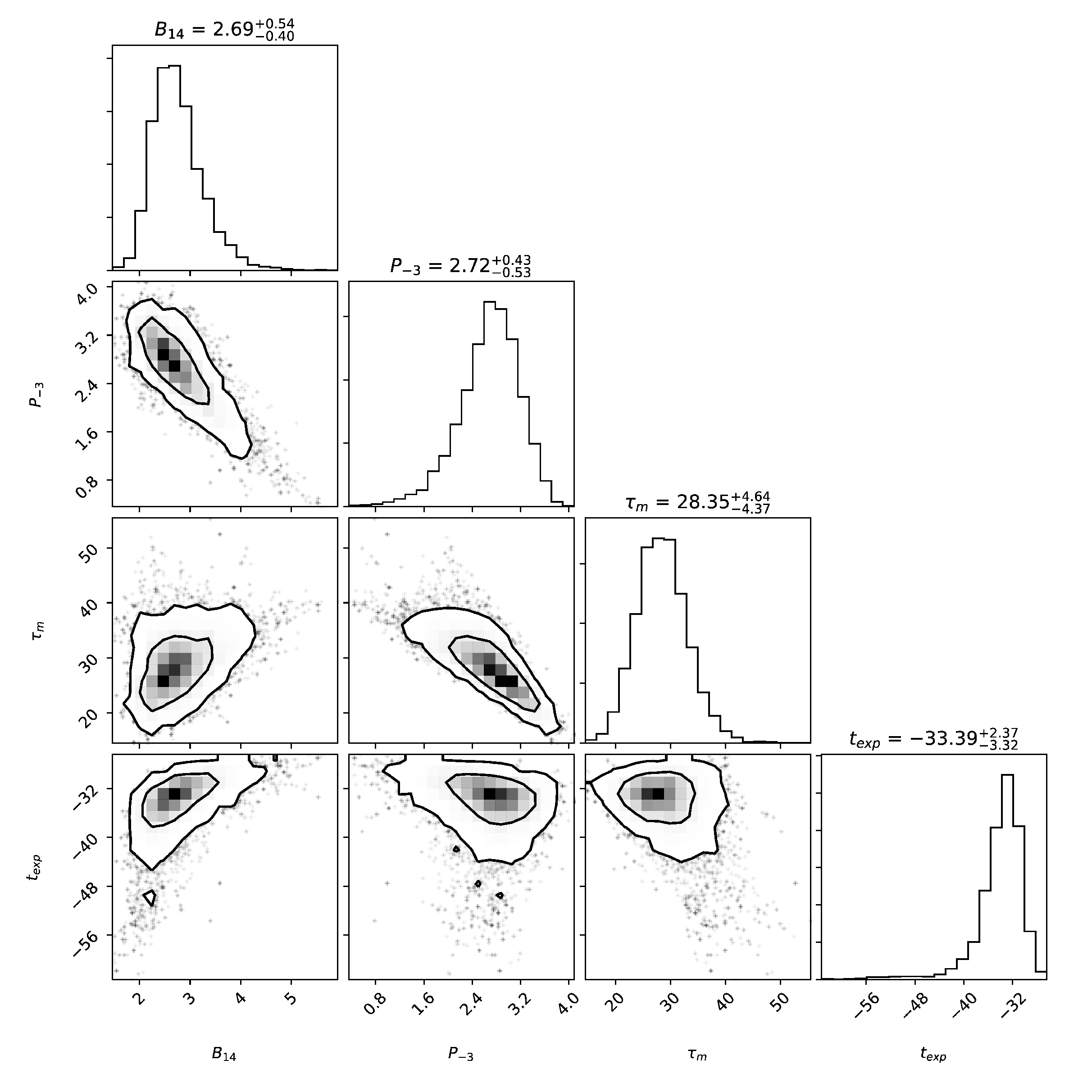}
\caption{Confidence levels of the best-fit parameters for the magnetar model of PTF~10vwg.\label{fig magnetar errors 10vwg}}
\end{figure*}
\newpage
\begin{figure*}[!ht]
\epsscale{0.7}
\plotone{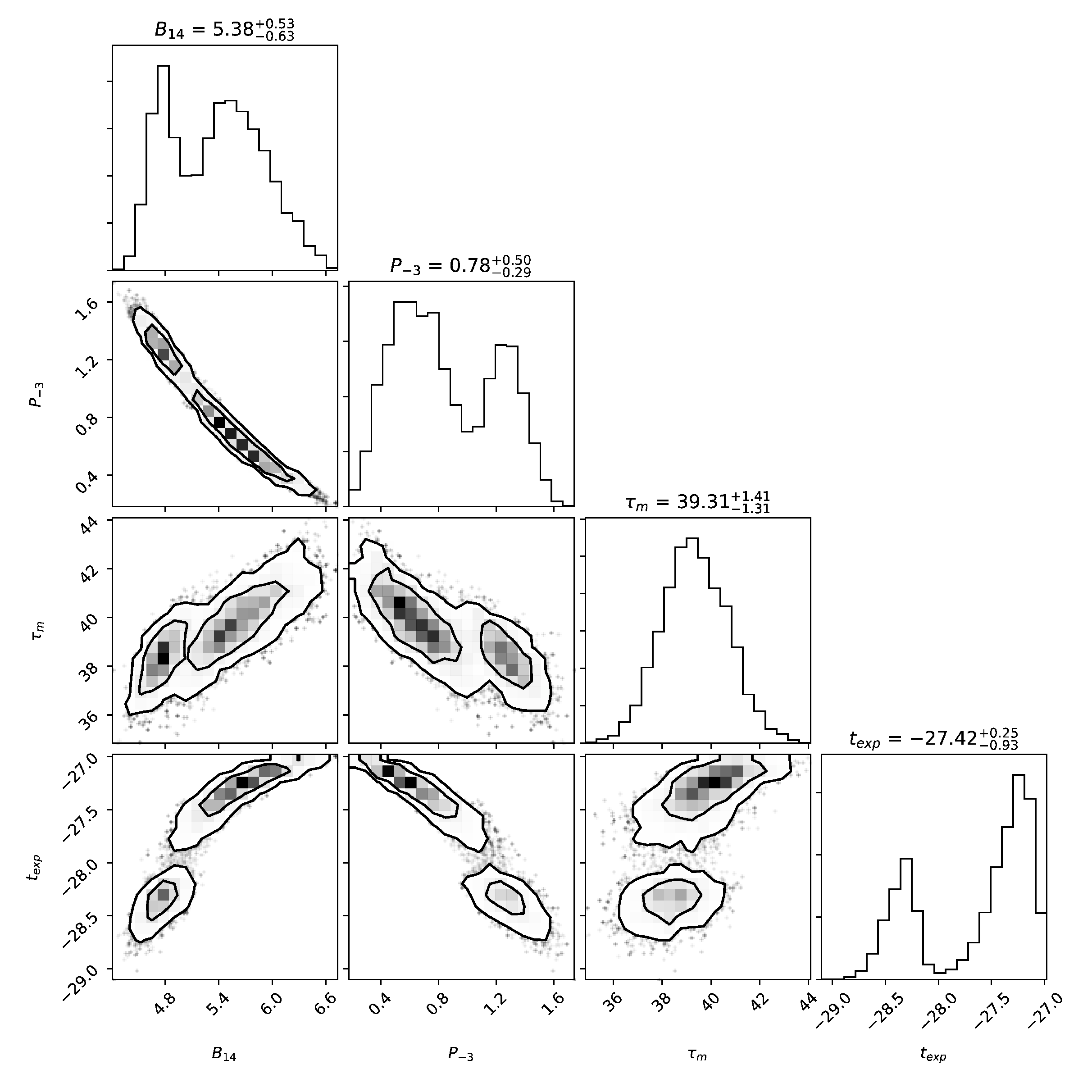}
\caption{Confidence levels of the best-fit parameters for the magnetar model of PTF~11dij.\label{fig magnetar errors 11dij}}
\end{figure*}
\begin{figure*}[!hb]
\epsscale{0.7}
\plotone{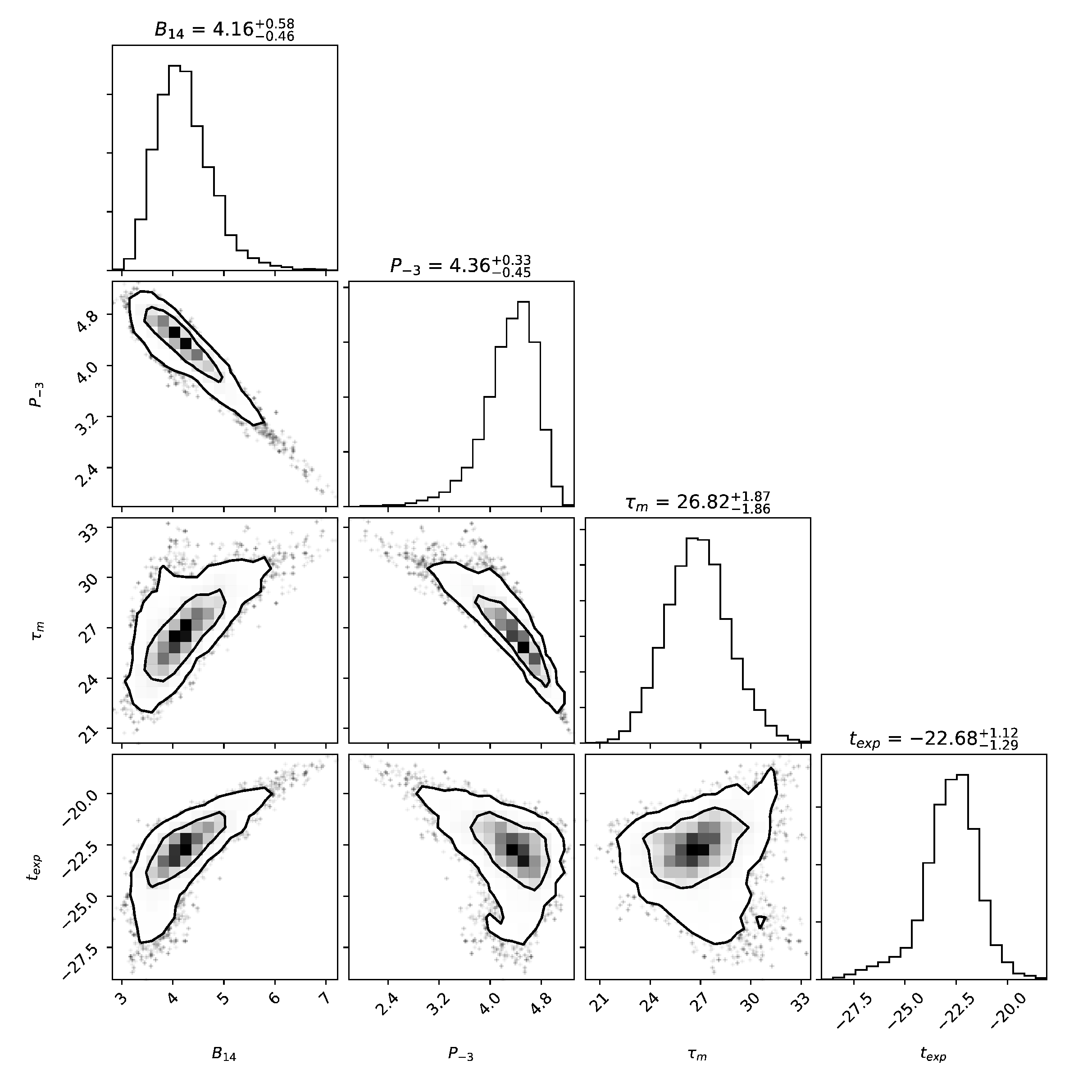}
\caption{Confidence levels of the best-fit parameters for the magnetar model of PTF~11rks.\label{fig magnetar errors 11rks}}
\end{figure*}
\newpage
\begin{figure*}[!ht]
\epsscale{0.7}
\plotone{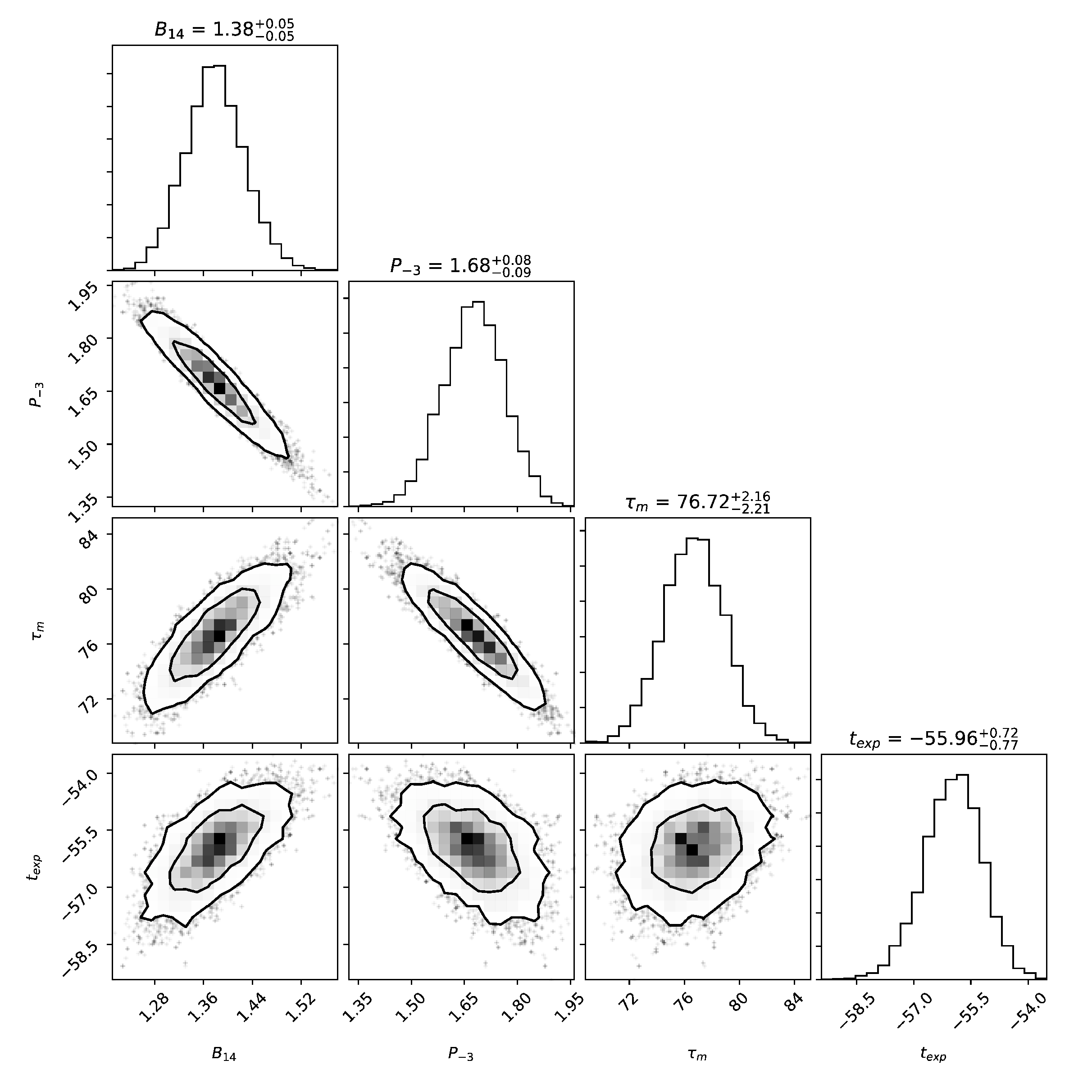}
\caption{Confidence levels of the best-fit parameters for the magnetar model of PTF~12dam.\label{fig magnetar errors 12dam}}
\end{figure*}
\begin{figure*}[!hb]
\epsscale{0.7}
\plotone{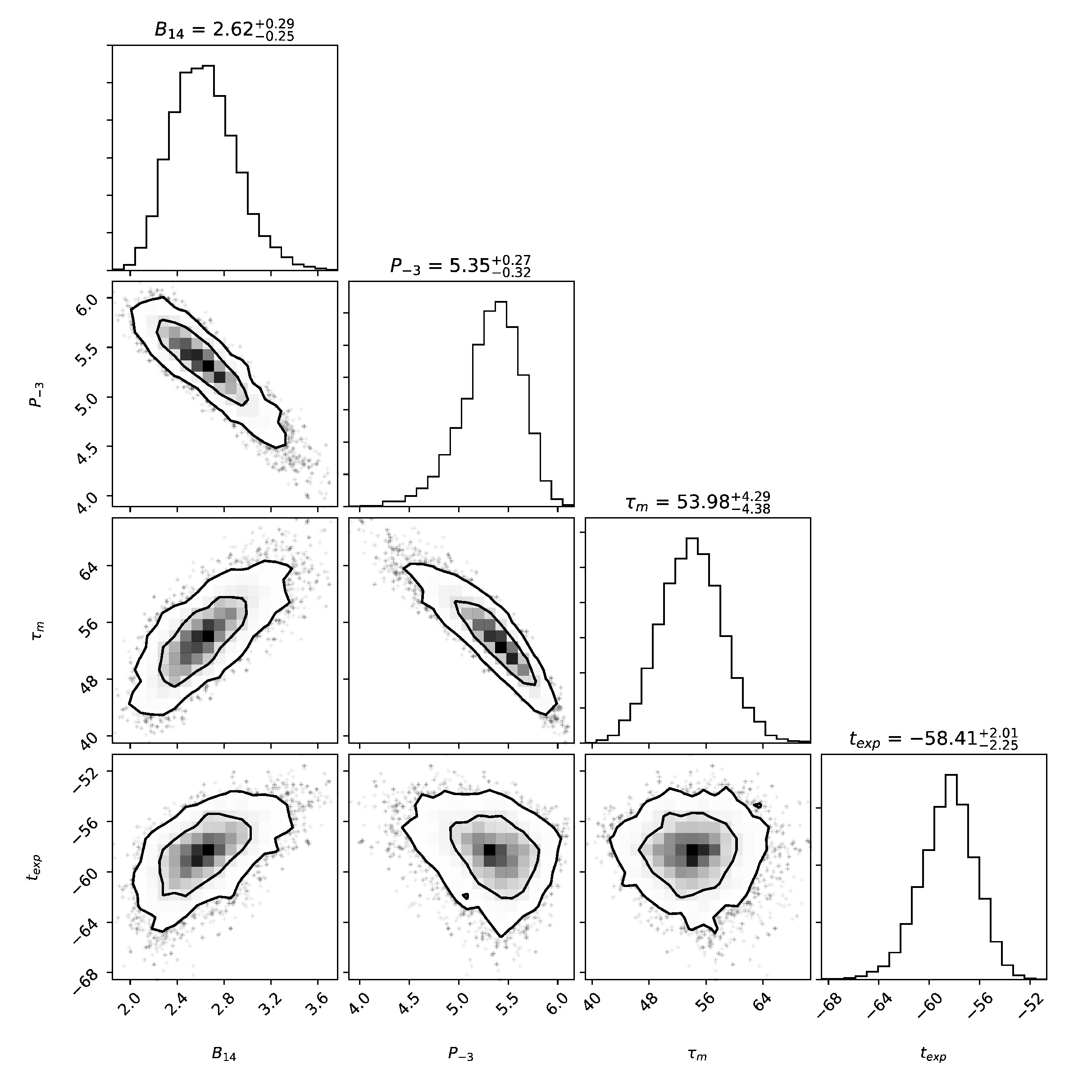}
\caption{Confidence levels of the best-fit parameters for the magnetar model of PTF~12gty.\label{fig magnetar errors 12gty}}
\end{figure*}
\newpage
\begin{figure*}[!ht]
\epsscale{0.7}
\plotone{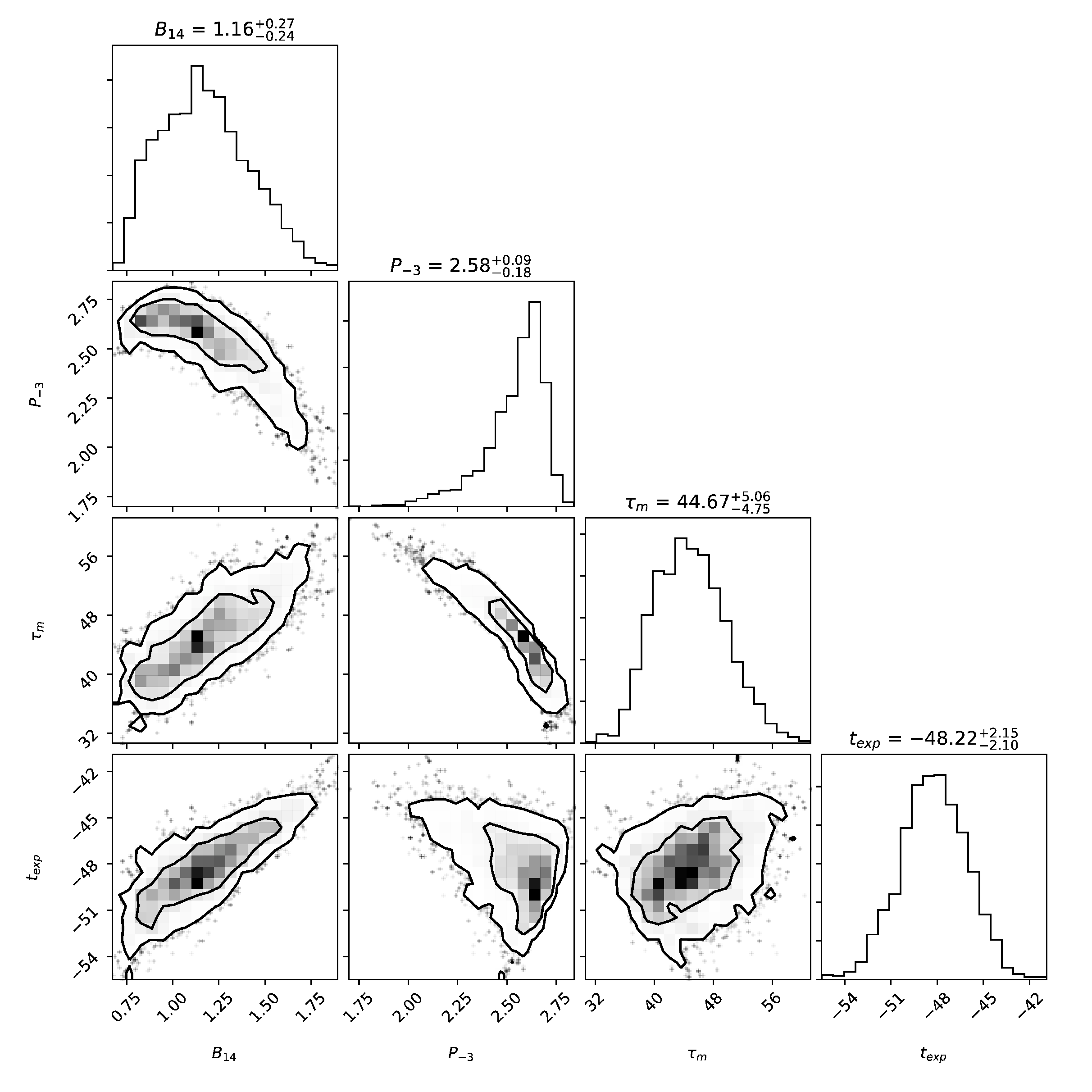}
\caption{Confidence levels of the best-fit parameters for the magnetar model of PTF~12mxx.\label{fig magnetar errors 12mxx}}
\end{figure*}
\begin{figure*}[!hb]
\epsscale{0.7}
\plotone{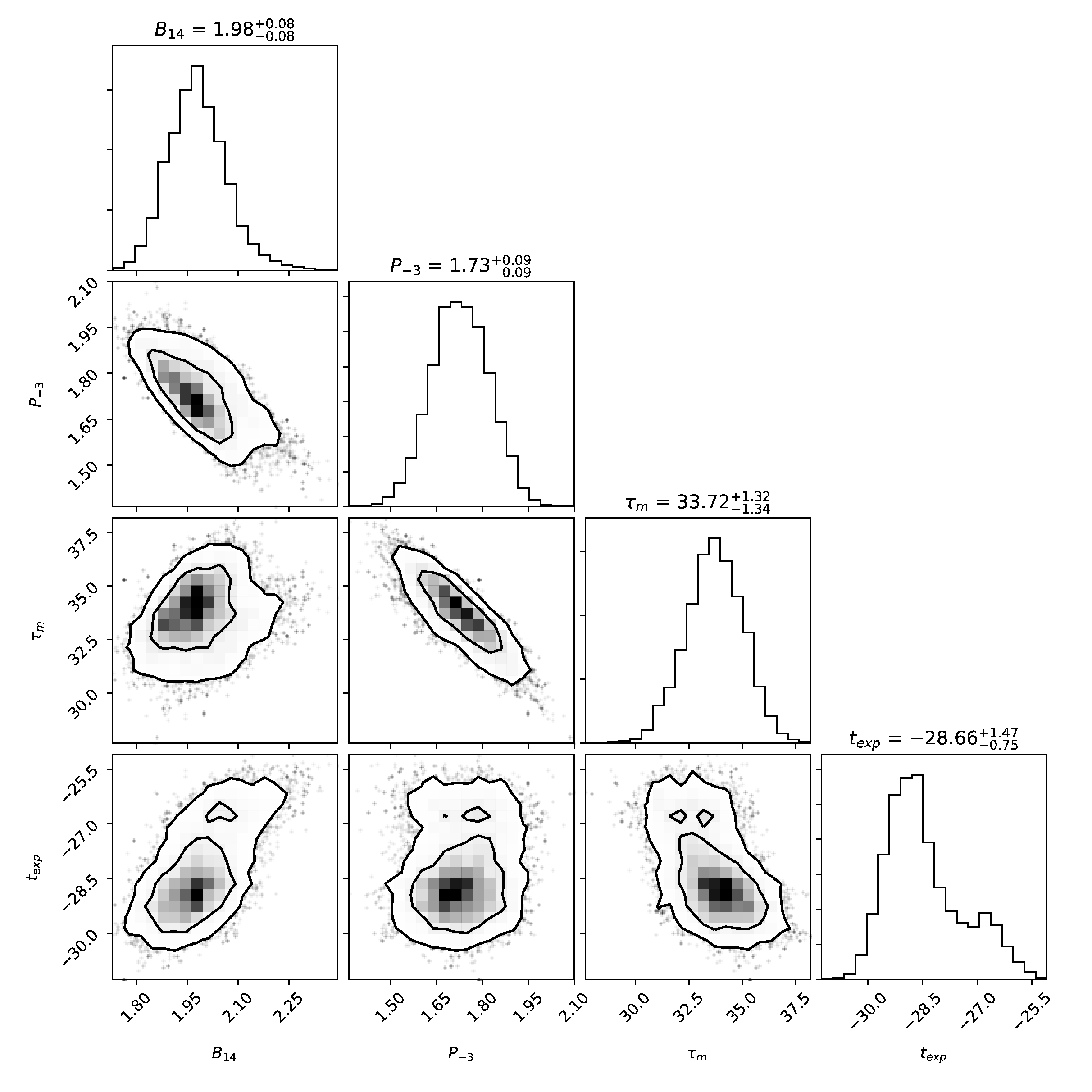}
\caption{Confidence levels of the best-fit parameters for the magnetar model of PTF~13ajg.\label{fig magnetar errors 13ajg}}
\end{figure*}

\clearpage
\newpage
\section{$k$-correction Tables}
The $k$-corrections from $r$ to rest-frame $g$ ($K_{gr}$) and from $i$ to rest-frame $r$ ($K_{ir}$) for the (i)PTF SLSN sample are listed in the tables below.
\floattable
\begin{deluxetable}{@{}r@{\hspace{1.0mm}} | @{}r@{\hspace{0.0mm}}r@{\hspace{0.0mm}}r@{\hspace{0.0mm}}r@{\hspace{0.0mm}}r@{\hspace{0.0mm}} r@{\hspace{0.0mm}}r@{\hspace{0.0mm}}r@{\hspace{0.0mm}}r@{\hspace{0.0mm}}r@{\hspace{0.0mm}} r@{\hspace{0.0mm}}r@{\hspace{0.0mm}} r@{}}[!h]
\tablecaption{$K_{gr}$ corrections\label{table k-corr}}
\tablewidth{0pt}
\tablehead{
\colhead{PTF ID}  &   \colhead{09as}  &   \colhead{09atu}  &   \colhead{09cnd}  &   \colhead{09cwl}  &   \colhead{10aagc}  &   \colhead{10bfz}  &   \colhead{10bjp}  &   \colhead{10cwr}  &   \colhead{10hgi}  &   \colhead{10nmn}  &   \colhead{10uhf}  &   \colhead{10vqv}  &   \colhead{10vwg} \\
 \colhead{$z$}  & \colhead{0.1864}  &   \colhead{  0.5014}  &   \colhead{  0.2585}  &   \colhead{  0.3502}  &   \colhead{  0.2067}  &   \colhead{  0.1699}  &   \colhead{  0.3585}  &   \colhead{  0.2301}  &   \colhead{  0.0982}  &   \colhead{  0.1236}  &   \colhead{  0.2879}  &   \colhead{  0.4520}  &   \colhead{  0.1901} \\ 
 \hline
\colhead{phase}  &   \colhead{$K_{gr}$} \\
}
\startdata
$   -25$ & $ 0.00$ & $-0.50$ & $-0.13$ & $-0.30$ & $-0.04$ & $ 0.04$ & $-0.31$ & $-0.08$ & $ 0.20$ & $ 0.16$ & $-0.19$ & $-0.42$ & $-0.01$ \\ 
$   -24$ & $-0.00$ & $-0.50$ & $-0.14$ & $-0.30$ & $-0.04$ & $ 0.04$ & $-0.32$ & $-0.09$ & $ 0.20$ & $ 0.15$ & $-0.19$ & $-0.42$ & $-0.01$ \\ 
$   -23$ & $-0.00$ & $-0.49$ & $-0.14$ & $-0.30$ & $-0.05$ & $ 0.03$ & $-0.32$ & $-0.09$ & $ 0.19$ & $ 0.15$ & $-0.19$ & $-0.42$ & $-0.01$ \\ 
$   -22$ & $-0.01$ & $-0.49$ & $-0.14$ & $-0.30$ & $-0.05$ & $ 0.03$ & $-0.32$ & $-0.09$ & $ 0.19$ & $ 0.14$ & $-0.20$ & $-0.42$ & $-0.02$ \\ 
$   -22$ & $-0.01$ & $-0.49$ & $-0.14$ & $-0.31$ & $-0.05$ & $ 0.03$ & $-0.32$ & $-0.10$ & $ 0.18$ & $ 0.14$ & $-0.20$ & $-0.42$ & $-0.02$ \\ 
$   -21$ & $-0.02$ & $-0.49$ & $-0.15$ & $-0.31$ & $-0.06$ & $ 0.02$ & $-0.32$ & $-0.10$ & $ 0.18$ & $ 0.13$ & $-0.20$ & $-0.42$ & $-0.02$ \\ 
$   -20$ & $-0.02$ & $-0.49$ & $-0.15$ & $-0.31$ & $-0.06$ & $ 0.02$ & $-0.32$ & $-0.10$ & $ 0.17$ & $ 0.13$ & $-0.20$ & $-0.42$ & $-0.03$ \\ 
$   -14$ & $-0.04$ & $-0.48$ & $-0.17$ & $-0.31$ & $-0.08$ & $-0.01$ & $-0.32$ & $-0.12$ & $ 0.14$ & $ 0.10$ & $-0.22$ & $-0.41$ & $-0.05$ \\ 
$    -8$ & $-0.06$ & $-0.46$ & $-0.19$ & $-0.32$ & $-0.10$ & $-0.03$ & $-0.33$ & $-0.14$ & $ 0.11$ & $ 0.07$ & $-0.24$ & $-0.41$ & $-0.07$ \\ 
$    -3$ & $-0.08$ & $-0.45$ & $-0.20$ & $-0.32$ & $-0.12$ & $-0.05$ & $-0.33$ & $-0.16$ & $ 0.09$ & $ 0.05$ & $-0.25$ & $-0.40$ & $-0.09$ \\ 
$     0$ & $-0.10$ & $-0.45$ & $-0.21$ & $-0.33$ & $-0.13$ & $-0.06$ & $-0.33$ & $-0.17$ & $ 0.07$ & $ 0.03$ & $-0.26$ & $-0.40$ & $-0.10$ \\ 
$     3$ & $-0.10$ & $-0.44$ & $-0.22$ & $-0.33$ & $-0.14$ & $-0.07$ & $-0.34$ & $-0.17$ & $ 0.06$ & $ 0.02$ & $-0.26$ & $-0.39$ & $-0.11$ \\ 
$     6$ & $-0.11$ & $-0.44$ & $-0.23$ & $-0.33$ & $-0.15$ & $-0.08$ & $-0.34$ & $-0.18$ & $ 0.04$ & $ 0.01$ & $-0.27$ & $-0.39$ & $-0.12$ \\ 
$    10$ & $-0.13$ & $-0.43$ & $-0.23$ & $-0.34$ & $-0.16$ & $-0.09$ & $-0.34$ & $-0.19$ & $ 0.03$ & $-0.01$ & $-0.28$ & $-0.39$ & $-0.13$ \\ 
$    16$ & $-0.14$ & $-0.42$ & $-0.25$ & $-0.34$ & $-0.17$ & $-0.11$ & $-0.34$ & $-0.21$ & $ 0.00$ & $-0.03$ & $-0.29$ & $-0.38$ & $-0.15$ \\ 
$    21$ & $-0.15$ & $-0.42$ & $-0.26$ & $-0.34$ & $-0.18$ & $-0.12$ & $-0.35$ & $-0.22$ & $-0.01$ & $-0.05$ & $-0.30$ & $-0.38$ & $-0.16$ \\ 
$    22$ & $-0.16$ & $-0.42$ & $-0.26$ & $-0.35$ & $-0.19$ & $-0.13$ & $-0.35$ & $-0.22$ & $-0.01$ & $-0.05$ & $-0.30$ & $-0.38$ & $-0.16$ \\ 
$    26$ & $-0.17$ & $-0.41$ & $-0.27$ & $-0.35$ & $-0.20$ & $-0.14$ & $-0.35$ & $-0.23$ & $-0.03$ & $-0.06$ & $-0.31$ & $-0.38$ & $-0.17$ \\ 
$    47$ & $-0.20$ & $-0.39$ & $-0.30$ & $-0.36$ & $-0.23$ & $-0.18$ & $-0.36$ & $-0.26$ & $-0.09$ & $-0.12$ & $-0.34$ & $-0.36$ & $-0.21$ \\ 
$    57$ & $-0.22$ & $-0.38$ & $-0.31$ & $-0.37$ & $-0.24$ & $-0.19$ & $-0.36$ & $-0.27$ & $-0.11$ & $-0.14$ & $-0.35$ & $-0.36$ & $-0.22$ \\ 
$    57$ & $-0.22$ & $-0.38$ & $-0.31$ & $-0.37$ & $-0.24$ & $-0.19$ & $-0.36$ & $-0.27$ & $-0.11$ & $-0.14$ & $-0.35$ & $-0.36$ & $-0.22$ \\ 
$    87$ & $-0.24$ & $-0.36$ & $-0.34$ & $-0.38$ & $-0.27$ & $-0.22$ & $-0.37$ & $-0.29$ & $-0.16$ & $-0.18$ & $-0.37$ & $-0.35$ & $-0.25$ \\ 
$   154$ & $-0.23$ & $-0.35$ & $-0.34$ & $-0.38$ & $-0.26$ & $-0.21$ & $-0.38$ & $-0.29$ & $-0.18$ & $-0.19$ & $-0.38$ & $-0.35$ & $-0.24$ \\ 
$   164$ & $-0.23$ & $-0.35$ & $-0.33$ & $-0.38$ & $-0.25$ & $-0.21$ & $-0.38$ & $-0.28$ & $-0.18$ & $-0.18$ & $-0.37$ & $-0.35$ & $-0.23$ \\ 
$   214$ & $-0.19$ & $-0.36$ & $-0.31$ & $-0.38$ & $-0.22$ & $-0.17$ & $-0.37$ & $-0.26$ & $-0.15$ & $-0.16$ & $-0.35$ & $-0.35$ & $-0.20$ \\ 
$   266$ & $-0.18$ & $-0.37$ & $-0.29$ & $-0.36$ & $-0.21$ & $-0.16$ & $-0.37$ & $-0.24$ & $-0.13$ & $-0.14$ & $-0.33$ & $-0.36$ & $-0.18$ \\ 
$   321$ & $-0.21$ & $-0.38$ & $-0.29$ & $-0.35$ & $-0.24$ & $-0.19$ & $-0.36$ & $-0.26$ & $-0.15$ & $-0.17$ & $-0.32$ & $-0.36$ & $-0.22$ \\ 
\enddata
\end{deluxetable}
\vspace{3cm}
\floattable
\begin{deluxetable}{@{}r@{\hspace{1.0mm}} | @{}r@{\hspace{0.0mm}}r@{\hspace{0.0mm}}r@{\hspace{0.0mm}}r@{\hspace{0.0mm}}r@{\hspace{0.0mm}} r@{\hspace{0.0mm}}r@{\hspace{0.0mm}}r@{\hspace{0.0mm}}r@{\hspace{0.0mm}}r@{\hspace{0.0mm}} r@{\hspace{0.0mm}}r@{\hspace{0.0mm}} r@{}}
\tablecaption{Countinuation of Table \ref{table k-corr}.\label{table k-corr 2}}
\tablewidth{0pt}
\tablehead{
\colhead{PTF ID}  &   \colhead{11dij}  &   \colhead{11hrq}  &   \colhead{11rks}  &   \colhead{12dam}  &   \colhead{12gty}  &   \colhead{12hni}  &   \colhead{12mxx}  &   \colhead{13ajg}  &   \colhead{13bdl}  &   \colhead{13bjz}  &   \colhead{13cjq}  &   \colhead{13dcc}  &   \colhead{13ehe} \\
\colhead{$z$}   &   \colhead{  0.1429}  &   \colhead{  0.0571}  &   \colhead{  0.1924}  &   \colhead{  0.1075}  &   \colhead{  0.1768}  &   \colhead{  0.1056}  &   \colhead{  0.3274}  &   \colhead{  0.7403}  &   \colhead{  0.4030}  &   \colhead{  0.2712}  &   \colhead{  0.3962}  &   \colhead{  0.4308}  &   \colhead{  0.3434} \\
 \hline
\colhead{phase}  &   \colhead{$K_{gr}$} \\
}
\startdata
$   -25$ & $ 0.11$ & $ 0.26$ & $-0.01$ & $ 0.19$ & $ 0.02$ & $ 0.19$ & $-0.26$ & $-0.81$ & $-0.37$ & $-0.16$ & $-0.36$ & $-0.40$ & $-0.29$ \\ 
$   -24$ & $ 0.11$ & $ 0.25$ & $-0.01$ & $ 0.18$ & $ 0.02$ & $ 0.19$ & $-0.27$ & $-0.80$ & $-0.37$ & $-0.16$ & $-0.36$ & $-0.39$ & $-0.29$ \\ 
$   -23$ & $ 0.10$ & $ 0.25$ & $-0.02$ & $ 0.18$ & $ 0.02$ & $ 0.18$ & $-0.27$ & $-0.79$ & $-0.37$ & $-0.16$ & $-0.36$ & $-0.39$ & $-0.29$ \\ 
$   -22$ & $ 0.10$ & $ 0.24$ & $-0.02$ & $ 0.17$ & $ 0.01$ & $ 0.18$ & $-0.27$ & $-0.78$ & $-0.36$ & $-0.17$ & $-0.36$ & $-0.39$ & $-0.29$ \\ 
$   -22$ & $ 0.09$ & $ 0.24$ & $-0.03$ & $ 0.17$ & $ 0.01$ & $ 0.17$ & $-0.27$ & $-0.77$ & $-0.36$ & $-0.17$ & $-0.36$ & $-0.39$ & $-0.30$ \\ 
$   -21$ & $ 0.09$ & $ 0.23$ & $-0.03$ & $ 0.16$ & $ 0.01$ & $ 0.17$ & $-0.27$ & $-0.76$ & $-0.36$ & $-0.17$ & $-0.36$ & $-0.39$ & $-0.30$ \\ 
$   -20$ & $ 0.08$ & $ 0.23$ & $-0.03$ & $ 0.16$ & $ 0.00$ & $ 0.16$ & $-0.27$ & $-0.75$ & $-0.36$ & $-0.17$ & $-0.36$ & $-0.39$ & $-0.30$ \\ 
$   -14$ & $ 0.06$ & $ 0.20$ & $-0.05$ & $ 0.13$ & $-0.02$ & $ 0.13$ & $-0.28$ & $-0.70$ & $-0.36$ & $-0.19$ & $-0.36$ & $-0.39$ & $-0.31$ \\ 
$    -8$ & $ 0.03$ & $ 0.17$ & $-0.08$ & $ 0.10$ & $-0.05$ & $ 0.10$ & $-0.29$ & $-0.64$ & $-0.36$ & $-0.21$ & $-0.36$ & $-0.38$ & $-0.31$ \\ 
$    -3$ & $ 0.01$ & $ 0.14$ & $-0.09$ & $ 0.07$ & $-0.06$ & $ 0.08$ & $-0.30$ & $-0.60$ & $-0.36$ & $-0.22$ & $-0.36$ & $-0.38$ & $-0.32$ \\ 
$     0$ & $-0.00$ & $ 0.13$ & $-0.11$ & $ 0.06$ & $-0.08$ & $ 0.06$ & $-0.31$ & $-0.57$ & $-0.36$ & $-0.23$ & $-0.35$ & $-0.38$ & $-0.32$ \\ 
$     3$ & $-0.01$ & $ 0.12$ & $-0.11$ & $ 0.04$ & $-0.09$ & $ 0.05$ & $-0.31$ & $-0.55$ & $-0.36$ & $-0.24$ & $-0.35$ & $-0.38$ & $-0.33$ \\ 
$     6$ & $-0.03$ & $ 0.10$ & $-0.13$ & $ 0.03$ & $-0.10$ & $ 0.03$ & $-0.32$ & $-0.52$ & $-0.36$ & $-0.25$ & $-0.35$ & $-0.37$ & $-0.33$ \\ 
$    10$ & $-0.04$ & $ 0.09$ & $-0.14$ & $ 0.02$ & $-0.11$ & $ 0.02$ & $-0.32$ & $-0.49$ & $-0.36$ & $-0.25$ & $-0.35$ & $-0.37$ & $-0.33$ \\ 
$    16$ & $-0.06$ & $ 0.06$ & $-0.15$ & $-0.01$ & $-0.13$ & $-0.01$ & $-0.33$ & $-0.45$ & $-0.35$ & $-0.27$ & $-0.35$ & $-0.37$ & $-0.34$ \\ 
$    21$ & $-0.07$ & $ 0.05$ & $-0.16$ & $-0.02$ & $-0.14$ & $-0.02$ & $-0.34$ & $-0.42$ & $-0.35$ & $-0.28$ & $-0.35$ & $-0.37$ & $-0.34$ \\ 
$    22$ & $-0.08$ & $ 0.04$ & $-0.16$ & $-0.03$ & $-0.14$ & $-0.02$ & $-0.34$ & $-0.42$ & $-0.35$ & $-0.28$ & $-0.35$ & $-0.37$ & $-0.34$ \\ 
$    26$ & $-0.09$ & $ 0.03$ & $-0.17$ & $-0.04$ & $-0.15$ & $-0.04$ & $-0.34$ & $-0.39$ & $-0.35$ & $-0.29$ & $-0.35$ & $-0.36$ & $-0.35$ \\ 
$    47$ & $-0.14$ & $-0.03$ & $-0.21$ & $-0.10$ & $-0.19$ & $-0.10$ & $-0.36$ & $-0.29$ & $-0.35$ & $-0.32$ & $-0.35$ & $-0.36$ & $-0.36$ \\ 
$    57$ & $-0.15$ & $-0.05$ & $-0.23$ & $-0.12$ & $-0.20$ & $-0.12$ & $-0.37$ & $-0.26$ & $-0.35$ & $-0.33$ & $-0.35$ & $-0.36$ & $-0.37$ \\ 
$    57$ & $-0.15$ & $-0.05$ & $-0.23$ & $-0.12$ & $-0.20$ & $-0.12$ & $-0.37$ & $-0.26$ & $-0.35$ & $-0.33$ & $-0.35$ & $-0.36$ & $-0.37$ \\ 
$    87$ & $-0.19$ & $-0.10$ & $-0.25$ & $-0.17$ & $-0.23$ & $-0.16$ & $-0.38$ & $-0.19$ & $-0.35$ & $-0.35$ & $-0.35$ & $-0.35$ & $-0.38$ \\ 
$   154$ & $-0.20$ & $-0.14$ & $-0.24$ & $-0.18$ & $-0.22$ & $-0.18$ & $-0.39$ & $-0.20$ & $-0.35$ & $-0.36$ & $-0.35$ & $-0.35$ & $-0.39$ \\ 
$   164$ & $-0.19$ & $-0.13$ & $-0.23$ & $-0.18$ & $-0.21$ & $-0.18$ & $-0.39$ & $-0.22$ & $-0.35$ & $-0.35$ & $-0.36$ & $-0.35$ & $-0.39$ \\ 
$   214$ & $-0.16$ & $-0.11$ & $-0.20$ & $-0.15$ & $-0.18$ & $-0.15$ & $-0.37$ & $-0.31$ & $-0.36$ & $-0.33$ & $-0.36$ & $-0.36$ & $-0.38$ \\ 
$   266$ & $-0.14$ & $-0.09$ & $-0.19$ & $-0.13$ & $-0.16$ & $-0.13$ & $-0.35$ & $-0.40$ & $-0.36$ & $-0.31$ & $-0.36$ & $-0.37$ & $-0.36$ \\ 
$   321$ & $-0.17$ & $-0.11$ & $-0.22$ & $-0.16$ & $-0.20$ & $-0.16$ & $-0.34$ & $-0.40$ & $-0.36$ & $-0.30$ & $-0.36$ & $-0.36$ & $-0.35$ \\ 
\enddata
\end{deluxetable}

\vspace{3cm}
\floattable
\begin{deluxetable}{@{}r@{\hspace{1.0mm}} | @{}r@{\hspace{0.0mm}}r@{\hspace{0.0mm}}r@{\hspace{0.0mm}}r@{\hspace{0.0mm}}r@{\hspace{0.0mm}} r@{\hspace{0.0mm}}r@{\hspace{0.0mm}}r@{\hspace{0.0mm}}r@{\hspace{0.0mm}}r@{\hspace{0.0mm}} r@{\hspace{0.0mm}}r@{\hspace{0.0mm}} r@{}}
\tablecaption{$K_{ri}$ corrections\label{table k-corr ir}}
\tablewidth{0pt}
\tablehead{
\colhead{PTF ID}  &   \colhead{09as}  &   \colhead{09atu}  &   \colhead{09cnd}  &   \colhead{09cwl}  &   \colhead{10aagc}  &   \colhead{10bfz}  &   \colhead{10bjp}  &   \colhead{10cwr}  &   \colhead{10hgi}  &   \colhead{10nmn}  &   \colhead{10uhf}  &   \colhead{10vqv}  &   \colhead{10vwg}  \\
 \colhead{z}  & \colhead{0.1864}  &   \colhead{  0.5014}  &   \colhead{  0.2585}  &   \colhead{  0.3502}  &   \colhead{  0.2067}  &   \colhead{  0.1699}  &   \colhead{  0.3585}  &   \colhead{  0.2301}  &   \colhead{  0.0982}  &   \colhead{  0.1236}  &   \colhead{  0.2879}  &   \colhead{  0.4520}  &   \colhead{  0.1901} \\ 
 \hline
\colhead{phase}  &   \colhead{$K_{ri}$} \\
}
\startdata
$   -25$ & $-0.16$ & $-0.70$ & $-0.26$ & $-0.44$ & $-0.20$ & $-0.13$ & $-0.46$ & $-0.23$ & $-0.05$ & $-0.08$ & $-0.30$ & $-0.62$ & $-0.17$ \\ 
$   -24$ & $-0.16$ & $-0.70$ & $-0.26$ & $-0.44$ & $-0.20$ & $-0.13$ & $-0.46$ & $-0.23$ & $-0.05$ & $-0.08$ & $-0.30$ & $-0.61$ & $-0.17$ \\ 
$   -23$ & $-0.16$ & $-0.70$ & $-0.26$ & $-0.44$ & $-0.20$ & $-0.13$ & $-0.46$ & $-0.23$ & $-0.06$ & $-0.08$ & $-0.30$ & $-0.61$ & $-0.17$ \\ 
$   -22$ & $-0.16$ & $-0.69$ & $-0.26$ & $-0.44$ & $-0.20$ & $-0.13$ & $-0.46$ & $-0.23$ & $-0.06$ & $-0.08$ & $-0.30$ & $-0.61$ & $-0.17$ \\ 
$   -22$ & $-0.16$ & $-0.69$ & $-0.26$ & $-0.44$ & $-0.20$ & $-0.13$ & $-0.46$ & $-0.23$ & $-0.06$ & $-0.09$ & $-0.30$ & $-0.61$ & $-0.17$ \\ 
$   -21$ & $-0.16$ & $-0.69$ & $-0.26$ & $-0.44$ & $-0.20$ & $-0.13$ & $-0.46$ & $-0.23$ & $-0.06$ & $-0.09$ & $-0.30$ & $-0.61$ & $-0.17$ \\ 
$   -20$ & $-0.16$ & $-0.68$ & $-0.26$ & $-0.44$ & $-0.20$ & $-0.13$ & $-0.46$ & $-0.23$ & $-0.06$ & $-0.09$ & $-0.30$ & $-0.60$ & $-0.17$ \\ 
$   -14$ & $-0.16$ & $-0.67$ & $-0.26$ & $-0.43$ & $-0.20$ & $-0.14$ & $-0.45$ & $-0.23$ & $-0.07$ & $-0.10$ & $-0.30$ & $-0.59$ & $-0.17$ \\ 
$    -8$ & $-0.16$ & $-0.65$ & $-0.26$ & $-0.43$ & $-0.20$ & $-0.14$ & $-0.45$ & $-0.23$ & $-0.08$ & $-0.11$ & $-0.30$ & $-0.58$ & $-0.17$ \\ 
$    -3$ & $-0.17$ & $-0.63$ & $-0.26$ & $-0.43$ & $-0.20$ & $-0.14$ & $-0.44$ & $-0.23$ & $-0.09$ & $-0.11$ & $-0.30$ & $-0.56$ & $-0.17$ \\ 
$     0$ & $-0.17$ & $-0.62$ & $-0.26$ & $-0.42$ & $-0.20$ & $-0.14$ & $-0.44$ & $-0.23$ & $-0.09$ & $-0.12$ & $-0.30$ & $-0.56$ & $-0.17$ \\ 
$     3$ & $-0.17$ & $-0.61$ & $-0.26$ & $-0.42$ & $-0.19$ & $-0.14$ & $-0.44$ & $-0.23$ & $-0.10$ & $-0.12$ & $-0.30$ & $-0.55$ & $-0.17$ \\ 
$     6$ & $-0.17$ & $-0.60$ & $-0.26$ & $-0.42$ & $-0.19$ & $-0.15$ & $-0.43$ & $-0.22$ & $-0.10$ & $-0.13$ & $-0.30$ & $-0.55$ & $-0.17$ \\ 
$    10$ & $-0.17$ & $-0.60$ & $-0.26$ & $-0.42$ & $-0.19$ & $-0.15$ & $-0.43$ & $-0.22$ & $-0.11$ & $-0.13$ & $-0.30$ & $-0.54$ & $-0.17$ \\ 
$    16$ & $-0.17$ & $-0.58$ & $-0.25$ & $-0.41$ & $-0.19$ & $-0.15$ & $-0.42$ & $-0.22$ & $-0.12$ & $-0.14$ & $-0.29$ & $-0.53$ & $-0.17$ \\ 
$    21$ & $-0.17$ & $-0.57$ & $-0.25$ & $-0.41$ & $-0.19$ & $-0.15$ & $-0.42$ & $-0.22$ & $-0.12$ & $-0.14$ & $-0.29$ & $-0.52$ & $-0.17$ \\ 
$    22$ & $-0.17$ & $-0.57$ & $-0.25$ & $-0.41$ & $-0.19$ & $-0.15$ & $-0.42$ & $-0.22$ & $-0.12$ & $-0.14$ & $-0.29$ & $-0.52$ & $-0.17$ \\ 
$    26$ & $-0.17$ & $-0.56$ & $-0.25$ & $-0.40$ & $-0.19$ & $-0.16$ & $-0.42$ & $-0.22$ & $-0.13$ & $-0.15$ & $-0.29$ & $-0.51$ & $-0.17$ \\ 
$    47$ & $-0.18$ & $-0.52$ & $-0.24$ & $-0.39$ & $-0.19$ & $-0.17$ & $-0.40$ & $-0.21$ & $-0.15$ & $-0.17$ & $-0.28$ & $-0.48$ & $-0.18$ \\ 
$    57$ & $-0.18$ & $-0.51$ & $-0.24$ & $-0.38$ & $-0.19$ & $-0.17$ & $-0.39$ & $-0.21$ & $-0.17$ & $-0.18$ & $-0.28$ & $-0.47$ & $-0.18$ \\ 
$    57$ & $-0.18$ & $-0.51$ & $-0.24$ & $-0.38$ & $-0.19$ & $-0.17$ & $-0.39$ & $-0.21$ & $-0.17$ & $-0.18$ & $-0.28$ & $-0.47$ & $-0.18$ \\ 
$    87$ & $-0.18$ & $-0.48$ & $-0.23$ & $-0.36$ & $-0.19$ & $-0.18$ & $-0.37$ & $-0.21$ & $-0.19$ & $-0.20$ & $-0.26$ & $-0.44$ & $-0.18$ \\ 
$   154$ & $-0.20$ & $-0.45$ & $-0.22$ & $-0.32$ & $-0.19$ & $-0.20$ & $-0.32$ & $-0.20$ & $-0.23$ & $-0.22$ & $-0.24$ & $-0.42$ & $-0.19$ \\ 
$   164$ & $-0.20$ & $-0.45$ & $-0.22$ & $-0.31$ & $-0.19$ & $-0.20$ & $-0.32$ & $-0.20$ & $-0.23$ & $-0.23$ & $-0.23$ & $-0.42$ & $-0.19$ \\ 
$   214$ & $-0.20$ & $-0.45$ & $-0.21$ & $-0.29$ & $-0.20$ & $-0.20$ & $-0.30$ & $-0.20$ & $-0.23$ & $-0.22$ & $-0.22$ & $-0.42$ & $-0.20$ \\ 
$   266$ & $-0.20$ & $-0.45$ & $-0.22$ & $-0.28$ & $-0.20$ & $-0.20$ & $-0.30$ & $-0.20$ & $-0.20$ & $-0.20$ & $-0.23$ & $-0.42$ & $-0.20$ \\ 
$   321$ & $-0.19$ & $-0.43$ & $-0.24$ & $-0.30$ & $-0.19$ & $-0.18$ & $-0.31$ & $-0.22$ & $-0.13$ & $-0.16$ & $-0.25$ & $-0.40$ & $-0.19$ \\ 
\enddata
\end{deluxetable}
\vspace{3cm}
\floattable
\begin{deluxetable}{@{}r@{\hspace{1.0mm}} | @{}r@{\hspace{0.0mm}}r@{\hspace{0.0mm}}r@{\hspace{0.0mm}}r@{\hspace{0.0mm}}r@{\hspace{0.0mm}} r@{\hspace{0.0mm}}r@{\hspace{0.0mm}}r@{\hspace{0.0mm}}r@{\hspace{0.0mm}}r@{\hspace{0.0mm}} r@{\hspace{0.0mm}}r@{\hspace{0.0mm}} r@{}}
\tablecaption{Continuation of Table \ref{table k-corr ir}.\label{table k-corr ir 2}}
\tablewidth{0pt}
\tablehead{
\colhead{PTF ID}  &  \colhead{11dij}  &   \colhead{11hrq}  &   \colhead{11rks}  &   \colhead{12dam}  &   \colhead{12gty}  &   \colhead{12hni}  &   \colhead{12mxx}  &   \colhead{13ajg}  &   \colhead{13bdl}  &   \colhead{13bjz}  &   \colhead{13cjq}  &   \colhead{13dcc}  &   \colhead{13ehe} \\
\colhead{z}   &   \colhead{  0.1429}  &   \colhead{  0.0571}  &   \colhead{  0.1924}  &   \colhead{  0.1075}  &   \colhead{  0.1768}  &   \colhead{  0.1056}  &   \colhead{  0.3274}  &   \colhead{  0.7403}  &   \colhead{  0.4030}  &   \colhead{  0.2712}  &   \colhead{  0.3962}  &   \colhead{  0.4308}  &   \colhead{  0.3434} \\
 \hline
\colhead{phase}  &   \colhead{$K_{ri}$} \\
}
\startdata
$   -25$ & $-0.09$ & $ 0.02$ & $-0.17$ & $-0.06$ & $-0.14$ & $-0.06$ & $-0.38$ & $-0.98$ & $-0.54$ & $-0.28$ & $-0.53$ & $-0.58$ & $-0.42$ \\ 
$   -24$ & $-0.09$ & $ 0.02$ & $-0.17$ & $-0.07$ & $-0.14$ & $-0.06$ & $-0.38$ & $-0.98$ & $-0.54$ & $-0.28$ & $-0.53$ & $-0.58$ & $-0.42$ \\ 
$   -23$ & $-0.09$ & $ 0.02$ & $-0.17$ & $-0.07$ & $-0.14$ & $-0.07$ & $-0.38$ & $-0.97$ & $-0.54$ & $-0.28$ & $-0.53$ & $-0.58$ & $-0.42$ \\ 
$   -22$ & $-0.09$ & $ 0.02$ & $-0.17$ & $-0.07$ & $-0.14$ & $-0.07$ & $-0.38$ & $-0.96$ & $-0.54$ & $-0.28$ & $-0.53$ & $-0.58$ & $-0.42$ \\ 
$   -22$ & $-0.09$ & $ 0.01$ & $-0.17$ & $-0.07$ & $-0.14$ & $-0.07$ & $-0.38$ & $-0.95$ & $-0.54$ & $-0.28$ & $-0.53$ & $-0.58$ & $-0.42$ \\ 
$   -21$ & $-0.10$ & $ 0.01$ & $-0.17$ & $-0.07$ & $-0.14$ & $-0.07$ & $-0.38$ & $-0.95$ & $-0.54$ & $-0.28$ & $-0.53$ & $-0.58$ & $-0.42$ \\ 
$   -20$ & $-0.10$ & $ 0.01$ & $-0.17$ & $-0.07$ & $-0.14$ & $-0.07$ & $-0.38$ & $-0.94$ & $-0.54$ & $-0.28$ & $-0.53$ & $-0.57$ & $-0.42$ \\ 
$   -14$ & $-0.10$ & $ 0.00$ & $-0.17$ & $-0.08$ & $-0.15$ & $-0.08$ & $-0.38$ & $-0.90$ & $-0.53$ & $-0.28$ & $-0.52$ & $-0.56$ & $-0.42$ \\ 
$    -8$ & $-0.11$ & $-0.01$ & $-0.17$ & $-0.09$ & $-0.15$ & $-0.09$ & $-0.38$ & $-0.86$ & $-0.52$ & $-0.28$ & $-0.51$ & $-0.55$ & $-0.41$ \\ 
$    -3$ & $-0.11$ & $-0.01$ & $-0.17$ & $-0.10$ & $-0.15$ & $-0.10$ & $-0.38$ & $-0.82$ & $-0.51$ & $-0.27$ & $-0.50$ & $-0.54$ & $-0.41$ \\ 
$     0$ & $-0.12$ & $-0.02$ & $-0.17$ & $-0.11$ & $-0.15$ & $-0.10$ & $-0.38$ & $-0.80$ & $-0.50$ & $-0.27$ & $-0.50$ & $-0.54$ & $-0.41$ \\ 
$     3$ & $-0.12$ & $-0.02$ & $-0.17$ & $-0.11$ & $-0.15$ & $-0.11$ & $-0.37$ & $-0.78$ & $-0.50$ & $-0.27$ & $-0.49$ & $-0.53$ & $-0.41$ \\ 
$     6$ & $-0.12$ & $-0.03$ & $-0.18$ & $-0.11$ & $-0.15$ & $-0.11$ & $-0.37$ & $-0.76$ & $-0.50$ & $-0.27$ & $-0.49$ & $-0.53$ & $-0.41$ \\ 
$    10$ & $-0.12$ & $-0.03$ & $-0.18$ & $-0.12$ & $-0.16$ & $-0.12$ & $-0.37$ & $-0.74$ & $-0.49$ & $-0.27$ & $-0.48$ & $-0.52$ & $-0.40$ \\ 
$    16$ & $-0.13$ & $-0.04$ & $-0.18$ & $-0.13$ & $-0.16$ & $-0.13$ & $-0.37$ & $-0.70$ & $-0.48$ & $-0.27$ & $-0.47$ & $-0.51$ & $-0.40$ \\ 
$    21$ & $-0.13$ & $-0.05$ & $-0.18$ & $-0.13$ & $-0.16$ & $-0.13$ & $-0.37$ & $-0.68$ & $-0.48$ & $-0.27$ & $-0.47$ & $-0.50$ & $-0.40$ \\ 
$    22$ & $-0.13$ & $-0.05$ & $-0.18$ & $-0.14$ & $-0.16$ & $-0.13$ & $-0.37$ & $-0.67$ & $-0.47$ & $-0.27$ & $-0.47$ & $-0.50$ & $-0.40$ \\ 
$    26$ & $-0.14$ & $-0.05$ & $-0.18$ & $-0.14$ & $-0.16$ & $-0.14$ & $-0.36$ & $-0.65$ & $-0.47$ & $-0.27$ & $-0.46$ & $-0.50$ & $-0.39$ \\ 
$    47$ & $-0.15$ & $-0.08$ & $-0.18$ & $-0.17$ & $-0.17$ & $-0.16$ & $-0.35$ & $-0.57$ & $-0.45$ & $-0.26$ & $-0.44$ & $-0.47$ & $-0.38$ \\ 
$    57$ & $-0.16$ & $-0.09$ & $-0.18$ & $-0.18$ & $-0.17$ & $-0.17$ & $-0.35$ & $-0.53$ & $-0.44$ & $-0.26$ & $-0.43$ & $-0.46$ & $-0.37$ \\ 
$    57$ & $-0.16$ & $-0.09$ & $-0.18$ & $-0.18$ & $-0.17$ & $-0.17$ & $-0.35$ & $-0.53$ & $-0.44$ & $-0.26$ & $-0.43$ & $-0.46$ & $-0.37$ \\ 
$    87$ & $-0.18$ & $-0.12$ & $-0.18$ & $-0.20$ & $-0.18$ & $-0.20$ & $-0.33$ & $-0.46$ & $-0.41$ & $-0.25$ & $-0.40$ & $-0.44$ & $-0.35$ \\ 
$   154$ & $-0.21$ & $-0.16$ & $-0.19$ & $-0.23$ & $-0.20$ & $-0.23$ & $-0.29$ & $-0.42$ & $-0.37$ & $-0.22$ & $-0.36$ & $-0.41$ & $-0.31$ \\ 
$   164$ & $-0.21$ & $-0.16$ & $-0.19$ & $-0.24$ & $-0.20$ & $-0.24$ & $-0.29$ & $-0.42$ & $-0.37$ & $-0.22$ & $-0.36$ & $-0.40$ & $-0.30$ \\ 
$   214$ & $-0.21$ & $-0.16$ & $-0.20$ & $-0.23$ & $-0.20$ & $-0.23$ & $-0.27$ & $-0.45$ & $-0.36$ & $-0.21$ & $-0.35$ & $-0.40$ & $-0.28$ \\ 
$   266$ & $-0.20$ & $-0.13$ & $-0.20$ & $-0.21$ & $-0.20$ & $-0.20$ & $-0.26$ & $-0.49$ & $-0.36$ & $-0.22$ & $-0.35$ & $-0.40$ & $-0.28$ \\ 
$   321$ & $-0.16$ & $-0.06$ & $-0.19$ & $-0.15$ & $-0.18$ & $-0.15$ & $-0.27$ & $-0.49$ & $-0.37$ & $-0.25$ & $-0.37$ & $-0.39$ & $-0.29$ \\ 
\enddata
\end{deluxetable}

\clearpage

\newpage
\section{Photometric data}
All magnitudes are in the AB system. The absolute magnitudes in rest-frame $g$-band are derived from $M_g = m_r - DM -K_{gr}$, where $DM$ is the distance modulus, $m_r$ is corrected for foreground Galactic extinction (Table~\ref{table sample}), and $K_{gr}$ is the $k$-correction (Table~\ref{table k-corr}), as described in Sect. \ref{sec k-corr}. The photometric data in Tables \ref{table phot09as} to \ref{table g phot13ehe} are available as machine-readable electronic tables in the online version of the paper (future link). The UV photometry from Swift was not corrected from host-galaxy contribution, but this should be minimal (see Sect. \ref{sec swift}). The table references are: [I] \citet{Quimby11}; [II] \citet{Pastorello10}; [III] \citet{Inserra13}; [IV] \citet{Nicholl13}, \citet{Chen15}; [V] \citet{Vreeswijk14}. 
\begin{table*}
\begin{center}
\caption{PTF~09as photometry data.\label{table phot09as}}
\begin{tabular}{c c c c c }
\tableline\tableline
MJD & Phase & $m$ & Filter & Telescope \\
\tableline
$54910.417$ & $  -6.56$ & $19.71\pm0.11$ &     g &             p48 \\
$54915.164$ & $  -2.56$ & $19.85\pm0.14$ &     g &             p48 \\
$54915.244$ & $  -2.49$ & $19.72\pm0.12$ &     g &             p48 \\
$54918.197$ & $   0.00$ & $20.13\pm0.11$ &     r &             p60 \\
$54918.199$ & $   0.00$ & $19.80\pm0.12$ &     i &             p60 \\
$54918.210$ & $   0.01$ & $20.19\pm0.15$ &     g &             p48 \\
$54919.195$ & $   0.84$ & $20.19\pm0.12$ &     r &             p60 \\
\end{tabular}
\end{center}
\end{table*}
\begin{table*}
\begin{center}
\caption{PTF~09as rest-frame $g$ photometry data.\label{table g phot09as}}
\begin{tabular}{c c c c c c}
\tableline\tableline
MJD & Phase & $m_g$ & $M_g$  & Telescope & Reference \\
\tableline
$54918.197$ & $   0.00$ & $20.29\pm0.11$ & $-19.50$ &             p60 &            \\
$54919.195$ & $   0.84$ & $20.36\pm0.12$ & $-19.43$ &             p60 &            \\
$54920.235$ & $   1.72$ & $20.54\pm0.22$ & $-19.24$ &             p60 &            \\
$54921.248$ & $   2.57$ & $20.45\pm0.13$ & $-19.33$ &             p60 &            \\
$54922.186$ & $   3.36$ & $20.49\pm0.19$ & $-19.30$ &             p60 &            \\
$54923.283$ & $   4.29$ & $20.57\pm0.13$ & $-19.21$ &             p60 &            \\
$54925.376$ & $   6.05$ & $20.64\pm0.26$ & $-19.14$ &             p60 &            \\
$54926.214$ & $   6.76$ & $21.00\pm0.22$ & $-18.78$ &             p60 &            \\
$54928.297$ & $   8.51$ & $20.90\pm0.22$ & $-18.88$ &             p60 &            \\
$54934.333$ & $  13.60$ & $21.62\pm0.27$ & $-18.16$ &             p60 &            \\
$54938.202$ & $  16.86$ & $21.59\pm0.24$ & $-18.19$ &             p60 &            \\
\tableline
\end{tabular}
\end{center}
\end{table*}
\clearpage
\begin{table*}
\begin{center}
\caption{PTF~09atu photometry data.\label{table phot09atu}}
\begin{tabular}{@{}c c c c c | ccccc@{}}
\tableline\tableline
MJD & Phase & $m$ & Filter & Tel & MJD & Phase & $m$ & Filter & Tel \\
\tableline
$54974.325$ & $ -58.61$ & $22.43\pm0.29$ &     r &             p48 & $55064.163$ & $   1.23$ & $20.09\pm0.10$ &     i &             p60 \\
$55016.222$ & $ -30.70$ & $20.58\pm0.15$ &     r &             p48 & $55064.165$ & $   1.23$ & $20.04\pm0.10$ &     i &             p60 \\
$55016.321$ & $ -30.64$ & $20.93\pm0.33$ &     r &             p48 & $55066.154$ & $   2.55$ & $20.32\pm0.08$ &     g &             p60 \\
$55025.227$ & $ -24.71$ & $20.45\pm0.11$ &     r &             p48 & $55066.157$ & $   2.56$ & $19.99\pm0.08$ &     r &             p60 \\
$55025.343$ & $ -24.63$ & $20.36\pm0.13$ &     r &             p48 & $55066.160$ & $   2.56$ & $20.01\pm0.09$ &     r &             p60 \\
$55029.337$ & $ -21.97$ & $20.39\pm0.11$ &     r &             p60 & $55066.162$ & $   2.56$ & $20.07\pm0.10$ &     i &             p60 \\
$55029.339$ & $ -21.97$ & $20.52\pm0.15$ &     i &             p60 & $55066.165$ & $   2.56$ & $19.98\pm0.10$ &     i &             p60 \\
\end{tabular}
\end{center}
\end{table*}
\begin{table*}
\begin{center}
\caption{PTF~09atu rest-frame $g$ photometry data.\label{table g phot09atu}}
\begin{tabular}{c c c c c c}
\tableline\tableline
MJD & Phase & $m_g$ & $M_g$  & Telescope & Reference \\
\tableline
$54974.325$ & $ -58.61$ & $23.00\pm0.29$ & $-19.27$ &             p48 &            \\
$55016.222$ & $ -30.70$ & $21.14\pm0.15$ & $-21.13$ &             p48 &            \\
$55016.321$ & $ -30.64$ & $21.50\pm0.33$ & $-20.77$ &             p48 &            \\
$55025.227$ & $ -24.71$ & $21.01\pm0.11$ & $-21.26$ &             p48 &            \\
$55025.343$ & $ -24.63$ & $20.93\pm0.13$ & $-21.34$ &             p48 &            \\
$55029.337$ & $ -21.97$ & $20.94\pm0.11$ & $-21.33$ &             p60 &            \\
$55035.250$ & $ -18.03$ & $20.74\pm0.09$ & $-21.53$ &             p48 &            \\
\end{tabular}
\end{center}
\end{table*}
\clearpage
\begin{table*}
\begin{center}
\caption{PTF~09cnd photometry data.\label{table phot09cnd}}
\begin{tabular}{@{}c c c c c | ccccc@{}}
\tableline\tableline
MJD & Phase & $m$ & Filter & Tel & MJD & Phase & $m$ & Filter & Tel \\
\tableline
$55025.378$ & $ -48.45$ & $20.66\pm0.26$ &     r &             p48 & $55064.148$ & $ -17.64$ & $18.54\pm0.08$ &     r &             p60 \\
$55025.418$ & $ -48.41$ & $20.49\pm0.29$ &     r &             p48 & $55064.151$ & $ -17.64$ & $18.76\pm0.11$ &     i &             p60 \\
$55035.218$ & $ -40.63$ & $20.15\pm0.12$ &     r &             p48 & $55064.153$ & $ -17.64$ & $18.74\pm0.11$ &     i &             p60 \\
$55035.313$ & $ -40.55$ & $20.31\pm0.13$ &     r &             p48 & $55065.829$ & $ -16.30$ & $19.41\pm0.07$ &  UVM2 &      Swift/UVOT \\
$55049.254$ & $ -29.47$ & $19.05\pm0.12$ &     r &             p48 & $55065.830$ & $ -16.30$ & $18.76\pm0.07$ &  UVW1 &      Swift/UVOT \\
$55049.360$ & $ -29.39$ & $18.93\pm0.14$ &     r &             p48 & $55065.830$ & $ -16.30$ & $18.13\pm0.06$ &     u &      Swift/UVOT \\
$55052.277$ & $ -27.07$ & $18.85\pm0.09$ &     r &             p48 & $55065.831$ & $ -16.30$ & $19.66\pm0.08$ &  UVW2 &      Swift/UVOT \\
\end{tabular}
\end{center}
\end{table*}
\begin{table*}
\begin{center}
\caption{Continuation of Table \ref{table phot09cnd}.\label{table phot09cnd 2}}
\begin{tabular}{c c c c c }
\tableline\tableline
MJD & Phase & $m$ & Filter & Telescope \\
\tableline
$55084.213$ & $  -1.70$ & $18.20\pm0.07$ &     r &             p48 \\
$55084.257$ & $  -1.66$ & $18.23\pm0.08$ &     r &             p48 \\
$55097.812$ & $   9.11$ & $20.47\pm0.15$ &  UVM2 &      Swift/UVOT \\
$55097.814$ & $   9.11$ & $19.71\pm0.11$ &  UVW1 &      Swift/UVOT \\
$55097.815$ & $   9.11$ & $18.37\pm0.07$ &     u &      Swift/UVOT \\
$55097.816$ & $   9.11$ & $20.56\pm0.12$ &  UVW2 &      Swift/UVOT \\
$55107.174$ & $  16.55$ & $18.25\pm0.08$ &     g &             p60 \\
\end{tabular}
\end{center}
\end{table*}
\begin{table*}
\begin{center}
\caption{PTF~09cnd rest-frame $g$ photometry data.\label{table g phot09cnd}}
\begin{tabular}{@{}c c c c c c | cccccc@{}}
\tableline\tableline
MJD & Phase & $m_g$ & $M_g$  & Tel & Ref & MJD & Phase & $m_g$ & $M_g$  & Tel & Ref \\
\tableline
$55025.378$ & $ -48.45$ & $20.85\pm0.26$ & $-19.73$ &             p48 &            & $55185.495$ & $  78.78$ & $19.86\pm0.18$ & $-20.73$ &             p60 &            \\
$55025.418$ & $ -48.41$ & $20.68\pm0.29$ & $-19.91$ &             p48 &            & $55189.486$ & $  81.95$ & $20.22\pm0.18$ & $-20.37$ &             p60 &            \\
$55035.218$ & $ -40.63$ & $20.34\pm0.12$ & $-20.24$ &             p48 &            & $55202.474$ & $  92.27$ & $20.73\pm0.23$ & $-19.85$ &             p60 &            \\
$55035.313$ & $ -40.55$ & $20.50\pm0.13$ & $-20.08$ &             p48 &            & $55203.448$ & $  93.05$ & $20.94\pm0.28$ & $-19.64$ &             p60 &            \\
$55049.254$ & $ -29.47$ & $19.25\pm0.12$ & $-21.34$ &             p48 &            & $55205.000$ & $  94.28$ & $21.44\pm0.20$ & $-19.14$ &            P200 &            \\
$55049.360$ & $ -29.39$ & $19.13\pm0.14$ & $-21.46$ &             p48 &            \\
$55052.277$ & $ -27.07$ & $19.04\pm0.09$ & $-21.54$ &             p48 &            \\
\end{tabular}
\end{center}
\end{table*}
\clearpage
\begin{table*}
\begin{center}
\caption{PTF~09cwl photometry data.\label{table phot09cwl}}
\begin{tabular}{c c c c c }
\tableline\tableline
MJD & Phase & $m$ & Filter & Telescope \\
\tableline
$55021.255$ & $ -34.07$ & $20.59\pm0.27$ &     r &             p48 \\
$55034.210$ & $ -24.47$ & $19.83\pm0.13$ &     r &             p48 \\
$55034.308$ & $ -24.40$ & $20.03\pm0.19$ &     r &             p48 \\
$55050.185$ & $ -12.64$ & $19.13\pm0.12$ &     r &             p48 \\
$55050.268$ & $ -12.58$ & $19.21\pm0.19$ &     r &             p48 \\
$55072.574$ & $   3.94$ & $22.04\pm0.26$ &  UVM2 &      Swift/UVOT \\
$55072.578$ & $   3.94$ & $20.86\pm0.14$ &  UVW1 &      Swift/UVOT \\
$55072.581$ & $   3.95$ & $19.50\pm0.09$ &     u &      Swift/UVOT \\
$55072.583$ & $   3.95$ & $22.85\pm0.33$ &  UVW2 &      Swift/UVOT \\
$55189.465$ & $  90.51$ & $21.72\pm0.29$ &     r &             p60 \\
$55207.416$ & $ 103.81$ & $22.28\pm0.32$ &     g &             p60 \\
$55208.416$ & $ 104.55$ & $21.57\pm0.35$ &     r &             p60 \\
$55211.405$ & $ 106.76$ & $20.76\pm0.23$ &     g &             p60 \\
$55330.000$ & $ 194.60$ & $23.93\pm0.35$ &     r &            Keck \\
\tableline
\end{tabular}
\end{center}
\end{table*}
\begin{table*}
\begin{center}
\caption{PTF~09cwl rest-frame $g$ photometry data.\label{table g phot09cwl}}
\begin{tabular}{c c c c c c}
\tableline\tableline
MJD & Phase & $m_g$ & $M_g$  & Telescope & Reference \\
\tableline
$55021.255$ & $ -34.07$ & $20.95\pm0.27$ & $-20.39$ &             p48 &            \\
$55034.210$ & $ -24.47$ & $20.20\pm0.13$ & $-21.15$ &             p48 &            \\
$55034.308$ & $ -24.40$ & $20.40\pm0.19$ & $-20.95$ &             p48 &            \\
$55050.185$ & $ -12.64$ & $19.51\pm0.12$ & $-21.83$ &             p48 &            \\
$55050.268$ & $ -12.58$ & $19.60\pm0.19$ & $-21.75$ &             p48 &            \\
$55081.227$ & $  10.35$ & $19.47\pm0.10$ & $-21.88$ &            Wise &          I \\
$55189.465$ & $  90.51$ & $22.16\pm0.29$ & $-19.18$ &             p60 &            \\
$55205.540$ & $ 102.42$ & $22.42\pm0.05$ & $-18.92$ &            p200 &          I \\
$55208.416$ & $ 104.55$ & $22.02\pm0.35$ & $-19.32$ &             p60 &            \\
$55330.000$ & $ 194.60$ & $24.39\pm0.35$ & $-16.96$ &            Keck &            \\
\tableline
\end{tabular}
\end{center}
\end{table*}
\clearpage
\begin{table*}
\begin{center}
\caption{PTF~10aagc photometry data.\label{table phot10aagc}}
\begin{tabular}{@{}c c c c c | ccccc@{}}
\tableline\tableline
MJD & Phase & $m$ & Filter & Tel & MJD & Phase & $m$ & Filter & Tel \\
\tableline
$55495.481$ & $  -3.31$ & $19.88\pm0.23$ &     r &             p48 & $55532.508$ & $  27.37$ & $21.44\pm0.29$ &     r &             p48 \\
$55495.506$ & $  -3.29$ & $19.96\pm0.16$ &     r &             p48 & $55534.353$ & $  28.90$ & $20.71\pm0.17$ &     i &             p60 \\
$55495.530$ & $  -3.27$ & $19.96\pm0.18$ &     r &             p48 & $55535.315$ & $  29.70$ & $20.49\pm0.33$ &     i &             p60 \\
$55497.524$ & $  -1.62$ & $19.87\pm0.19$ &     r &             p48 & $55535.318$ & $  29.70$ & $21.55\pm0.24$ &     r &             p60 \\
$55503.502$ & $   3.33$ & $19.60\pm0.11$ &     r &             p48 & $55535.394$ & $  29.76$ & $21.37\pm0.33$ &     r &             p48 \\
$55503.546$ & $   3.37$ & $19.68\pm0.14$ &     r &             p48 & $55538.307$ & $  32.18$ & $21.58\pm0.24$ &     i &             p60 \\
$55504.434$ & $   4.11$ & $19.54\pm0.09$ &     r &             p60 & $55538.309$ & $  32.18$ & $21.86\pm0.23$ &     r &             p60 \\
\end{tabular}
\end{center}
\end{table*}
\begin{table*}
\begin{center}
\caption{PTF~10aagc rest-frame $g$ photometry data.\label{table g phot10aagc}}
\begin{tabular}{c c c c c c}
\tableline\tableline
MJD & Phase & $m_g$ & $M_g$  & Telescope & Reference \\
\tableline
$55495.481$ & $  -3.31$ & $20.14\pm0.23$ & $-19.90$ &             p48 &            \\
$55495.506$ & $  -3.29$ & $20.22\pm0.16$ & $-19.81$ &             p48 &            \\
$55495.530$ & $  -3.27$ & $20.22\pm0.18$ & $-19.81$ &             p48 &            \\
$55497.524$ & $  -1.62$ & $20.14\pm0.19$ & $-19.90$ &             p48 &            \\
$55503.502$ & $   3.33$ & $19.88\pm0.11$ & $-20.16$ &             p48 &            \\
$55503.546$ & $   3.37$ & $19.96\pm0.14$ & $-20.08$ &             p48 &            \\
$55504.434$ & $   4.11$ & $19.82\pm0.09$ & $-20.21$ &             p60 &            \\
\end{tabular}
\end{center}
\end{table*}
\clearpage
\begin{table*}
\begin{center}
\caption{PTF~10bfz photometry data.\label{table phot10bfz}}
\begin{tabular}{@{}c c c c c | ccccc@{}}
\tableline\tableline
MJD & Phase & $m$ & Filter & Tel & MJD & Phase & $m$ & Filter & Tel \\
\tableline
$55227.463$ & $  -0.00$ & $18.52\pm0.09$ &     r &             p48 & $55283.303$ & $  47.73$ & $20.38\pm0.29$ &     r &             p60 \\
$55227.507$ & $   0.04$ & $18.49\pm0.08$ &     r &             p48 & $55292.501$ & $  55.59$ & $21.01\pm0.36$ &     r &             p48 \\
$55242.264$ & $  12.65$ & $18.74\pm0.06$ &     g &             p60 & $55292.502$ & $  55.59$ & $20.97\pm0.34$ &     r &             p48 \\
$55242.266$ & $  12.65$ & $18.69\pm0.06$ &     r &             p60 & $55295.494$ & $  58.15$ & $20.88\pm0.21$ &     i &             p60 \\
$55242.267$ & $  12.65$ & $18.90\pm0.08$ &     i &             p60 & $55296.146$ & $  58.71$ & $21.20\pm0.26$ &     r &             p48 \\
$55244.452$ & $  14.52$ & $18.73\pm0.07$ &     r &             p48 & $55296.189$ & $  58.74$ & $21.47\pm0.25$ &     r &             p48 \\
$55244.496$ & $  14.56$ & $18.80\pm0.07$ &     r &             p48 & $55296.452$ & $  58.97$ & $21.47\pm0.31$ &     r &             p60 \\
\end{tabular}
\end{center}
\end{table*}
\begin{table*}
\begin{center}
\caption{PTF~10bfz rest-frame $g$ photometry data.\label{table g phot10bfz}}
\begin{tabular}{c c c c c c}
\tableline\tableline
MJD & Phase & $m_g$ & $M_g$  & Telescope & Reference \\
\tableline
$55227.463$ & $  -0.00$ & $18.72\pm0.09$ & $-20.84$ &             p48 &            \\
$55227.507$ & $   0.04$ & $18.70\pm0.08$ & $-20.87$ &             p48 &            \\
$55242.266$ & $  12.65$ & $18.94\pm0.06$ & $-20.63$ &             p60 &            \\
$55244.452$ & $  14.52$ & $18.98\pm0.07$ & $-20.58$ &             p48 &            \\
$55244.496$ & $  14.56$ & $19.05\pm0.07$ & $-20.51$ &             p48 &            \\
$55250.438$ & $  19.64$ & $19.14\pm0.08$ & $-20.42$ &             p60 &            \\
$55251.464$ & $  20.51$ & $19.14\pm0.07$ & $-20.42$ &             p48 &            \\
\end{tabular}
\end{center}
\end{table*}
\clearpage
\begin{table*}
\begin{center}
\caption{PTF~10bjp photometry data.\label{table phot10bjp}}
\begin{tabular}{c c c c c }
\tableline\tableline
MJD & Phase & $m$ & Filter & Telescope \\
\tableline
$55205.547$ & $ -34.57$ & $21.37\pm0.35$ &     r &             p48 \\
$55207.341$ & $ -33.25$ & $21.32\pm0.30$ &     r &             p48 \\
$55207.386$ & $ -33.22$ & $21.65\pm0.28$ &     r &             p48 \\
$55209.503$ & $ -31.66$ & $21.37\pm0.34$ &     r &             p48 \\
$55209.548$ & $ -31.63$ & $21.08\pm0.28$ &     r &             p48 \\
$55221.515$ & $ -22.82$ & $20.77\pm0.20$ &     r &             p48 \\
$55221.560$ & $ -22.79$ & $21.15\pm0.32$ &     r &             p48 \\
$55227.346$ & $ -18.53$ & $20.69\pm0.34$ &     r &             p48 \\
$55227.391$ & $ -18.49$ & $20.46\pm0.27$ &     r &             p48 \\
$55243.366$ & $  -6.74$ & $20.30\pm0.16$ &     r &             p48 \\
$55243.412$ & $  -6.70$ & $20.39\pm0.17$ &     r &             p48 \\
$55246.213$ & $  -4.64$ & $20.59\pm0.23$ &     r &             p48 \\
$55246.309$ & $  -4.57$ & $20.46\pm0.15$ &     r &             p48 \\
$55246.353$ & $  -4.54$ & $20.42\pm0.20$ &     r &             p48 \\
$55250.405$ & $  -1.55$ & $20.51\pm0.27$ &     r &             p48 \\
$55250.449$ & $  -1.52$ & $20.52\pm0.24$ &     r &             p48 \\
$55271.409$ & $  13.91$ & $20.49\pm0.25$ &     r &             p48 \\
$55275.363$ & $  16.82$ & $21.34\pm0.32$ &     r &             p48 \\
$55275.407$ & $  16.85$ & $20.88\pm0.25$ &     r &             p48 \\
$55284.156$ & $  23.29$ & $20.34\pm0.35$ &     r &             p48 \\
$55296.199$ & $  32.16$ & $21.20\pm0.28$ &     r &             p48 \\
$55297.395$ & $  33.04$ & $21.22\pm0.33$ &     r &             p48 \\
\tableline
\end{tabular}
\end{center}
\end{table*}
\begin{table*}
\begin{center}
\caption{PTF~10bjp rest-frame $g$ photometry data.\label{table g phot10bjp}}
\begin{tabular}{c c c c c c}
\tableline\tableline
MJD & Phase & $m_g$ & $M_g$  & Telescope & Reference \\
\tableline
$55205.547$ & $ -34.57$ & $21.83\pm0.35$ & $-19.57$ &             p48 &            \\
$55207.341$ & $ -33.25$ & $21.78\pm0.30$ & $-19.63$ &             p48 &            \\
$55207.386$ & $ -33.22$ & $22.10\pm0.28$ & $-19.30$ &             p48 &            \\
$55209.503$ & $ -31.66$ & $21.83\pm0.34$ & $-19.58$ &             p48 &            \\
$55209.548$ & $ -31.63$ & $21.54\pm0.28$ & $-19.87$ &             p48 &            \\
$55221.515$ & $ -22.82$ & $21.23\pm0.20$ & $-20.17$ &             p48 &            \\
$55221.560$ & $ -22.79$ & $21.61\pm0.32$ & $-19.80$ &             p48 &            \\
\end{tabular}
\end{center}
\end{table*}
\clearpage
\begin{table*}
\begin{center}
\caption{PTF~10cwr photometry data.\label{table phot10cwr}}
\begin{tabular}{@{}c c c c c | ccccc@{}}
\tableline\tableline
MJD & Phase & $m$ & Filter & Tel & MJD & Phase & $m$ & Filter & Tel \\
\tableline
$55268.233$ & $ -10.57$ & $18.69\pm0.09$ &     r &             p48 & $55319.275$ & $  30.92$ & $19.49\pm0.12$ &     r &             p60 \\
$55268.277$ & $ -10.53$ & $18.56\pm0.07$ &     r &             p48 & $55360.000$ & $  64.03$ & $22.20\pm0.19$ &     r &            P200 \\
$55271.264$ & $  -8.11$ & $18.52\pm0.09$ &    r &             p48 \\
$55271.315$ & $  -8.06$ & $18.53\pm0.08$ &    r &             p48 \\
$55272.168$ & $  -7.37$ & $18.14\pm0.10$ &    g &             p60 \\
$55272.169$ & $  -7.37$ & $18.49\pm0.09$ &    r &             p60 \\
$55272.170$ & $  -7.37$ & $18.68\pm0.12$ &    i &             p60 \\
\end{tabular}
\end{center}
\end{table*}
\begin{table*}
\begin{center}
\caption{PTF~10cwr rest-frame $g$ photometry data.\label{table g phot10cwr}}
\begin{tabular}{c c c c c c}
\tableline\tableline
MJD & Phase & $m_g$ & $M_g$  & Telescope & Reference \\
\tableline
$55268.233$ & $ -10.57$ & $18.97\pm0.09$ & $-21.33$ &             p48 &            \\
$55268.277$ & $ -10.53$ & $18.84\pm0.07$ & $-21.46$ &             p48 &            \\
$55271.264$ & $  -8.11$ & $18.80\pm0.09$ & $-21.49$ &             p48 &            \\
$55271.315$ & $  -8.06$ & $18.82\pm0.08$ & $-21.48$ &             p48 &            \\
$55272.169$ & $  -7.37$ & $18.78\pm0.09$ & $-21.52$ &             p60 &            \\
$55274.163$ & $  -5.75$ & $18.71\pm0.08$ & $-21.59$ &             p60 &            \\
$55276.157$ & $  -4.13$ & $18.81\pm0.08$ & $-21.49$ &             p60 &            \\
$55277.345$ & $  -3.16$ & $18.79\pm0.08$ & $-21.50$ &             p60 &            \\
$55280.147$ & $  -0.88$ & $18.62\pm0.08$ & $-21.68$ &             p60 &            \\
$55286.211$ & $   4.05$ & $18.75\pm0.08$ & $-21.55$ &             p60 &            \\
$55293.228$ & $   9.75$ & $18.85\pm0.12$ & $-21.44$ &             p60 &            \\
$55297.177$ & $  12.96$ & $18.99\pm0.08$ & $-21.31$ &             p60 &            \\
$55303.211$ & $  17.87$ & $19.26\pm0.09$ & $-21.04$ &             p60 &            \\
$55319.275$ & $  30.92$ & $19.87\pm0.12$ & $-20.43$ &             p60 &            \\
$55360.000$ & $  64.03$ & $22.61\pm0.19$ & $-17.68$ &            P200 &            \\
\tableline
\end{tabular}
\end{center}
\end{table*}
\clearpage
\begin{table*}
\begin{center}
\caption{PTF~10hgi photometry data.\label{table phot10hgi}}
\begin{tabular}{@{}c c c c c | ccccc@{}}
\tableline\tableline
MJD & Phase & $m$ & Filter & Tel & MJD & Phase & $m$ & Filter & Tel \\
\tableline
$55320.381$ & $ -42.85$ & $21.40\pm0.34$ &     r &             p48 & $55372.277$ & $   4.41$ & $17.89\pm0.08$ &     r &             p48 \\
$55331.441$ & $ -32.77$ & $19.10\pm0.08$ &     r &             p48 & $55372.289$ & $   4.42$ & $18.15\pm0.15$ &     i &             p60 \\
$55331.485$ & $ -32.73$ & $19.15\pm0.09$ &     r &             p48 & $55372.290$ & $   4.42$ & $17.88\pm0.10$ &     r &             p60 \\
$55332.190$ & $ -32.09$ & $19.36\pm0.18$ &     i &             p60 & $55372.295$ & $   4.43$ & $18.03\pm0.13$ &     B &             p60 \\
$55332.192$ & $ -32.09$ & $19.15\pm0.12$ &     r &             p60 & $55375.368$ & $   7.22$ & $17.94\pm0.09$ &     r &             p60 \\
$55332.195$ & $ -32.09$ & $19.35\pm0.18$ &     i &             p60 & $55375.368$ & $   7.22$ & $18.22\pm0.13$ &     i &             p60 \\
$55332.197$ & $ -32.09$ & $19.08\pm0.12$ &     r &             p60 & $55375.370$ & $   7.23$ & $18.02\pm0.13$ &     B &             p60 \\
\end{tabular}
\end{center}
\end{table*}
\begin{table*}
\begin{center}
\caption{PTF~10hgi rest-frame $g$ photometry data.\label{table g phot10hgi}}
\begin{tabular}{@{}c c c c c c | cccccc@{}}
\tableline\tableline
MJD & Phase & $m_g$ & $M_g$  & Tel & Ref & MJD & Phase & $m_g$ & $M_g$  & Tel & Ref \\
\tableline
$55320.381$ & $ -42.85$ & $21.34\pm0.34$ & $-16.94$ &             p48 &            & $55415.296$ & $  43.58$ & $19.67\pm0.14$ & $-18.61$ &             p60 &            \\
$55331.441$ & $ -32.77$ & $19.04\pm0.08$ & $-19.24$ &             p48 &            & $55416.177$ & $  44.38$ & $19.58\pm0.12$ & $-18.69$ &             p48 &            \\
$55331.485$ & $ -32.73$ & $19.09\pm0.09$ & $-19.18$ &             p48 &            & $55416.221$ & $  44.42$ & $19.48\pm0.11$ & $-18.79$ &             p48 &            \\
$55332.192$ & $ -32.09$ & $19.09\pm0.12$ & $-19.19$ &             p60 &            & $55419.182$ & $  47.12$ & $19.60\pm0.12$ & $-18.67$ &             p48 &            \\
$55332.197$ & $ -32.09$ & $19.01\pm0.12$ & $-19.26$ &             p60 &            & $55419.226$ & $  47.16$ & $19.60\pm0.13$ & $-18.68$ &             p48 &            \\
$55333.221$ & $ -31.15$ & $18.93\pm0.13$ & $-19.34$ &             p60 &            & $55421.272$ & $  49.02$ & $19.73\pm0.15$ & $-18.55$ &             p60 &            \\
$55334.570$ & $ -29.92$ & $18.68\pm0.07$ & $-19.59$ &             PS1 &        III & $55422.224$ & $  49.89$ & $19.69\pm0.10$ & $-18.58$ &             p48 &            \\
\end{tabular}
\end{center}
\end{table*}
\clearpage
\begin{table*}
\begin{center}
\caption{PTF~10nmn photometry data.\label{table phot10nmn}}
\begin{tabular}{c c c c c }
\tableline\tableline
MJD & Phase & $m$ & Filter & Telescope \\
\tableline
$55267.413$ & $-103.94$ & $19.79\pm0.14$ &     r &             p48 \\
$55267.456$ & $-103.90$ & $19.70\pm0.12$ &     r &             p48 \\
$55384.200$ & $   0.00$ & $18.54\pm0.07$ &     r &             p48 \\
$55384.245$ & $   0.04$ & $18.52\pm0.07$ &     r &             p48 \\
$55391.275$ & $   6.30$ & $18.72\pm0.15$ &     i &             p60 \\
$55391.277$ & $   6.30$ & $18.67\pm0.08$ &     r &             p60 \\
$55392.213$ & $   7.13$ & $18.63\pm0.14$ &     i &             p60 \\
\end{tabular}
\end{center}
\end{table*}
\begin{table*}
\begin{center}
\caption{PTF~10nmn rest-frame $g$ photometry data.\label{table g phot10nmn}}
\begin{tabular}{c c c c c c}
\tableline\tableline
MJD & Phase & $m_g$ & $M_g$  & Telescope & Reference \\
\tableline
$55267.413$ & $-103.94$ & $19.77\pm0.14$ & $-19.04$ &             p48 &            \\
$55267.456$ & $-103.90$ & $19.68\pm0.12$ & $-19.13$ &             p48 &            \\
$55384.200$ & $   0.00$ & $18.65\pm0.07$ & $-20.16$ &             p48 &            \\
$55384.245$ & $   0.04$ & $18.63\pm0.07$ & $-20.18$ &             p48 &            \\
$55391.277$ & $   6.30$ & $18.81\pm0.08$ & $-20.00$ &             p60 &            \\
$55392.215$ & $   7.13$ & $18.83\pm0.08$ & $-19.98$ &             p60 &            \\
$55404.165$ & $  17.77$ & $19.33\pm0.12$ & $-19.48$ &             p60 &            \\
\end{tabular}
\end{center}
\end{table*}
\clearpage
\begin{table*}
\begin{center}
\caption{PTF~10uhf photometry data.\label{table phot10uhf}}
\begin{tabular}{c c c c c }
\tableline\tableline
MJD & Phase & $m$ & Filter & Telescope \\
\tableline
$55433.249$ & $ -14.75$ & $20.75\pm0.36$ &     r &             p48 \\
$55436.275$ & $ -12.40$ & $20.38\pm0.25$ &     r &             p48 \\
$55436.319$ & $ -12.37$ & $20.34\pm0.28$ &     r &             p48 \\
$55444.149$ & $  -6.29$ & $20.14\pm0.15$ &     r &             p48 \\
$55444.196$ & $  -6.25$ & $20.22\pm0.17$ &     r &             p48 \\
$55447.135$ & $  -3.97$ & $20.45\pm0.32$ &     r &             p60 \\
$55447.136$ & $  -3.97$ & $20.12\pm0.26$ &     g &             p60 \\
$55448.900$ & $  -2.60$ & $19.76\pm0.08$ &     g &              LT \\
$55448.900$ & $  -2.60$ & $19.78\pm0.10$ &     i &              LT \\
$55448.900$ & $  -2.60$ & $19.74\pm0.09$ &     r &              LT \\
$55450.900$ & $  -1.05$ & $19.81\pm0.10$ &     i &              LT \\
$55450.900$ & $  -1.05$ & $19.78\pm0.08$ &     g &              LT \\
$55450.900$ & $  -1.05$ & $19.78\pm0.09$ &     r &              LT \\
$55453.149$ & $   0.70$ & $20.10\pm0.15$ &     r &             p48 \\
$55453.193$ & $   0.73$ & $19.90\pm0.14$ &     r &             p48 \\
$55457.136$ & $   3.79$ & $20.00\pm0.16$ &     r &             p48 \\
$55457.179$ & $   3.83$ & $20.09\pm0.18$ &     r &             p48 \\
$55459.800$ & $   5.86$ & $19.83\pm0.10$ &     r &              LT \\
$55459.800$ & $   5.86$ & $19.96\pm0.11$ &     g &              LT \\
$55460.138$ & $   6.12$ & $20.19\pm0.19$ &     r &             p48 \\
$55460.196$ & $   6.17$ & $20.19\pm0.21$ &     r &             p48 \\
$55485.900$ & $  26.13$ & $21.54\pm0.22$ &     g &              LT \\
$55485.900$ & $  26.13$ & $20.82\pm0.16$ &     r &              LT \\
$55485.900$ & $  26.13$ & $20.57\pm0.15$ &     i &              LT \\
\tableline
\end{tabular}
\end{center}
\end{table*}
\begin{table*}
\begin{center}
\caption{PTF~10uhf rest-frame $g$ photometry data.\label{table g phot10uhf}}
\begin{tabular}{c c c c c c}
\tableline\tableline
MJD & Phase & $m_g$ & $M_g$  & Telescope & Reference \\
\tableline
$55433.249$ & $ -14.75$ & $21.11\pm0.36$ & $-19.74$ &             p48 &            \\
$55436.275$ & $ -12.40$ & $20.75\pm0.25$ & $-20.10$ &             p48 &            \\
$55436.319$ & $ -12.37$ & $20.71\pm0.28$ & $-20.14$ &             p48 &            \\
$55444.149$ & $  -6.29$ & $20.52\pm0.15$ & $-20.33$ &             p48 &            \\
$55444.196$ & $  -6.25$ & $20.61\pm0.17$ & $-20.24$ &             p48 &            \\
$55447.135$ & $  -3.97$ & $20.84\pm0.32$ & $-20.01$ &             p60 &            \\
$55448.900$ & $  -2.60$ & $20.12\pm0.09$ & $-20.73$ &              LT &            \\
$55450.900$ & $  -1.05$ & $20.17\pm0.09$ & $-20.69$ &              LT &            \\
$55453.149$ & $   0.70$ & $20.50\pm0.15$ & $-20.35$ &             p48 &            \\
$55453.193$ & $   0.73$ & $20.30\pm0.14$ & $-20.55$ &             p48 &            \\
$55457.136$ & $   3.79$ & $20.41\pm0.16$ & $-20.45$ &             p48 &            \\
$55457.179$ & $   3.83$ & $20.50\pm0.18$ & $-20.36$ &             p48 &            \\
$55459.800$ & $   5.86$ & $20.23\pm0.10$ & $-20.62$ &              LT &            \\
$55460.138$ & $   6.12$ & $20.61\pm0.19$ & $-20.25$ &             p48 &            \\
$55460.196$ & $   6.17$ & $20.61\pm0.21$ & $-20.24$ &             p48 &            \\
$55485.900$ & $  26.13$ & $21.27\pm0.16$ & $-19.59$ &              LT &            \\
\tableline
\end{tabular}
\end{center}
\end{table*}
\clearpage
\begin{table*}
\begin{center}
\caption{PTF~10vqv photometry data.\label{table phot10vqv}}
\begin{tabular}{c c c c c }
\tableline\tableline
MJD & Phase & $m$ & Filter & Telescope \\
\tableline
$55455.517$ & $ -10.33$ & $20.02\pm0.12$ &     r &             p48 \\
$55456.498$ & $  -9.65$ & $20.02\pm0.10$ &     r &             p48 \\
$55457.502$ & $  -8.96$ & $20.06\pm0.10$ &     r &             p48 \\
$55458.295$ & $  -8.42$ & $20.23\pm0.18$ &     r &             p60 \\
$55458.296$ & $  -8.42$ & $19.68\pm0.14$ &     g &             p60 \\
$55459.293$ & $  -7.73$ & $20.21\pm0.20$ &     r &             p60 \\
$55459.518$ & $  -7.57$ & $19.91\pm0.15$ &     r &             p48 \\
\end{tabular}
\end{center}
\end{table*}
\begin{table*}
\begin{center}
\caption{PTF~10vqv rest-frame $g$ photometry data.\label{table g phot10vqv}}
\begin{tabular}{c c c c c c}
\tableline\tableline
MJD & Phase & $m_g$ & $M_g$  & Telescope & Reference \\
\tableline
$55455.517$ & $ -10.33$ & $20.57\pm0.12$ & $-21.43$ &             p48 &            \\
$55456.498$ & $  -9.65$ & $20.57\pm0.10$ & $-21.43$ &             p48 &            \\
$55457.502$ & $  -8.96$ & $20.61\pm0.10$ & $-21.39$ &             p48 &            \\
$55458.295$ & $  -8.42$ & $20.78\pm0.18$ & $-21.22$ &             p60 &            \\
$55459.293$ & $  -7.73$ & $20.76\pm0.20$ & $-21.24$ &             p60 &            \\
$55459.518$ & $  -7.57$ & $20.46\pm0.15$ & $-21.54$ &             p48 &            \\
$55460.291$ & $  -7.04$ & $20.52\pm0.14$ & $-21.48$ &             p60 &            \\
\end{tabular}
\end{center}
\end{table*}
\clearpage
\begin{table*}
\begin{center}
\caption{PTF~10vwg photometry data.\label{table phot10vwg}}
\begin{tabular}{c c c c c }
\tableline\tableline
MJD & Phase & $m$ & Filter & Telescope \\
\tableline
$55400.401$ & $ -46.12$ & $19.46\pm0.32$ &     r &             p48 \\
$55401.359$ & $ -45.32$ & $19.11\pm0.35$ &     r &             p48 \\
$55401.403$ & $ -45.28$ & $19.30\pm0.34$ &     r &             p48 \\
$55425.290$ & $ -25.21$ & $18.40\pm0.09$ &     r &             p48 \\
$55427.211$ & $ -23.59$ & $18.56\pm0.11$ &     r &             p48 \\
$55427.261$ & $ -23.55$ & $18.34\pm0.10$ &     r &             p48 \\
$55428.257$ & $ -22.71$ & $18.34\pm0.10$ &     r &             p48 \\
\end{tabular}
\end{center}
\end{table*}
\begin{table*}
\begin{center}
\caption{PTF~10vwg rest-frame $g$ photometry data.\label{table g phot10vwg}}
\begin{tabular}{c c c c c c}
\tableline\tableline
MJD & Phase & $m_g$ & $M_g$  & Telescope & Reference \\
\tableline
$55400.401$ & $ -46.12$ & $19.61\pm0.32$ & $-20.22$ &             p48 &            \\
$55401.359$ & $ -45.32$ & $19.26\pm0.35$ & $-20.58$ &             p48 &            \\
$55401.403$ & $ -45.28$ & $19.45\pm0.34$ & $-20.39$ &             p48 &            \\
$55425.290$ & $ -25.21$ & $18.55\pm0.09$ & $-21.29$ &             p48 &            \\
$55427.211$ & $ -23.59$ & $18.71\pm0.11$ & $-21.12$ &             p48 &            \\
$55427.261$ & $ -23.55$ & $18.49\pm0.10$ & $-21.34$ &             p48 &            \\
$55428.257$ & $ -22.71$ & $18.50\pm0.10$ & $-21.34$ &             p48 &            \\
\end{tabular}
\end{center}
\end{table*}
\clearpage
\begin{table*}
\begin{center}
\caption{PTF~11dij photometry data.\label{table phot11dij}}
\begin{tabular}{c c c c c }
\tableline\tableline
MJD & Phase & $m$ & Filter & Telescope \\
\tableline
$55652.370$ & $ -28.00$ & $20.01\pm0.11$ &     r &             p48 \\
$55652.429$ & $ -27.95$ & $19.85\pm0.11$ &     r &             p48 \\
$55666.197$ & $ -15.90$ & $18.08\pm0.08$ &     r &             p48 \\
$55666.240$ & $ -15.86$ & $18.04\pm0.07$ &     r &             p48 \\
$55694.324$ & $   8.71$ & $17.95\pm0.06$ &     r &             p60 \\
$55694.327$ & $   8.71$ & $17.83\pm0.06$ &     g &             p60 \\
$55694.329$ & $   8.71$ & $18.05\pm0.06$ &     i &             p60 \\
$55694.330$ & $   8.71$ & $17.90\pm0.06$ &     r &             p60 \\
$55694.334$ & $   8.72$ & $17.98\pm0.06$ &     g &             p60 \\
$55695.205$ & $   9.48$ & $17.89\pm0.08$ &     r &             p48 \\
$55695.249$ & $   9.52$ & $17.96\pm0.07$ &     r &             p48 \\
$55708.423$ & $  21.04$ & $18.42\pm0.09$ &     i &             p60 \\
$55708.425$ & $  21.05$ & $18.42\pm0.08$ &     r &             p60 \\
$55713.378$ & $  25.38$ & $18.88\pm0.11$ &     g &             p60 \\
$55744.294$ & $  52.43$ & $20.32\pm0.14$ &     i &             p60 \\
$55744.296$ & $  52.43$ & $20.68\pm0.15$ &     r &             p60 \\
$55744.299$ & $  52.43$ & $21.30\pm0.20$ &     g &             p60 \\
$55754.299$ & $  61.18$ & $21.02\pm0.17$ &     i &             p60 \\
$55754.301$ & $  61.19$ & $21.55\pm0.19$ &     r &             p60 \\
$55755.301$ & $  62.06$ & $20.68\pm0.17$ &     i &             p60 \\
$55766.273$ & $  71.66$ & $21.00\pm0.24$ &     i &             p60 \\
$55779.196$ & $  82.97$ & $21.72\pm0.24$ &     i &             p60 \\
\tableline
\end{tabular}
\end{center}
\end{table*}
\begin{table*}
\begin{center}
\caption{PTF~11dij rest-frame $g$ photometry data.\label{table g phot11dij}}
\begin{tabular}{c c c c c c}
\tableline\tableline
MJD & Phase & $m_g$ & $M_g$  & Telescope & Reference \\
\tableline
$55652.370$ & $ -28.00$ & $19.98\pm0.11$ & $-19.17$ &             p48 &            \\
$55652.429$ & $ -27.95$ & $19.82\pm0.11$ & $-19.33$ &             p48 &            \\
$55653.570$ & $ -26.95$ & $19.29\pm0.05$ & $-19.86$ &             PS1 &        III \\
$55653.580$ & $ -26.94$ & $19.30\pm0.05$ & $-19.85$ &             PS1 &        III \\
$55666.197$ & $ -15.90$ & $18.11\pm0.08$ & $-21.05$ &             p48 &            \\
$55666.240$ & $ -15.86$ & $18.06\pm0.07$ & $-21.09$ &             p48 &            \\
$55666.800$ & $ -15.37$ & $17.94\pm0.06$ & $-21.21$ &             PS1 &        III \\
\end{tabular}
\end{center}
\end{table*}
\clearpage
\begin{table*}
\begin{center}
\caption{PTF~11hrq photometry data.\label{table phot11hrq}}
\begin{tabular}{@{}c c c c c | ccccc@{}}
\tableline\tableline
MJD & Phase & $m$ & Filter & Tel & MJD & Phase & $m$ & Filter & Tel \\
\tableline
$55753.477$ & $   0.00$ & $17.21\pm0.07$ &     r &             p48 & $55820.294$ & $  63.21$ & $18.07\pm0.10$ &     r &             p48 \\
$55754.486$ & $   0.95$ & $17.23\pm0.07$ &     r &             p48 & $55820.338$ & $  63.25$ & $18.25\pm0.09$ &     r &             p48 \\
$55755.474$ & $   1.89$ & $17.25\pm0.07$ &     r &             p48 & $55821.324$ & $  64.18$ & $18.15\pm0.09$ &     r &             p48 \\
$55757.473$ & $   3.78$ & $17.37\pm0.09$ &     r &             p48 & $55822.289$ & $  65.10$ & $18.27\pm0.09$ &     r &             p48 \\
$55761.458$ & $   7.55$ & $17.43\pm0.08$ &     r &             p48 & $55822.332$ & $  65.14$ & $18.22\pm0.09$ &     r &             p48 \\
$55762.492$ & $   8.53$ & $17.47\pm0.07$ &     r &             p48 & $55823.301$ & $  66.05$ & $18.31\pm0.08$ &     r &             p48 \\
$55763.451$ & $   9.44$ & $17.50\pm0.07$ &     r &             p48 & $55823.345$ & $  66.09$ & $18.25\pm0.07$ &     r &             p48 \\
\end{tabular}
\end{center}
\end{table*}
\begin{table*}
\begin{center}
\caption{Continuation of Table \ref{table phot11hrq}.\label{table phot11hrq 2}}
\begin{tabular}{c c c c c }
\tableline\tableline
MJD & Phase & $m$ & Filter & Telescope \\
\tableline
$55901.130$ & $ 139.68$ & $18.96\pm0.28$ &     r &             p48 \\
$55902.093$ & $ 140.59$ & $18.94\pm0.24$ &     r &             p48 \\
$55902.138$ & $ 140.63$ & $19.21\pm0.22$ &     r &             p48 \\
$55903.088$ & $ 141.53$ & $19.64\pm0.25$ &     r &             p48 \\
$55903.114$ & $ 141.55$ & $19.21\pm0.18$ &     r &             p48 \\
$55904.100$ & $ 142.49$ & $19.47\pm0.27$ &     r &             p48 \\
$55904.146$ & $ 142.53$ & $19.30\pm0.27$ &     r &             p48 \\
$55906.147$ & $ 144.42$ & $19.62\pm0.17$ &     r &             p48 \\
$55916.115$ & $ 153.85$ & $19.46\pm0.15$ &     r &             p48 \\
$55916.167$ & $ 153.90$ & $19.33\pm0.20$ &     r &             p48 \\
\tableline
\end{tabular}
\end{center}
\end{table*}
\begin{table*}
\begin{center}
\caption{PTF~11hrq rest-frame $g$ photometry data.\label{table g phot11hrq}}
\begin{tabular}{@{}c c c c c c | cccccc@{}}
\tableline\tableline
MJD & Phase & $m_g$ & $M_g$  & Tel & Ref & MJD & Phase & $m_g$ & $M_g$  & Tel & Ref \\
\tableline
$55753.477$ & $   0.00$ & $17.17\pm0.07$ & $-19.87$ &             p48 &            & $55820.294$ & $  63.21$ & $18.22\pm0.10$ & $-18.82$ &             p48 &            \\
$55754.486$ & $   0.95$ & $17.19\pm0.07$ & $-19.84$ &             p48 &            & $55820.338$ & $  63.25$ & $18.40\pm0.09$ & $-18.63$ &             p48 &            \\
$55755.474$ & $   1.89$ & $17.21\pm0.07$ & $-19.82$ &             p48 &            & $55821.324$ & $  64.18$ & $18.31\pm0.09$ & $-18.73$ &             p48 &            \\
$55757.473$ & $   3.78$ & $17.35\pm0.09$ & $-19.69$ &             p48 &            & $55822.289$ & $  65.10$ & $18.43\pm0.09$ & $-18.61$ &             p48 &            \\
$55761.458$ & $   7.55$ & $17.42\pm0.08$ & $-19.62$ &             p48 &            & $55822.332$ & $  65.14$ & $18.37\pm0.09$ & $-18.66$ &             p48 &            \\
$55762.492$ & $   8.53$ & $17.46\pm0.07$ & $-19.57$ &             p48 &            & $55823.301$ & $  66.05$ & $18.46\pm0.08$ & $-18.57$ &             p48 &            \\
$55763.451$ & $   9.44$ & $17.49\pm0.07$ & $-19.54$ &             p48 &            & $55823.345$ & $  66.09$ & $18.41\pm0.07$ & $-18.62$ &             p48 &            \\
\end{tabular}
\end{center}
\end{table*}
\clearpage
\begin{table*}
\begin{center}
\caption{Continuation of Table \ref{table g phot11hrq}.\label{table g phot11hrq 2}}
\begin{tabular}{c c c c c c }
\tableline\tableline
MJD & Phase & $m_g$ & $M_g$  & Telescope & Reference \\
\tableline
$55901.130$ & $ 139.68$ & $19.18\pm0.28$ & $-17.86$ &             p48 &            \\
$55902.093$ & $ 140.59$ & $19.16\pm0.24$ & $-17.87$ &             p48 &            \\
$55902.138$ & $ 140.63$ & $19.43\pm0.22$ & $-17.60$ &             p48 &            \\
$55903.088$ & $ 141.53$ & $19.86\pm0.25$ & $-17.17$ &             p48 &            \\
$55903.114$ & $ 141.55$ & $19.43\pm0.18$ & $-17.61$ &             p48 &            \\
$55904.100$ & $ 142.49$ & $19.70\pm0.27$ & $-17.34$ &             p48 &            \\
$55904.146$ & $ 142.53$ & $19.52\pm0.27$ & $-17.52$ &             p48 &            \\
$55906.147$ & $ 144.42$ & $19.84\pm0.17$ & $-17.19$ &             p48 &            \\
$55916.115$ & $ 153.85$ & $19.68\pm0.15$ & $-17.35$ &             p48 &            \\
$55916.167$ & $ 153.90$ & $19.55\pm0.20$ & $-17.48$ &             p48 &            \\
\tableline
\end{tabular}
\end{center}
\end{table*}
\clearpage
\begin{table*}
\begin{center}
\caption{PTF~11rks photometry data.\label{table phot11rks}}
\begin{tabular}{@{}c c c c c | ccccc@{}}
\tableline\tableline
MJD & Phase & $m$ & Filter & Tel & MJD & Phase & $m$ & Filter & Tel \\
\tableline
$55906.082$ & $ -24.37$ & $20.18\pm0.23$ &     r &             p48 & $55946.092$ & $   9.19$ & $18.65\pm0.31$ &     r &             p48 \\
$55906.140$ & $ -24.32$ & $20.57\pm0.20$ &     r &             p48 & $55946.127$ & $   9.22$ & $19.33\pm0.20$ &     i &             p60 \\
$55916.092$ & $ -15.97$ & $19.43\pm0.16$ &     r &             p48 & $55946.129$ & $   9.22$ & $19.45\pm0.17$ &     r &             p60 \\
$55916.136$ & $ -15.93$ & $19.63\pm0.11$ &     r &             p48 & $55946.130$ & $   9.22$ & $19.69\pm0.16$ &     B &             p60 \\
$55922.137$ & $ -10.90$ & $19.16\pm0.10$ &     r &             p48 & $55946.134$ & $   9.22$ & $18.91\pm0.14$ &     i &             p60 \\
$55923.094$ & $ -10.10$ & $19.23\pm0.11$ &     r &             p48 & $55946.135$ & $   9.22$ & $18.86\pm0.15$ &     r &             p48 \\
$55923.138$ & $ -10.06$ & $19.16\pm0.09$ &     r &             p48 & $55946.136$ & $   9.22$ & $19.28\pm0.13$ &     r &             p60 \\
\end{tabular}
\end{center}
\end{table*}
\begin{table*}
\begin{center}
\caption{Continuation of Table \ref{table phot11rks}.\label{table phot11rks 2}}
\begin{tabular}{c c c c c }
\tableline\tableline
MJD & Phase & $m$ & Filter & Telescope \\
\tableline
$55975.134$ & $  33.54$ & $19.99\pm0.09$ &     i &             p60 \\
$55975.135$ & $  33.54$ & $20.33\pm0.09$ &     r &             p60 \\
$55975.137$ & $  33.55$ & $21.68\pm0.28$ &     B &             p60 \\
$55975.140$ & $  33.55$ & $21.60\pm0.16$ &     g &             p60 \\
$55979.121$ & $  36.89$ & $22.06\pm0.21$ &     g &             p60 \\
$55990.150$ & $  46.14$ & $20.94\pm0.34$ &     i &             p60 \\
$55990.153$ & $  46.14$ & $21.05\pm0.35$ &     r &             p60 \\
$56079.481$ & $ 121.05$ & $21.58\pm0.22$ &     r &             p60 \\
$56082.478$ & $ 123.57$ & $21.09\pm0.22$ &     g &             p60 \\
$56086.460$ & $ 126.91$ & $22.19\pm0.29$ &     r &             p60 \\
$56089.452$ & $ 129.42$ & $22.04\pm0.26$ &     r &             p60 \\
$56098.432$ & $ 136.95$ & $21.42\pm0.32$ &     i &             p60 \\
$56099.447$ & $ 137.80$ & $22.22\pm0.32$ &     i &             p60 \\
$56105.407$ & $ 142.80$ & $21.54\pm0.29$ &     i &             p60 \\
$56138.399$ & $ 170.46$ & $22.19\pm0.30$ &     r &             p60 \\
$56151.333$ & $ 181.31$ & $22.33\pm0.35$ &     i &             p60 \\
\tableline
\end{tabular}
\end{center}
\end{table*}
\begin{table*}
\begin{center}
\caption{PTF~11rks rest-frame $g$ photometry data.\label{table g phot11rks}}
\begin{tabular}{@{}c c c c c c | cccccc@{}}
\tableline\tableline
MJD & Phase & $m_g$ & $M_g$  & Tel & Ref & MJD & Phase & $m_g$ & $M_g$  & Tel & Ref \\
\tableline
$55906.082$ & $ -24.37$ & $20.28\pm0.23$ & $-19.58$ &             p48 &            & $55949.148$ & $  11.75$ & $19.26\pm0.09$ & $-20.60$ &             p48 &            \\
$55906.140$ & $ -24.32$ & $20.67\pm0.20$ & $-19.20$ &             p48 &            & $55949.195$ & $  11.79$ & $19.31\pm0.12$ & $-20.55$ &             p48 &            \\
$55916.092$ & $ -15.97$ & $19.56\pm0.16$ & $-20.30$ &             p48 &            & $55951.730$ & $  13.92$ & $19.33\pm0.03$ & $-20.53$ &             FTN &        III \\
$55916.136$ & $ -15.93$ & $19.76\pm0.11$ & $-20.10$ &             p48 &            & $55952.106$ & $  14.23$ & $19.38\pm0.10$ & $-20.48$ &             p48 &            \\
$55922.137$ & $ -10.90$ & $19.31\pm0.10$ & $-20.56$ &             p48 &            & $55952.118$ & $  14.24$ & $19.45\pm0.08$ & $-20.42$ &             p60 &            \\
$55923.094$ & $ -10.10$ & $19.39\pm0.11$ & $-20.48$ &             p48 &            & $55952.151$ & $  14.27$ & $19.29\pm0.09$ & $-20.57$ &             p48 &            \\
$55923.138$ & $ -10.06$ & $19.31\pm0.09$ & $-20.55$ &             p48 &            & $55953.097$ & $  15.06$ & $19.25\pm0.10$ & $-20.61$ &             p48 &            \\
\end{tabular}
\end{center}
\end{table*}
\clearpage
\begin{table*}
\begin{center}
\caption{PTF~12dam photometry data.\label{table phot12dam}}
\begin{tabular}{@{}c c c c c | ccccc@{}}
\tableline\tableline
MJD & Phase & $m$ & Filter & Tel & MJD & Phase & $m$ & Filter & Tel \\
\tableline
$56025.328$ & $ -60.50$ & $20.03\pm0.22$ &     r &             p48 & $56070.436$ & $ -19.77$ & $17.20\pm0.06$ &     r &             p48 \\
$56027.257$ & $ -58.75$ & $19.93\pm0.17$ &     r &             p48 & $56070.469$ & $ -19.74$ & $17.21\pm0.06$ &     r &             p48 \\
$56027.286$ & $ -58.73$ & $20.02\pm0.16$ &     r &             p48 & $56072.340$ & $ -18.05$ & $17.78\pm0.03$ &     i &           LCOGT \\
$56027.315$ & $ -58.70$ & $19.95\pm0.15$ &     r &             p48 & $56072.342$ & $ -18.05$ & $17.43\pm0.03$ &     g &           LCOGT \\
$56033.489$ & $ -53.13$ & $19.80\pm0.13$ &     r &             p48 & $56072.343$ & $ -18.05$ & $17.42\pm0.04$ &     g &           LCOGT \\
$56034.162$ & $ -52.52$ & $19.87\pm0.13$ &     r &             p48 & $56075.234$ & $ -15.43$ & $17.24\pm0.06$ &     i &             p60 \\
$56034.194$ & $ -52.49$ & $19.87\pm0.13$ &     r &             p48 & $56075.235$ & $ -15.43$ & $17.13\pm0.06$ &     r &             p60 \\
\end{tabular}
\end{center}
\end{table*}
\begin{table*}
\begin{center}
\caption{Continuation of Table \ref{table phot12dam}.\label{table phot12dam 2}}
\begin{tabular}{@{}c c c c c | ccccc@{}}
\tableline\tableline
MJD & Phase & $m$ & Filter & Tel & MJD & Phase & $m$ & Filter & Tel \\
\tableline
$56089.338$ & $  -2.70$ & $17.23\pm0.03$ &     i &           LCOGT & $56118.257$ & $  23.41$ & $17.00\pm0.06$ &     r &             p60 \\
$56090.200$ & $  -1.92$ & $16.99\pm0.06$ &     i &             p60 & $56118.258$ & $  23.41$ & $17.02\pm0.06$ &     B &             p60 \\
$56090.201$ & $  -1.92$ & $16.84\pm0.06$ &     r &             p60 & $56118.259$ & $  23.41$ & $16.86\pm0.06$ &     g &             p60 \\
$56090.202$ & $  -1.92$ & $16.40\pm0.06$ &     B &             p60 & $56127.277$ & $  31.56$ & $17.35\pm0.07$ &     g &             p60 \\
$56090.203$ & $  -1.92$ & $16.61\pm0.06$ &     g &             p60 & $56128.378$ & $  32.55$ & $17.33\pm0.04$ &     g &           LCOGT \\
$56090.204$ & $  -1.92$ & $17.12\pm0.06$ &     i &             p60 & $56128.379$ & $  32.55$ & $17.35\pm0.04$ &     g &           LCOGT \\
$56090.205$ & $  -1.92$ & $16.83\pm0.06$ &     r &             p60 & $56128.380$ & $  32.55$ & $17.12\pm0.02$ &     r &           LCOGT \\
\end{tabular}
\end{center}
\end{table*}
\begin{table*}
\begin{center}
\caption{Continuation of Table \ref{table phot12dam}.\label{table phot12dam 2}}
\begin{tabular}{@{}c c c c c | ccccc@{}}
\tableline\tableline
MJD & Phase & $m$ & Filter & Tel & MJD & Phase & $m$ & Filter & Tel \\
\tableline
$56177.150$ & $  76.59$ & $17.86\pm0.06$ &     g &             p60 & $56266.472$ & $ 157.24$ & $19.08\pm0.09$ &     r &             p60 \\
$56181.231$ & $  80.27$ & $18.18\pm0.07$ &     g &           LCOGT & $56266.473$ & $ 157.24$ & $19.50\pm0.12$ &     g &             p60 \\
$56181.233$ & $  80.28$ & $17.79\pm0.04$ &     r &           LCOGT & $56268.466$ & $ 159.04$ & $19.98\pm0.15$ &     B &             p60 \\
$56181.234$ & $  80.28$ & $17.82\pm0.03$ &     r &           LCOGT & $56273.452$ & $ 163.54$ & $19.44\pm0.10$ &     r &             p60 \\
$56181.235$ & $  80.28$ & $18.05\pm0.04$ &     i &           LCOGT & $56273.454$ & $ 163.54$ & $19.74\pm0.11$ &     g &             p60 \\
$56185.159$ & $  83.82$ & $17.84\pm0.07$ &     i &             p60 & $56281.429$ & $ 170.75$ & $19.57\pm0.20$ &     i &             p60 \\
$56185.160$ & $  83.82$ & $17.84\pm0.07$ &     r &             p60 & $56281.431$ & $ 170.75$ & $19.42\pm0.14$ &     r &             p60 \\
\end{tabular}
\end{center}
\end{table*}
\begin{table*}
\begin{center}
\caption{Continuation of Table \ref{table phot12dam}.\label{table phot12dam 2}}
\begin{tabular}{@{}c c c c c | ccccc@{}}
\tableline\tableline
MJD & Phase & $m$ & Filter & Tel & MJD & Phase & $m$ & Filter & Tel \\
\tableline
$56362.281$ & $ 243.75$ & $20.84\pm0.24$ &     B &             p60 & $56507.216$ & $ 374.62$ & $22.01\pm0.28$ &     r &             p60 \\
$56362.283$ & $ 243.75$ & $20.70\pm0.15$ &    g &             p60 \\
$56362.479$ & $ 243.93$ & $20.62\pm0.18$ &    r &             p48 \\
$56362.509$ & $ 243.96$ & $20.34\pm0.15$ &    r &             p48 \\
$56362.539$ & $ 243.98$ & $20.50\pm0.18$ &    r &             p48 \\
$56364.203$ & $ 245.49$ & $20.18\pm0.14$ &    i &             p60 \\
$56366.489$ & $ 247.55$ & $20.32\pm0.22$ &    r &             p48 \\
\end{tabular}
\end{center}
\end{table*}
\begin{table*}
\begin{center}
\caption{PTF~12dam rest-frame $g$ photometry data.\label{table g phot12dam}}
\begin{tabular}{@{}c c c c c c | cccccc@{}}
\tableline\tableline
MJD & Phase & $m_g$ & $M_g$  & Tel & Ref & MJD & Phase & $m_g$ & $M_g$  & Tel & Ref \\
\tableline
$56025.328$ & $ -60.50$ & $19.92\pm0.22$ & $-18.57$ &             p48 &            & $56080.204$ & $ -10.95$ & $16.91\pm0.06$ & $-21.57$ &             p60 &            \\
$56027.257$ & $ -58.75$ & $19.82\pm0.17$ & $-18.66$ &             p48 &            & $56080.207$ & $ -10.94$ & $16.93\pm0.06$ & $-21.55$ &             p60 &            \\
$56027.286$ & $ -58.73$ & $19.91\pm0.16$ & $-18.57$ &             p48 &            & $56080.940$ & $ -10.28$ & $16.93\pm0.01$ & $-21.55$ &       LT+RATCam &         IV \\
$56027.315$ & $ -58.70$ & $19.84\pm0.15$ & $-18.64$ &             p48 &            & $56081.940$ & $  -9.38$ & $16.91\pm0.01$ & $-21.57$ &       LT+RATCam &         IV \\
$56030.480$ & $ -55.84$ & $19.25\pm0.17$ & $-19.24$ &             PS1 &         IV & $56083.360$ & $  -8.10$ & $16.89\pm0.02$ & $-21.59$ &           LCOGT &            \\
$56033.489$ & $ -53.13$ & $19.69\pm0.13$ & $-18.79$ &             p48 &            & $56085.211$ & $  -6.43$ & $16.89\pm0.06$ & $-21.59$ &             p60 &            \\
$56034.162$ & $ -52.52$ & $19.76\pm0.13$ & $-18.72$ &             p48 &            & $56085.214$ & $  -6.42$ & $16.96\pm0.06$ & $-21.52$ &             p60 &            \\
\end{tabular}
\end{center}
\end{table*}
\clearpage
\begin{table*}
\begin{center}
\caption{Continuation of Table \ref{table g phot12dam}.\label{table g phot12dam 2}}
\begin{tabular}{@{}c c c c c c | cccccc@{}}
\tableline\tableline
MJD & Phase & $m_g$ & $M_g$  & Tel & Ref & MJD & Phase & $m_g$ & $M_g$  & Tel & Ref \\
\tableline
$56191.860$ & $  89.87$ & $18.37\pm0.01$ & $-20.11$ &             WHT &         IV & $56374.513$ & $ 254.79$ & $20.62\pm0.17$ & $-17.87$ &             p48 &            \\
$56196.215$ & $  93.80$ & $18.34\pm0.02$ & $-20.14$ &           LCOGT &            & $56375.293$ & $ 255.50$ & $21.47\pm0.23$ & $-17.02$ &             p60 &            \\
$56196.216$ & $  93.80$ & $18.19\pm0.03$ & $-20.29$ &           LCOGT &            & $56376.170$ & $ 256.29$ & $21.02\pm0.30$ & $-17.46$ &             p60 &            \\
$56197.224$ & $  94.71$ & $18.28\pm0.02$ & $-20.20$ &           LCOGT &            & $56376.506$ & $ 256.59$ & $21.15\pm0.25$ & $-17.34$ &             p48 &            \\
$56197.225$ & $  94.72$ & $18.39\pm0.02$ & $-20.09$ &           LCOGT &            & $56385.515$ & $ 264.73$ & $20.61\pm0.32$ & $-17.87$ &             p48 &            \\
$56200.213$ & $  97.41$ & $18.36\pm0.02$ & $-20.12$ &           LCOGT &            & $56388.496$ & $ 267.42$ & $21.32\pm0.29$ & $-17.16$ &             p60 &            \\
$56200.214$ & $  97.41$ & $18.40\pm0.02$ & $-20.09$ &           LCOGT &            & $56390.279$ & $ 269.03$ & $20.85\pm0.28$ & $-17.63$ &             p60 &            \\
\end{tabular}
\end{center}
\end{table*}
\clearpage
\begin{table*}
\begin{center}
\caption{PTF~12gty photometry data.\label{table phot12gty}}
\begin{tabular}{@{}c c c c c | ccccc@{}}
\tableline\tableline
MJD & Phase & $m$ & Filter & Tel & MJD & Phase & $m$ & Filter & Tel \\
\tableline
$56086.286$ & $ -48.50$ & $21.20\pm0.36$ &     r &             p48 & $56159.279$ & $  13.53$ & $20.23\pm0.13$ &     g &             p60 \\
$56090.440$ & $ -44.97$ & $20.91\pm0.35$ &     r &             p48 & $56166.255$ & $  19.46$ & $19.61\pm0.10$ &     i &             p60 \\
$56092.374$ & $ -43.32$ & $21.30\pm0.36$ &     r &             p48 & $56166.257$ & $  19.46$ & $19.90\pm0.11$ &     r &             p60 \\
$56092.404$ & $ -43.30$ & $20.95\pm0.32$ &     r &             p48 & $56166.259$ & $  19.46$ & $20.54\pm0.29$ &     g &             p60 \\
$56097.373$ & $ -39.07$ & $20.92\pm0.35$ &     r &             p48 & $56166.262$ & $  19.47$ & $19.52\pm0.09$ &     i &             p60 \\
$56097.431$ & $ -39.02$ & $20.30\pm0.21$ &     r &             p48 & $56166.263$ & $  19.47$ & $19.83\pm0.10$ &     r &             p60 \\
$56099.391$ & $ -37.36$ & $20.59\pm0.25$ &     r &             p48 & $56166.266$ & $  19.47$ & $20.25\pm0.12$ &     g &             p60 \\
\end{tabular}
\end{center}
\end{table*}
\begin{table*}
\begin{center}
\caption{PTF~12gty rest-frame $g$ photometry data.\label{table g phot12gty}}
\begin{tabular}{@{}c c c c c c | cccccc@{}}
\tableline\tableline
MJD & Phase & $m_g$ & $M_g$  & Tel & Ref & MJD & Phase & $m_g$ & $M_g$  & Tel & Ref \\
\tableline
$56086.286$ & $ -48.50$ & $21.26\pm0.36$ & $-18.40$ &             p48 &            & $56186.137$ & $  36.35$ & $20.32\pm0.12$ & $-19.34$ &             p60 &            \\
$56090.440$ & $ -44.97$ & $20.96\pm0.35$ & $-18.70$ &             p48 &            & $56187.138$ & $  37.20$ & $20.44\pm0.12$ & $-19.22$ &             p60 &            \\
$56092.374$ & $ -43.32$ & $21.35\pm0.36$ & $-18.31$ &             p48 &            & $56199.120$ & $  47.39$ & $20.95\pm0.20$ & $-18.71$ &             p60 &            \\
$56092.404$ & $ -43.30$ & $21.00\pm0.32$ & $-18.65$ &             p48 &            & $56200.117$ & $  48.23$ & $20.83\pm0.13$ & $-18.83$ &             p60 &            \\
$56097.373$ & $ -39.07$ & $20.97\pm0.35$ & $-18.68$ &             p48 &            & $56201.116$ & $  49.08$ & $21.02\pm0.18$ & $-18.64$ &             p60 &            \\
$56097.431$ & $ -39.02$ & $20.36\pm0.21$ & $-19.30$ &             p48 &            & $56202.114$ & $  49.93$ & $21.17\pm0.18$ & $-18.49$ &             p60 &            \\
$56099.391$ & $ -37.36$ & $20.64\pm0.25$ & $-19.02$ &             p48 &            & $56214.110$ & $  60.12$ & $20.87\pm0.17$ & $-18.79$ &             p60 &            \\
\end{tabular}
\end{center}
\end{table*}
\clearpage
\begin{table*}
\begin{center}
\caption{PTF~12hni photometry data.\label{table phot12hni}}
\begin{tabular}{@{}c c c c c | ccccc@{}}
\tableline\tableline
MJD & Phase & $m$ & Filter & Tel & MJD & Phase & $m$ & Filter & Tel \\
\tableline
$56146.255$ & $  -7.23$ & $18.45\pm0.09$ &     r &             p48 & $56228.108$ & $  66.80$ & $21.01\pm0.25$ &     r &             p60 \\
$56146.319$ & $  -7.18$ & $18.52\pm0.08$ &     r &             p48 & $56228.177$ & $  66.86$ & $21.16\pm0.25$ &     r &             p48 \\
$56147.229$ & $  -6.35$ & $18.49\pm0.08$ &     r &             p48 & $56229.188$ & $  67.78$ & $20.89\pm0.27$ &     r &             p48 \\
$56147.294$ & $  -6.30$ & $18.53\pm0.08$ &     r &             p48 & $56231.104$ & $  69.51$ & $20.47\pm0.16$ &     r &             p48 \\
$56148.232$ & $  -5.45$ & $18.52\pm0.08$ &     r &             p48 & $56231.168$ & $  69.57$ & $20.88\pm0.25$ &     r &             p48 \\
$56148.296$ & $  -5.39$ & $18.74\pm0.07$ &     r &             p48 & $56232.155$ & $  70.46$ & $20.43\pm0.17$ &     r &             p48 \\
$56151.485$ & $  -2.50$ & $18.67\pm0.07$ &     r &             p60 & $56234.128$ & $  72.25$ & $20.58\pm0.17$ &     i &             p60 \\
\end{tabular}
\end{center}
\end{table*}
\begin{table*}
\begin{center}
\caption{PTF~12hni rest-frame $g$ photometry data.\label{table g phot12hni}}
\begin{tabular}{c c c c c c}
\tableline\tableline
MJD & Phase & $m_g$ & $M_g$  & Telescope & Reference \\
\tableline
$56146.255$ & $  -7.23$ & $18.44\pm0.09$ & $-20.00$ &             p48 &            \\
$56146.319$ & $  -7.18$ & $18.51\pm0.08$ & $-19.93$ &             p48 &            \\
$56147.229$ & $  -6.35$ & $18.48\pm0.08$ & $-19.96$ &             p48 &            \\
$56147.294$ & $  -6.30$ & $18.52\pm0.08$ & $-19.92$ &             p48 &            \\
$56148.232$ & $  -5.45$ & $18.51\pm0.08$ & $-19.93$ &             p48 &            \\
$56148.296$ & $  -5.39$ & $18.74\pm0.07$ & $-19.70$ &             p48 &            \\
$56151.485$ & $  -2.50$ & $18.68\pm0.07$ & $-19.76$ &             p60 &            \\
\end{tabular}
\end{center}
\end{table*}
\clearpage
\begin{table*}
\begin{center}
\caption{PTF~12mxx photometry data.\label{table phot12mxx}}
\begin{tabular}{@{}c c c c c | ccccc@{}}
\tableline\tableline
MJD & Phase & $m$ & Filter & Tel & MJD & Phase & $m$ & Filter & Tel \\
\tableline
$56239.096$ & $ -39.96$ & $21.19\pm0.31$ &     r &             p48 & $56283.800$ & $  -6.28$ & $19.13\pm0.07$ &     g &              LT \\
$56243.141$ & $ -36.91$ & $20.56\pm0.20$ &     g &             p48 & $56283.800$ & $  -6.28$ & $19.29\pm0.09$ &     r &              LT \\
$56243.178$ & $ -36.89$ & $20.74\pm0.22$ &     g &             p48 & $56283.800$ & $  -6.28$ & $19.36\pm0.10$ &     i &              LT \\
$56245.207$ & $ -35.36$ & $20.49\pm0.17$ &     g &             p48 & $56284.180$ & $  -6.00$ & $21.06\pm0.62$ &  UVW2 &      Swift/UVOT \\
$56245.242$ & $ -35.33$ & $20.38\pm0.17$ &     g &             p48 & $56284.186$ & $  -5.99$ & $19.23\pm0.14$ &     r &             p60 \\
$56252.227$ & $ -30.07$ & $20.17\pm0.19$ &     r &             p48 & $56284.188$ & $  -5.99$ & $19.26\pm0.29$ &     u &      Swift/UVOT \\
$56252.259$ & $ -30.04$ & $20.34\pm0.24$ &     r &             p48 & $56284.191$ & $  -5.99$ & $19.14\pm0.19$ &     B &      Swift/UVOT \\
\end{tabular}
\end{center}
\end{table*}
\begin{table*}
\begin{center}
\caption{Continuation of Table \ref{table phot12mxx}.\label{table phot12mxx 2}}
%
\end{center}
\end{table*}
\begin{table*}
\begin{center}
\caption{PTF~12mxx rest-frame $g$ photometry data.\label{table g phot12mxx}}
%
\end{center}
\end{table*}
\clearpage
\begin{table*}
\begin{center}
\caption{PTF~13ajg photometry data.\label{table phot13ajg}}
%
\end{center}
\end{table*}
\begin{table*}
\begin{center}
\caption{Continuation of Table \ref{table phot13ajg}.\label{table phot13ajg 2}}
%
\end{center}
\end{table*}
\begin{table*}
\begin{center}
\caption{PTF~13ajg rest-frame $g$ photometry data.\label{table g phot13ajg}}
%
\end{center}
\end{table*}
\clearpage
\begin{table*}
\begin{center}
\caption{Continuation of Table \ref{table g phot13ajg}.\label{table g phot13ajg 2}}
%
\end{center}
\end{table*}
\clearpage
\begin{table*}
\begin{center}
\caption{Continuation of Table \ref{table g phot13ajg}.\label{table g phot13ajg 2}}
%
\end{center}
\end{table*}
\clearpage
\begin{table*}
\begin{center}
\caption{PTF~13bdl photometry data.\label{table phot13bdl}}
%
\end{center}
\end{table*}
\begin{table*}
\begin{center}
\caption{PTF~13bdl rest-frame $g$ photometry data.\label{table g phot13bdl}}
%
\end{center}
\end{table*}
\clearpage
\begin{table*}
\begin{center}
\caption{PTF~13bjz photometry data.\label{table phot13bjz}}
%
\end{center}
\end{table*}
\begin{table*}
\begin{center}
\caption{PTF~13bjz rest-frame $g$ photometry data.\label{table g phot13bjz}}
%
\end{center}
\end{table*}
\clearpage
\begin{table*}
\begin{center}
\caption{PTF~13cjq photometry data.\label{table phot13cjq}}
%
\end{center}
\end{table*}
\begin{table*}
\begin{center}
\caption{PTF~13cjq rest-frame $g$ photometry data.\label{table g phot13cjq}}
%
\end{center}
\end{table*}
\clearpage
\begin{table*}
\begin{center}
\caption{PTF~13dcc photometry data.\label{table phot13dcc}}
%
\end{center}
\end{table*}
\begin{table*}
\begin{center}
\caption{Continuation of Table \ref{table phot13dcc}.\label{table phot13dcc 2}}
%
\end{center}
\end{table*}
\begin{table*}
\begin{center}
\caption{PTF~13dcc rest-frame $g$ photometry data.\label{table g phot13dcc}}
%
\end{center}
\end{table*}
\clearpage
\begin{table*}
\begin{center}
\caption{Continuation of Table \ref{table g phot13dcc}.\label{table g phot13dcc 2}}
%
\end{center}
\end{table*}
\clearpage
\begin{table*}
\begin{center}
\caption{PTF~13ehe photometry data.\label{table phot13ehe}}
%
\end{center}
\end{table*}
\begin{table*}
\begin{center}
\caption{PTF~13ehe rest-frame $g$ photometry data.\label{table g phot13ehe}}
%
\end{center}
\end{table*}
\clearpage

\end{document}